\documentclass [11 pt, twoside, a4paper]{article}

\usepackage[utf8]{inputenc}

\usepackage[autostyle]{csquotes}
\usepackage[english]{babel}
\usepackage[T1]{fontenc}

\usepackage[sorting=none, bibstyle=numeric,backend=biber]{biblatex}
\newcommand{\A}{\mathbf{A}}
\newcommand{\X}{\mathbf{X}}
\newcommand{\G}{\mathbf{\Gamma}}

\usepackage{makecell}
\usepackage{amsfonts}
\usepackage{graphicx}
\usepackage{tabularx}
\usepackage{adjustbox}
\usepackage{amsmath}
\usepackage{amsthm}
\usepackage{cases}
\usepackage{epigraph}
\usepackage{amssymb} 
\usepackage{hyperref}
\usepackage{epstopdf}
\usepackage{latexsym}
\usepackage{eucal}
\usepackage{float}
\usepackage{verbatim}
\usepackage{fancyhdr}
\usepackage{setspace}
\usepackage{graphicx}
\usepackage{url}
\usepackage{bm}
\usepackage{siunitx}
\usepackage{caption}
\usepackage{subfig}
\DeclareCaptionLabelFormat{onlynumber}{Figure (\Alph)}
\DeclareCaptionLabelFormat{Anumber}{A#2}
\DeclareCaptionLabelFormat{3number}{3#2}
\DeclareCaptionLabelFormat{Hnumber}{H#2}

\fancyhfoffset[L]{0.7cm} 
\fancyhfoffset[R]{0.7cm} 
\theoremstyle{plain}

\theoremstyle{definition}

\theoremstyle{remark}

\usepackage[paper=a4paper,margin=1in]{geometry}

\usepackage{listings}
\usepackage{color}

\definecolor{dkgreen}{rgb}{0,0.6,0}
\definecolor{gray}{rgb}{0.5,0.5,0.5}
\definecolor{mauve}{rgb}{0.58,0,0.82}

\usepackage{nameref}
\usepackage{varioref}
\usepackage{hyperref}
\usepackage{cleveref}

\begin{document}
\title{\MakeUppercase{\textbf{Classification of BCI-EEG based on Augmented Covariance Matrix}}}
\author{Igor Carrara$^{1, 2}$ Th\'eodore Papadopoulo$^{1, 2}$ \\
\small $^1$ Université Côte d'Azur (UCA)\\
\small $^2$ Centre Inria d'Université Côte d'Azur, Cronos Team \\
\small igor.carrara@inria.fr and theodore.papadopoulo@inria.fr}
\date{}

\maketitle

\begin{abstract}
\textit{Objective}: Electroencephalography signals are recorded as a multidimensional dataset. We propose a new framework based on the augmented covariance extracted from an autoregressive model to improve motor imagery classification.
\textit{Methods}: From the autoregressive model can be derived the Yule-Walker equations, which show the emergence of a symmetric positive definite matrix: the augmented covariance matrix. The state-of the art for classifying covariance matrices is based on Riemannian Geometry. A fairly natural idea is therefore to extend the standard approach using these augmented covariance matrices.
The methodology for creating the augmented covariance matrix shows a natural connection with the delay embedding theorem proposed by Takens for dynamical systems. Such an embedding method is based on the knowledge of two parameters: the delay and the embedding dimension, respectively related to the lag and the order of the autoregressive model. This approach provides new methods to compute the hyper-parameters in addition to standard grid search.
\textit{Results}: The augmented covariance matrix performed noticeably better than any state-of-the-art methods. We will test our approach on several datasets and several subjects using the MOABB framework, using both within-session and cross-session evaluation.
\textit{Conclusion}: The improvement in results is due to the fact that the augmented covariance matrix incorporates not only spatial but also temporal information, incorporating nonlinear components of the signal through an embedding procedure, which allows the leveraging of dynamical systems algorithms.
\textit{Significance}: These results extend the concepts and the results of the Riemannian distance based classification algorithm.
\end{abstract}


\section{Introduction}
\label{sec:introduction}
Electroencephalography (EEG) is a modality that allows the passive measurement outside the head of the electrical potential arising mainly from the electrical activity of the brain.
The measurements obtained at sensors at time \textit{t} are linear combinations of the electrical activities at this same time \textit{t} of a set of “sources” located at the level of the cerebral cortex. These sources are time dependent on each other according to a complex biophysical model that is poorly understood (because it depends on many parameters) and which is therefore not easy to establish. The temporal dependence between cortical sources remains poorly understood and difficult to obtain, at least if we are trying to stick closely to biophysical models. There are, however, simple models which,
even if they are not completely justified by biology, are general enough to be used with EEG measurements. An example of those are auto regressive models (AR).
These models are based on the hypothesis that the signal can be explained by a combination of its past values plus some random factor called innovation. The AR model has been used on several occasions and in different ways. For example they allow to model the interactions between the different zones of the brain during cognitive tasks~\cite{belaoucha-papadopoulo:20}, to extract information from signals~\cite{ding-bressler-etal:00} or detect state changes~\cite{kirch-muhsal-etal:15}. Finally, autoregressive models are sometimes used in conjunction with machine learning methods for performing classification tasks~\cite{faradji-ward-etal:19}. 

One promising application of EEG signal analysis is Brain Computer Interfaces (BCI). BCI can be defined as a technology that measures brain activity and translates that signal into instructions or commands for a digital system.
We focus on electroencephalography (EEG)-based BCI (BCI-EEG), which is non-invasive, possess a high time resolution and is relatively inexpensive.  Because of these characteristics, BCI-EEG interfaces are candidates to becoming the main BCI technology in a mass deployment perspective for everyday use.
In particular, we focus on motor imagery BCI paradigms, where the user changes his brain activity by imagining the movement of a body part.

Even though BCI was created for medical purposes, the use of this technology has expanded, and found its way into non-medical applications ranging from drone control~\cite{duan-xie-etal:19}, rehabilitation in neurological diseases to applications in the context of Virtual and Augmented reality~\cite{wen-fan-etal:21} or video gaming. 

However, despite the incredible progress made in the last ten years, the BCI field is still in its infancy. In fact, we have not yet obtained a level of accuracy that would allow an application in everyday life. Moreover, the classification performance are particularly sub-optimal for some particular subjects suffering from 'BCI illiteracy':
some research estimates that about $30\%$ of all BCI users are unable to reach at least $70\%$ of accuracy with the current technology, despite the fact that they undergo the same training process as normal subjects~\cite{blankertz-sannelli-etal:10}. This problem is extremely limiting for possible future applications because it compromises the universality of BCI systems.

Over the past 10 years, the Riemann distance-based classification paradigm has shown interesting performances for BCI-EEG classification~\cite{barachant-bonnet-etal:10}. 
The key idea behind this method is to map covariance matrices into an appropriate geometrical space: the space of Symmetric Positive Definite (SPD) matrices, which is actually a differentiable manifold with a natural Riemann structure~\cite{forstner-moonen:03}. It is therefore natural to try to extend this Riemannian approach with ideas from auto-regressive models in order to optimize the performance even for the most difficult subjects.

In order to create a new algorithm for the BCI-EEG classification, we must focus on the following basic characteristics: be computationally efficient, improve accuracy compared to the state of the art, keep improvements applicable to all possible BCI paradigms, do not require excessive time for training the hyper-parameters related to the model. So far, there have been several attempts to improve over the results obtained with the Riemann distance-based algorithm~\cite{barachant-bonnet-etal:11b}. However, these attempts, while showing superior performance, obtained such results by introducing additional problems, which requires complex optimization of hyper-parameters. 

The goal of this work is to improve the performance of the classification algorithms, exploiting the nonlinear components of the EEG signal and combining this approach with Riemann distance-based classifiers. The idea is thus to push further the logic of the Riemann algorithm by defining an appropriate SPD matrix that maximizes the amount of information related to a particular task, contained in the EEG signal. 

The AR model can be computed using the Yule-Walker equations, which involve a SPD matrix i.e. the augmented covariance matrix with lags. Thus, a fairly natural idea is to extend the standard Riemannian approach by using these augmented covariance matrices with lags.
This idea of using an augmented covariance matrix with delays was also recently introduced~\cite{sosulski-tangermann:22} where it was used in combination with a Linear Discriminant Analysis algorithm and applied on Event Related Potential datasets. To the authors' knowledge the approach of combing the augmented covariance with the Riemannian classification is completely a novelty.

This approach can also be seen from the point of view of dynamical systems: since the AR Yule-Walker matrix is a matrix of delayed covariance, we can extract the same matrix using the standard covariance matrix stemming from an embedding of the original system in a high dimensional space. Thus such a matrix incorporates the nonlinear properties of the EEG signal.
Hence, it is natural to connect our approach with the delay embedding theorem proposed by Takens~\cite{takens:81} in the context of non-linear dynamical systems.
Most EEG classification algorithms are based on the assumption that the signal can be described through a linear model. However, studies have shown that the EEG signal follows a nonlinear model~\cite{casdagli-iasemidis-etal:97} and because of this nonlinear component, the use of signal analysis techniques based on linearity assumptions could lose important characteristics. Only recently some classification algorithms are being proposed based on the use of quantities defined in the context of nonlinear systems theory such as correlation dimension, Lyapunov exponents, mutual information and the minimum embedding dimension~\cite{fang-chen-etal:15, hosseinifard-moradi-etal:13}. Such features, however, are extremely computationally demanding.
The basic idea of our approach is to embed the EEG data in a high dimensional space using the theory of non-linear dynamics, and then to classify the EEG signal using the Riemannian framework. In this sense, we are thus combining the two strengths of both approaches, the use of the nonlinear component of the signal and the performance of the Riemannian distance-based algorithm.

In order to validate the methods, we apply our approach both with a classification on the Riemann surface using the Minimum Distance to the Mean and another classification on the Tangent Space using Support Vector Machine. We will test our approach on several datasets and several subjects using the \href{http://moabb.neurotechx.com/docs/index.html}{MOABB} framework~\cite{jayaram-barachant:18}, with both within-session and cross-session evaluation.

The article is structured as follows: in section~\ref{sec:materials}, we first describe the theoretical approach that underlies the model, and then the datasets considered and the performed statistical analysis. In section~\ref{sec:results}, we lists the results obtained using within-session and cross-session evaluation. Finally, in section~\ref{sec:discussion}, we analyze the implications of the method and its current limitations. Section~\ref{sec:conclusion} summarizes the results of our study.

\section{Materials and Methods}
\label{sec:materials}

\subsection{BCI classification}
The EEG signal can be represented using a multivariate time-serie $\mathbf{X} \in \mathbb{R}^{d \times T}$ where $T$ is the number of sampled data point and $d$ is the number of electrodes.

\begin{figure}[!ht]
\centering
\includegraphics[width=\linewidth]{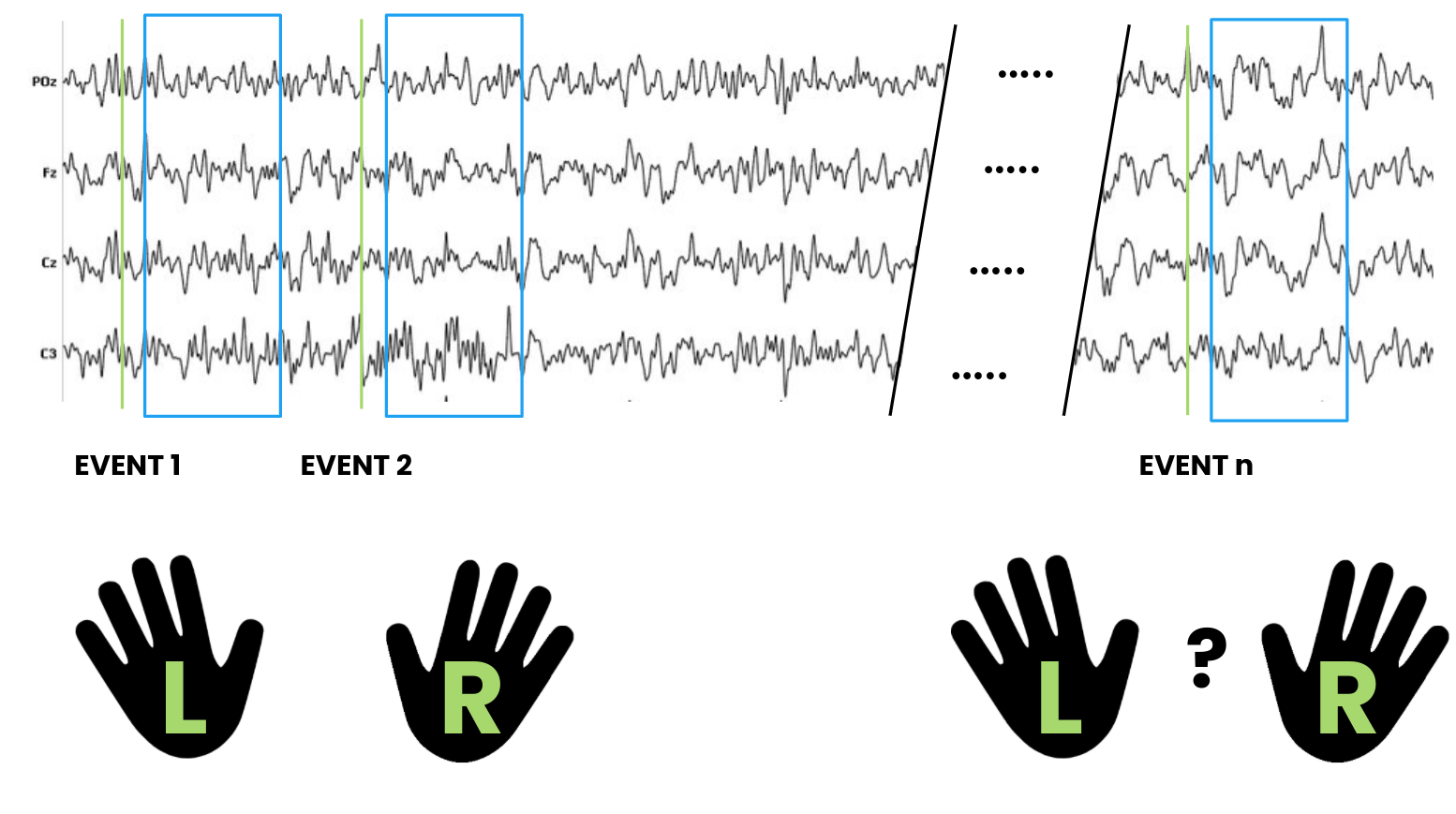}
\caption{Example of an EEG dataset composed of 4 electrodes. For a motor imagery task for the classification between right and left hand. The experimental paradigm consists in asking the subject to imagine the movement of a body part (left or right hand in this case), at specific times instants marked with green lines. For each green line, an epoch of data (blue boxes) is collected and associated with its left/right hand label. A subset of these epochs forms the training dataset. Later when presented with a new epoch (not contained in the training dataset), we would like to be able to classify whether the subject is thinking about moving his right or the left hand.}
\label{fig1:Figure_dataset_BCI}
\end{figure}

This work is focused on motor imagery classification tasks, so for example we are interested in predicting whether a subject is thinking about moving the right hand or the left hand (see
Fig.~\ref{fig1:Figure_dataset_BCI}).
We segment the EEG datasets into different short-time windows or epochs corresponding to a single motor task. Each such epoch is labelled by the actual motor task achieved by the subject. So for example to classify right vs left hand, we will get some epochs associated with the left hand label and other with the right hand one. We apply a feature extraction on each epoch in order to create disjoint training and testing datasets. The training dataset is used to train a classification algorithm. After the training procedure, we use the classifier to predict the outcome on the testing dataset.

\subsection{Riemann geometry}
Over the past 10 years, Riemannian distance-based classification paradigms have shown interesting performances for SPD matrix classification~\cite{barachant-bonnet-etal:10}. The key idea behind this method is to map the covariance matrix into an appropriate geometrical space. The space of SPD matrix is a differentiable manifold with a natural Riemann structure~\cite{forstner-moonen:03}.

The spaces of $n \times n$ real square 
 and real symmetric matrices are denoted the eigenvalues of  by $M(n)$ and $S(n)$, respectively: $S(n) = \left\{\mathbf{S} \in M(n), \;\mathbf{S}^{T}=\mathbf{S}\right\}$. The space of SPD matrices is defined as $P(n)=\{\mathbf{P} \in S(n), \mathbf{P}>0\}$. 
In this way, we can represent a matrix $\mathbf{P} \in S(n)$ as a point on a Riemann surface of dimension $n(n+1)/2$. 
Since the space of SPD matrix is a manifold of non zero curvature, we cannot use the concepts of Euclidean geometry since they are not reliable and have instead to rely on Riemannian geometry. It is possible to define a distance between two SPD matrix $\mathbf{P}_1$ and $\mathbf{P}_2$ in $P(n)$ as the length of the geodesic that connects $\mathbf{P}_1$ and $\mathbf{P}_2$ on the Riemann surface. In this paper, we use the affine-invariant metric~\cite{moakher:05}
\begin{equation}
    \delta_{R}\left(\mathbf{P}_{1}, \mathbf{P}_{2}\right)=\left\|\log \left(\mathbf{P}_{1}^{-\frac{1}{2}} \mathbf{P}_{2} \mathbf{P}_{1}^{-\frac{1}{2}}\right)\right\|_{F}=\left[\sum_{i=1}^{n} \log ^{2} \lambda_{i}\right]^{1 / 2}
    \label{Riemann metric}
\end{equation}
where $\|\cdot\|_{F}$ is the Frobenius norm, $Log()$ is the logarithm of a matrix and $\lambda_i$ are the eigenvalues of  $\mathbf{P}_1^{-\frac{1}{2}}\mathbf{P}_2\mathbf{P}_1^{-\frac{1}{2}}$. 
The choice of the metric is not unique but we choose that metric because it possesses relevant properties for BCI applications: invariance under reordering, invariance under congruent transformation and invariance under inversion~\cite{congedo-barachant-etal:17}.
To define the mean of $m$ SPD matrices $(\mathbf{P}_1, ..., \mathbf{P}_m)$, we use the concept of Fréchet mean~\cite{moakher:05}
\begin{equation}
    \mathbf{\overline{P}} =\underset{\mathbf{P} \in P(n)}{\operatorname{argmin}} \sum_{i=1}^{m} \delta_{R}^{2}\left(\mathbf{P}, \mathbf{P}_{i}\right)
\end{equation}
$\mathbf{\overline{P}}$ is called the geometric mean in the Riemannian sense or the center of mass of $(\mathbf{P}_1, ..., \mathbf{P}_m)$. Note that, for $m>2$, no closed-form expression is know, so we need to employ an iterative algorithm~\cite{pennec-fillard-etal:04}. Due to the properties of Eq.~(\ref{Riemann metric}), the geometric mean inherits the properties of congruence invariance~\cite{congedo-barachant-etal:17}. The geometric mean is also robust, both with respect to outliers and for cross-session and across-subject applications.

\begin{figure}[!ht]
 \centering
 \includegraphics[width=0.8\linewidth]{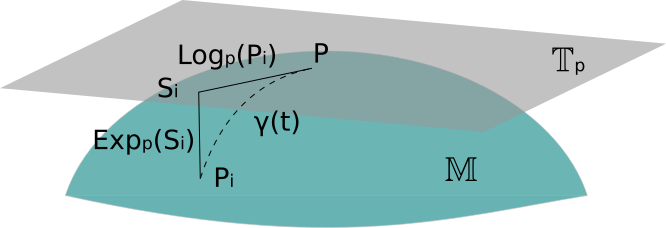}
 \caption{A Riemann manifold and the maps from/to the tangent plane.}
  \label{fig:figure_Riemann}
\end{figure}

For every point $\mathbf{P}_i \in P(n)$, in a complete Riemann space, we can define a point $\mathbf{S}_i \in S(n)$ in the tangent space $\mathbf{T}_P$ to the Riemann surface at point $\mathbf{P} \in P(n)$. The tangent space has zero curvature by definition and allows us to use standard machine learning algorithms and other classical tools. Mappings from the Riemann surface to the tangent space and from the tangent space to the Riemann surface, are provided respectively by the Riemannian Exp and Log maps~\cite{moakher:05} (see Fig.~\ref{fig:figure_Riemann}):
\begin{equation}
    \operatorname{Exp}_{\mathbf{P}}\left(\mathbf{S}_{i}\right)  =  \mathbf{P}^{1 / 2} \operatorname{Exp}\left(\mathbf{P}^{-1 / 2} \mathbf{S}_{i} \mathbf{P}^{-1 / 2}\right) \mathbf{P}^{1 / 2}\\
\end{equation}

    \begin{equation}            \operatorname{Log}_{\mathbf{P}}\left(\mathbf{P}_{i}\right)= \mathbf{P}^{1 / 2}  \operatorname{Log}\left(\mathbf{P}^{-1 / 2} \mathbf{P}_{i} \mathbf{P}^{-1 / 2}\right) \mathbf{P}^{1 / 2}
\end{equation}

In the context of a Riemann distance based classification algorithm, we estimate a spatial covariance matrix for each epoch with the standard approach of the spatial sample covariance estimator
\begin{equation}
    \operatorname{Cov}(\mathbf{X})=\frac{1}{Time-1} \sum_{i=1}^{Time}\mathbf{X}_{i}\mathbf{X}_{i}^{\textit{T}} \;
\end{equation}
where $T$ is the transpose operator. This is not the only option for estimating the covariance. We could have used other estimators, but this one has the advantage of being unbiased when the number of electrodes is much smaller than the temporal instants contained in an epoch.

As the covariance matrix belongs to the SPD space, it is possible to classify the different mental states using the Riemannian framework.
Note that the classification can be perform either on the Riemann surface, using algorithms specifically defined on the Riemann manifold as Minimum Distance to the Mean (MDM), or by moving points to tangent space and using standard classification algorithms such as Support Vector Machine (SVM) or Logistic Regression~\cite{barachant-bonnet-etal:11b}. 

\subsection{Augmented covariance}
This section discusses the autoregressive model and the motivation to create the augmented covariance method.
Our approach consists in considering the recorded EEG signal as a realization of a multivariate random process whose law is unknown. With certain assumptions, we can assure that this random process follows an autoregressive model.

EEG time series are not stationary: the signal properties change over time according to the underlying mental processes, but for sufficiently short time windows where the mental task does not vary, we can consider the signal as weakly stationary which means that $\mathbb{E}\left(\widetilde{\mathbf{X}}\right)$ and $\mathbb{E}\left(\widetilde{\mathbf{X}_{t+i}}, \widetilde{\mathbf{X}_t}^T\right)$ are independent of time $t$ in each window $i$.

Under the assumption of weak stationary, we can model the signal as a random process that follows an autoregressive model: this means that future values are expressed as a function of past ones plus a random component called innovation. Thus, we can model the multivariate EEG signal using an autoregressive model of order $p$~\cite{lutkepohl:05}
\begin{equation}
    \X_{t}=\sum_{i=1}^{p} \A_{i} \X_{t-i\tau}+\bm{\varepsilon}_{\mathbf{t}}
    \label{Autoregressive_eq}
\end{equation}
where $\X_t \in \mathbb{R}^{d \times 1}$ is the EEG signal at time $t$, $\A_i \in \mathbb{R}^{d \times d}$ are the autoregressive coefficients , $\tau$ is the lag and $\bm{\varepsilon}_t \in \mathbb{R}^{d \times 1}$ is the innovation component.

The order $p$ of the process, the delay as well as the autoregressive coefficients $\mathbf{A}_i$ are unknowns and must be estimated from the signal. The classic approach to estimate the autoregressive coefficients is using the Yule-Walker equation.
In order to derive it, we simply post multiply Equation (\ref{Autoregressive_eq}) by $\X_{t-k}$ and compute the expectation value. Denoting by $\G(i) = \mathbb{E}(\X_t, \X_{t-k}^T)\in \mathbb{R}^{d \times d}$ the matrix of auto-covariance with lag $i$ and by $\mathbf{U}$ the auto-covariance matrix of the innovation considered as independent from past values, we get the following equations for $i \in [1, p]$
\begin{equation}
    \left\{\begin{array}{l}
\G(\mathbf{0})=\sum_{k=1}^{p} \A_{k} \G(-k\tau)+\mathbf{U} \; \; \;  \text { for } i=0 \\
\G(\mathbf{i}) \; =\sum_{k=1}^{p} \A_{k} \G(i-k\tau)  \; \; \; \; \; \; \; \;  \text { for } i \neq 0
\end{array}\right.
\label{YW_eq}
\end{equation}
The second Equation (\ref{YW_eq}) in matrix notation yields
\begin{equation}
    \left[\begin{array}{cccc}
\G_{0} & \G_{-1} & \G_{-2} & \cdots \\
\G_{1} & \G_{0} & \G_{-1} & \cdots \\
\G_{2} & \G_{1} & \G_{0} & \cdots \\
\vdots & \vdots & \vdots & \ddots \\
\G_{p-1} & \G_{p-2} & \G_{p-3} & \cdots
\end{array}\right]
\left[\begin{array}{c}
\A_{1} \\
\A_{2} \\
\A_{3} \\
\vdots \\
\A_{p}
\end{array}\right]
= 
\left[\begin{array}{c}
\G_{1} \\
\G_{2} \\
\G_{3} \\
\vdots \\
\G_{p}
\end{array}\right]
\end{equation}
The autoregressive coefficients $\A_i$ are solutions of a linear system. We define the augmented covariance Matrix $\G_{aug}$ as 
\begin{equation}
        \G_{Aug} = \left[\begin{array}{cccc}
\G_{0} & \G_{-1} & \G_{-2} & \cdots \\
\G_{1} & \G_{0} & \G_{-1} & \cdots \\
\G_{2} & \G_{1} & \G_{0} & \cdots \\
\vdots & \vdots & \vdots & \ddots \\
\G_{p-1} & \G_{p-2} & \G_{p-3} & \cdots
\end{array}\right]
\end{equation}
$\G_{aug}$ is symmetric by construction since $\G_{i} = \G_{-i}$. A fairly natural idea is therefore to use these $\G_{aug}$ matrices for classification algorithms using the Riemannian framework.

\begin{figure*}[!ht]
\centering
  \subfloat[]{%
       \includegraphics[width=0.5\linewidth]{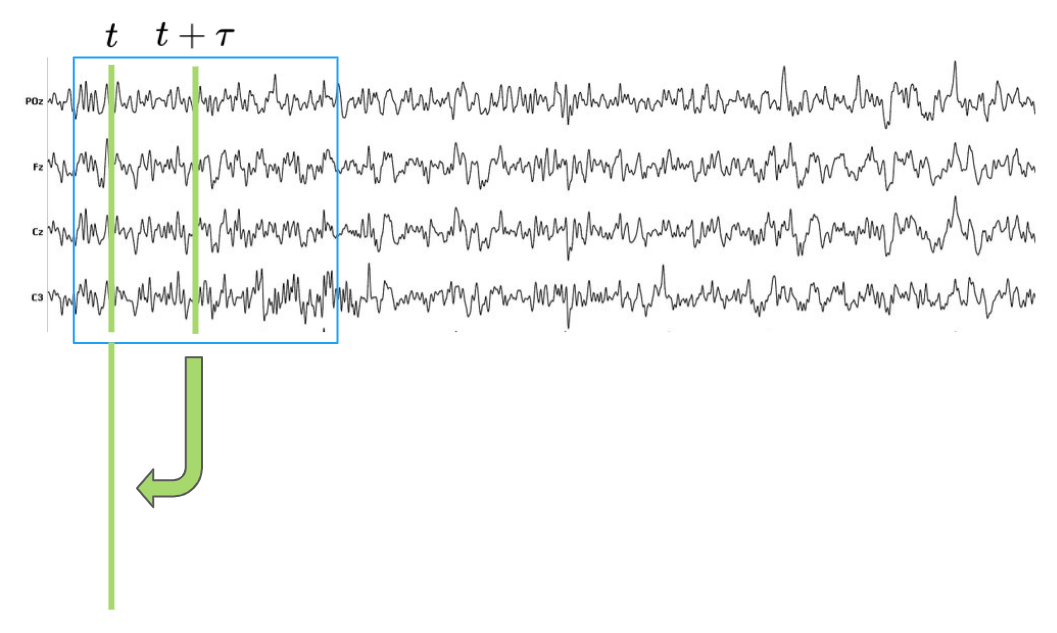}}
    \hfill
  \subfloat[]{%
        \includegraphics[width=0.5\linewidth]{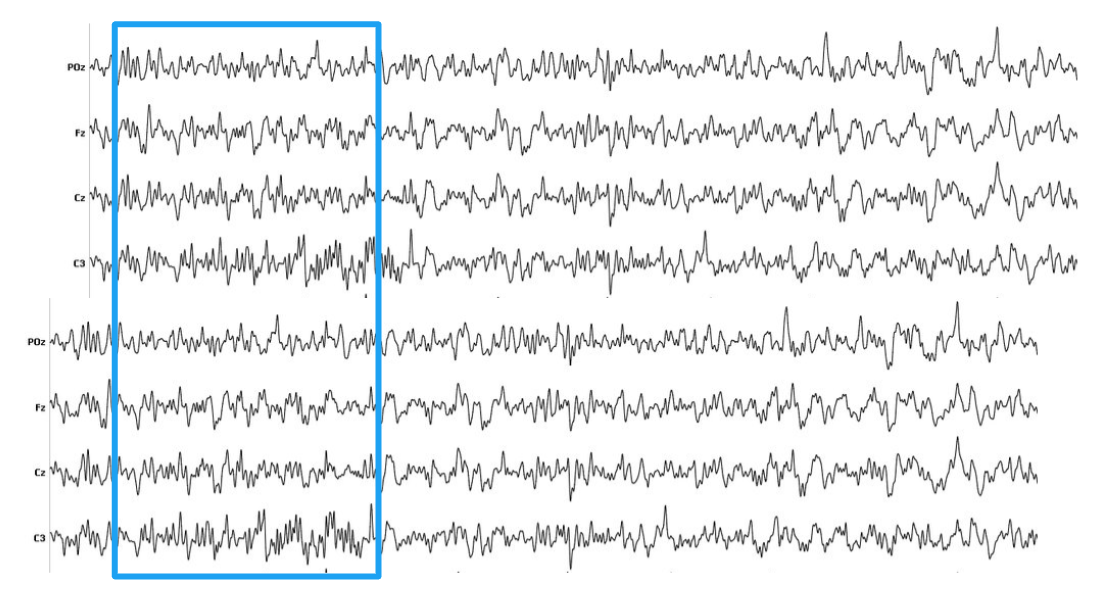}}
\caption{(a) Consider the EEG signal with 4 channels and an epoch (blue box), and create a new dataset by concatenating its values at time $t$ with those at time $t+\tau$.
(b) Result of the previous procedure at every point of the dataset. This is a higher dimensional dataset for a fixed delay $\tau$ and with dimension $d \times p$ (here $d=4$ and $p=2$).
The augmented covariance matrix of the original signal $\G_{aug}$ is exactly the standard covariance matrix of this concatenated dataset.
}
\label{Figure_dataset}
\end{figure*}

$\G_{aug}$ combines spatial covariance with some temporal information on the signal.
It is equivalent to consider an augmented dataset by embedding our EEG dataset in a dimension equal to $d \times p$, with a fixed delay $\tau$ (Fig. \ref{Figure_dataset}). 
Segmenting this new augmented dataset in epochs and computing the epoch covariance matrices results in exactly the same $\G_{aug}$ matrix.

The sample covariance matrix is an effcient and unbiased estimator of the real covariance matrix~\cite{hastie-tibshirani-etal:09}.  
In the case of BCI applications, we are in a Finite Observation Large-Dimensional Limit condition (FOLDL)~\cite{bartz:16}. In the case of $\G_{aug}$ the number of time sample $T$ is kept fixed while the dimensionality tends to be multiplied by the embedding dimension. The performance of the sample covariance is no more reliable and the estimator is no longer consistent.
A reliable estimator is obtained using a shrinkage estimator defined as
\begin{equation}
    \widehat{\mathbf{C}}_{shrink} = (1-\lambda)\widehat{\mathbf{C}}_{sample} + \lambda \frac{Tr\left(\widehat{\mathbf{C}}_{sample}\right)}{d} I
\end{equation}
Specifically, we are going to use the Shrinkage estimator introduced by Ledoit-Wolf~\cite{ledoit-wolf:04}, who developed an analytical formula to find the parameter $\lambda$ that minimizes the mean squared error between the estimated and the real covariance matrices.

\subsection{Non Linear Dynamical approach}

The formulation of the problem using the augmented dataset shows a natural connection with the delay-embedding theorem attributed to Takens~\cite{takens:81} in the context of non-linear dynamical systems.

Most EEG classification algorithms are based on the assumption that the signal is describable through a linear theory. However, recent studies have proven that the EEG signal follows a nonlinear model~\cite{casdagli-iasemidis-etal:97}. Because of this nonlinear component, the use of signal analysis techniques based on linearity assumptions could lose important characteristics. 
Recently some algorithms have been proposed that use quantities defined in the context of nonlinear systems theory as classification features, however, these quantities are extremely time consuming.

\begin{figure}[!ht]
 \centering
 \includegraphics[width=\linewidth]{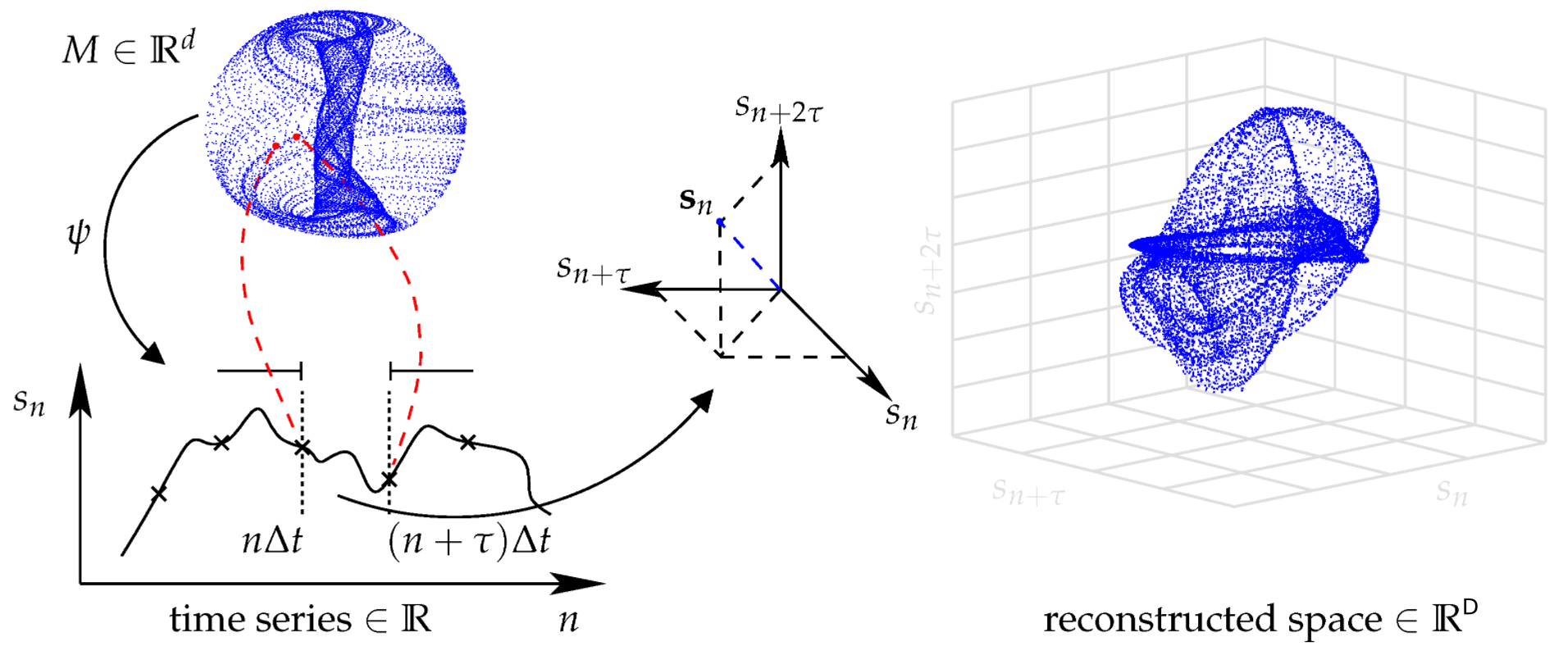}
 \caption{Illustration of the concept of the Delay-Embedding theorem. Reproduced with permission from Ana González-Marcos~\cite{pedro-carracedo-fuentes-jimenez-etal:20}}
  \label{fig:fig1}
\end{figure}

Let us now try to understand the idea behind phase space reconstruction based on Takens's theorem. Usually what we observe in an experiment is not a phase space object but a time series, most likely only a sequence of scalar measurements, so that not all the dynamic variables involved in the dynamics of the system under study are made available through the measurement process. So in most cases a time series can be seen as the values taken by the observed variables of a partially observable dynamical system. It is obvious that even with a precise knowledge of the measurement process it may be impossible to reconstruct the state space of the original system from the data. Fortunately, a reconstruction of the original phase space is not really necessary for data analysis and sometimes
not even desirable. It is sufficient to construct a new space that is dynamically equivalent to the original one. The process that we have just enunciated is very well summarized by Figure~\ref{fig:fig1}.

Considering that a process of measurement create a time series ${s(n)}$.
According to Takens’s theorem, the geometric structure of the multi variable dynamics of the system can be unfolded from the observable $s(n)$ in a $D$-dimensional space constructed from the delay vector
\begin{equation}
    \textbf{s}_E(n) = [s(n), s(n - \tau), ..., s(n - (D - 1)\tau)]^T
\end{equation}
where $\tau$ is a positive integer called the embedding delay and $D$ embedding dimension. So $\textbf{s}_E(n) \in \mathbb{R}^{D}$ is an embedding of the original phase space.
As seen previously, the augmented covariance matrix is equivalent to using a standard covariance matrix on an embedding the EEG data in a high dimensional space. The approach is thus related to the theory of non-linear dynamics, but then, instead of trying to identify the dynamics's parameters, we classify the mental state using the Riemann framework. In this sense, the two strengths of both approaches are thus combined: the use of the nonlinear component of the signal and the performance of the Riemann distance-based algorithms. Our approach turns out to be a natural extension of the Riemann distance-based classification that considers SPD matrices that contain a greatest amount of information extracted from the signal for dynamical systems theory. Thus the classification method based on the Riemann framework becomes even more effective by accounting for some non-linearities of the signal measured through the embedding of the data.

\subsection{Datasets Considered}
To test the performance of our method, we used datasets implemented in the MOABB framework, which are open-access. 
We consider five motor imagery BCI datasets consisting of several subjects for each dataset and several sessions for each subject.
Table~\ref{table:dataset} contains all the details about the considered datasets.

\begin{table}[!ht]
\resizebox{\linewidth}{!}{\begin{tabular}{c|c|c|c|c|c|c|c}
\text { Dataset } & \text { subjects } & \text { channels } & \text {sampling rate } & \text { sessions } & \text { tasks } & \text { trials/class } & \text{Epoch (s)} \\
\hline \text { Zhou2016~\cite{ zhou-wu-etal:16}} & 3 & 14 & 250 Hz & 3 & 3 & 160 & [0, 5] \\
\hline \text {BNCI2014001~\cite{tangermann-muller-etal:12}} & 9 & 22 & 250 Hz & 2 & 4 & 144 & [2, 6] \\
\hline \text {BNCI2015001~\cite{faller-vidaurre-etal:12}} & 12 & 13 & 512 Hz & 1 & 2 & 200 & [0, 5] \\
\hline \text {BNCI2014002~\cite{steyrl-scherer-etal:16}} & 14 & 15 & 512 Hz & 5 & 2 & 80 & [3, 8] \\
\hline \text {BNCI2014004~\cite{leeb-lee-etal:07}} & 9 & 3 & 250 Hz & 1 & 2 & 360 & [3, 7.5]\\
\hline 
\end{tabular}}
\caption{Dataset considered during this study}
\label{table:dataset}
\end{table}

On all datasets, we apply a standard band-pass filter with range [8; 35] Hz for the motor imagery task. We consider epochs with lengths equal to the duration of task conditions which varies with the datasets under consideration. All classes are balanced in these datasets.

\subsection{Parameter Estimation}
Our model depends on two hyper-parameters, the order of the autoregressive model (equivalently the embedding dimension $D$) and the delay $\tau$. In order to compute these two parameters for every dataset and every subject, we developed two different strategies: one based on the grid search algorithm and another one based on the concept of non-linear dynamical systems. We select hyper-parameters for each subject in the case of performance evaluation using the within-session and cross-session evaluation methodology.

\subsubsection{Grid Search}
In order to find the hyper-parameters, we allow the grid search on two parameters to run on the domain defined by: $D \in [1, 10]$ and $\tau \in [1, 10]$. 
The grid search are performed using the exact same procedure used for the evaluation procedure as we will outline later. Note that Grid-Search should always provide the best solution in the explored domain (multiple minima are possible).

\subsubsection{Non Linear Dynamical approach}
The parameters $D$ and $\tau$ can also be estimated using the concepts of non-linear dynamical systems. There are several algorithms for estimating optimal parameters in noisy conditions that can be grouped in two categories: traditional and unified approaches. Both approaches were implemented using the \href{https://juliadynamics.github.io/DynamicalSystems.jl/dev/}{DynamicalSystem.jl} library~\cite{datseris:18}.

\paragraph{Traditional approach}
The traditional approach tries to find the best value of the time delay $\tau$ and then an optimal embedding dimension $D$ separately. 
Our approach is based on computing the Average Mutual Information (AMI)~\cite{fraser-swinney:86} and the number of neighbours using Cao's algorithm~\cite{cao:97}.
The embedding theorem places no restrictions on the choice of time delay in the reconstruction process. 
In theory, with an unlimited number of measurements without noise, any delay is equally valid, except certain multiples of the precise period of a periodic signal. In practice, the amount of data is finite, which means that some constraint is needed to identify the best value of time delay $\tau$. The AMI algorithm is using the position of the first minimum of the mutual information as the optimal value of $\tau$.
There is no rule of thumb to impose a minimum reconstruction dimension $D$ but, among all possible techniques, the False Nearest Neighbors (FNN) with Cao’s modification stands out. This algorithm is based on the consideration that in a reconstruction space of very low dimension, two points appear to be closer to each other than they actually are. Thus, two points are considered as real neighbors if their distance remains constant as the reconstruction dimension increases. The embedding dimension is selected as the value where the FNN saturates around 1.

Since that algorithms were created for dealing with univariate data we accumulate the results of each channel and epoch and we select $\tau$ as the first minimum of the overall Mutual Information. With analogous consideration we compute the embedding dimension $D$ is the value for which the overall false nearest neighbour saturates around 1.

\paragraph{Unified approach}
The unified approach finds at the same time an optimal combination of $\tau$ and $D$. For this approach,  we use the MDOP algorithm~\cite{nichkawde:13}.
The MDOP algorithm works iteratively by finding the best delay and embedding dimensions together. The algorithm starts with the current dataset and adds a lagged version of that dataset to it by finding the correct value of $\tau$ such that the redundancy between coordinates is minimized, i.e. by maximizing the beta statistics~\cite{nichkawde:13}. After each embedding cycle the FNN-statistic~\cite{kennel-brown-etal:92} is being checked and as soon as this statistic drops below a threshold, the algorithm terminates.
It should be noted that this algorithm produces different values of $\tau$ for each embedding cycle, so we decide to select the optimal value of $\tau$ as an average of these values, while we select $D$ as the number of cycles.

\subsection{Pipeline}
We intend to evaluate how the augmented covariance method (ACM) as a general technique can increase the performance in different settings. In particular, we consider several classification methods on the Riemann surface and on the tangent space as shown in Table~\ref{table:pipeline}.

\begin{table*}[!ht]
\resizebox{\linewidth}{!}{
    \begin{tabular}{c|c|*{3}{c}}
        \hline
        \textbf{Surface} &  \textbf{Standard algorithm}  &  \multicolumn{3}{c}{\bfseries Augmented algorithm} \\
        \hline
        &   & \multicolumn{1}{c}{\bfseries Grid Search} & \multicolumn{1}{c}{\bfseries Traditional Approach} & \multicolumn{1}{c}{\bfseries Unified Approach} \\
        \hline
        Riemann Surface & MDM & ACM+MDM & ACM+MDM AMI aFNN & ACM+MDM MDOP \\
        Tangent Space & TANG+SVM & ACM+TANG+SVM & ACM+TANG+SVM AMI aFNN & ACM+TANG+SVM MDOP \\
        \hline
    \end{tabular}
    }
\caption{Considered pipelines with classification both on Riemann surface and on Tangent Space.}
\label{table:pipeline}
\end{table*}

We compare our best method "ACM+TANG+SVM" (classification on tangent space using a SVM with grid search), against several algorithms in the literature of BCI.
\begin{enumerate}
    \item "CSP+LDA", a combination of the Common spatial pattern (CSP) algorithm followed by a classification performed on a shrinkage Linear Discriminant Analysis (LDA)~\cite{lotte-guan:10b}.
    \item "FgMDM", a classification method by Minimum Distance to the Mean after having applied a geodesic filtering~\cite{yger-berar-etal:16}.
    \item "COV+EN", a classification method in the tangent space using an Elastic Network as classifier~\cite{corsi-chevallier-etal:22}.
\end{enumerate}

For classification scores, we use different model validation techniques depending on the considered paradigm: for within-session evaluations (WS), we used a 5-fold cross-validation while for cross-session evaluations (CS), we used a leave-one-session-out paradigm.

All the hyper parameters of the classifiers used in this study are optimized with grid search. Table~\ref{table:pipeline_parameter} summarizes the domains over which these grid searches are performed.

\begin{table}[!ht]
\centering
\begin{tabular}{c|c|c}
  \hline
  \textbf{Pipeline} &  \textbf{Parameter}  &  \textbf{Value} \\ \hline
   "COV+EN" & l1 ratio & [0.15, 0.30, 0.45, 0.60, 0.75] \\ \hline
   "TANG+SVM" & C & [0.5, 1, 1.5] \\ 
    & Kernel & ["linear", "rbf"] \\ \hline
    "CSP+LDA" & nfilter & [1, 2, 3, 4, 5, 6, 7, 8] \\ \hline
    "ACM+TANG+SVM" & C & [0.5, 1, 1.5] \\ 
    & Kernel & ["linear", "rbf"] \\
    & Order & [1, 2, 3, 4, 5, 6, 7, 8, 9, 10] \\
    & Lag & [1, 2, 3, 4, 5, 6, 7, 8, 9, 10] \\\hline
    "ACM+MDM" & Order & [1, 2, 3, 4, 5, 6, 7, 8, 9, 10] \\
    & Lag & [1, 2, 3, 4, 5, 6, 7, 8, 9, 10] \\\hline
\end{tabular}

\caption{Parameter considered for each pipelines}
\label{table:pipeline_parameter}
\end{table}

\subsection{Statistical Analysis}
In order to validate the best pipeline from a statistical point of view, we use the instrument provided by the MOABB framework. This algorithm use a mixture of permutation and non-parametric tests in order to keep the algorithm fast.
The MOABB framework uses a one-tailed permutation-based paired t-test~\cite{student:08} for datasets with less than 20 subjects, or a Wilcoxon signed-rank test~\cite{wilcoxon:92} otherwise.
This statistical test is used to generate a p-value that compares two pipelines for each pair of pipelines. Then, the p-values are combined using Stouffer’s method~\cite{stouffer-suchman-etal:49} in order to get a final p-value for each hypothesis.
In order to prevent the problem of false positive, we apply the Bonferroni correction~\cite{bonferroni:36}.
In this way, we can validate the results using several subjects and several datasets.

The score are obtained using the Area Under (AUC) Receiver Operating Characteristic (ROC) curve. AUC ranges in value from 0 to 1 and a higher AUC value is a better classifier. In case of a multi-class classification the metric used is the accuracy.

\section{Results}
\label{sec:results}
In this section, we report the results of the augmented covariance method over the classification on both the Riemann surface and the Tangent Space. To validate the robustness and validity of our approach, we test the algorithm on different datasets, subjects and tasks using the MOABB framework. By definition Grid Search will always give the best solution for the Augmented approaches provided the optimum is in the search domain. Thus, the comparison between Grid-Search and Non Linear Dynamics approaches will be always in favor of the former because we chose a Gird-Search domain that is large enough.

\subsection{Right hand vs Left hand}
We consider 3 different datasets: BNCI2014001, BNCI2014004 and Zhou2016 to classify a right vs left hand task classification using within-session and cross-session evaluation procedures.

\subsubsection{Riemann surface classification}
All methods using ACM show an improvement with respect to the standard MDM algorithm as shown in Table~\ref{table:MDM-rhlh-whithinsession}. Using ACM and selecting hyper parameters with grid search gives performance improvements in the range 3\%-7\%, while selecting those with the traditional and unified approaches give close results (but with decreases of AUC of up to 3\% for the traditional approach and 5\% for the unified one). These tendencies are true in both the within-session (WS) and cross-session (CS) frameworks. Figures~\ref{fig:MDM-rhlh-whithinsession} and~\ref{fig:MDM-rhlh-crosssession} provide a detailed statistical study of the significance of these results.

\begin{table}[!ht]
\resizebox{\linewidth}{!}{\begin{tabular}{c|c|c|c|c|c}
\text { Dataset } & Eval & \text {MDM} & \text {ACM+Grid Search} & \text {ACM+Traditional} & \text {ACM+Unified} \\
\hline \text {BNCI2014001} & WS & $0.84 \pm 0.14$ & \textbf{0.91} $\pm$ \textbf{0.11} & 0.89 $\pm$ 0.12 & $0.86 \pm 0.14$  \\
\hline \text {BNCI2014004} & WS & $0.78 \pm 0.16$ & \textbf{0.83} $\pm$ \textbf{0.14} & $0.82 \pm 0.15$ & $0.81 \pm 0.16$  \\
\hline \text { Zhou2016} & WS & $0.89 \pm 0.08$ & \textbf{0.93} $\pm$ \textbf{0.07} & $0.90 \pm 0.07$ & $0.91 \pm 0.08$  \\ \hline
\hline \text {BNCI2014001} & CS & $0.83 \pm 0.15$ & \textbf{0.91} $\pm$ \textbf{0.11} & $0.88 \pm 0.13$ & $0.86 \pm 0.13$  \\
\hline \text {BNCI2014004} & CS & $0.79 \pm 0.15$ & \textbf{0.83} $\pm$ \textbf{0.14} & \textbf{0.83} $\pm$ \textbf{0.14} & $0.81 \pm 0.15$  \\
\hline \text {Zhou2016} & CS & $0.92 \pm 0.05$ & \textbf{0.95} $\pm$ \textbf{0.04} & $0.93 \pm 0.04$ & $0.93 \pm 0.04$  \\
\hline 
\end{tabular}}
\caption{Performance (AUC) of Right hand vs Left hand using MDM based algorithms.}
\label{table:MDM-rhlh-whithinsession}
\end{table}

\subsubsection{Tangent space classification}
The tangent space classification with SVM is in general more effective than with the Riemann surface classification approaches as shown in Table~\ref{table:TANG+SVM-rhlh-whithinsession}. This is true for the baseline method as well as for the ACM ones. The ACM based methods with parameter estimation based on dynamical system theory shows results comparable to those obtained with the standard covariance matrix (AUC changes from -1\% to +2\%). Yet, the use of the ACM with grid search shows some notable improvements with respect to all other methods which behave similarly on this dataset (3-7\% depending on the dataset). Again, these improvements occur similarly for both the WS and CS frameworks. Figures~\ref{fig:TANG+SVM-rhlh-whithinsession} and~\ref{fig:TANG+SVM-rhlh-crosssession} provide a detailed study of the statistical significance of these results.

\begin{table}[!ht]
\resizebox{\linewidth}{!}{\begin{tabular}{c|c|c|c|c|c}
\text {Dataset} & Eval & \text {TANG+SVM} & \text {ACM+Grid Search} & \text {ACM+Traditional} & \text {ACM+Unified } \\
\hline \text {BNCI2014001} & WS & $0.89 \pm 0.11$ & \textbf{0.95} $\pm$ \textbf{0.07} & $0.89 \pm 0.12$ & $0.88 \pm 0.13$  \\
\hline \text {BNCI2014004} & WS & $0.81 \pm 0.15$ & \textbf{0.88} $\pm$ \textbf{0.12} & $0.81 \pm 0.16$ & $0.81 \pm 0.15$  \\
\hline \text {Zhou2016} & WS & $0.93 \pm 0.07$ & \textbf{0.96} $\pm$ \textbf{0.05} & $0.94 \pm 0.06$ & $0.93 \pm 0.07$  \\ \hline
\hline \text {BNCI2014001} & CS & $0.86 \pm 0.13$ & \textbf{0.93} $\pm$ \textbf{0.09} & $0.88 \pm 0.12$ & $0.88 \pm 0.12$  \\
\hline \text {BNCI2014004} & CS & $0.81 \pm 0.14$ & \textbf{0.85} $\pm$ \textbf{0.14} & $0.80 \pm 0.16$ & $0.81 \pm 0.15$  \\
\hline \text {Zhou2016} & CS & $0.92 \pm 0.07$ & \textbf{0.95} $\pm$ \textbf{0.04} & $0.93 \pm 0.05$ & $0.93 \pm 0.06$  \\
\hline 
\end{tabular}}
\caption{Performance (AUC) of Right hand vs Left hand using TANG+SVM based algorithm.}
\label{table:TANG+SVM-rhlh-whithinsession}
\end{table}

\subsubsection{State of the Art} Our best approach (ACM with TANG+SVM classification and hyper-parameters found with grid search) is compared to advanced state of the art methods (usually showing better results than the baseline method) in Table~\ref{table:TANG+SVM-rhlh-whithinsession-stateart}. It brings a notable improvements over all these methods with an average improvement of AUC by 2-9\% and also a significant decrease in the standard deviation of these values meaning that the results are more consistent across subjects. A detailed study of the statistical significance of these results is provided in Figures~\ref{fig:TANG+SVM-rhlh-whithinsession-stateart} and~\ref{fig:TANG+SVM-rhlh-crosssession-stateart}.

\begin{table}[!ht]
\resizebox{\linewidth}{!}{\begin{tabular}{c|c|c|c|c|c}
\text { Dataset } & Eval & \text {CSP+LDA} & \text {ACM+TANG+SVM(Grid Search)} & \text {COV+EN} & \text {FgMDM} \\
\hline \text {BNCI2014001} & WS & $0.86 \pm 0.13$ & \textbf{0.95} $\pm$ \textbf{0.07} & $0.89 \pm 0.11$ & $0.88 \pm 0.11$  \\
\hline \text {BNCI2014004} & WS & $0.80 \pm 0.15$ & \textbf{0.88} $\pm$ \textbf{0.12} & $0.80 \pm 0.15$ & $0.79 \pm 0.16$  \\
\hline \text {Zhou2016} & WS & $0.93 \pm 0.08$ & \textbf{0.96} $\pm$ \textbf{0.05} & $0.94 \pm 0.07$ & $0.91 \pm 0.08$  \\ \hline
\hline \text {BNCI2014001} & CS & $0.86 \pm 0.14$ & \textbf{0.93} $\pm$ \textbf{0.09} & $0.87 \pm 0.13$ & $0.85 \pm 0.14$  \\
\hline \text {BNCI2014004} & CS  & $0.81 \pm 0.14$ & \textbf{0.85} $\pm$ \textbf{0.14} & $0.81 \pm 0.14$ & $0.80 \pm 0.14$  \\
\hline \text {Zhou2016} & CS & $0.93 \pm 0.07$ & \textbf{0.95} $\pm$ \textbf{0.04} & $0.93 \pm 0.07$ & $0.90 \pm 0.09$  \\
\hline 
\end{tabular}}
\caption{Performance (AUC) of Right hand vs Left hand using ACM+TANG+SVM algorithm against the state of the art.}
\label{table:TANG+SVM-rhlh-whithinsession-stateart}
\end{table}

\subsection{Right hand vs Feet}
This classification task was performed on 4 different datasets: BNCI2014001, BNCI2014002, BNCI2015001 and Zhou2016 to classify a right hand vs feet task classification using a within-session and cross-session evaluation procedure.

\subsubsection{Riemann surface classification}
All methods using ACM show an improvement with respect to the standard MDM algorithm as shown in Table~\ref{table:MDM-rf-whithinsession}. The best performance is obtained with using ACM and selecting hyper parameters with grid search (3-6\% improvements in AUC). Selecting those obtained with the traditional approach gives close results (with decrease of performance of up-to 3\%). The unified approach is usually worse than the traditional one. As before, the tendency is similar for both WS and CS frameworks.
Figures~\ref{fig:MDM-rf-whithinsession} and~\ref{fig:MDM-rf-crosssession} show a detailed study of the statistical significance.

\begin{table}[!ht]
\resizebox{\linewidth}{!}{\begin{tabular}{c|c|c|c|c|c}
\text {Dataset} & Eval & \text {MDM} & \text {ACM+Grid Search} & \text {ACM+Traditional} & \text {ACM+Unified} \\
\hline \text {BNCI2014001} & WS & $0.91 \pm 0.10$ & \textbf{0.96} $\pm$ \textbf{0.05} & $0.95 \pm 0.06$ & $0.93 \pm 0.10$  \\
\hline \text {BNCI2014002} & WS & $0.77 \pm 0.15$ & \textbf{0.83} $\pm$ \textbf{0.15} & $0.81 \pm 0.15$ & $0.78 \pm 0.17$  \\
\hline \text {BNCI2015001} & WS & $0.86 \pm 0.13$ & \textbf{0.91} $\pm$ \textbf{0.11} & 0.90 $\pm$ 0.11 & $0.89 \pm 0.12$  \\
\hline \text {Zhou2016} & WS & $0.92 \pm 0.05$ & \textbf{0.96} $\pm$ \textbf{0.04} & $0.93 \pm 0.05$ & $0.94 \pm 0.05$  \\ \hline
\hline \text {BNCI2014001} & CS & $0.90 \pm 0.12$ & \textbf{0.95} $\pm$ \textbf{0.06} & $0.93 \pm 0.07$ & $0.92 \pm 0.10$  \\
\hline \text {BNCI2015001} & CS  & $0.87 \pm 0.12$ & \textbf{0.91} $\pm$ \textbf{0.09} & \textbf{0.91} $\pm$ \textbf{0.10} & $0.89 \pm 0.11$  \\
\hline \text {Zhou2016} & CS & $0.93 \pm 0.06$ & \textbf{0.96} $\pm$ \textbf{0.04} & $0.95 \pm 0.05$ & $0.95 \pm 0.05$  \\
\hline 
\end{tabular}}
\caption{Performance (AUC) of Right hand vs Feet using MDM based algorithms.}
\label{table:MDM-rf-whithinsession}
\end{table}

\subsubsection{Tangent space classification}
As in the Left vs Right hand case, the tangent space classification with SVM is in general more effective than with the Riemann surface classification approaches as shown in Table~\ref{table:TANG+SVM-rf-whithinsession}. The ACM based methods with parameter estimation based on dynamical system theory shows comparable results to the standard covariance matrix (in one dataset, performance decreases slightly). Yet, the use of the ACM with grid search shows some notable improvements (1-4\%) with respect to all other methods which behave similarly on these datasets, which is notable given the very high AUC values already obtained with the baseline method. Figures~\ref{fig:TANG+SVM-rf-whithinsession} and~\ref{fig:TANG+SVM-rf-crosssession} provide a detailed study of the statistical significance of these results.

\begin{table}[!ht]
\resizebox{\linewidth}{!}{\begin{tabular}{c|c|c|c|c|c}
\text {Dataset} & Eval & \text {TANG+SVM} & \text {ACM+Grid Search} & \text {ACM+Traditional} & \text {ACM+Unified} \\
\hline \text {BNCI2014001} & WS & $0.95 \pm 0.07$ & \textbf{0.98} $\pm$ \textbf{0.02} & $0.95 \pm 0.06$ & $0.95 \pm 0.07$  \\
\hline \text {BNCI2014002} & WS & $0.87 \pm 0.12$ & \textbf{0.91} $\pm$ \textbf{0.10} & $0.86 \pm 0.13$ & $0.85 \pm 0.15$  \\
\hline \text {BNCI2015001} & WS & $0.92 \pm 0.08$ & \textbf{0.95} $\pm$ \textbf{0.06} & \textbf{0.93} $\pm$ \textbf{0.09} & $0.92 \pm 0.09$  \\
\hline \text {Zhou2016} & WS & $0.97 \pm 0.03$ & \textbf{0.99} $\pm$ \textbf{0.01} & $0.97 \pm 0.02$ & $0.97 \pm 0.04$  \\ \hline
\hline \text {BNCI2014001} & CS & $0.94 \pm 0.07$ & \textbf{0.97} $\pm$ \textbf{0.04} & $0.94 \pm 0.07$ & $0.93 \pm 0.08$  \\
\hline \text {BNCI2015001} & CS & $0.90 \pm 0.10$ & \textbf{0.94} $\pm$ \textbf{0.07} & $0.92 \pm 0.08$ & $0.91 \pm 0.09$  \\
\hline \text {Zhou2016} & CS & $0.96 \pm 0.05$ & \textbf{0.99} $\pm$ \textbf{0.01} & $0.98 \pm 0.02$ & $0.98 \pm 0.03$  \\
\hline 
\end{tabular}}
\caption{Performance (AUC) of Right hand vs Feet using TANG+SVM based algorithms.}
\label{table:TANG+SVM-rf-whithinsession}
\end{table}

\subsubsection{State of the Art}
Our best approach (ACM with TANG+SVM classification and hyper-parameters found with grid search) is compared to other state of the art in Table~\ref{table:TANG+SVM-rf-whithinsession-stateart}. It brings a notable improvement over all these methods. A detailed study of the statistical significance of these results is provided in Figures~\ref{fig:TANG+SVM-rf-whithinsession-stateart} and~\ref{fig:TANG+SVM-rf-crosssession-stateart}.

\begin{table}[!ht]
\resizebox{\linewidth}{!}{\begin{tabular}{c|c|c|c|c|c}
\text {Dataset} & Eval & \text {CSP+LDA} & \text {ACM+TANG+SVM(Grid Search)} & \text {COV+EN} & \text {FgMDM} \\
\hline \text {BNCI2014001} & WS & $0.93 \pm 0.10$ & \textbf{0.98} $\pm$ \textbf{0.02} & $0.95 \pm 0.06$ & $0.94 \pm 0.08$  \\
\hline \text {BNCI2014002} & WS & $0.86 \pm 0.12$ & \textbf{0.91} $\pm$ \textbf{0.10} & $0.87 \pm 0.12$ & $0.85 \pm 0.12$  \\
\hline \text {BNCI2015001} & WS & $0.90 \pm 0.10$ & \textbf{0.95} $\pm$ \textbf{0.06} & $0.92 \pm 0.08$ & $0.90 \pm 0.10$  \\
\hline \text {Zhou2016} & WS & $0.95 \pm 0.04$ & \textbf{0.99} $\pm$ \textbf{0.01} & $0.97 \pm 0.03$ & $0.96 \pm 0.03$  \\ \hline
\hline \text {BNCI2014001} & CS & $0.90 \pm 0.12$ & \textbf{0.97} $\pm$ \textbf{0.05} & $0.94 \pm 0.07$ & $0.93 \pm 0.08$  \\
\hline \text {BNCI2015001} & CS & $0.89 \pm 0.11$ & \textbf{0.94} $\pm$ \textbf{0.07} & $0.91 \pm 0.09$ & $0.90 \pm 0.10$  \\
\hline \text {Zhou2016} & CS & $0.95 \pm 0.05$ & \textbf{0.99} $\pm$ \textbf{0.01} & $0.96 \pm 0.04$ & $0.95 \pm 0.06$  \\
\hline 
\end{tabular}}
\caption{Performance (AUC) of Right hand vs Feet using ACM+TANG+SVM algorithm against the state of the art.}
\label{table:TANG+SVM-rf-whithinsession-stateart}
\end{table}

\subsection{Right hand vs Left hand vs Feet}
In this case, we use 2 different datasets BNCI2014001 and Zhou2016 to classify a right vs left hand vs feet tasks using a within-session and cross-session evaluation procedure.

\subsubsection{Riemann surface classification}
In all cases, methods using ACM show an improvement with respect to the standard MDM algorithm as shown in Table~\ref{table:MDM-3class-whithinsession}. Grid search parameter optimisation leads to 6-9\% improvements.
The improvements obtained with the other two methods (standard or unified) are lower 1-6\%. Again, in all cases,
the standard deviations are lower which indicate more consistent results across subjects. Figures~\ref{fig:MDM-3class-whithinsession} and~\ref{fig:MDM-3class-crosssession} show a detailed study of the statistical significance.

\begin{table}[!ht]
\resizebox{\linewidth}{!}{\begin{tabular}{c|c|c|c|c|c}
\text {Dataset} & Eval & \text {MDM} & \text {ACM+Grid Search} & \text {ACM+Traditional} & \text {ACM+Unified} \\
\hline \text {BNCI2014001} & WS & $0.72 \pm 0.15$ & \textbf{0.81} $\pm$ \textbf{0.11} & $0.78 \pm 0.13$ & $0.76 \pm 0.14$  \\
\hline \text {Zhou2016} & WS & $0.75 \pm 0.07$ & \textbf{0.81} $\pm$ \textbf{0.04} & $0.76 \pm 0.06$ & 0.79 $\pm$ 0.05  \\ \hline
\hline \text {BNCI2014001} & CS & $0.66 \pm 0.13$ & \textbf{0.75} $\pm$ \textbf{0.11} & $0.74 \pm 0.12$ & $0.70 \pm 0.12$  \\
\hline \text {Zhou2016} & CS & $0.70 \pm 0.09$ & \textbf{0.75} $\pm$ \textbf{0.06} & $0.72 \pm 0.09$ & 0.72 $\pm$ 0.09  \\
\hline 
\end{tabular}}
\caption{Performance (Accuracy) of Right hand vs Left hand vs Feet using MDM based algorithms.}
\label{table:MDM-3class-whithinsession}
\end{table}

\subsubsection{Tangent space classification}
As in the previous results, the tangent space classification with SVM is more effective than with the Riemann surface classification approaches as shown in Table~\ref{table:TANG+SVM-3class-whithinsession}. In this case, the traditional method for computing the ACM hyper-parameters gives worse results than the baseline, failing to select a good set of parameters. The unified approach does a little bit better with improvements of 1-2\%. Using of the ACM with grid search gives the best results and shows some notable (7\%-8\%) improvements with respect to the baseline method.
Figures~\ref{fig:TANG+SVM-3class-whithinsession} and~\ref{fig:TANG+SVM-3class-crosssession} show a detailed study of the statistical significance.

\begin{table}[!ht]
\resizebox{\linewidth}{!}{\begin{tabular}{c|c|c|c|c|c}
\text {Dataset} & Eval & \text {TANG+SVM} & \text {ACM+Grid Search} & \text {ACM+Traditional} & \text {ACM+Unified} \\
\hline \text {BNCI2014001} & WS & $0.79 \pm 0.13$ & \textbf{0.87} $\pm$ \textbf{0.09} & $0.79 \pm 0.13$ & $0.80 \pm 0.14$  \\
\hline \text {Zhou2016} & WS & $0.84 \pm 0.05$ & \textbf{0.88} $\pm$ \textbf{0.03} & $0.83 \pm 0.03$ & $0.86 \pm 0.04$  \\ \hline
\hline \text {BNCI2014001} & CS & $0.69 \pm 0.13$ & \textbf{0.80} $\pm$ \textbf{0.10} & $0.74 \pm 0.12$ & $0.73 \pm 0.14$  \\
\hline \text {Zhou2016} & CS & $0.74 \pm 0.09$ & \textbf{0.81} $\pm$ \textbf{0.07} & $0.79 \pm 0.08$ & 0.76 $\pm$ 0.08  \\
\hline 
\end{tabular}}
\caption{Performance (Accuracy) of Right hand vs Left hand vs Feet using TANG+SVM based algorithms.}
\label{table:TANG+SVM-3class-whithinsession}
\end{table}

\subsubsection{State of the Art}
Our best approach (ACM with TANG+SVM classification and hyper-parameters found with grid search) is compared to the state of the art in Table~\ref{table:TANG+SVM-3class-whithinsession-stateart}. It brings a notable improvement (7\%-11\%) over all these methods. A detailed study of the statistical significance of these results is provided in Figures~\ref{fig:TANG+SVM-3class-whithinsession-stateart} and~\ref{fig:TANG+SVM-3class-crosssession-stateart}.

\begin{table}[!ht]
\resizebox{\linewidth}{!}{\begin{tabular}{c|c|c|c|c|c}
\text {Dataset} & Eval & \text {CSP+LDA} & \text {ACM+TANG+SVM(Grid Search)} & \text {COV+EN} & \text {FgMDM} \\
\hline \text {BNCI2014001} & WS & $0.76 \pm 0.15$ & \textbf{0.87} $\pm$ \textbf{0.09} & $0.80 \pm 0.14$ & $0.77 \pm 0.14$  \\
\hline \text {Zhou2016} & WS & $0.83 \pm 0.06$ & \textbf{0.88} $\pm$ \textbf{0.03} & $0.85 \pm 0.05$ & $0.81 \pm 0.06$  \\ \hline
\hline \text {BNCI2014001} & CS & $0.68 \pm 0.15$ & \textbf{0.80} $\pm$ \textbf{0.10} & $0.70 \pm 0.11$ & $0.69 \pm 0.13$  \\
\hline \text {Zhou2016} & CS & $0.74 \pm 0.13$ & \textbf{0.81} $\pm$ \textbf{0.07} & $0.75 \pm 0.10$ & 0.72 $\pm$ 0.10  \\
\hline 
\end{tabular}}
\caption{Performance (Accuracy) of Right hand vs Left hand vs Feet using ACM+TANG+SVM algorithm against the state of the art.}
\label{table:TANG+SVM-3class-whithinsession-stateart}
\end{table}

\section{Discussion}
\label{sec:discussion}
The numerous tests on different classification tasks, both binary and multi class, have shown that the augmented covariance approach is always superior in performance in all situations considered. This result is also confirmed through comparison with the state of the art algorithms and this is true in both the within- and cross- session frameworks. The performance improvements are naturally less important when the baseline methods are performing better initially, but this definitely shows that the augmented covariance approach brings some valuable information into the classification problem for these datasets. Naturally, the selection of hyper parameters that produces the best performance is through the use of grid search (because we used a search domain large enough). However, the modality based on non-linear dynamics concepts is also shown to produce in general better performance than standard algorithms (there are a few exceptions), with the traditional method being in general slightly better performing than the unified approach. 

\begin{figure*}[!ht]
\centering
  \subfloat[]{%
       \includegraphics[width=0.25\linewidth]{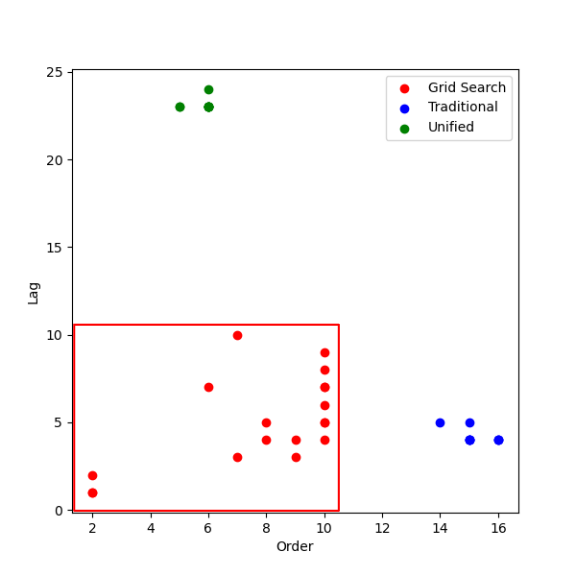}}
    \hfill
  \subfloat[]{%
        \includegraphics[width=0.31\linewidth]{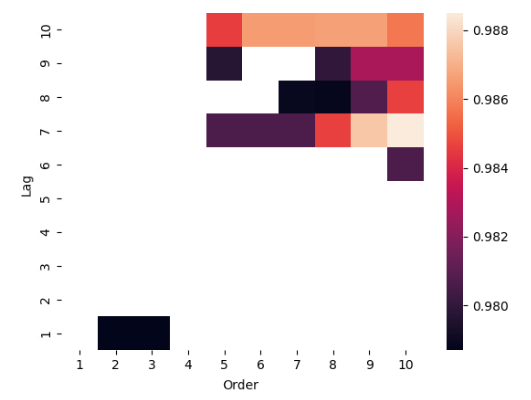}}
    \hfill
  \subfloat[]{%
        \includegraphics[width=0.29\linewidth]{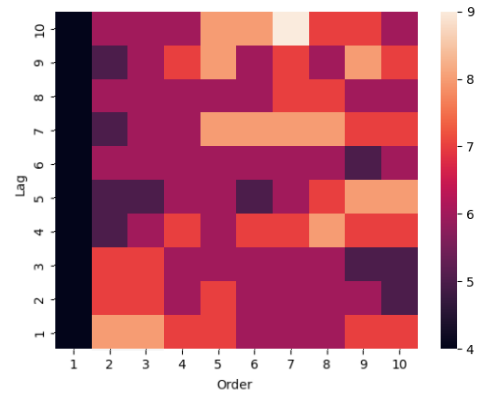}}

\caption{(a) Plot of the best hyper parameter for the ACM+MDM algorithm for each session and subject in the left hand right hand task, using the dataset BNCI2014001. Red box define the limit where we run the grid search.
(b) Heatmap that show the area that contain the hyper parameter that provide a classification score of $99\%$ of the maximum score obtained. This results is shown for Subject 1 in BNCI2014001 Session E in a right hand vs left hand task.
(c) General Heatmap that sow the maximum occurrence of best area of hyper parameter. It is obtained by counting the number of non empty boxes in plots (b) over all sessions and subjects.
}
\label{Figure_hyper}
\end{figure*}

However, the selection of the hyper parameters through grid search is time consuming in the training phase of the algorithm to search for hyper parameters, and since these values are shown to be subject dependent it is necessary to search for the correct hyper parameters for each subject. Fig.~\ref{Figure_hyper} (a) shows one example of the spread of the order and lag hyper parameters over subjects and sessions. There are 15 sets of parameters found by grid search for 18 subjects $\times$ sessions. Comparatively the unified and traditional approaches give more condensed point sets. It can be seen that they are able to correctly identify only one of the two hyper parameters, the lag and the order respectively, but not both. Furthermore, these algorithms select parameters outside the grid search domain and create extreme high dimensional ACM, that need to be regularized using a Ledoit-Wolf~\cite{ledoit-wolf:04} procedure, which is probably the reason for their lower performances. Yet, these methods based on nonlinear systems theory place an important theoretical foundation for finding efficiently proper hyper parameters which generally achieve performance improvements for both within-session and cross-session evaluation, even if far from optimal. Figure~\ref{Figure_hyper} (b) shows the set of parameters that provide an accuracy within 99\% of the best one. On this example, we see that there are numerous almost equivalent sets of hyper parameters, which shows that focusing on the actual optimal value is maybe too strict. Figure~\ref{Figure_hyper} (c) is a voting map which counts for each couple of hyper parameters when they are found in this 99\% region over subjects $\times$ sessions. It can be seen that the best vote reaches the value of 9 which is only 50\% of the maximum possible value of 18. This shows that the distribution of optimal hyper parameters is not unimodal and that it will be difficult to find a fixed value that would work in every situation.
Other classification tasks and datasets behave similarly.

While our approach has been shown to achieve improvement in accuracy performance in many BCI applications in the Motor Imagery task, it has some limitations. First of all, like other methods, it relies on the assumption of the validity of the sample covariance estimator. When the used time windows are way greater than the dimension of covariance matrix, this assumption remains valid. But this dimension increases quickly with higher AR orders. In that case, the matrix is still a symmetric positive matrix but is not definite and Riemannian metrics can no longer be used. To solve this issue, it is possible to adopt shrinkage or other methods to estimate covariance~\cite{ledoit-wolf:04,bartz-muller:14,sabbagh-ablin-etal:19}. 

\begin{figure}[!ht]
 \centering
 \includegraphics[width=0.6\linewidth]{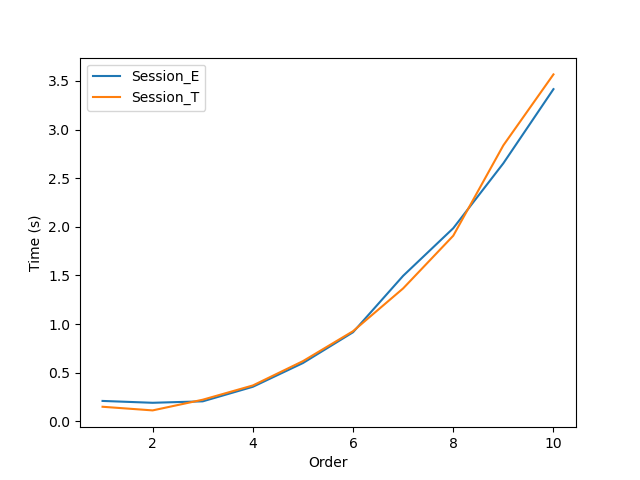}
 \caption{Total evaluation computational time for the MOABB within-session evaluation for subject 1 of BNCI2014001 with the task right hand vs feet. Computational times at order 1 correspond to standard covariance. Similar behaviours occur in the other evaluation procedures, subjects, tasks and datasets.}
  \label{fig:figure_Time}
\end{figure}

Finally, note that the augmented method increases computational time because of the increased dimensionality of the augmented covariance compared to the standard covariance~\cite{you-park:22}, as shown in Figure~\ref{fig:figure_Time}. Yet, even in the higher orders, the total time divided by the number of classified windows shows that the augmented approach is compatible with real time computation.

\section{Conclusion}
\label{sec:conclusion}
In this research, we explored the use of the augmented covariance matrix for the classification of Motor Imagery task in EEG-BCI. 
This methodology amounts to integrate temporal information into the normal spatial covariance, and improves quite dramatically classification results over the state of the art when used with Riemann distance based classification algorithms. There remain some issues with the definiteness of the augmented covariance matrix, which stems from the augmentation of the dimensionality of the covariance matrix. Another issue is the choice of the hyper parameters which is costly. Yet, we show that methods inspired by the Takens' theorem can provide some reasonable parameter values at the cost of a decrease of the improvements (but with improvements still in most cases).

The current study focused on offline classification. Our next direction will be to extend the approach for online classification. To do this, it will be necessary to test the algorithms on narrower epoch windows and integrate them with a sliding window approach.  Another important development of the augmented approach would be to see whether it is also efficient with smaller training dataset contexts i.e. with either a low number of electrodes or with fewer training samples. This would provide an important usability improvement for BCIs, potentially reducing the setup and/or the calibration times. We have already seen that performance over the state of the art is improved in the case of BNCI2014004, which possesses only 3 electrodes, but this needs to be generalized to the other datasets and to a smaller set of training samples. 

Another work direction is to analyse more in depth the distribution of hyper parameters with the objective of identifying a methodology to make the classifier parameter free (e.g. by using ensemble classifiers methodology).

\section*{Acknowledgment}
This work has been partially financed by a EUR DS4H/Neuromod fellowship. The authors are grateful to the OPAL infrastructure from Université Côte d'Azur for providing resources and support. 

\section*{Data and Code Availability}
The codes used to produce the results of this study are publicly available in this Github repository: 
\url{https://github.com/carraraig/Augmented_Covariance_BCI}.

\printbibliography

\appendix
\begin{figure*}[ht]  
    \centering
    \centering
     \subfloat[]{%
            \includegraphics[width=0.45\linewidth]{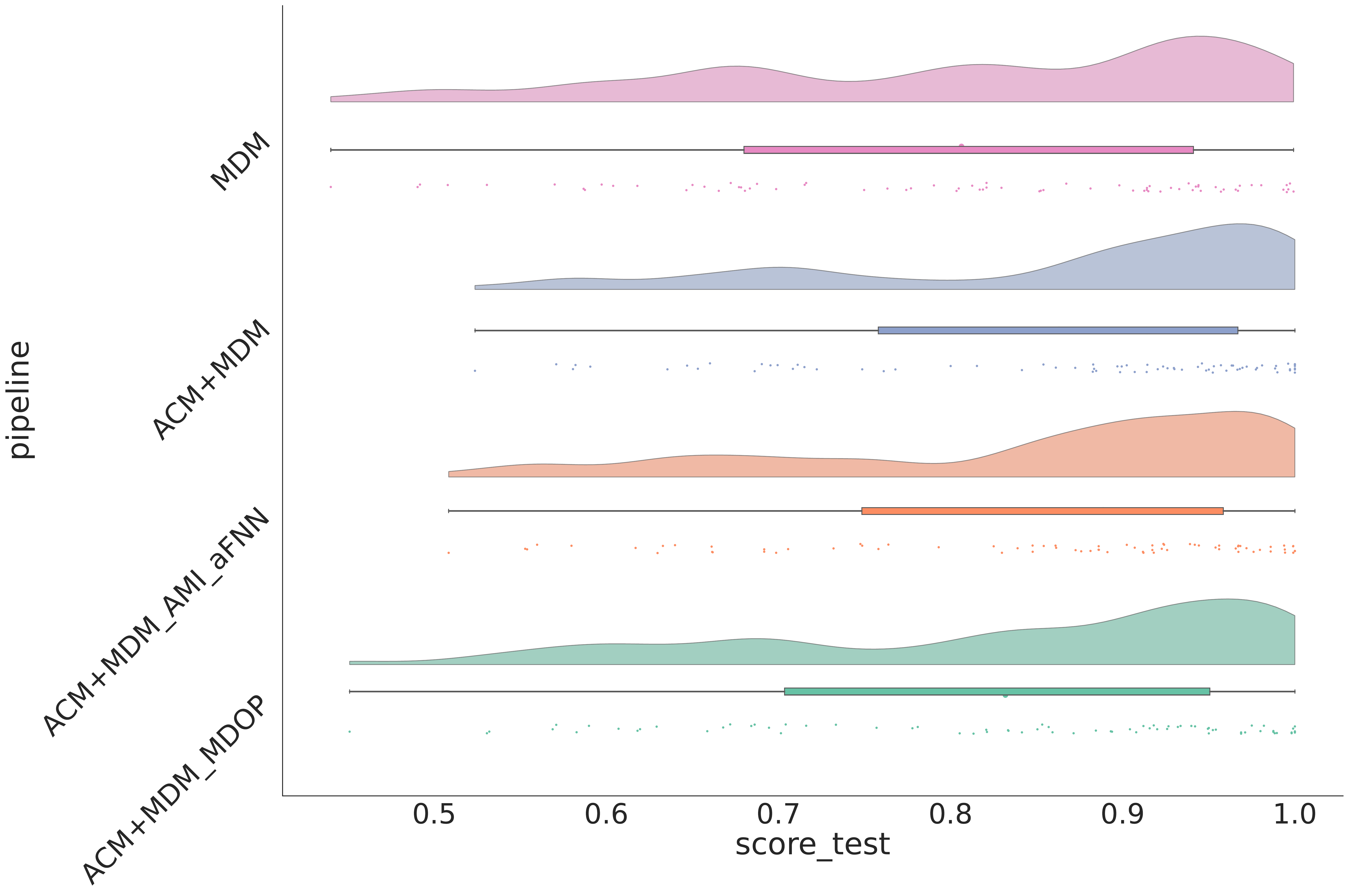}}
            \hfill
     \subfloat[]{%
            \includegraphics[width=0.45\linewidth]{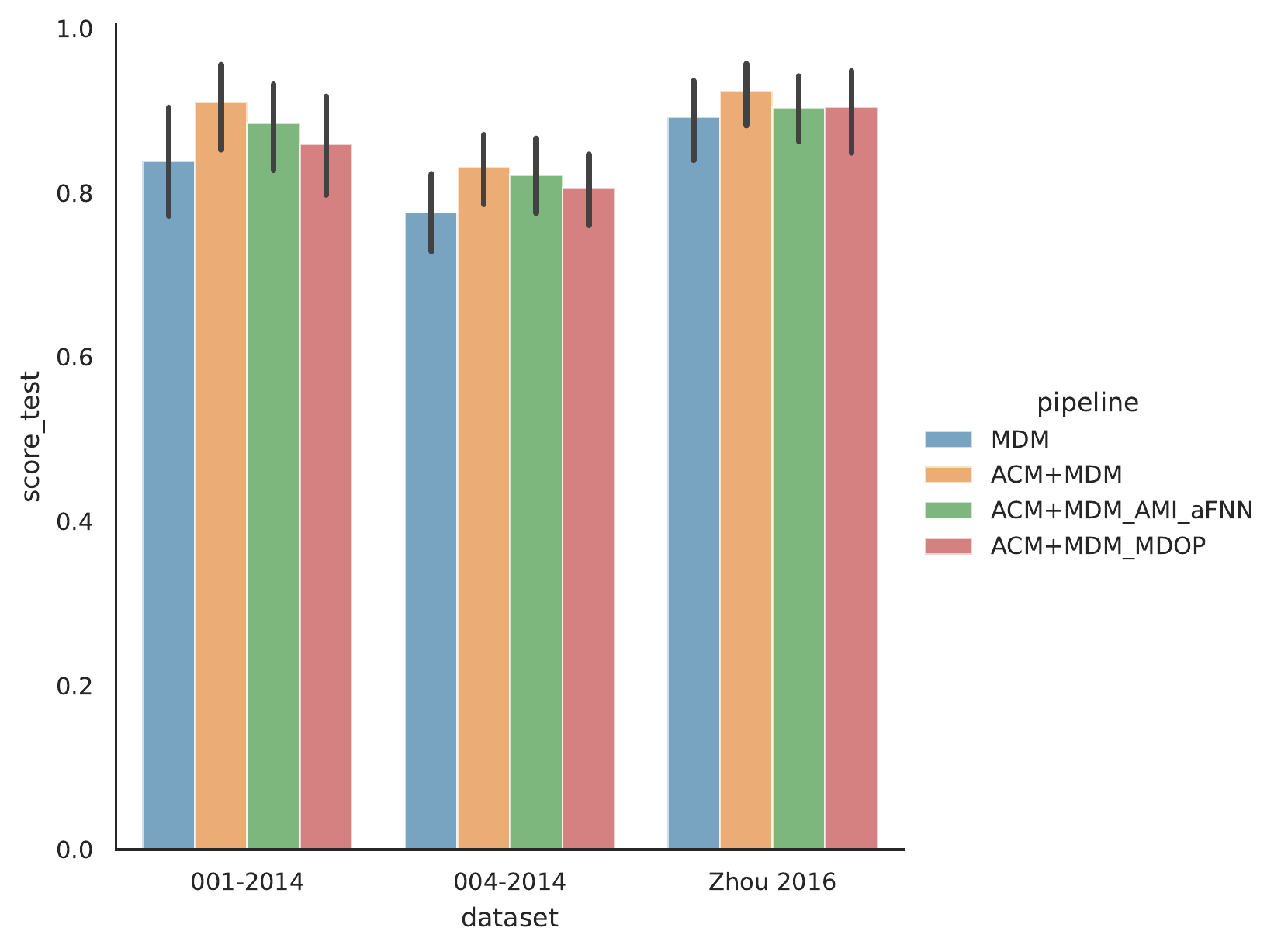}}
    \\
    \subfloat[]{%
        \includegraphics[width=0.5\linewidth]{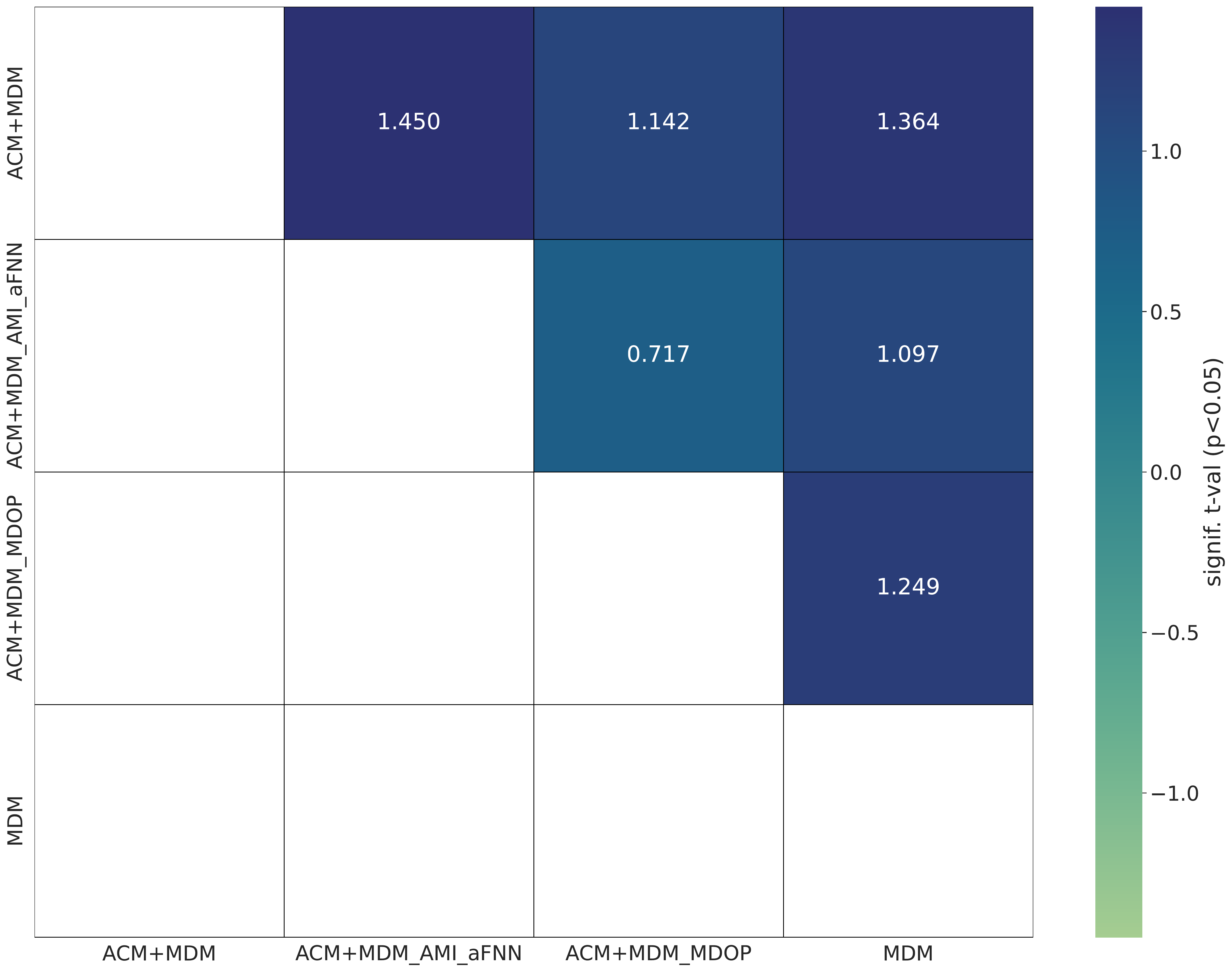}}
        \\
   \subfloat[]{%
            \includegraphics[width=0.30\linewidth]{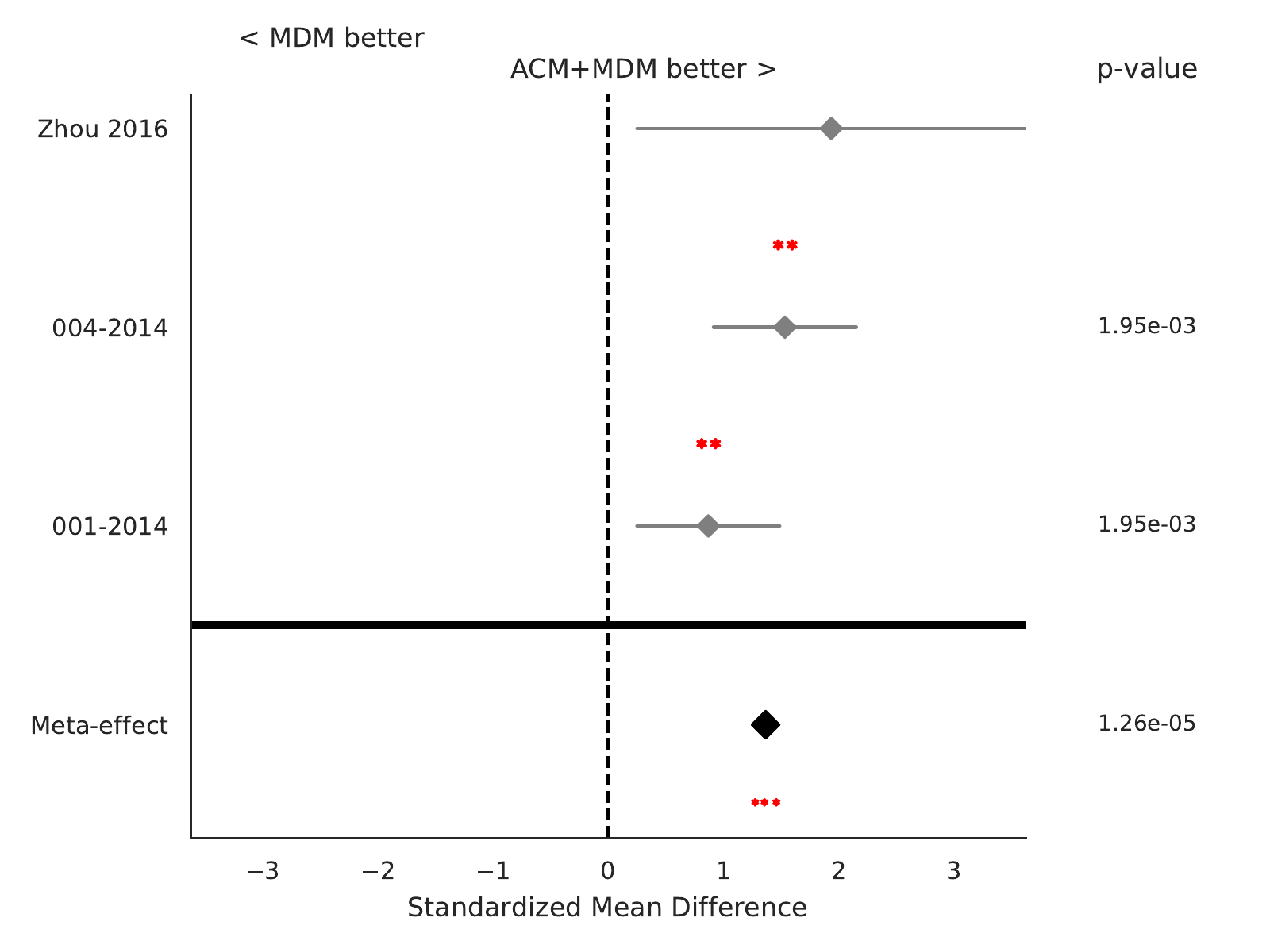}}
            \hfill
   \subfloat[]{%
            \includegraphics[width=0.30\linewidth]{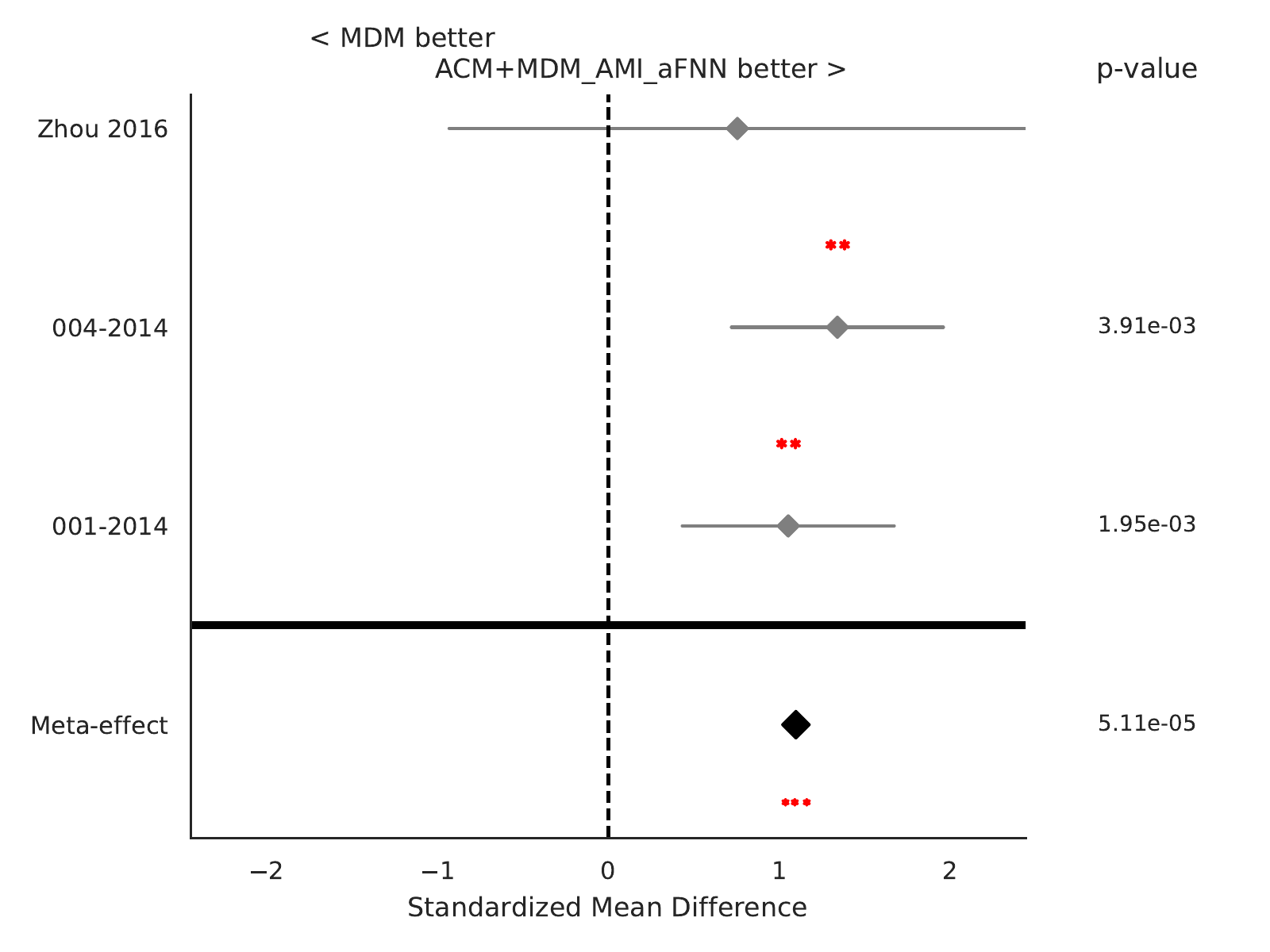}}
            \hfill
   \subfloat[]{%
            \includegraphics[width=0.30\linewidth]{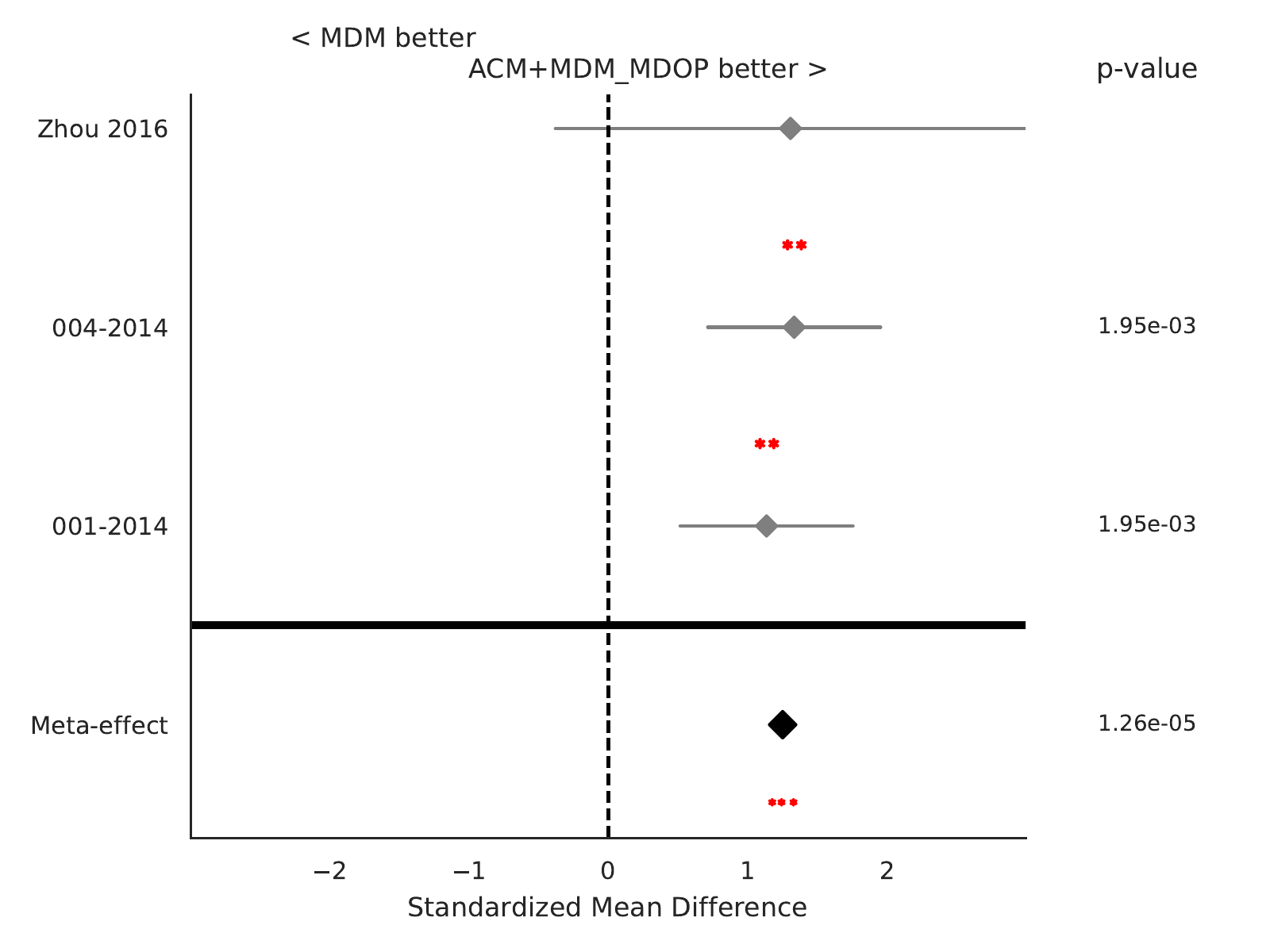}}
            \hfill

    \caption{Result for right hand vs left hand classification using the MDM algorithm, using withing-session evaluation. (a) show the rain clouds plots for each pipeline, showing the distribution of the score of every subject. (b) show the bar plot of the score withe the error of the different pipeline and for every dataset considered. (c) show the meta analysis of the different pipeline considered. This plot the significance that the algorithm on the y-axis is better than the one on the x-axis. The color represents the significance level of the difference of accuracy, in terms of t-values, and we show only the significant interactions ($p < 0.05$). (d) (e) (f) show the meta analysis of the standard MDM algorithm against the augmented covariance method with the selection of the hyper-parameter based on grid search, traditional and unified Takens approach respectively. We show the standardized mean differences, while p-values are computed as one-tailed Wilcoxon signed-rank test for the hypothesis given as title of the plot and the gray bar  denote $95\%$ interval. Here, * stands for $p < 0.05$, ** for $p < 0.01$, and *** for $p < 0.001$.
    }
    \label{fig:MDM-rhlh-whithinsession}
\end{figure*}

\begin{figure*}[ht]  
    \centering
    \centering
     \subfloat[]{%
            \includegraphics[width=0.45\linewidth]{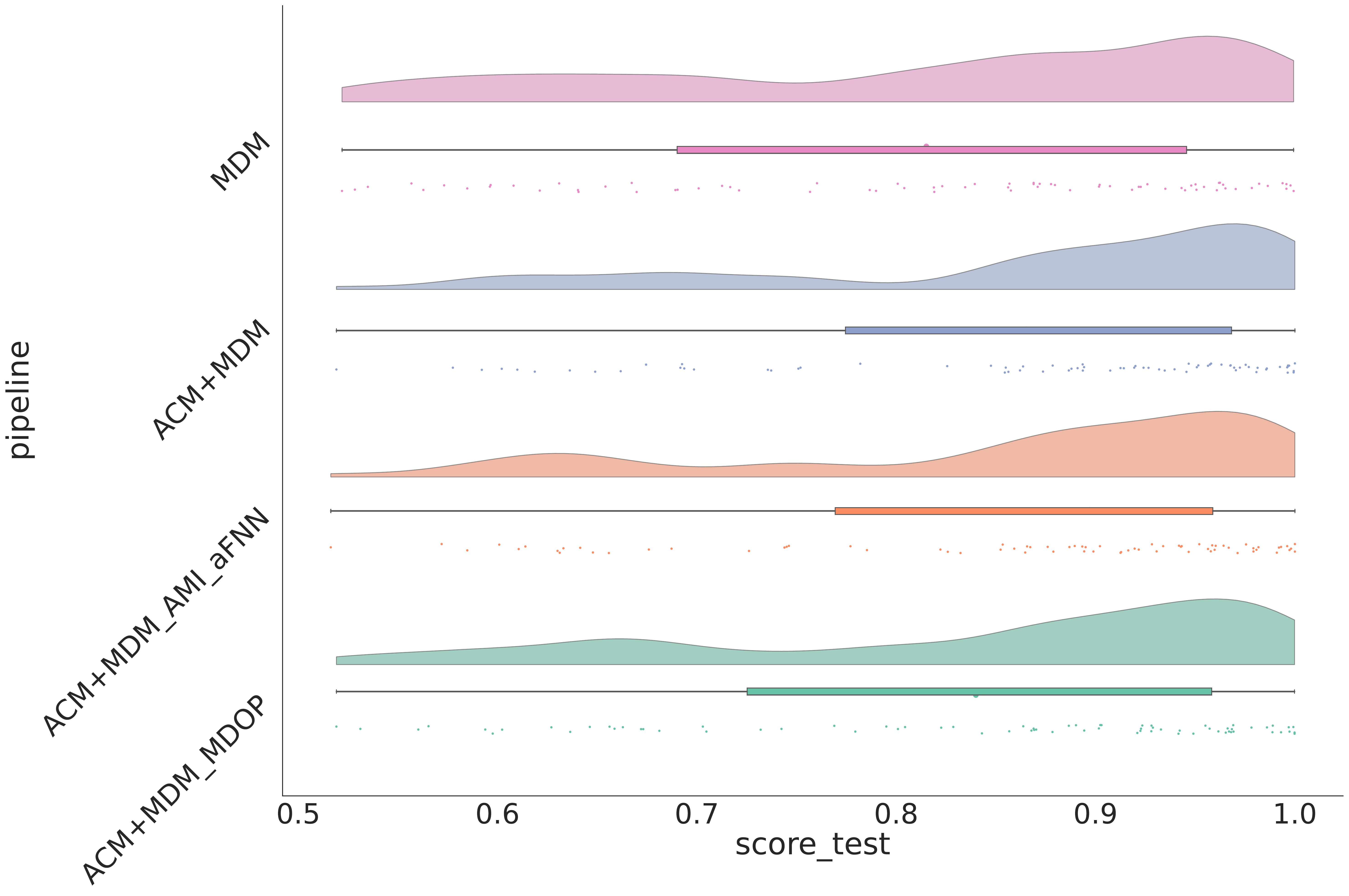}}
            \hfill
     \subfloat[]{%
            \includegraphics[width=0.45\linewidth]{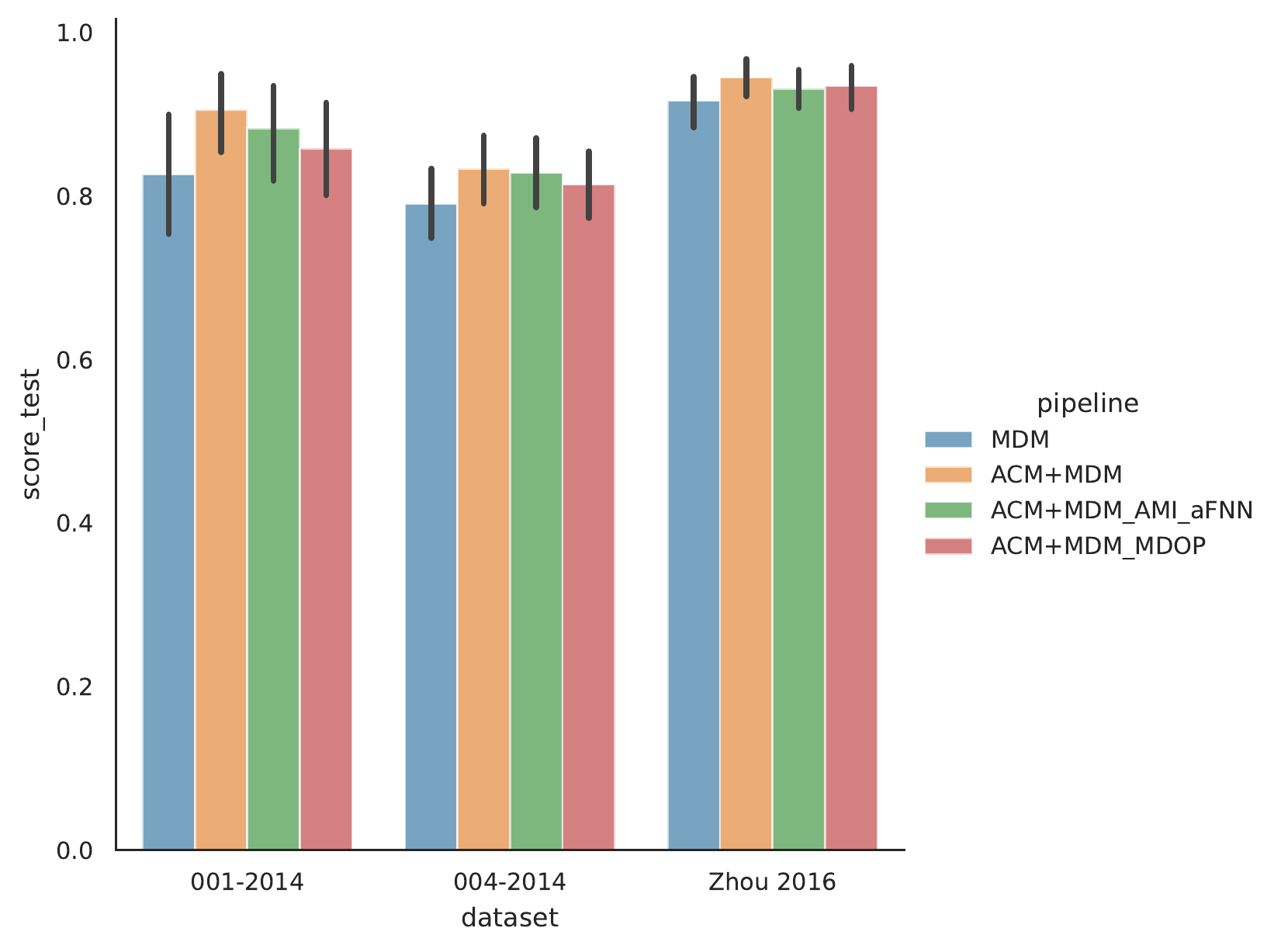}}
    \\
    \subfloat[]{%
        \includegraphics[width=0.5\linewidth]{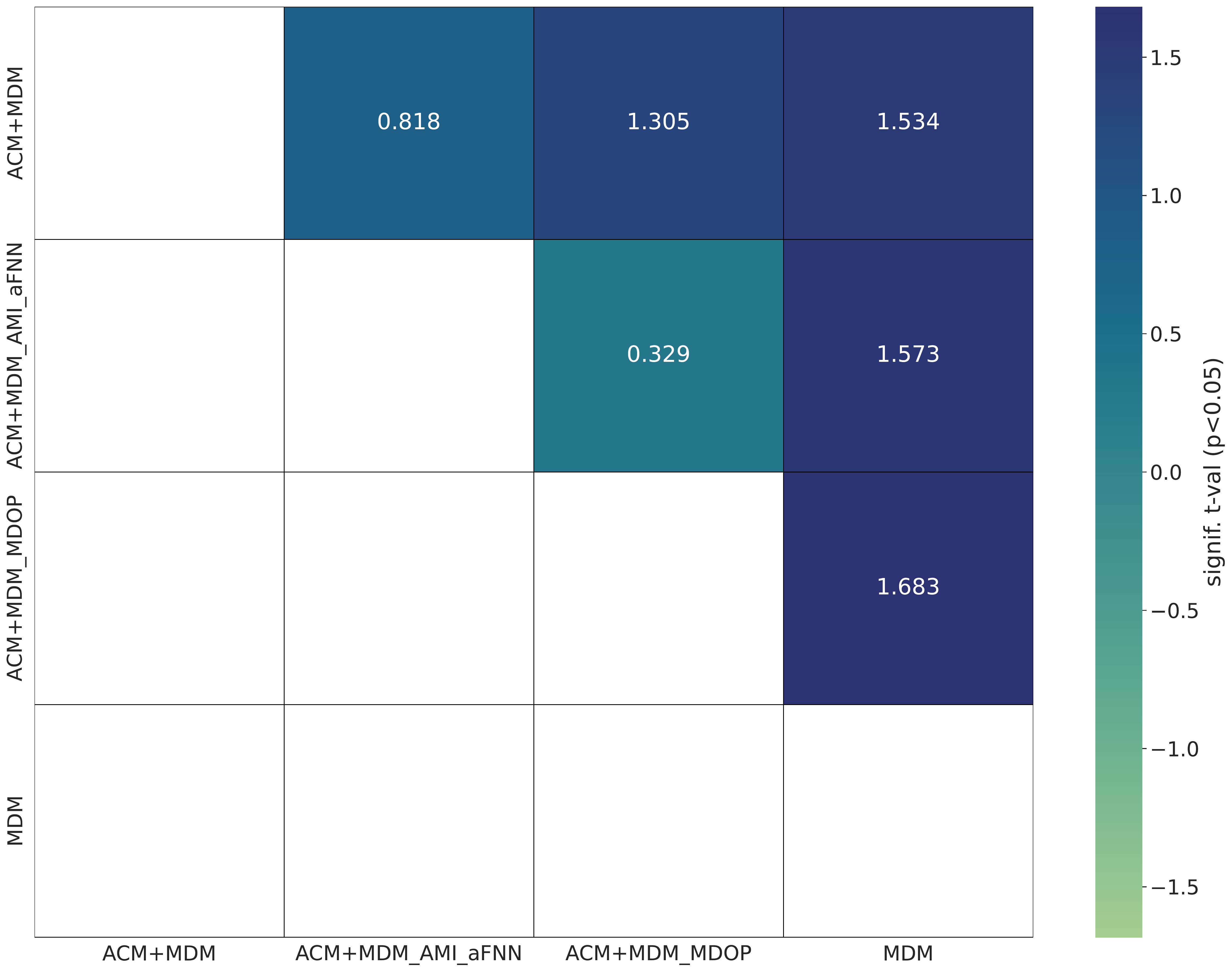}}
        \\
   \subfloat[]{%
            \includegraphics[width=0.30\linewidth]{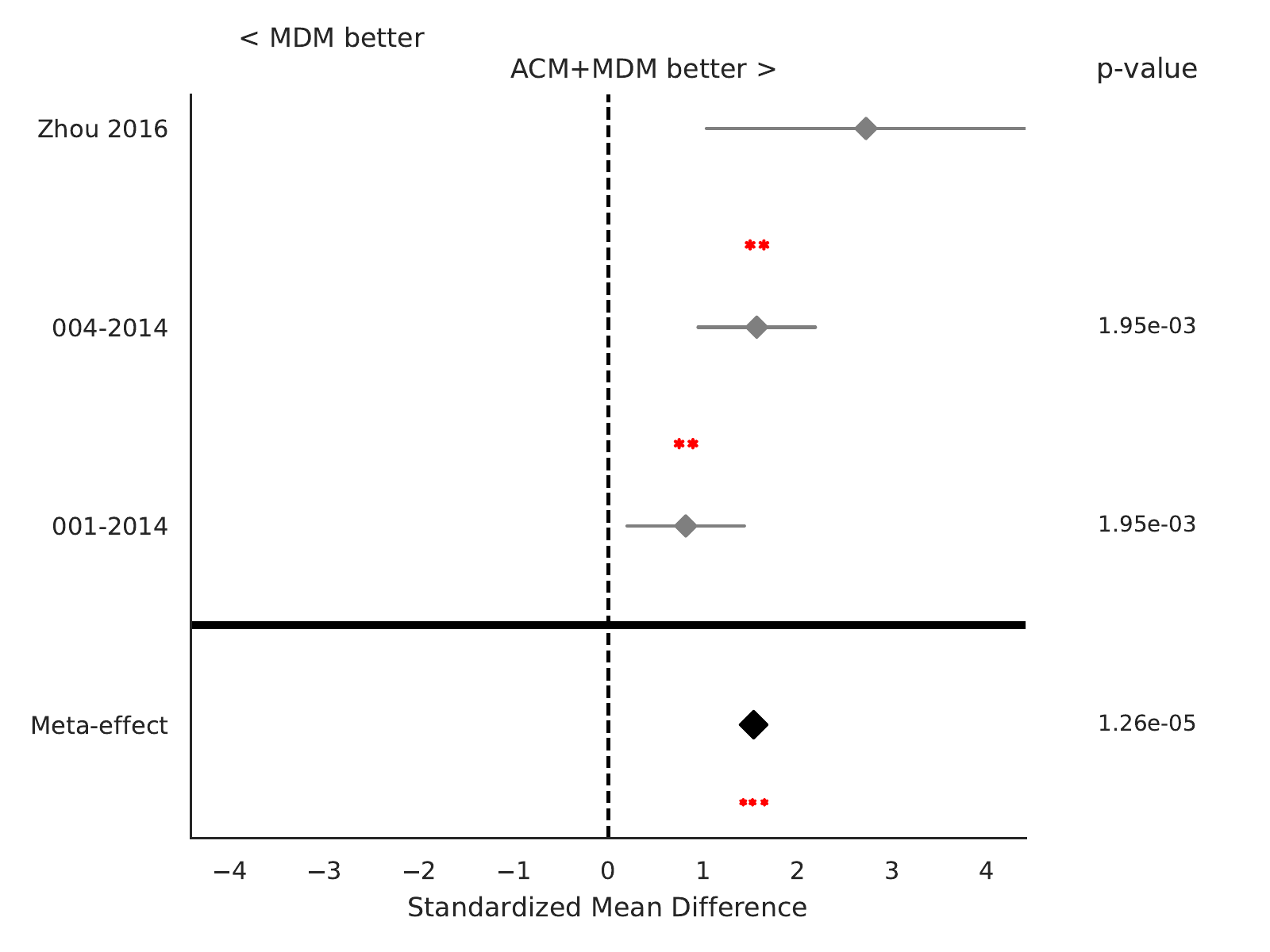}}
            \hfill
   \subfloat[]{%
            \includegraphics[width=0.30\linewidth]{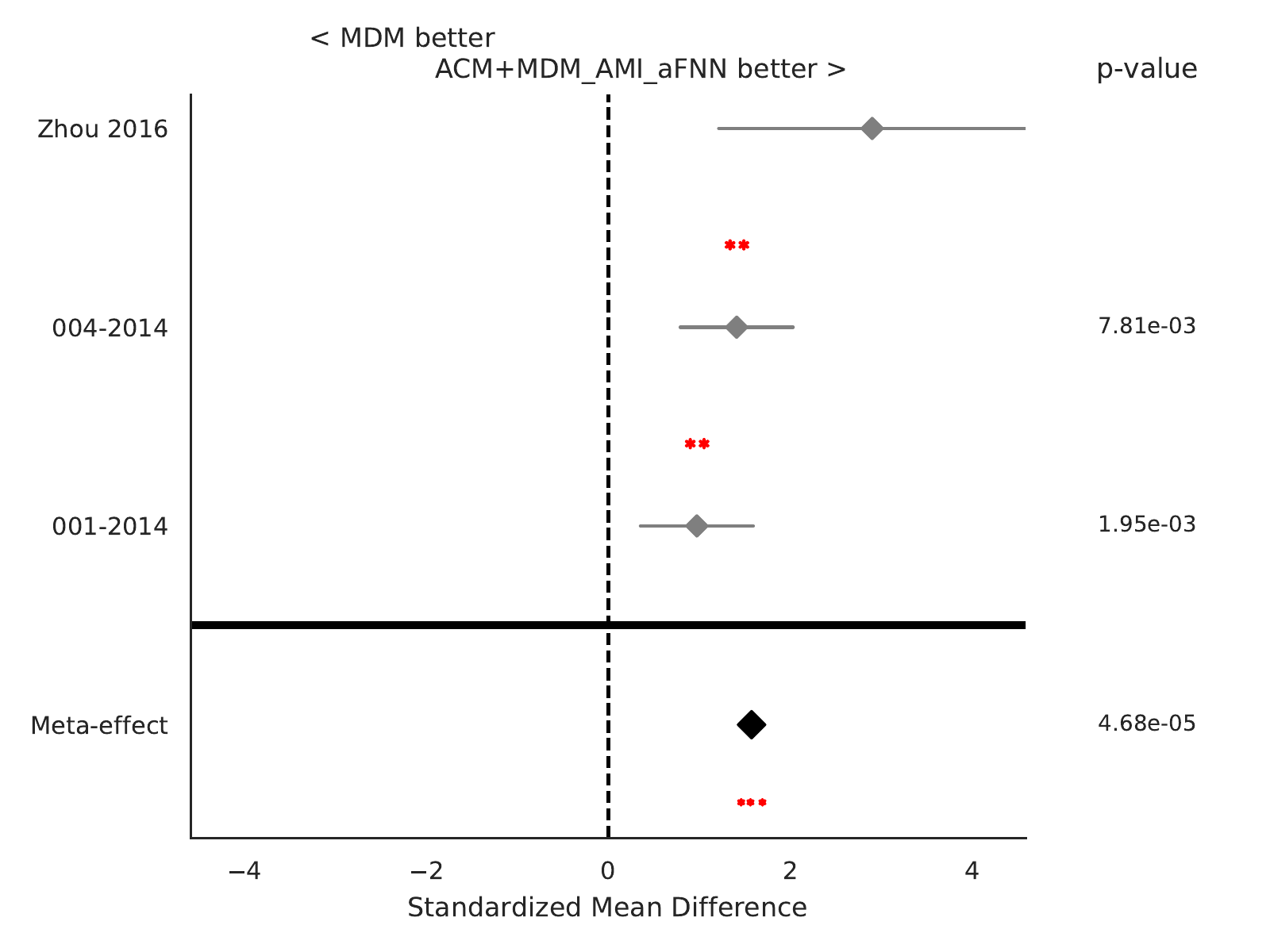}}
            \hfill
   \subfloat[]{%
            \includegraphics[width=0.30\linewidth]{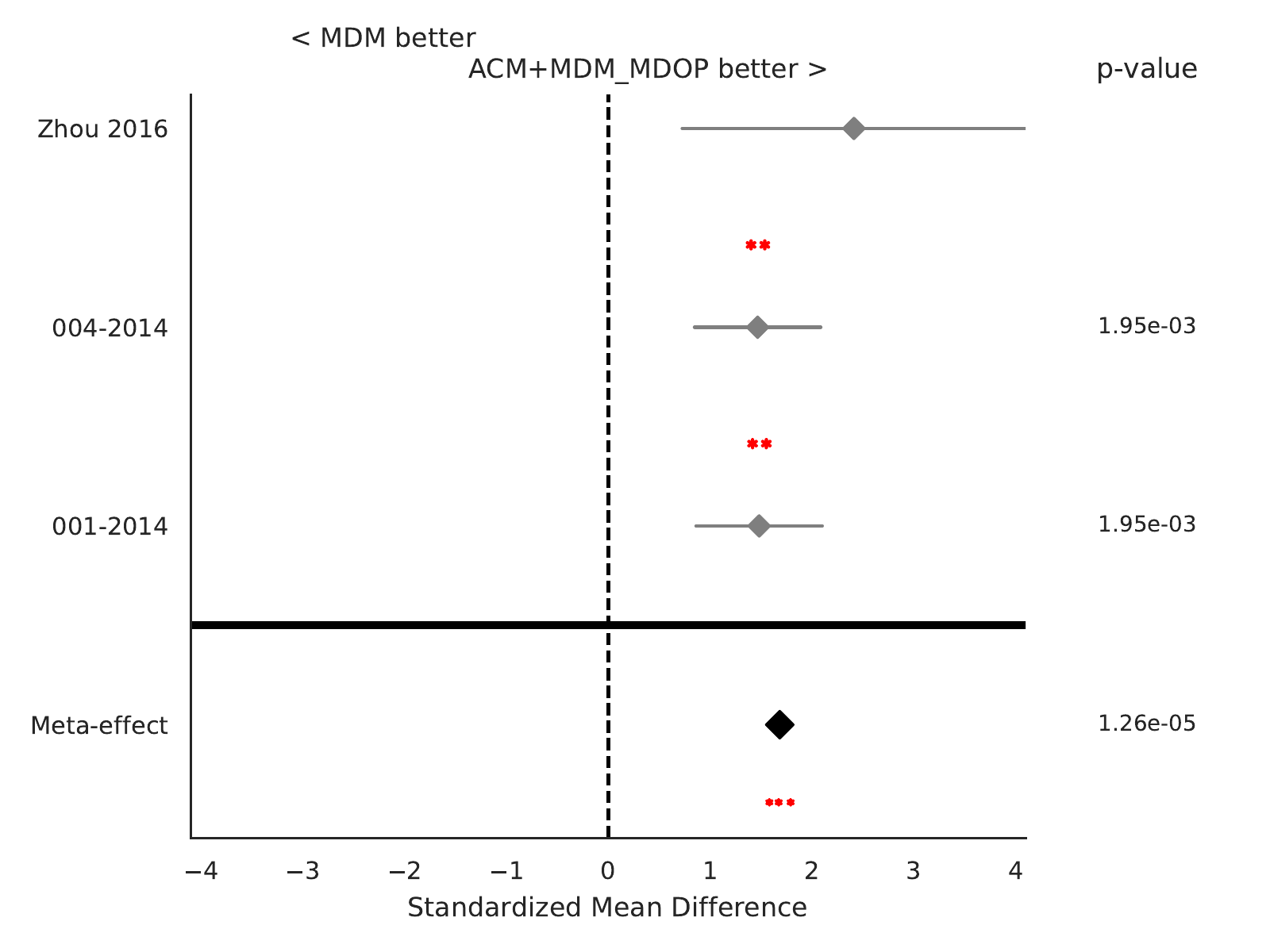}}
            \hfill

    \caption{Result for right hand vs left hand classification using the MDM algorithm, using cross-session evaluation. (a) show the rain clouds plots for each pipeline, showing the distribution of the score of every subject. (b) show the bar plot of the score withe the error of the different pipeline and for every dataset considered. (c) show the meta analysis of the different pipeline considered. This plot the significance that the algorithm on the y-axis is better than the one on the x-axis. The color represents the significance level of the difference of accuracy, in terms of t-values, and we show only the significant interactions ($p < 0.05$). (d) (e) (f) show the meta analysis of the standard MDM algorithm against the augmented covariance method with the selection of the hyper-parameter based on grid search, traditional and unified Takens approach respectively. We show the standardized mean differences, while p-values are computed as one-tailed Wilcoxon signed-rank test for the hypothesis given as title of the plot and the gray bar  denote $95\%$ interval. Here, * stands for $p < 0.05$, ** for $p < 0.01$, and *** for $p < 0.001$.
    }
    \label{fig:MDM-rhlh-crosssession}
\end{figure*}

\begin{figure*}[ht]  
    \centering
    \centering
     \subfloat[]{%
            \includegraphics[width=0.45\linewidth]{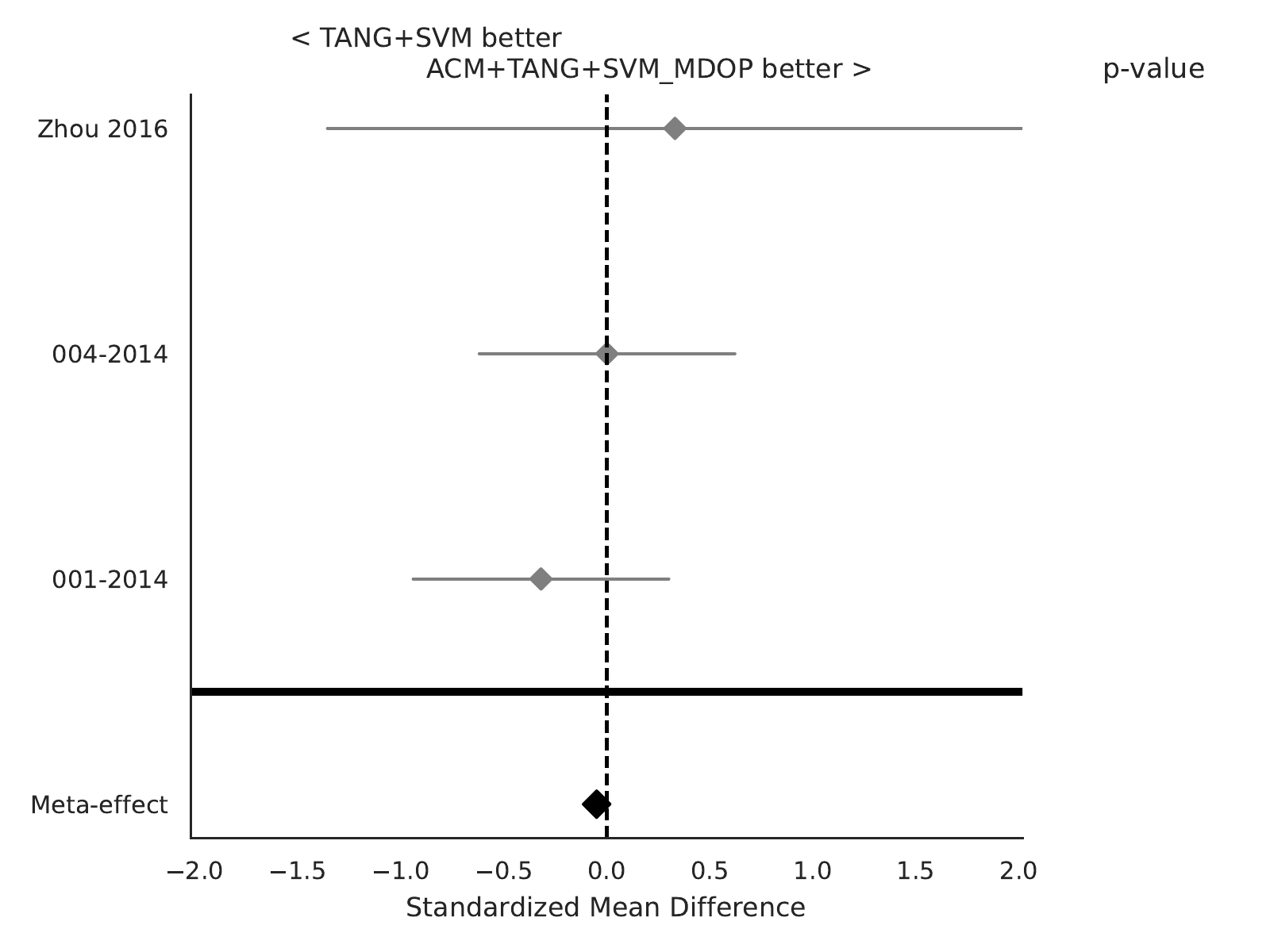}}
            \hfill
     \subfloat[]{%
            \includegraphics[width=0.45\linewidth]{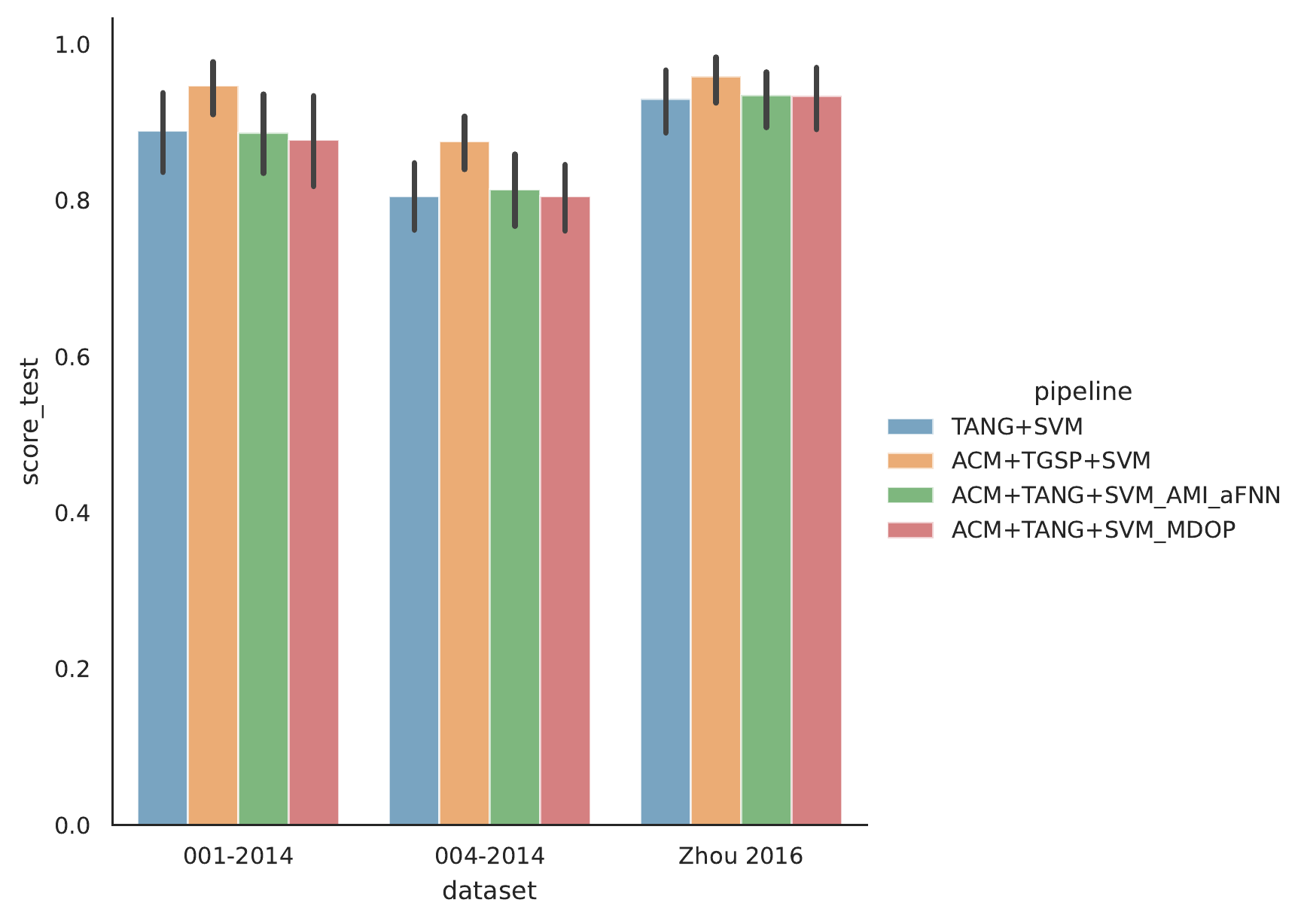}}
    \\
    \subfloat[]{%
        \includegraphics[width=0.5\linewidth]{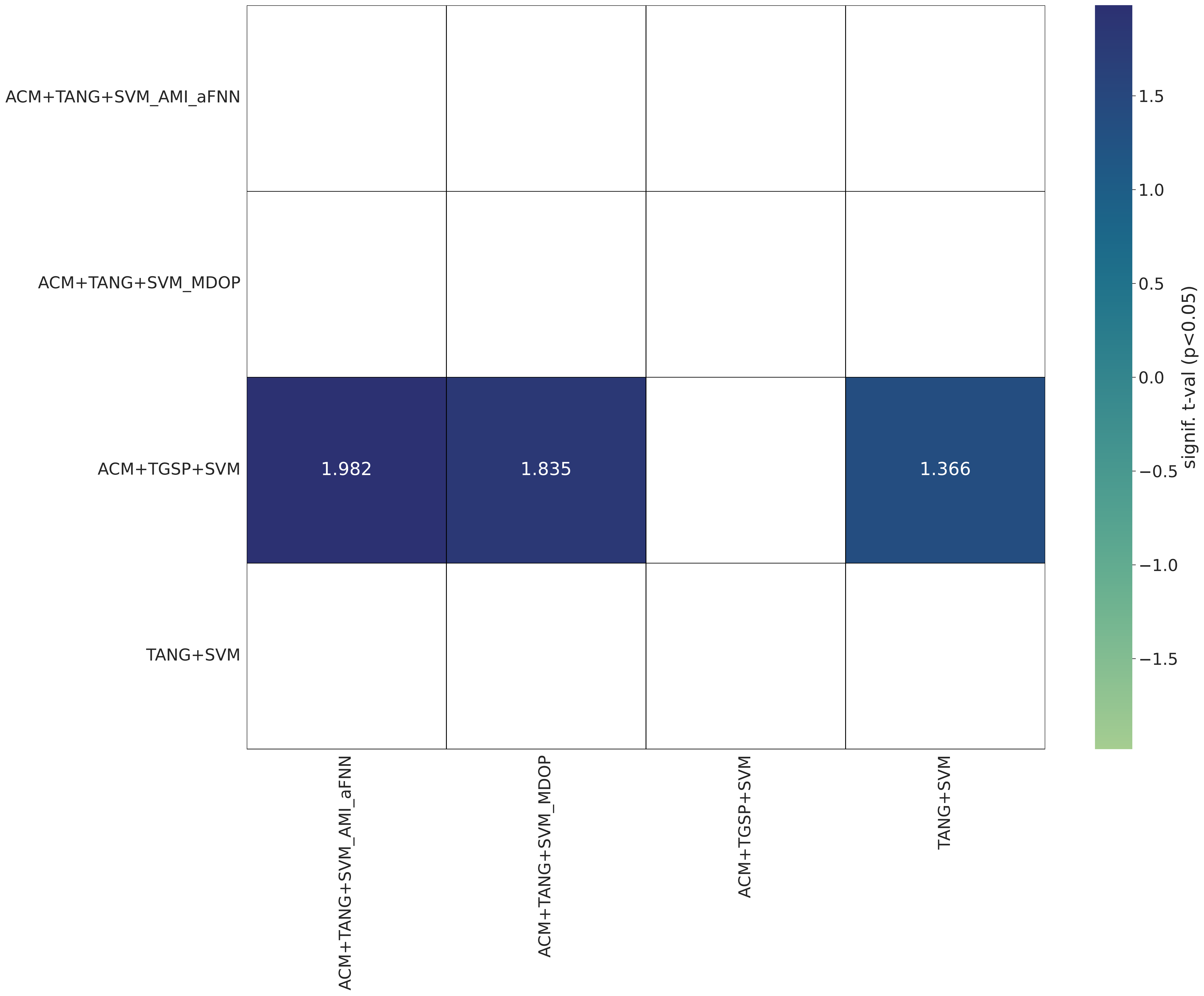}}
        \\
   \subfloat[]{%
            \includegraphics[width=0.30\linewidth]{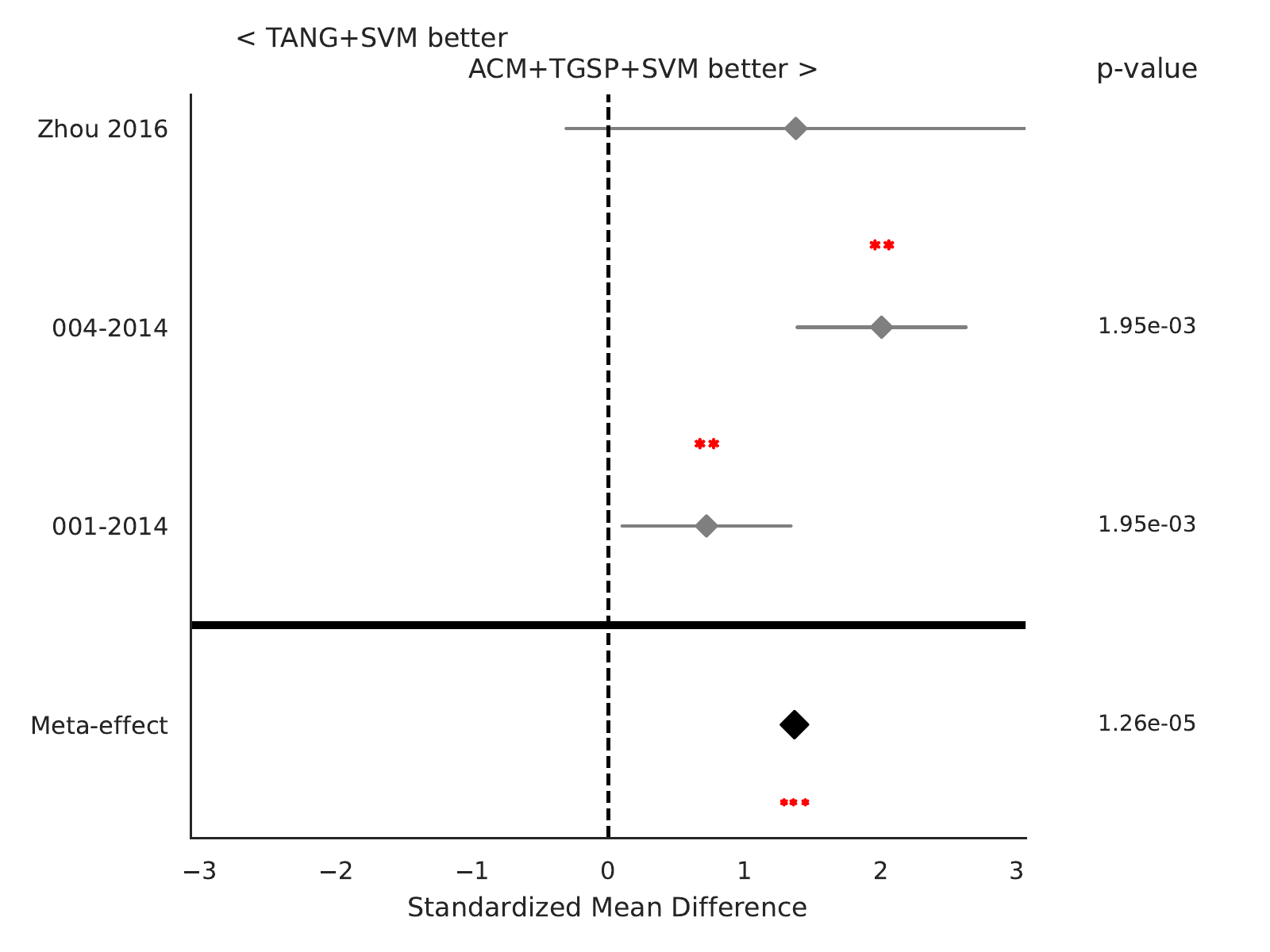}}
            \hfill
   \subfloat[]{%
            \includegraphics[width=0.30\linewidth]{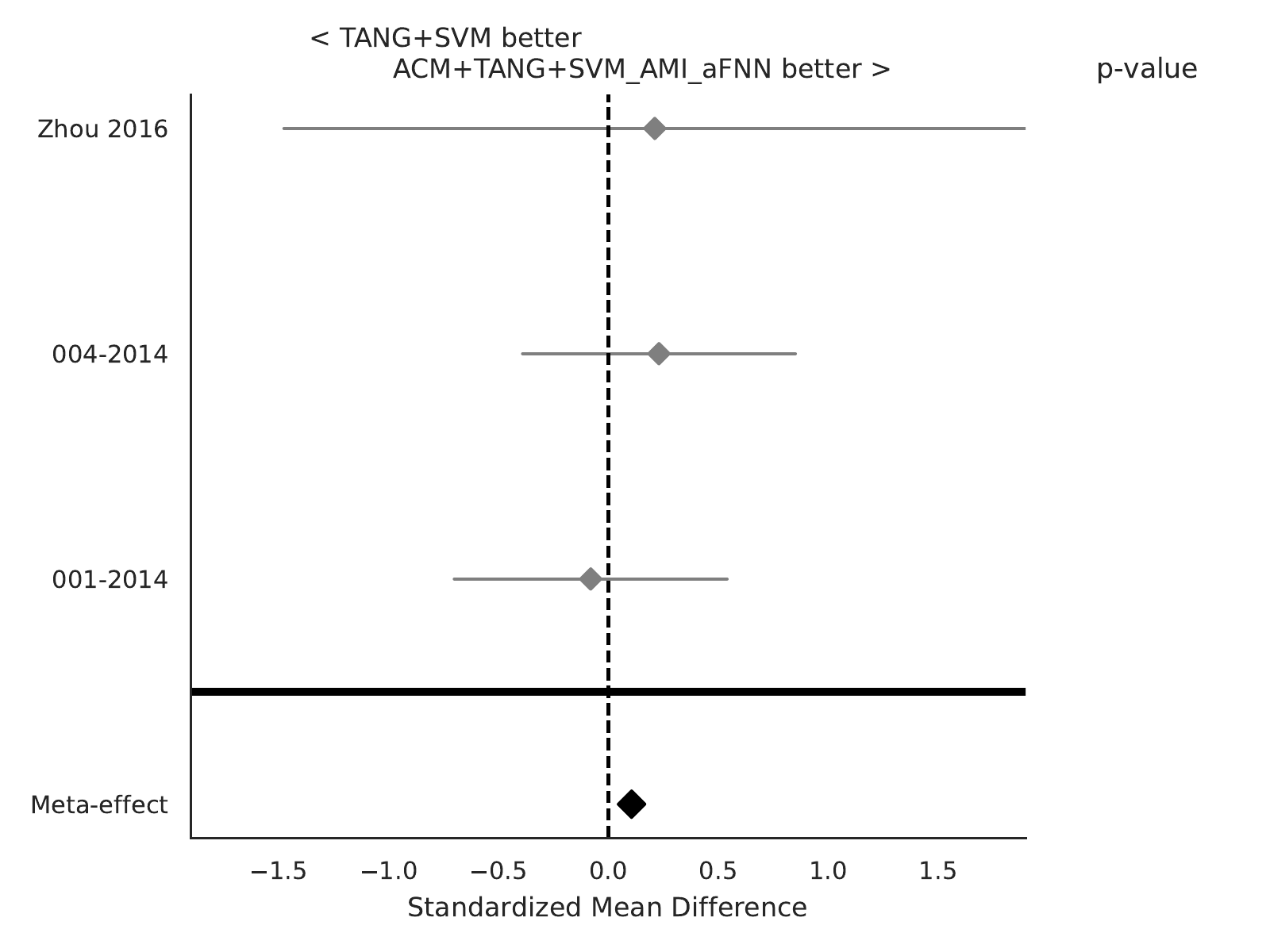}}
            \hfill
   \subfloat[]{%
            \includegraphics[width=0.30\linewidth]{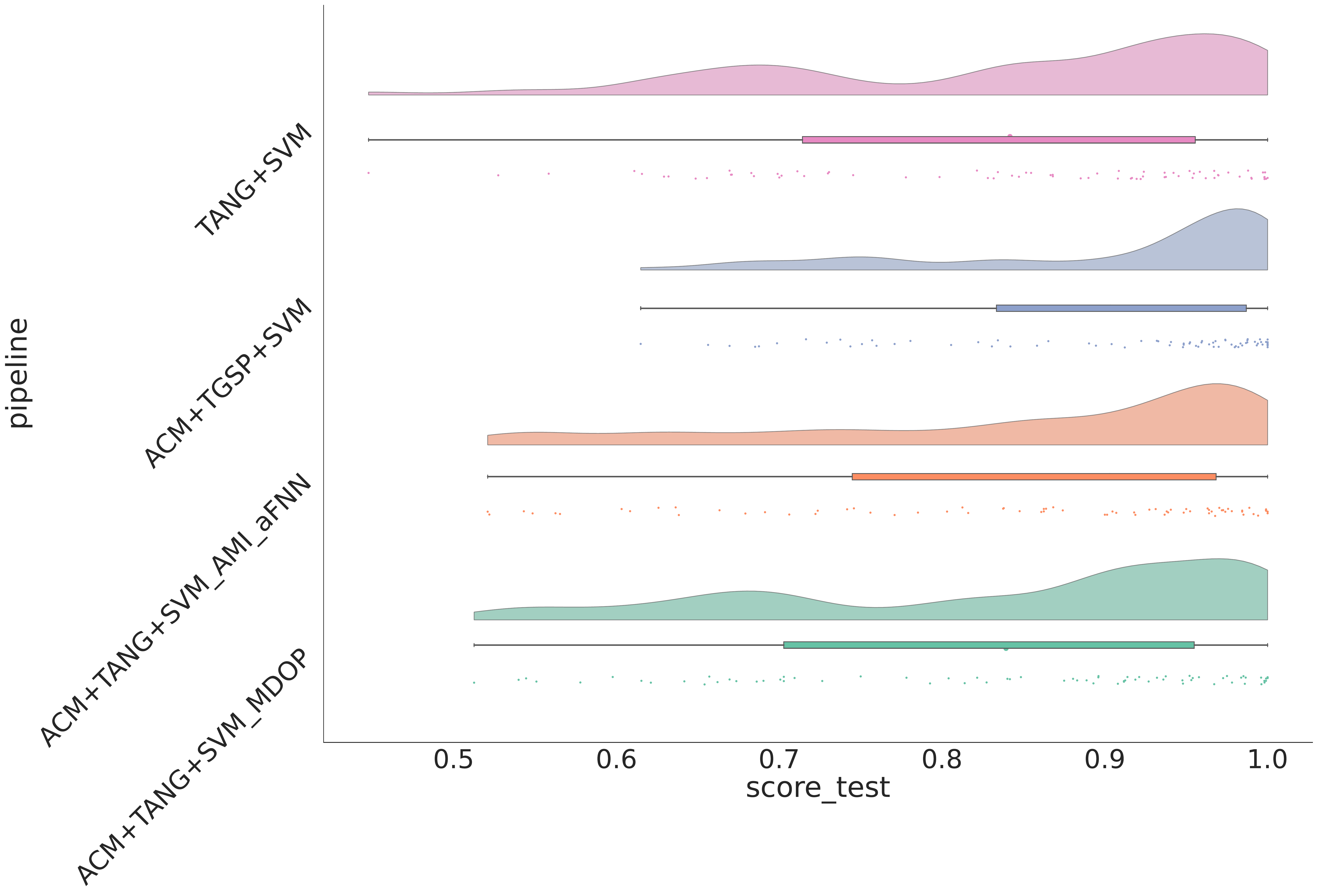}}
            \hfill

    \caption{Result for right hand vs left hand classification using the TANG algorithm, using withing-session evaluation. (a) show the rain clouds plots for each pipeline, showing the distribution of the score of every subject. (b) show the bar plot of the score withe the error of the different pipeline and for every dataset considered. (c) show the meta analysis of the different pipeline considered. This plot the significance that the algorithm on the y-axis is better than the one on the x-axis. The color represents the significance level of the difference of accuracy, in terms of t-values, and we show only the significant interactions ($p < 0.05$). (d) (e) (f) show the meta analysis of the standard TANG algorithm against the augmented covariance method with the selection of the hyper-parameter based on grid search, traditional and unified Takens approach respectively. We show the standardized mean differences, while p-values are computed as one-tailed Wilcoxon signed-rank test for the hypothesis given as title of the plot and the gray bar  denote $95\%$ interval. Here, * stands for $p < 0.05$, ** for $p < 0.01$, and *** for $p < 0.001$.
    }
    \label{fig:TANG+SVM-rhlh-whithinsession}
\end{figure*}

\begin{figure*}[ht]  
    \centering
    \centering
     \subfloat[]{%
            \includegraphics[width=0.45\linewidth]{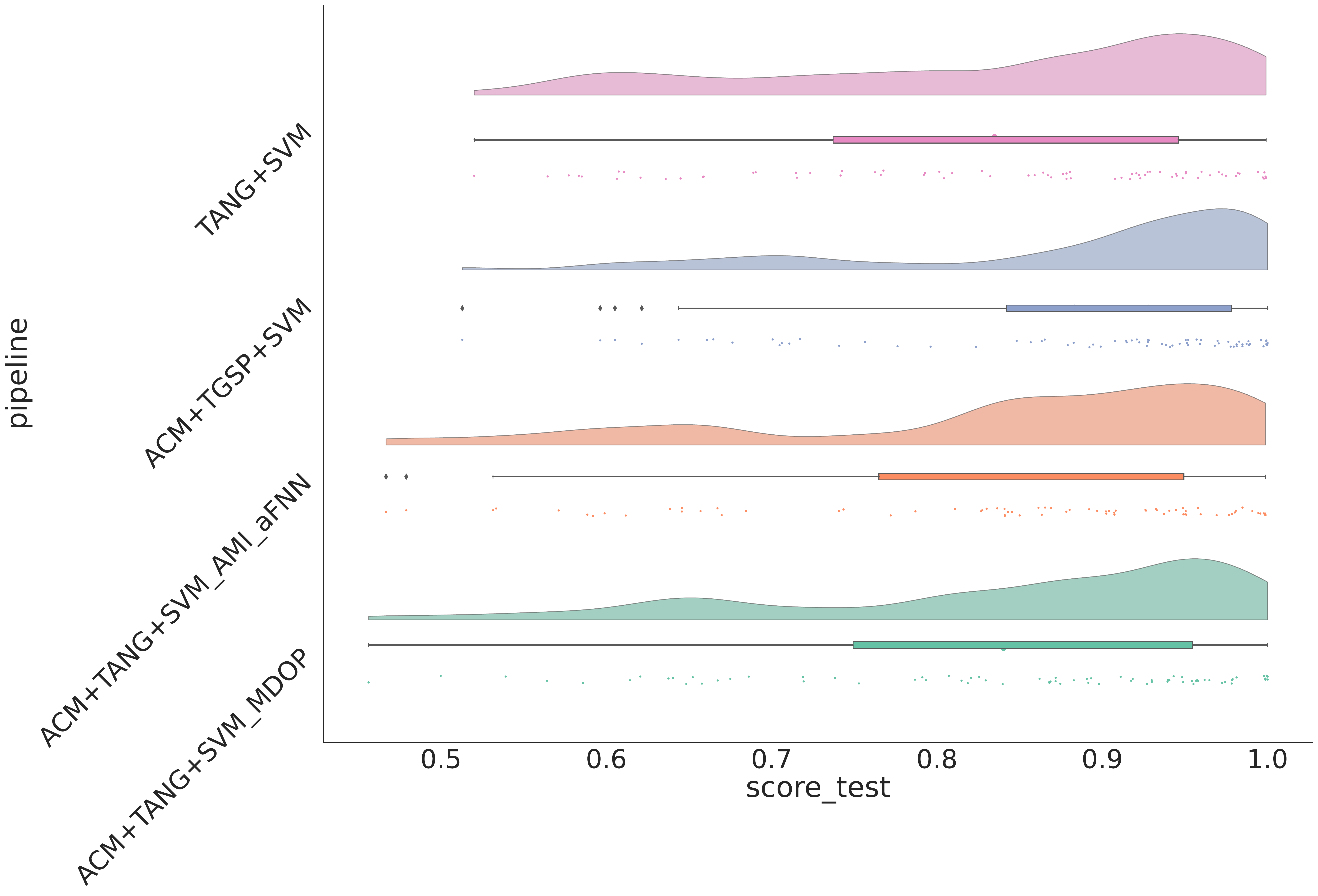}}
            \hfill
     \subfloat[]{%
            \includegraphics[width=0.45\linewidth]{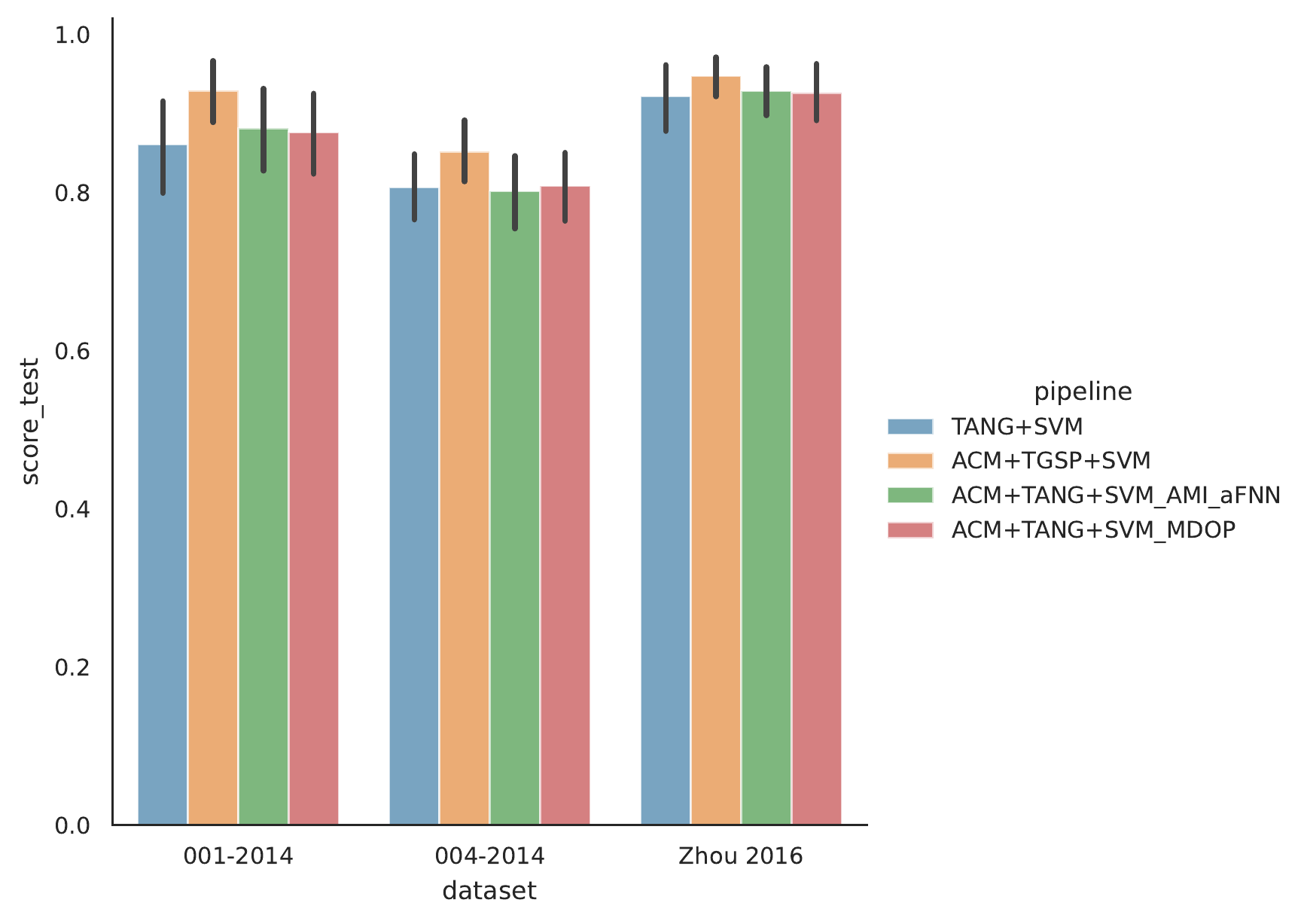}}
    \\
    \subfloat[]{%
        \includegraphics[width=0.5\linewidth]{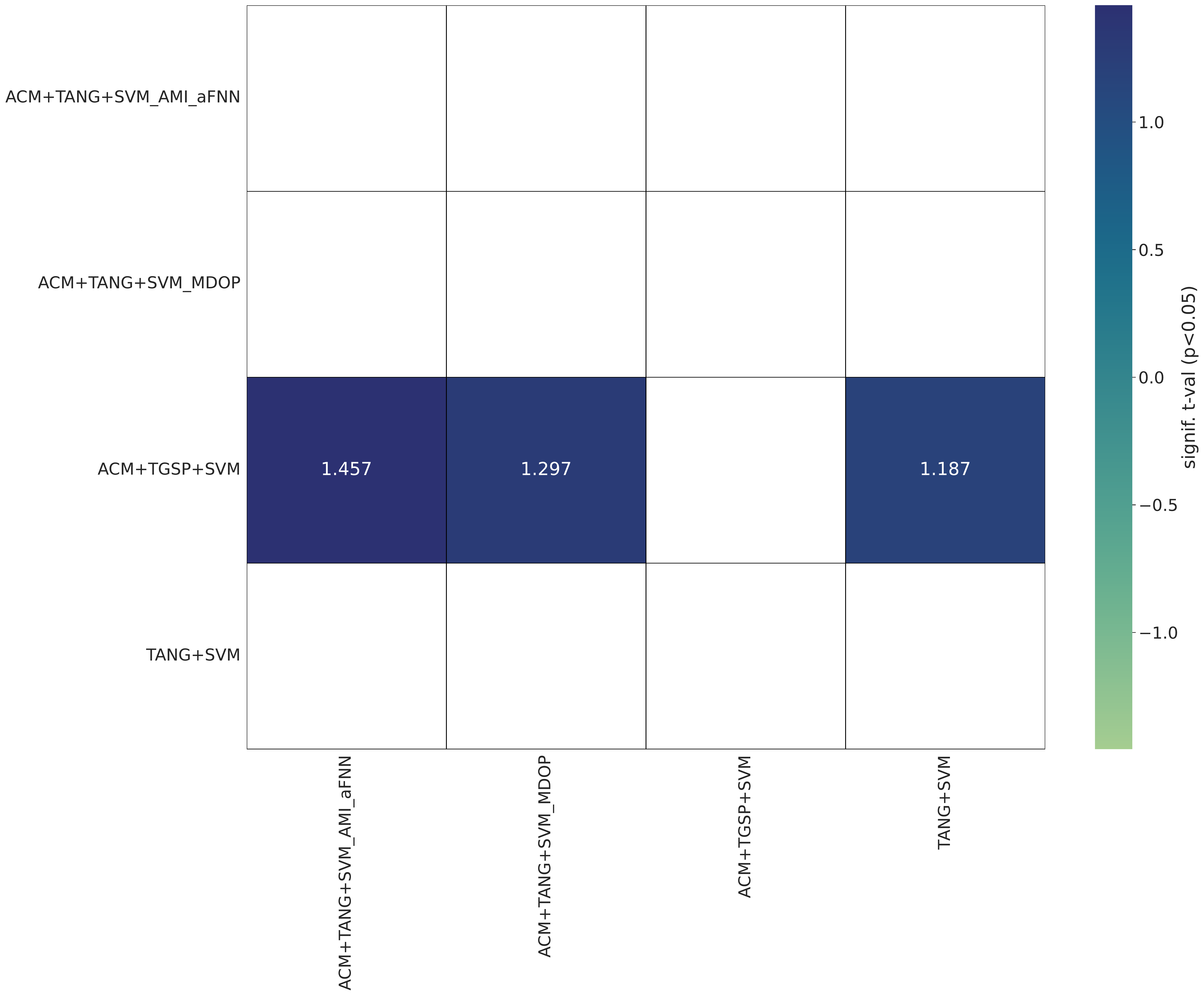}}
        \\
   \subfloat[]{%
            \includegraphics[width=0.30\linewidth]{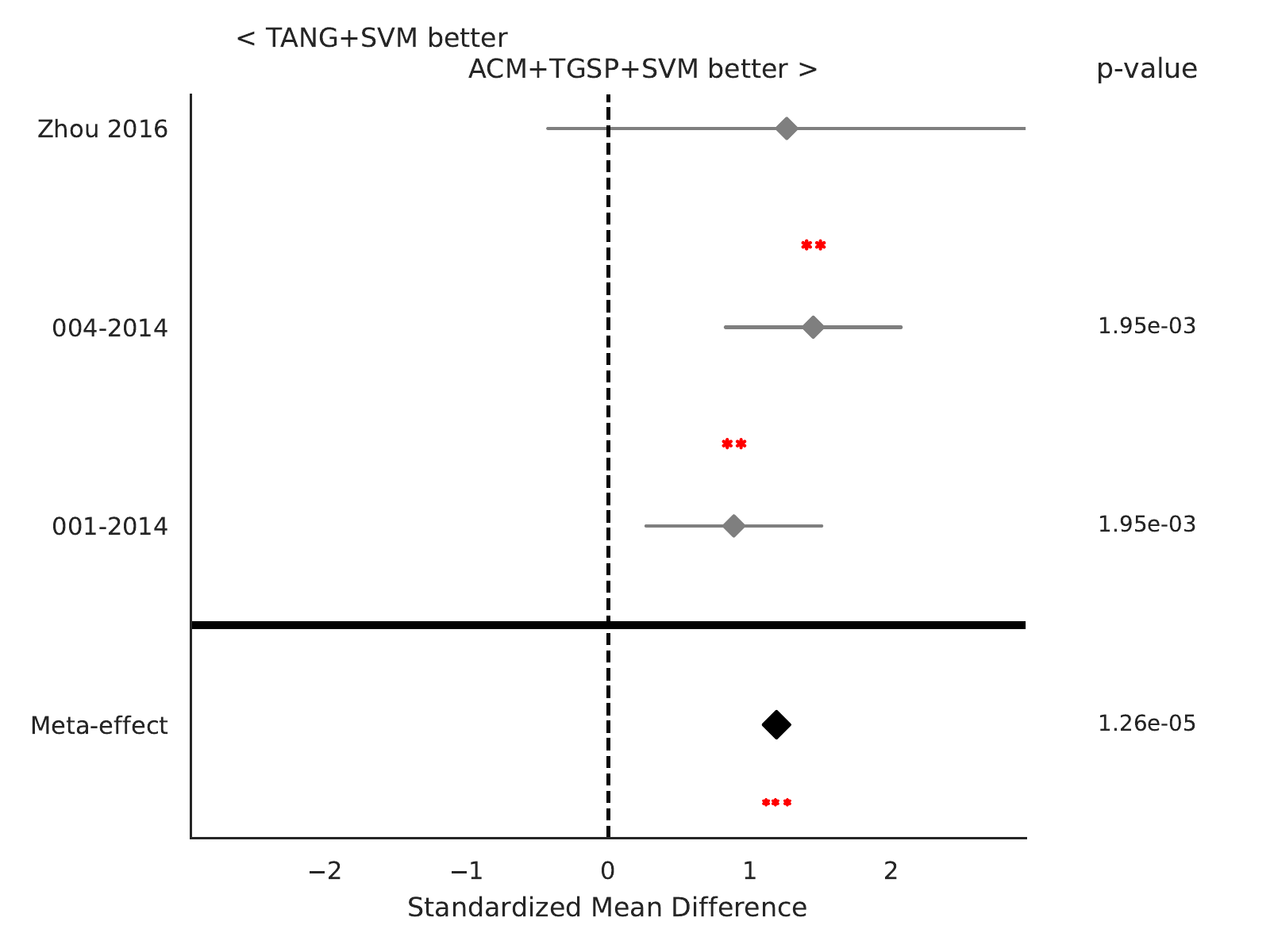}}
            \hfill
   \subfloat[]{%
            \includegraphics[width=0.30\linewidth]{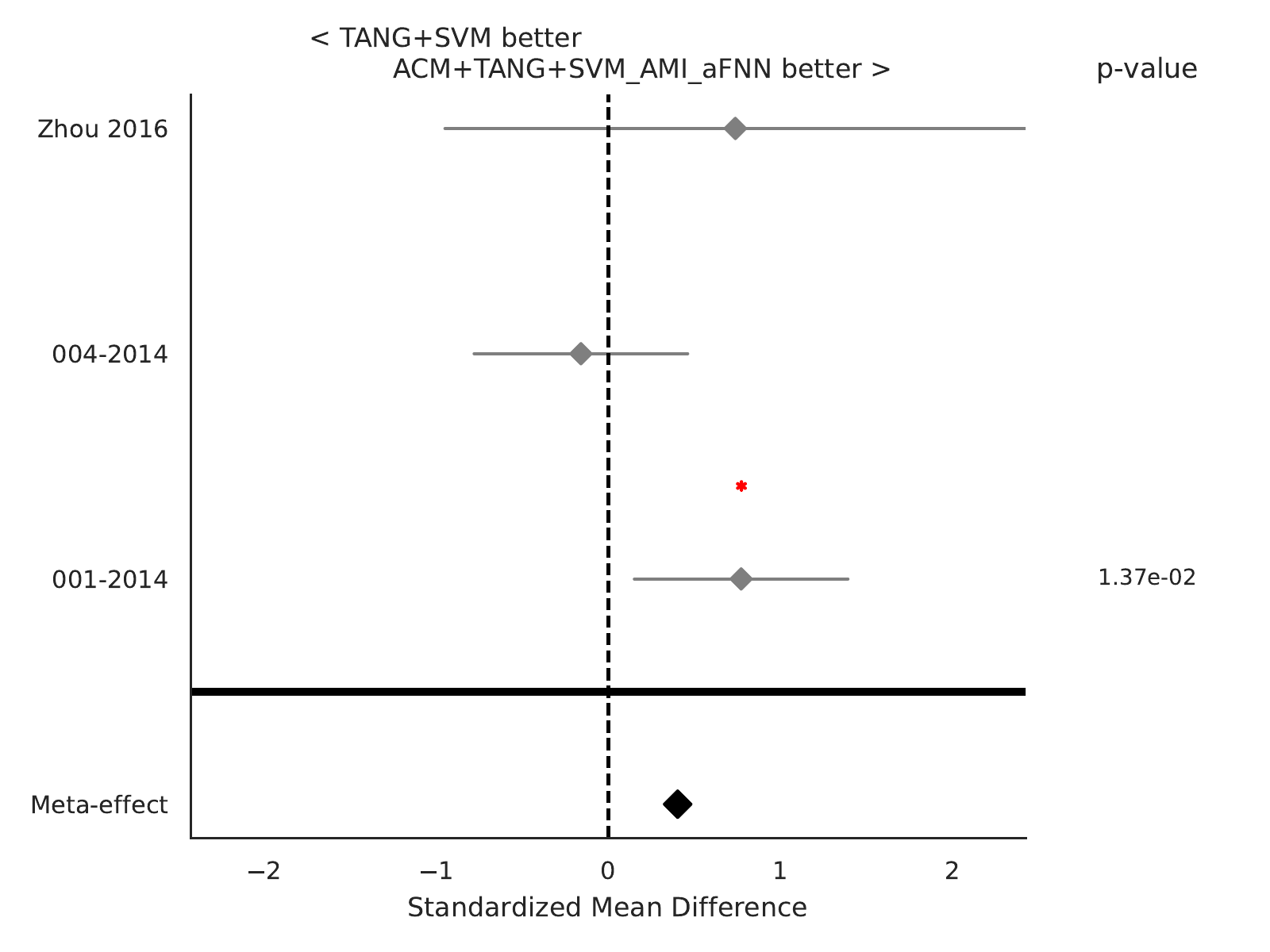}}
            \hfill
   \subfloat[]{%
            \includegraphics[width=0.30\linewidth]{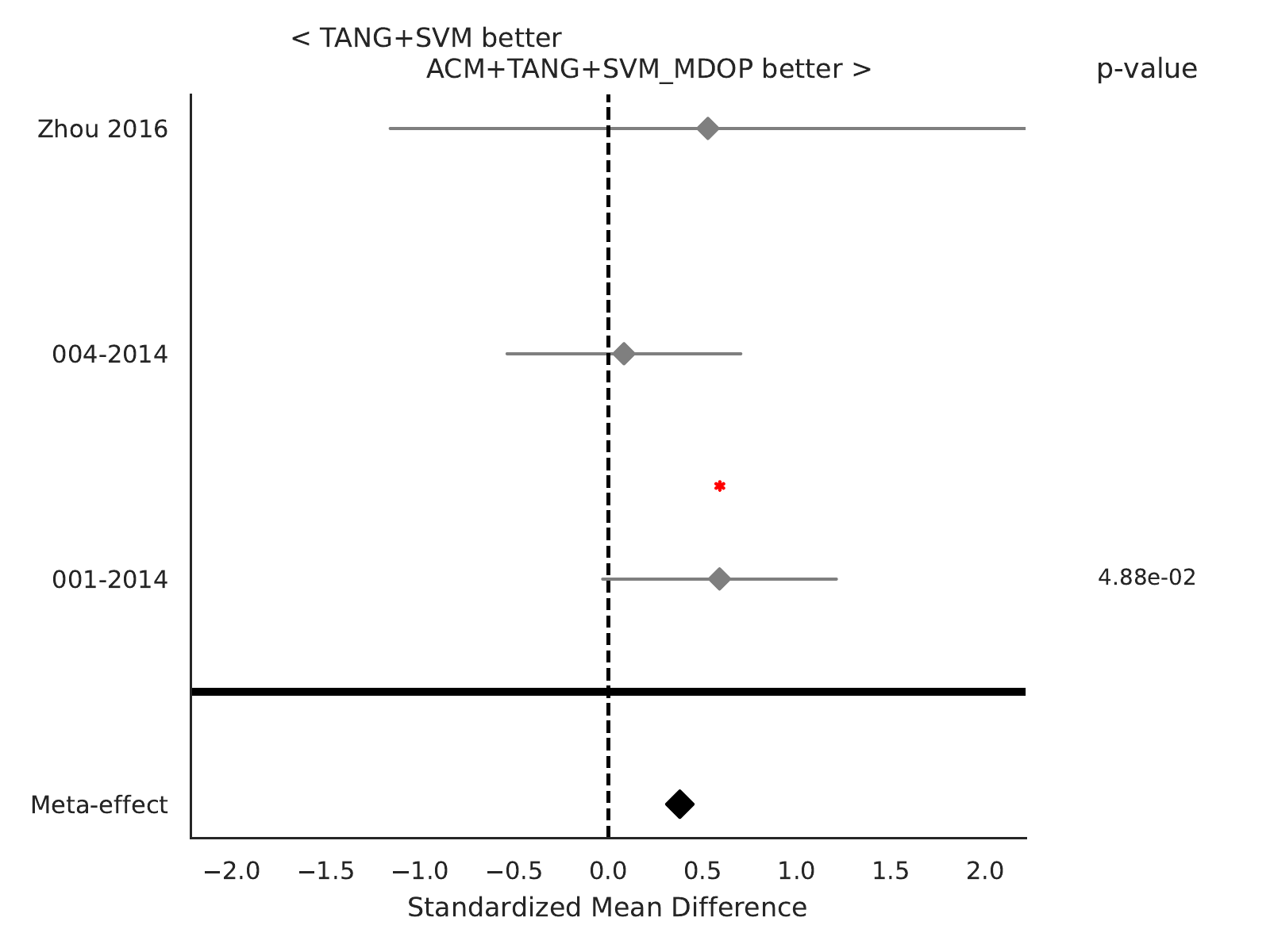}}
            \hfill

    \caption{Result for right hand vs left hand classification using the TANG algorithm, using cross-session evaluation. (a) show the rain clouds plots for each pipeline, showing the distribution of the score of every subject. (b) show the bar plot of the score withe the error of the different pipeline and for every dataset considered. (c) show the meta analysis of the different pipeline considered. This plot the significance that the algorithm on the y-axis is better than the one on the x-axis. The color represents the significance level of the difference of accuracy, in terms of t-values, and we show only the significant interactions ($p < 0.05$). (d) (e) (f) show the meta analysis of the standard TANG algorithm against the augmented covariance method with the selection of the hyper-parameter based on grid search, traditional and unified Takens approach respectively. We show the standardized mean differences, while p-values are computed as one-tailed Wilcoxon signed-rank test for the hypothesis given as title of the plot and the gray bar  denote $95\%$ interval. Here, * stands for $p < 0.05$, ** for $p < 0.01$, and *** for $p < 0.001$.
    }
    \label{fig:TANG+SVM-rhlh-crosssession}
\end{figure*}

\begin{figure*}[ht]  
    \centering
    \centering
     \subfloat[]{%
            \includegraphics[width=0.45\linewidth]{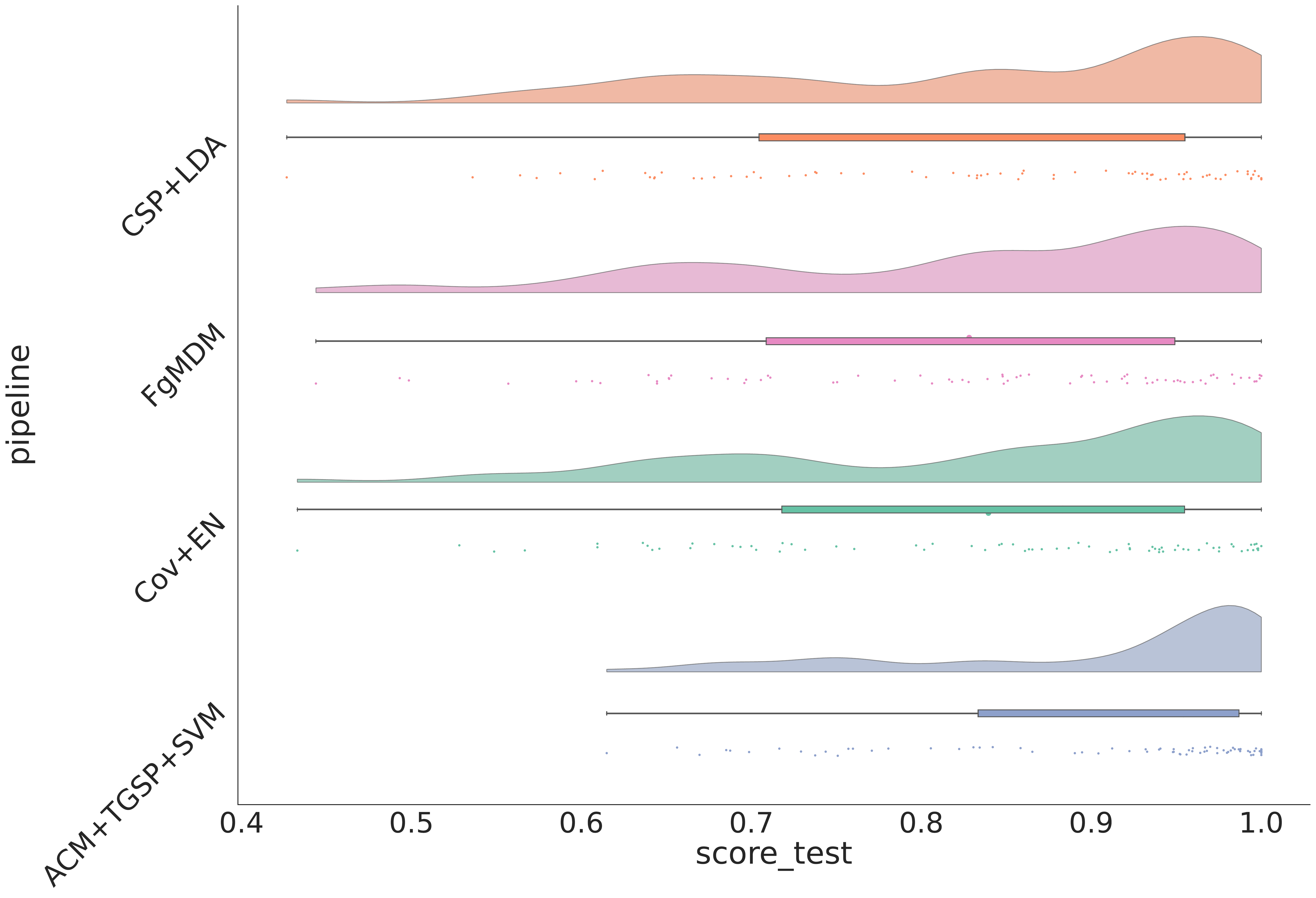}}
            \hfill
     \subfloat[]{%
            \includegraphics[width=0.45\linewidth]{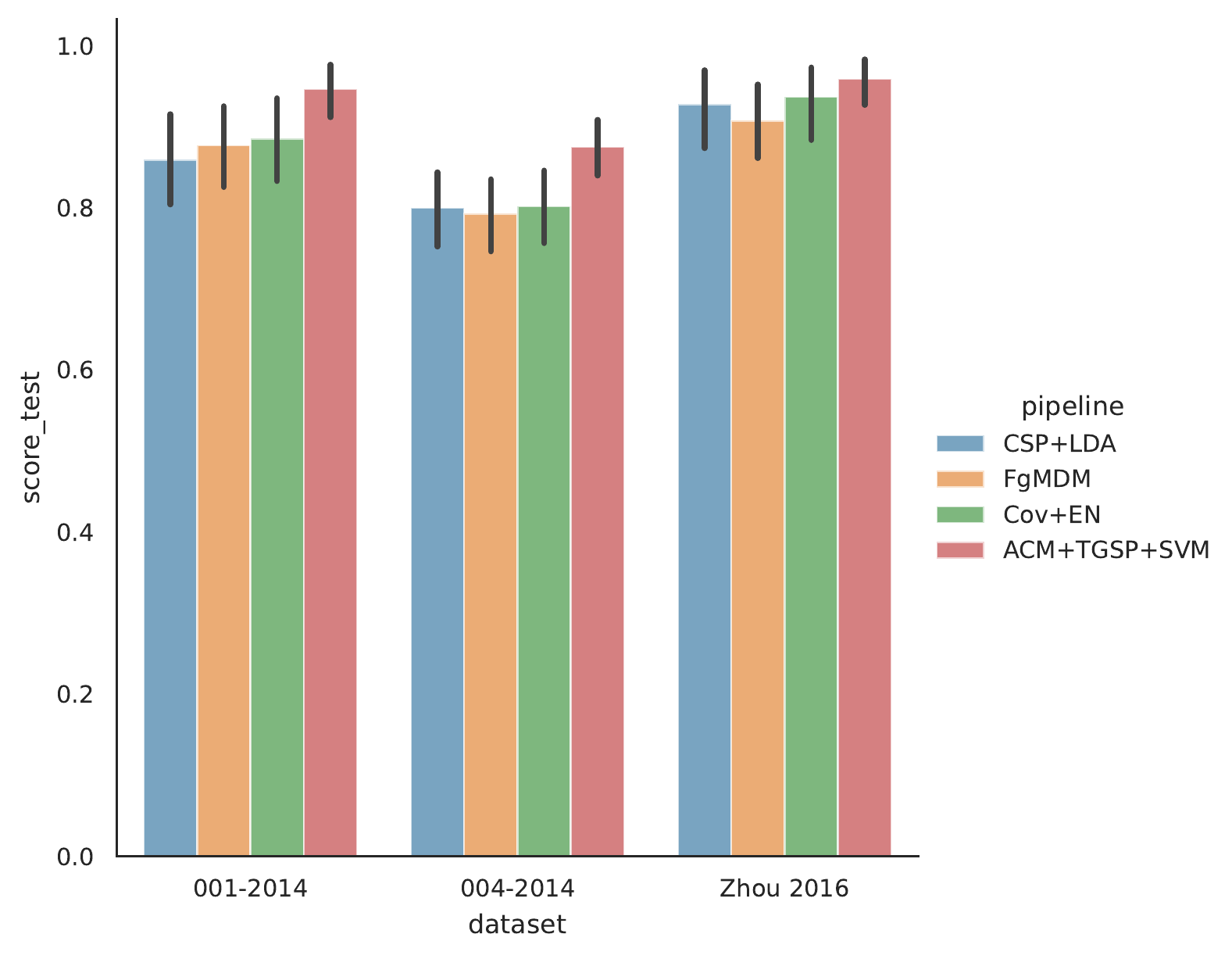}}
    \\
    \subfloat[]{%
        \includegraphics[width=0.5\linewidth]{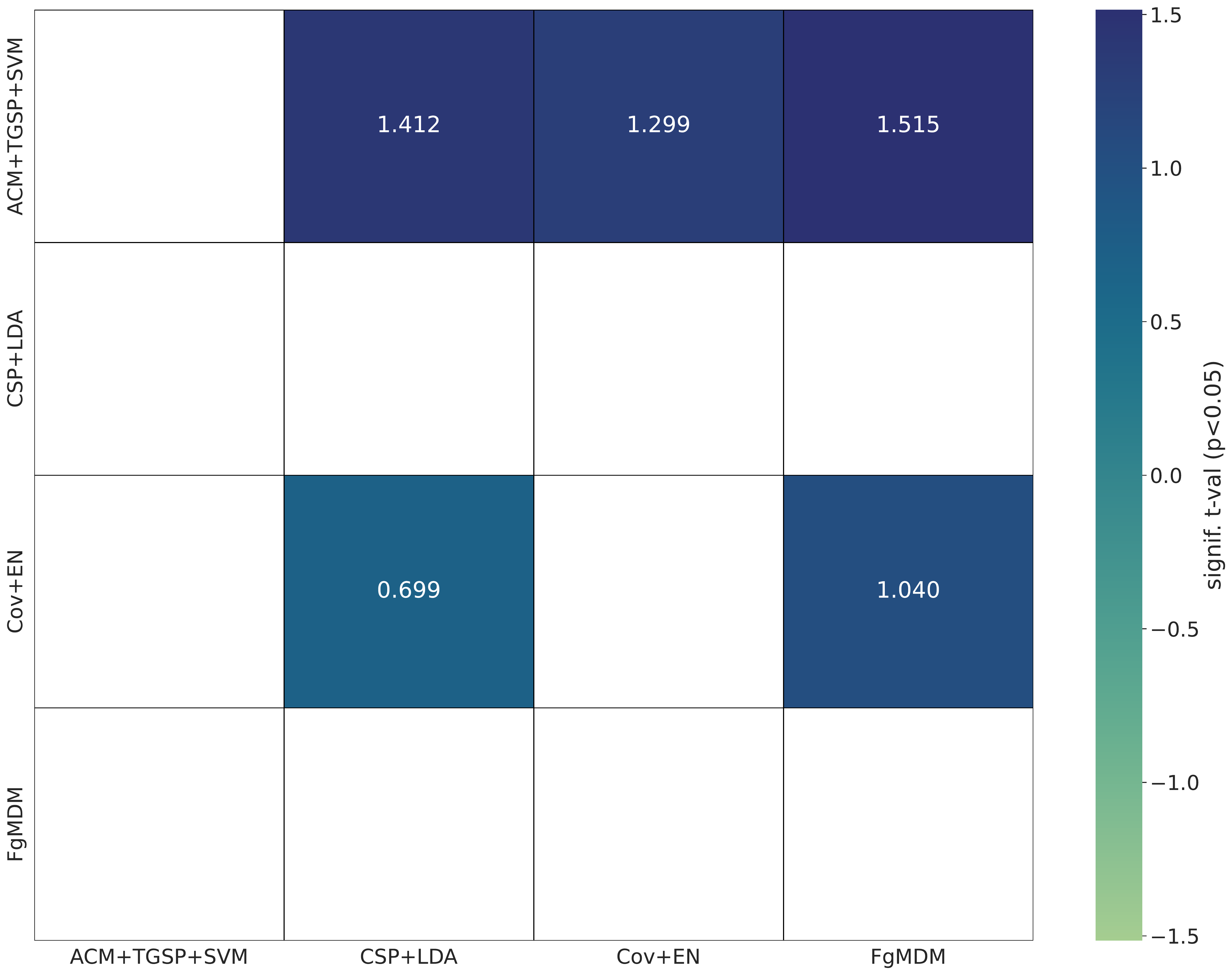}}
        \\
   \subfloat[]{%
            \includegraphics[width=0.30\linewidth]{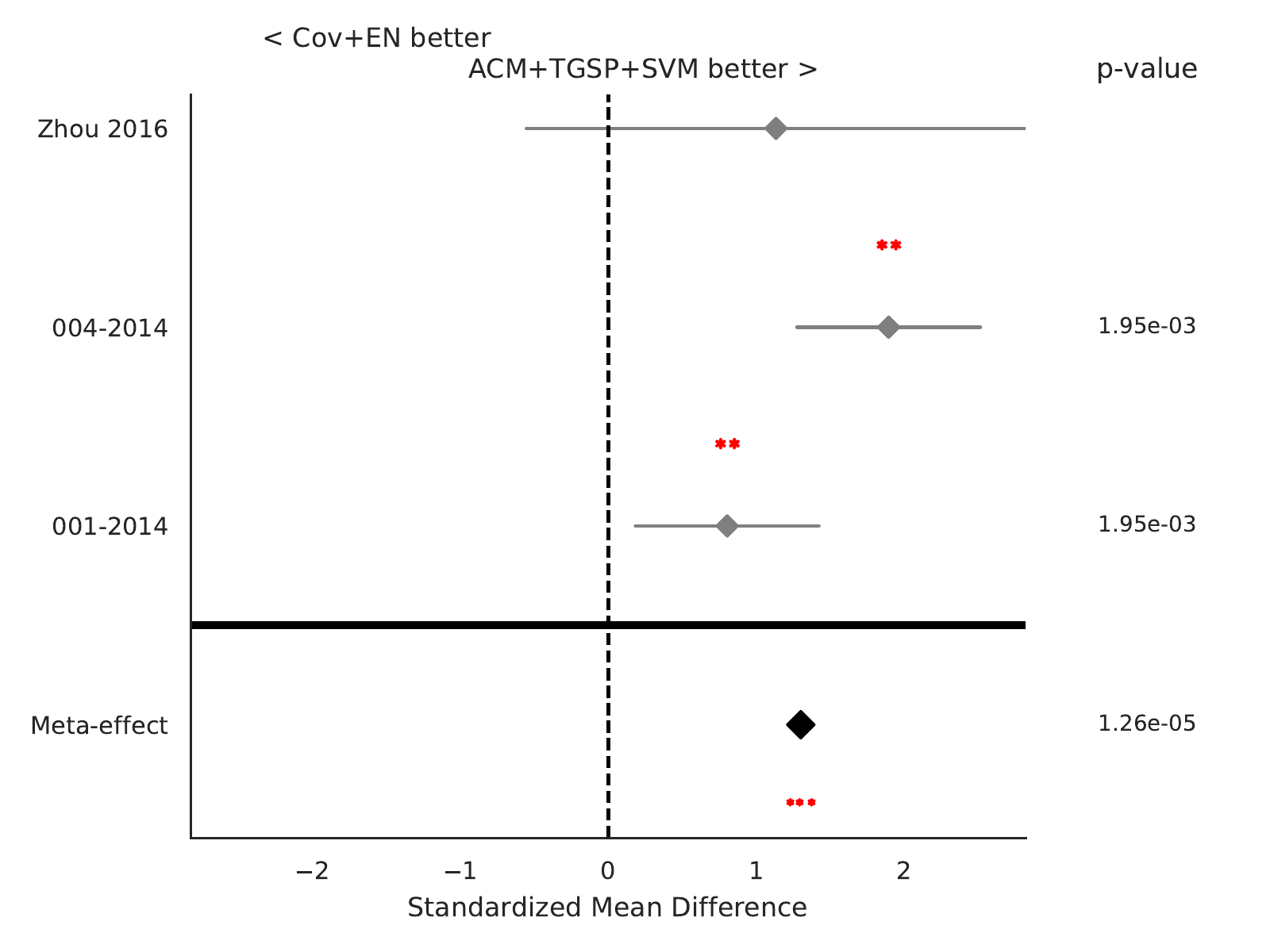}}
            \hfill
   \subfloat[]{%
            \includegraphics[width=0.30\linewidth]{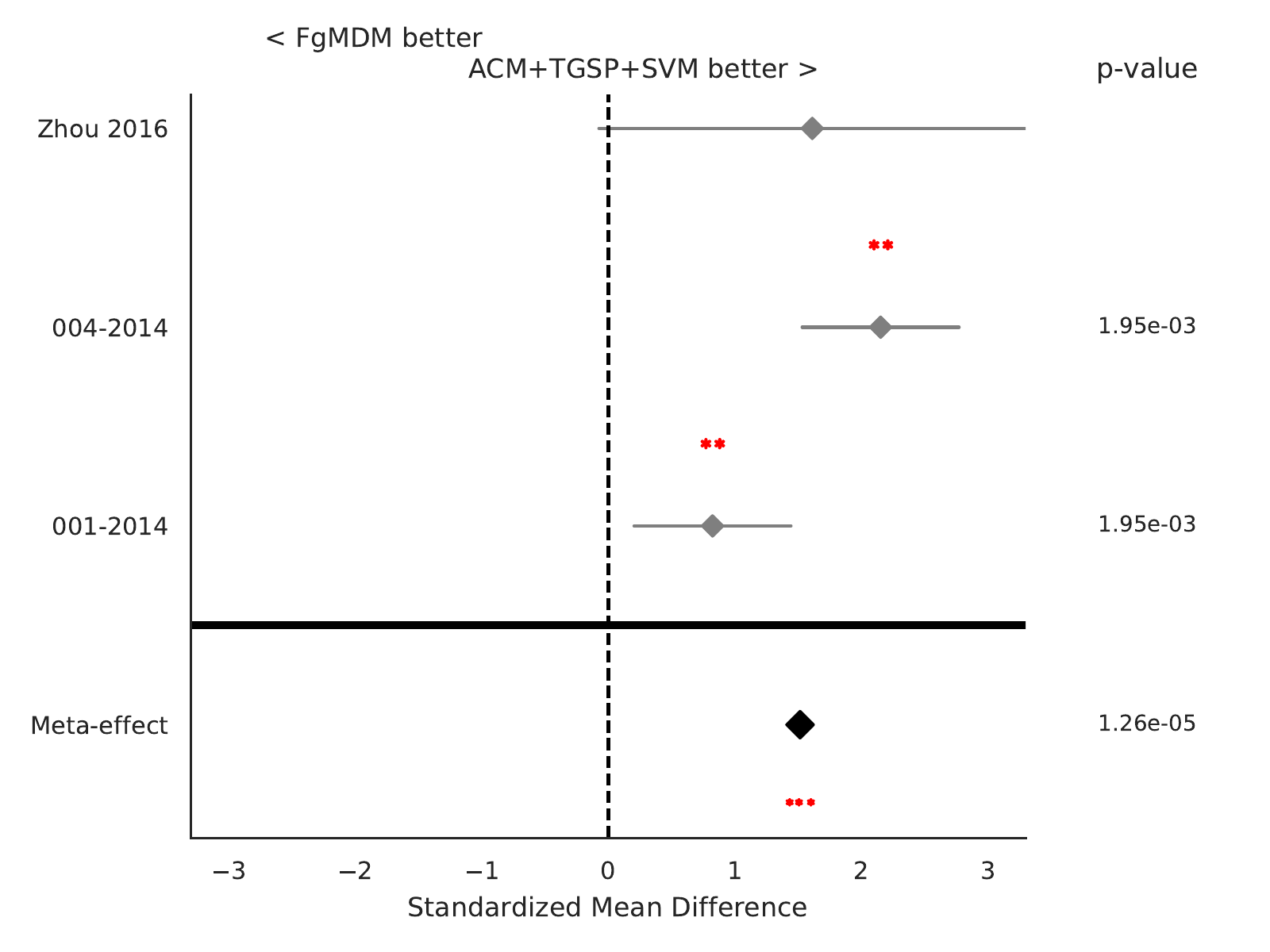}}
            \hfill
   \subfloat[]{%
            \includegraphics[width=0.30\linewidth]{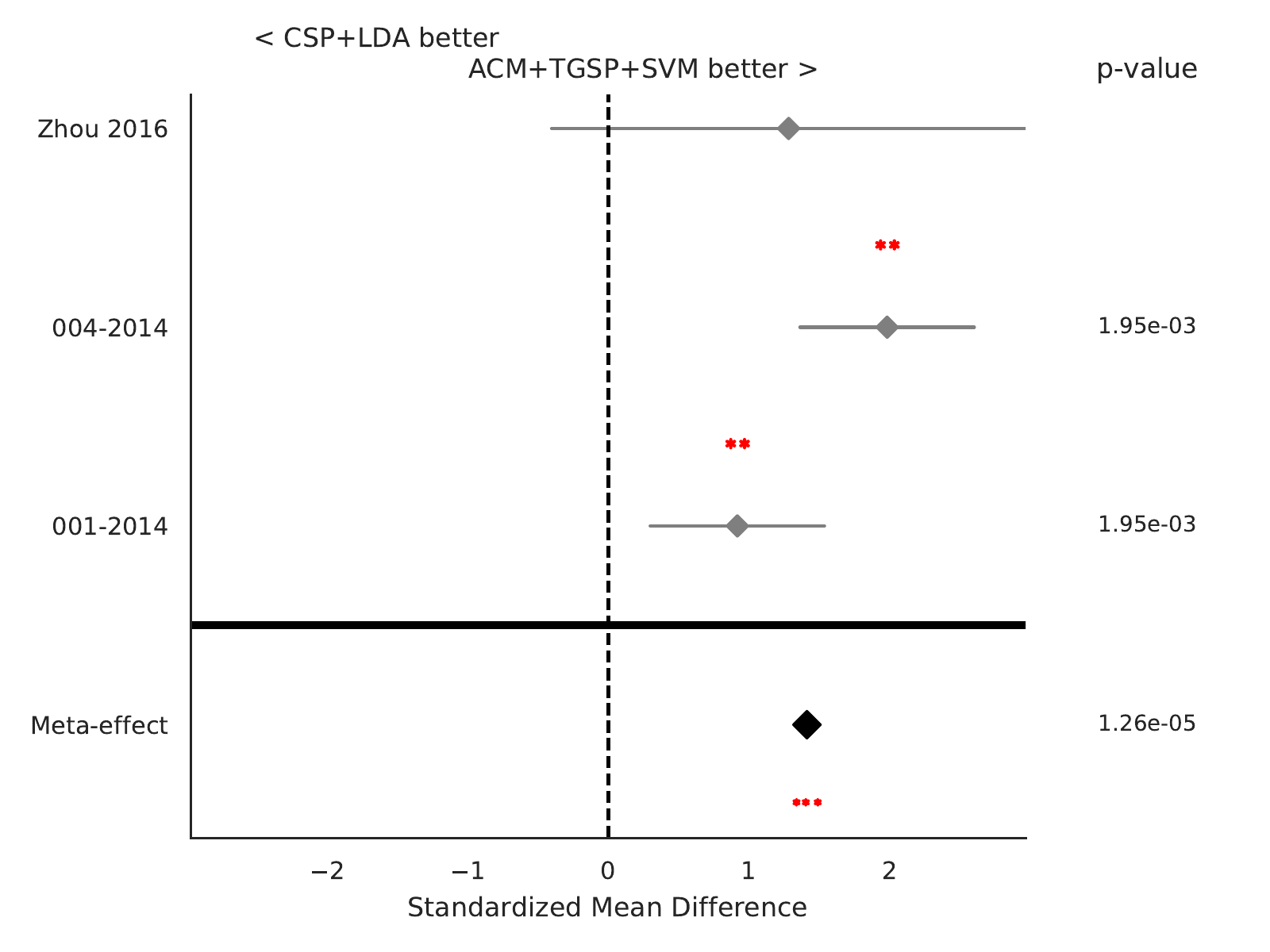}}
            \hfill

    \caption{Result for right hand vs left hand classification, using withing-session evaluation. (a) show the rain clouds plots for each pipeline, showing the distribution of the score of every subject. (b) show the bar plot of the score withe the error of the different pipeline and for every dataset considered. (c) show the meta analysis of the different pipeline considered. This plot the significance that the algorithm on the y-axis is better than the one on the x-axis. The color represents the significance level of the difference of accuracy, in terms of t-values, and we show only the significant interactions ($p < 0.05$). (d) (e) (f) show the meta analysis of augmented method with SVM against the state of the art. We show the standardized mean differences, while p-values are computed as one-tailed Wilcoxon signed-rank test for the hypothesis given as title of the plot and the gray bar  denote $95\%$ interval. Here, * stands for $p < 0.05$, ** for $p < 0.01$, and *** for $p < 0.001$.
    }
    \label{fig:TANG+SVM-rhlh-whithinsession-stateart}
\end{figure*}

\begin{figure*}[ht]  
    \centering
    \centering
     \subfloat[]{%
            \includegraphics[width=0.45\linewidth]{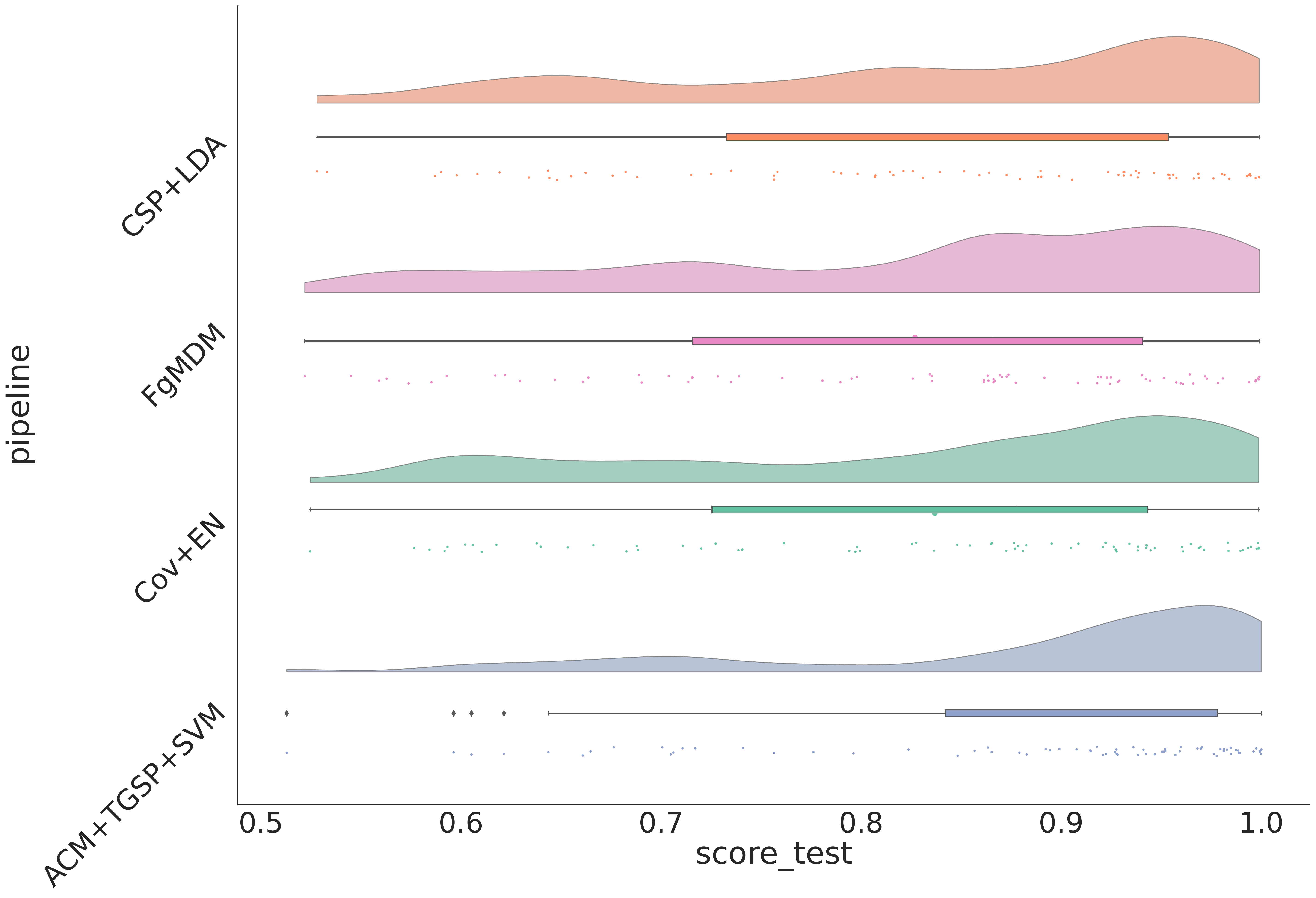}}
            \hfill
     \subfloat[]{%
            \includegraphics[width=0.45\linewidth]{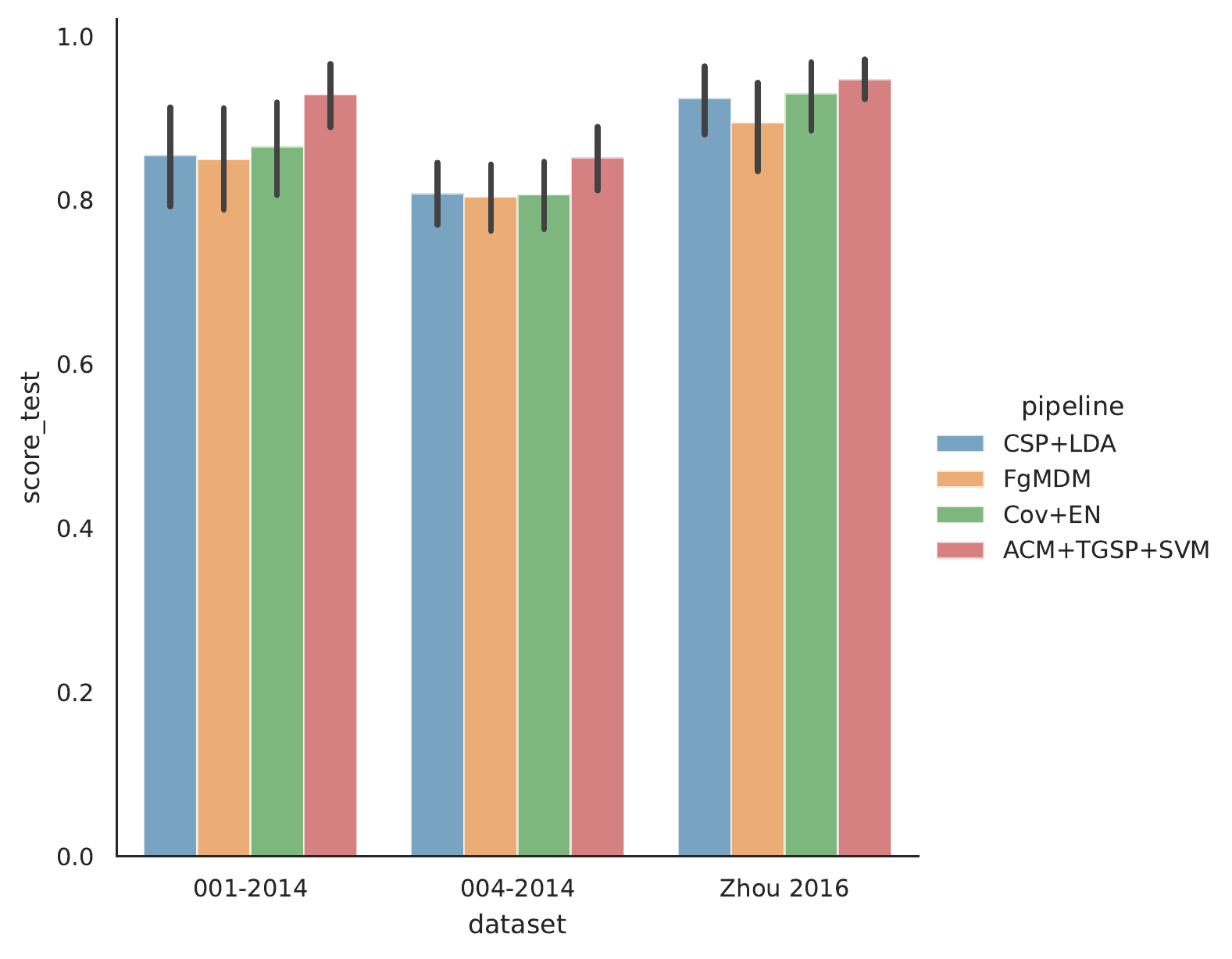}}
    \\
    \subfloat[]{%
        \includegraphics[width=0.5\linewidth]{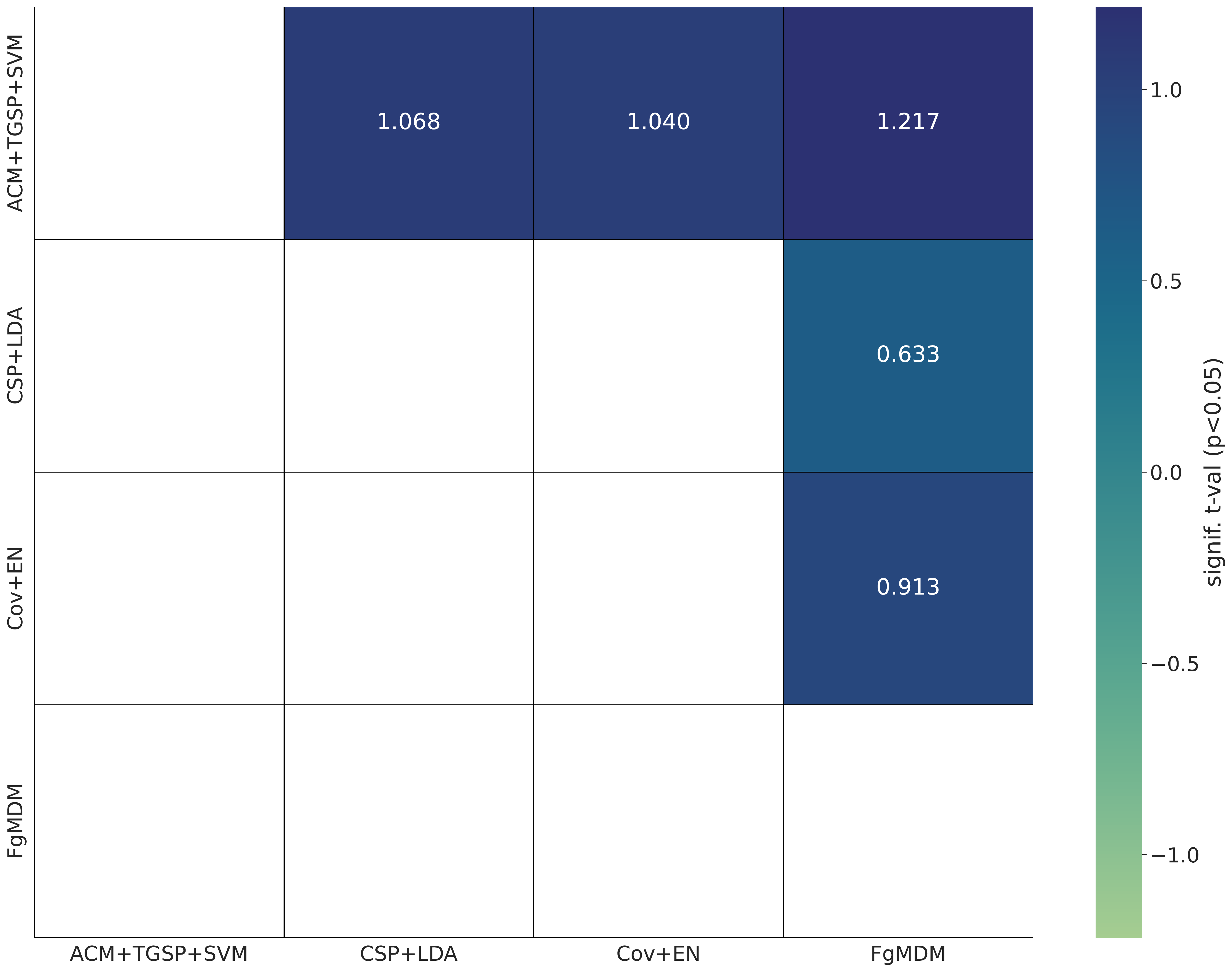}}
        \\
   \subfloat[]{%
            \includegraphics[width=0.30\linewidth]{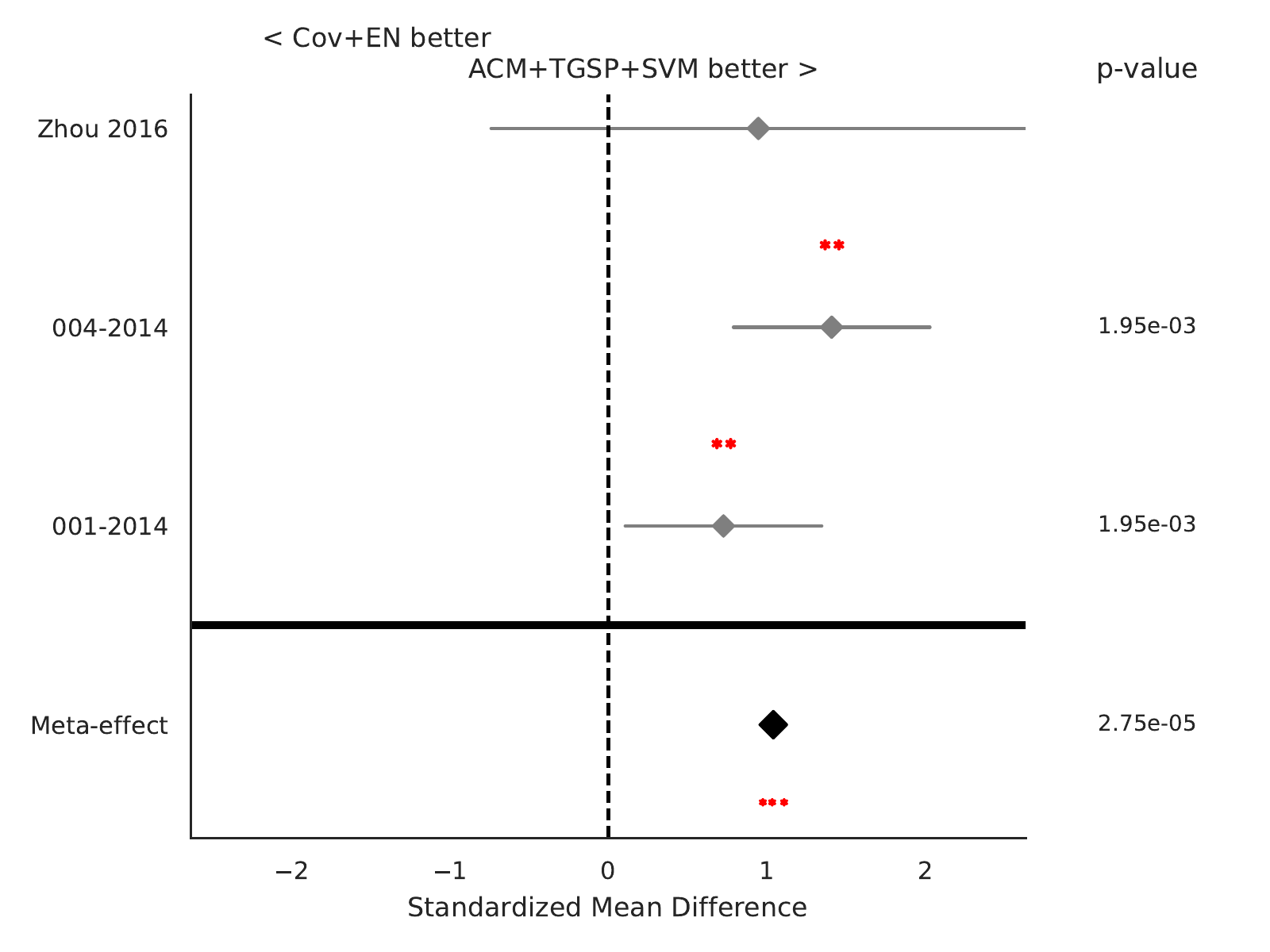}}
            \hfill
   \subfloat[]{%
            \includegraphics[width=0.30\linewidth]{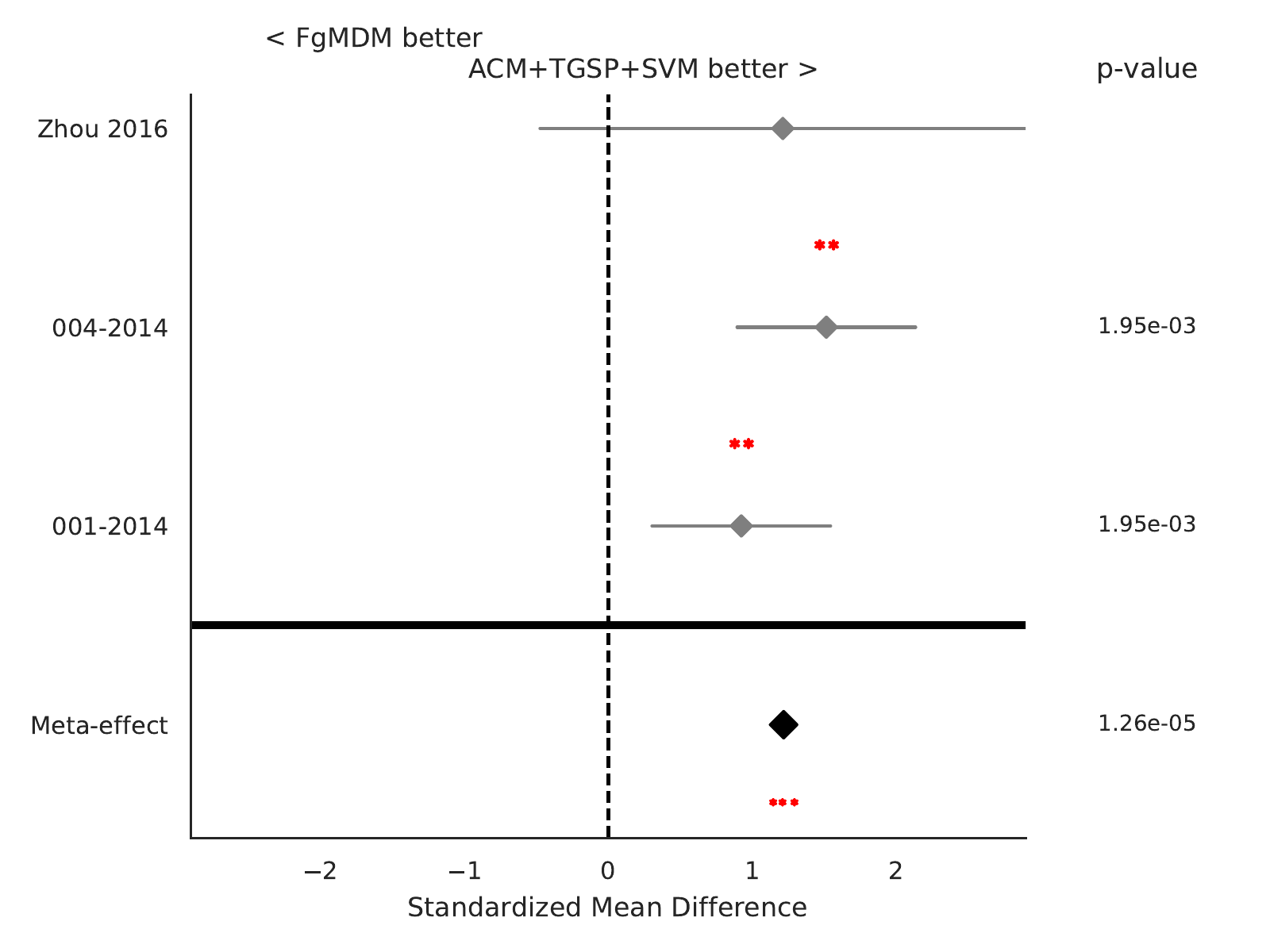}}
            \hfill
   \subfloat[]{%
            \includegraphics[width=0.30\linewidth]{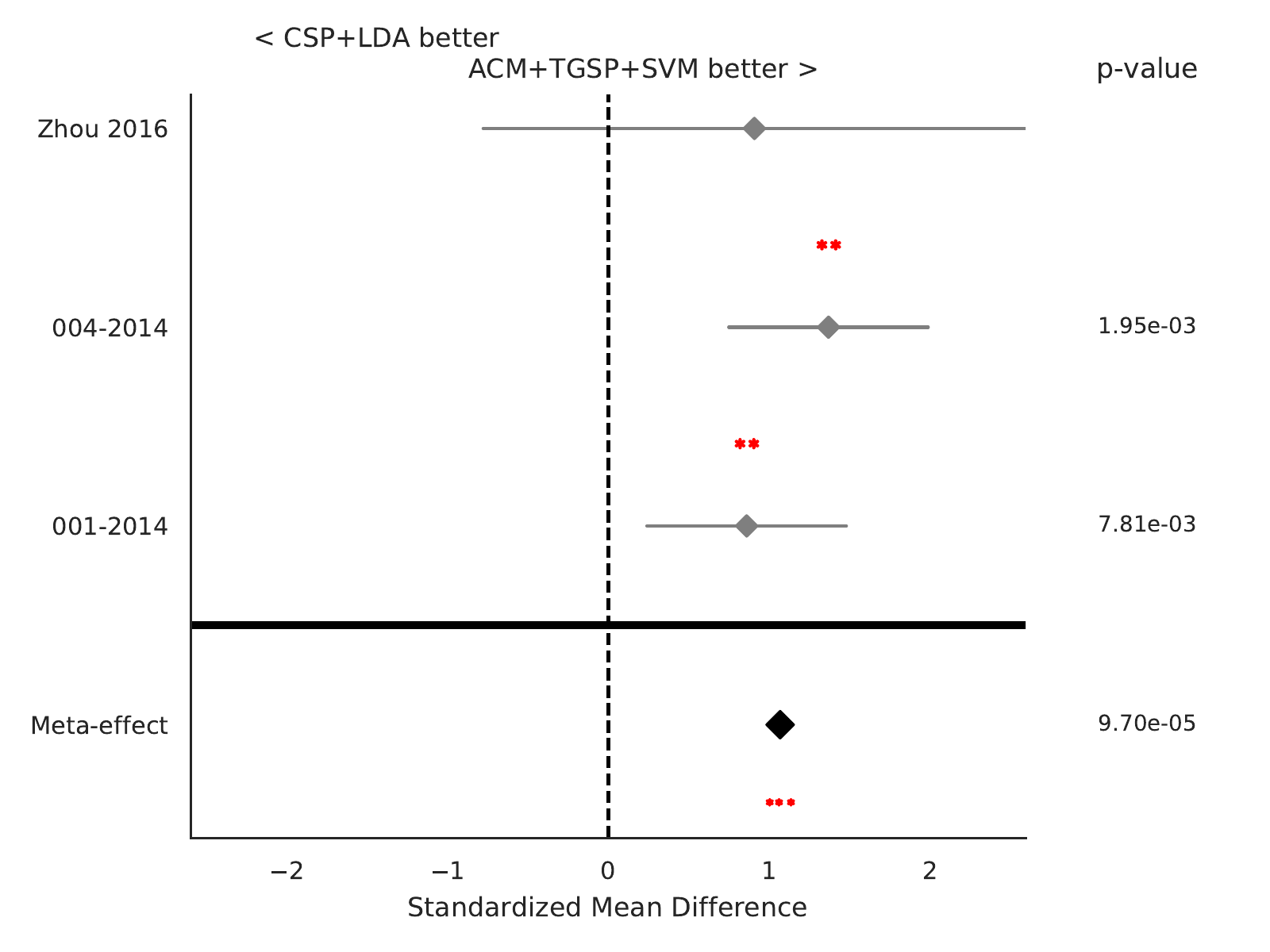}}
            \hfill

    \caption{Result for right hand vs left hand classification, using cross-session evaluation. (a) show the rain clouds plots for each pipeline, showing the distribution of the score of every subject. (b) show the bar plot of the score withe the error of the different pipeline and for every dataset considered. (c) show the meta analysis of the different pipeline considered. This plot the significance that the algorithm on the y-axis is better than the one on the x-axis. The color represents the significance level of the difference of accuracy, in terms of t-values, and we show only the significant interactions ($p < 0.05$). (d) (e) (f) show the meta analysis of augmented method with SVM against the state of the art. We show the standardized mean differences, while p-values are computed as one-tailed Wilcoxon signed-rank test for the hypothesis given as title of the plot and the gray bar  denote $95\%$ interval. Here, * stands for $p < 0.05$, ** for $p < 0.01$, and *** for $p < 0.001$.
    }
    \label{fig:TANG+SVM-rhlh-crosssession-stateart}
\end{figure*}

\begin{figure*}[ht]  
    \centering
    \centering
     \subfloat[]{%
            \includegraphics[width=0.45\linewidth]{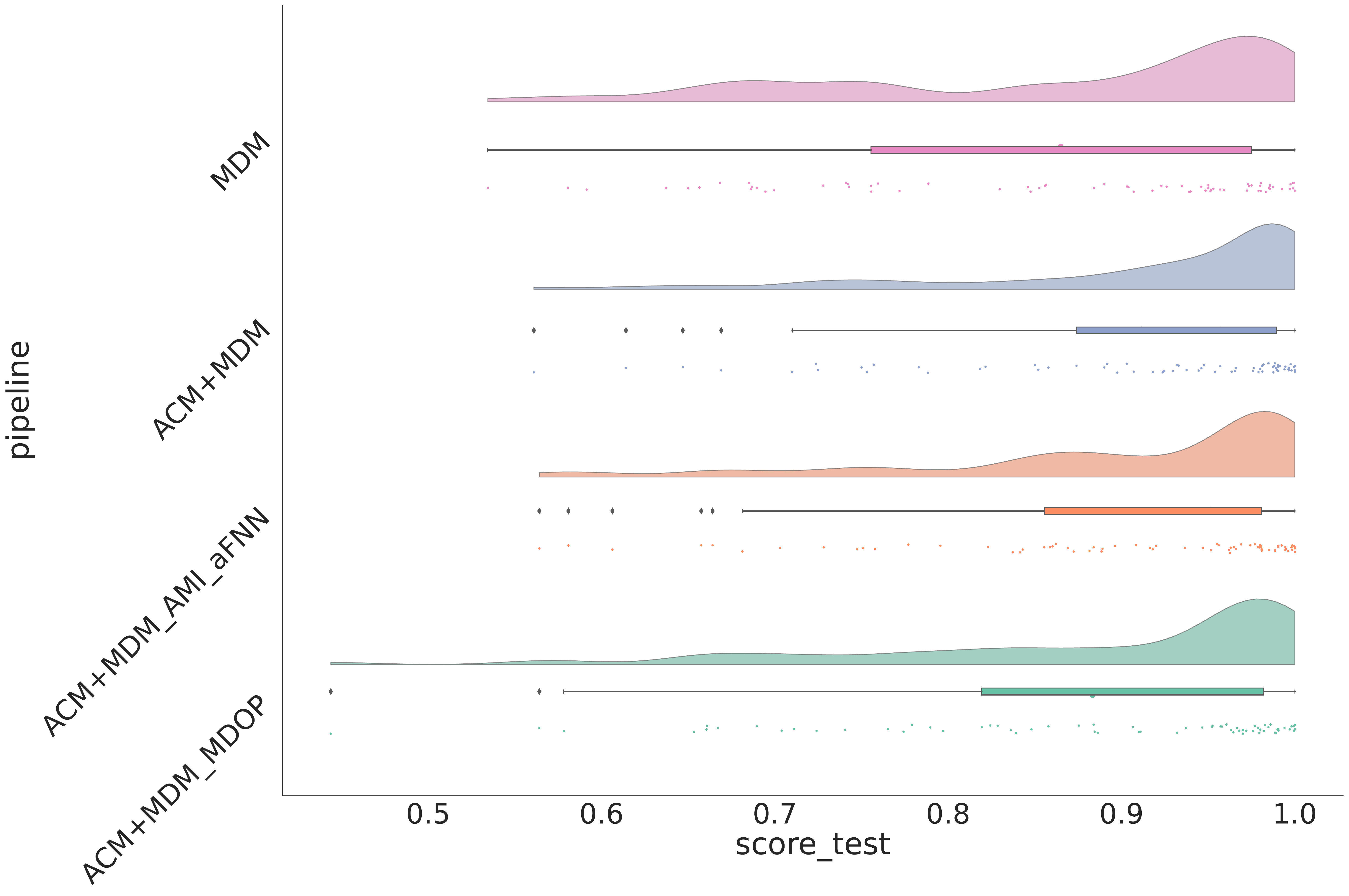}}
            \hfill
     \subfloat[]{%
            \includegraphics[width=0.45\linewidth]{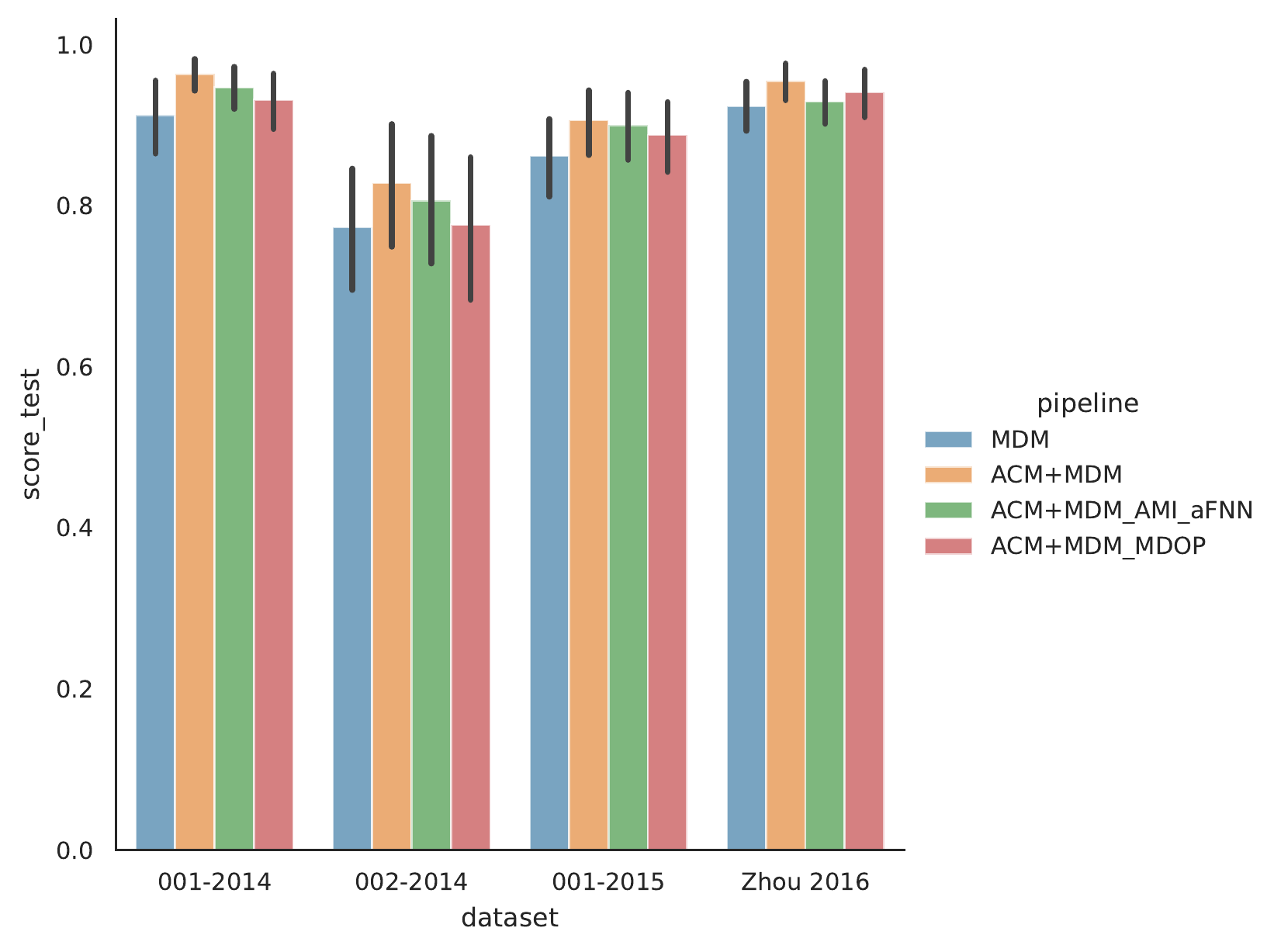}}
    \\
    \subfloat[]{%
        \includegraphics[width=0.5\linewidth]{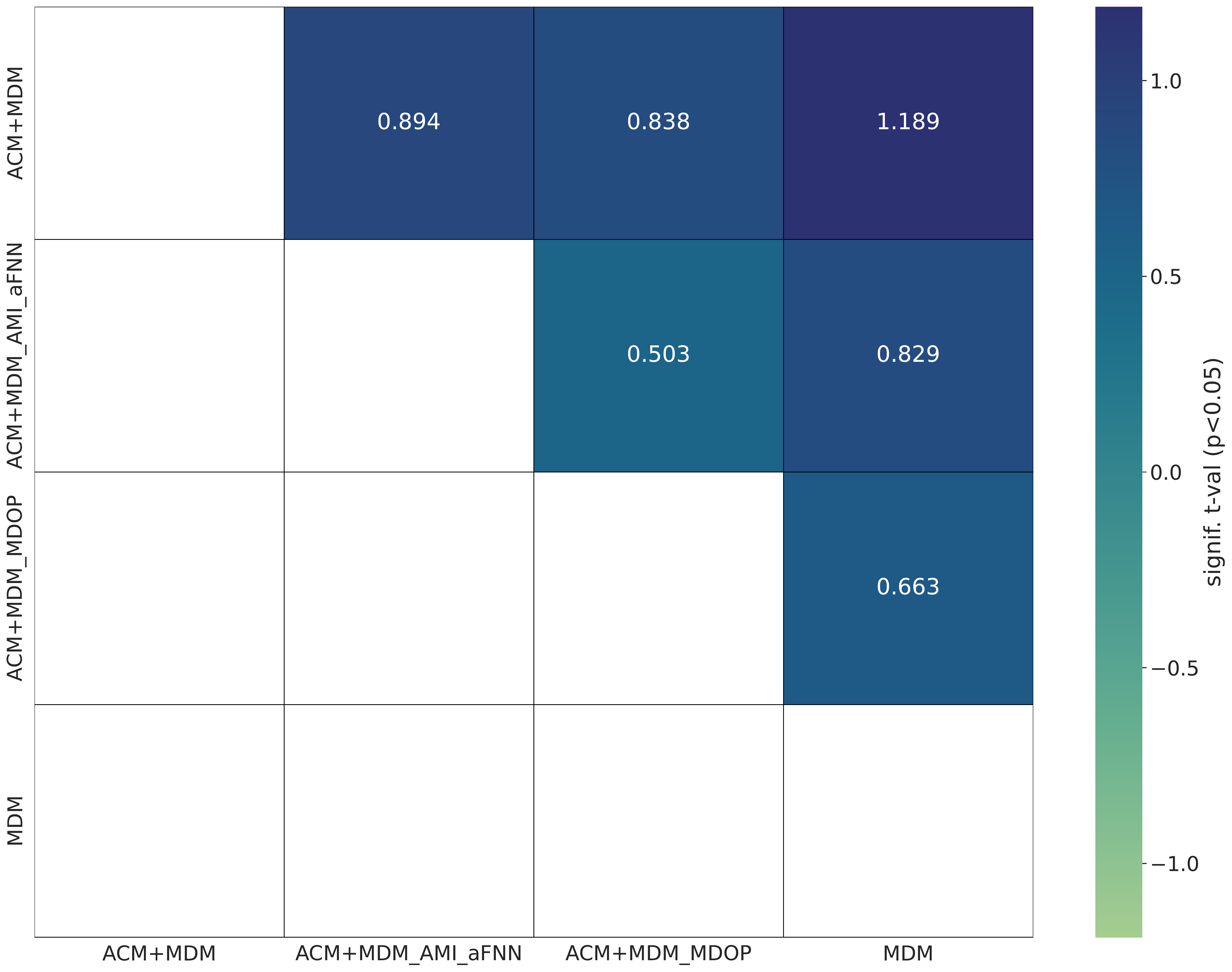}}
        \\
   \subfloat[]{%
            \includegraphics[width=0.30\linewidth]{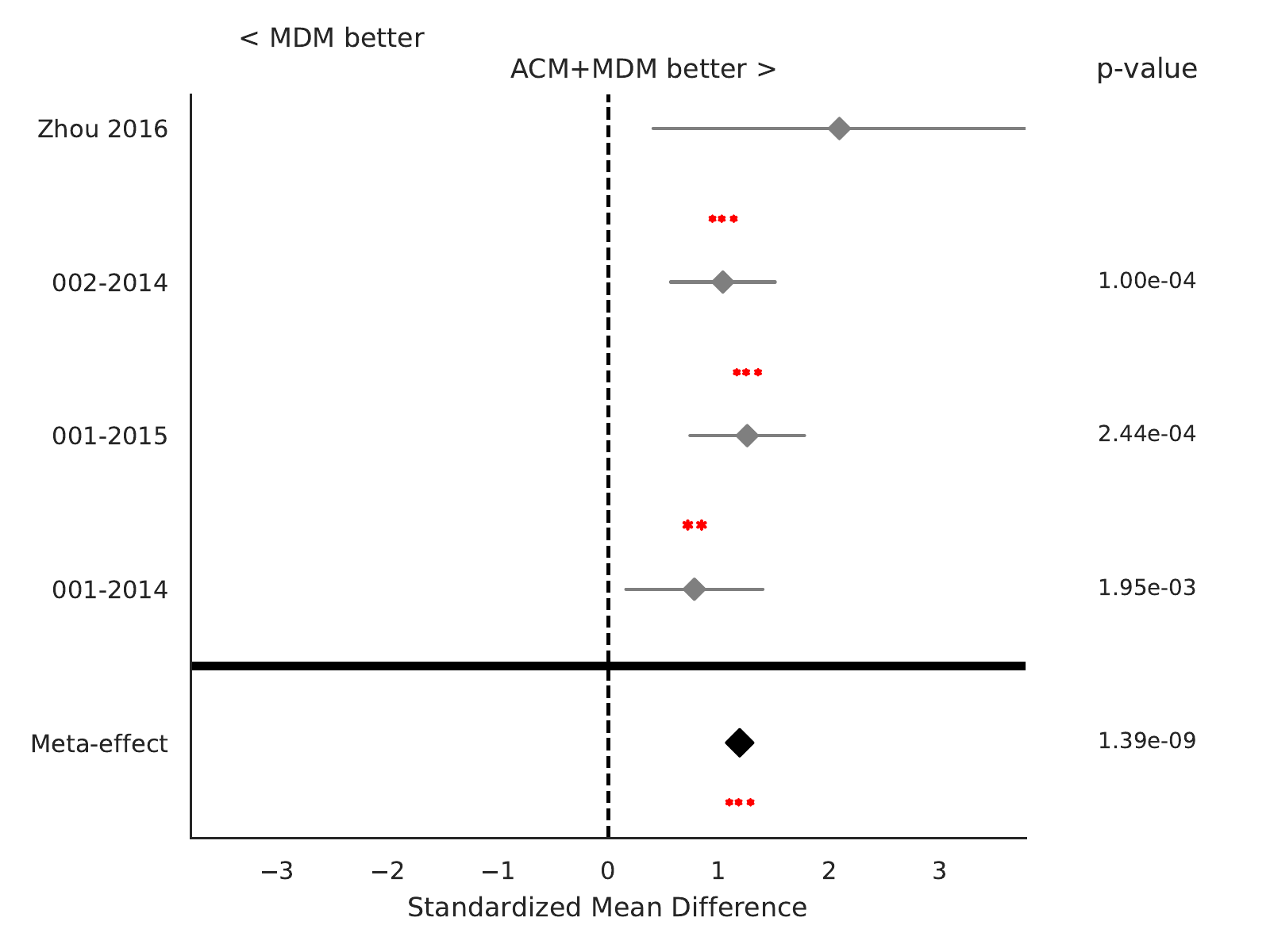}}
            \hfill
   \subfloat[]{%
            \includegraphics[width=0.30\linewidth]{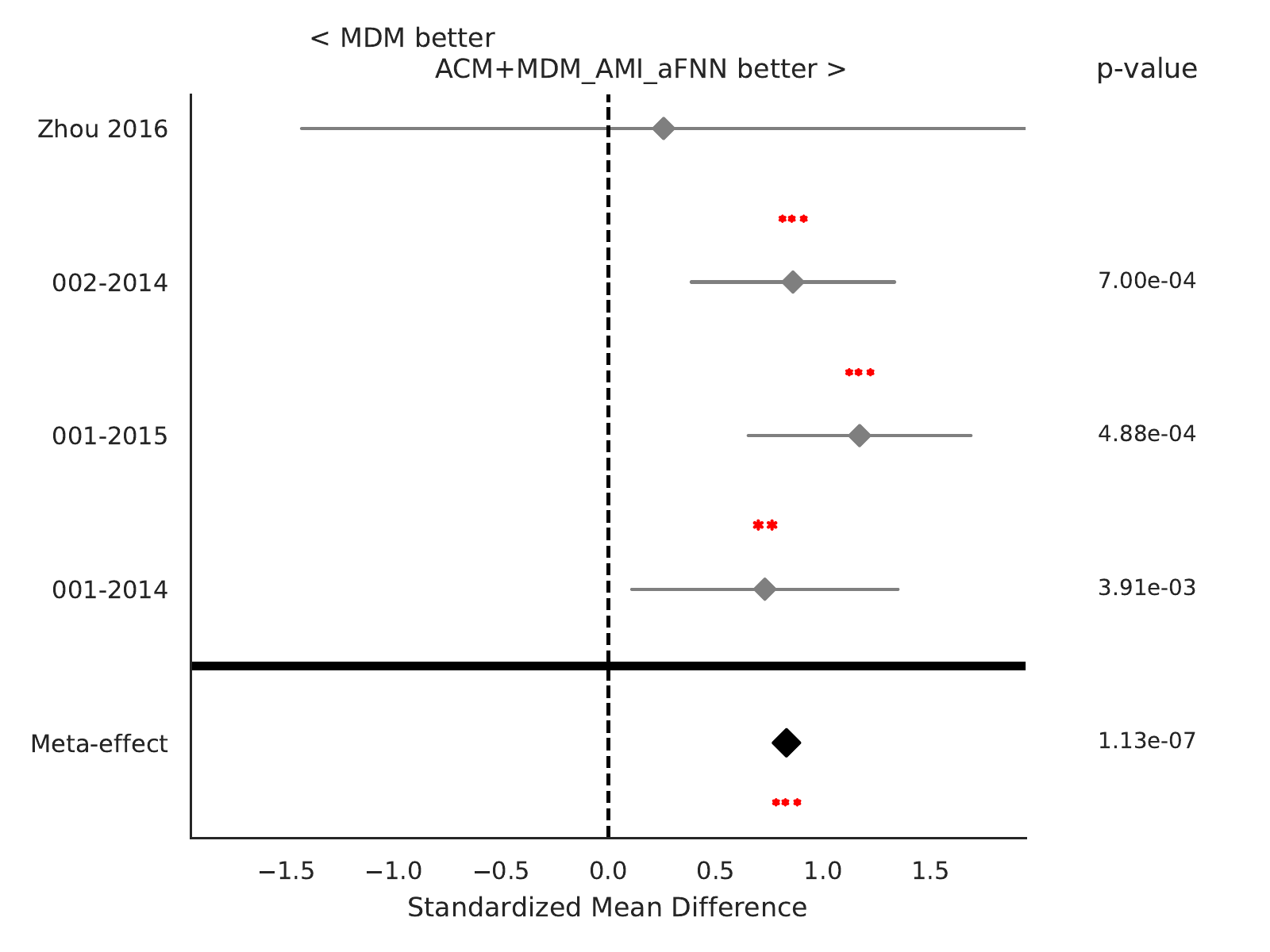}}
            \hfill
   \subfloat[]{%
            \includegraphics[width=0.30\linewidth]{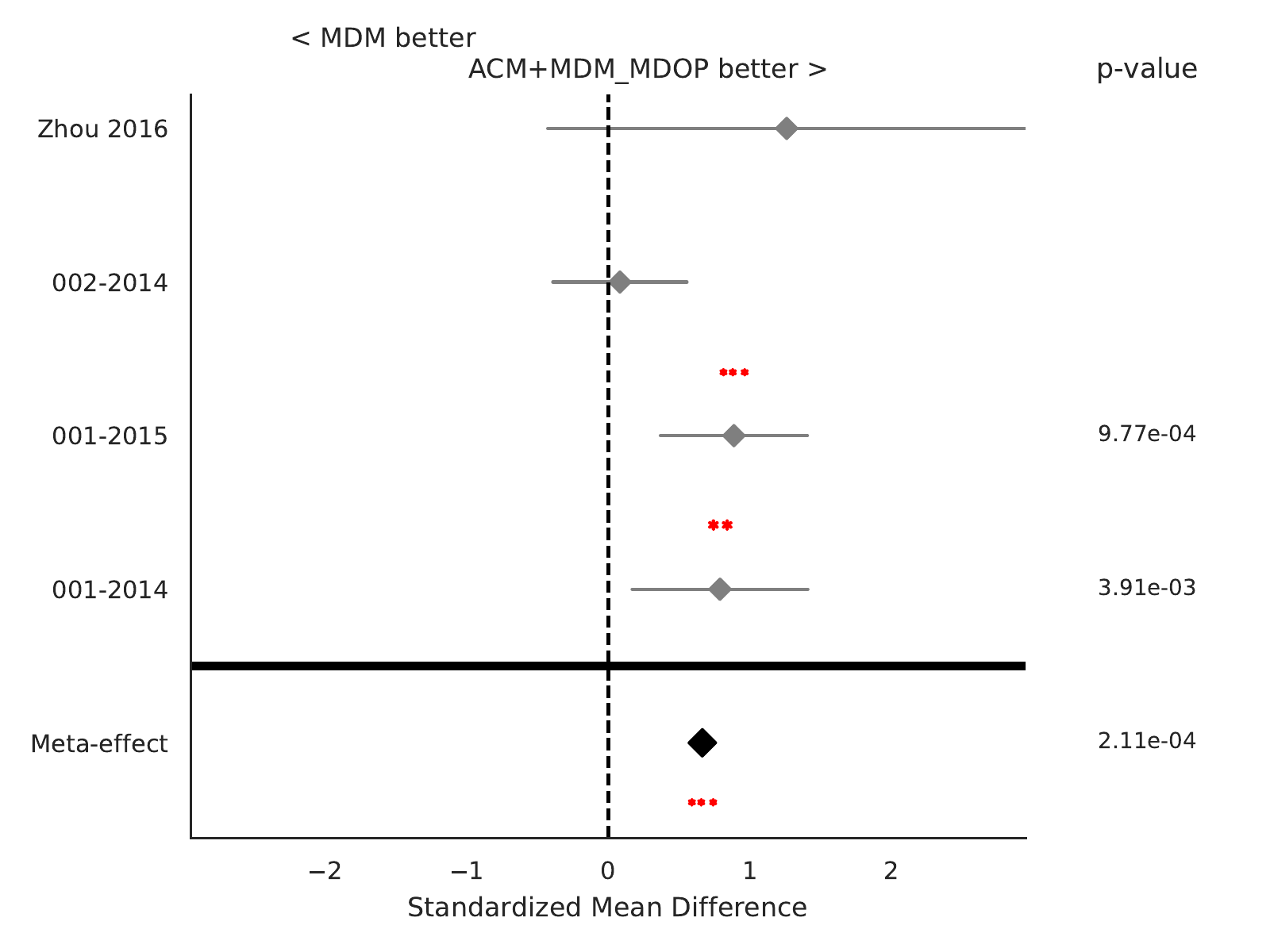}}
            \hfill

    \caption{Result for right hand vs feet classification using the MDM algorithm, using withing-session evaluation. (a) show the rain clouds plots for each pipeline, showing the distribution of the score of every subject. (b) show the bar plot of the score withe the error of the different pipeline and for every dataset considered. (c) show the meta analysis of the different pipeline considered. This plot the significance that the algorithm on the y-axis is better than the one on the x-axis. The color represents the significance level of the difference of accuracy, in terms of t-values, and we show only the significant interactions ($p < 0.05$). (d) (e) (f) show the meta analysis of the standard MDM algorithm against the augmented covariance method with the selection of the hyper-parameter based on grid search, traditional and unified Takens approach respectively. We show the standardized mean differences, while p-values are computed as one-tailed Wilcoxon signed-rank test for the hypothesis given as title of the plot and the gray bar  denote $95\%$ interval. Here, * stands for $p < 0.05$, ** for $p < 0.01$, and *** for $p < 0.001$.
    }
    \label{fig:MDM-rf-whithinsession}
\end{figure*}

\begin{figure*}[ht]  
    \centering
    \centering
     \subfloat[]{%
            \includegraphics[width=0.45\linewidth]{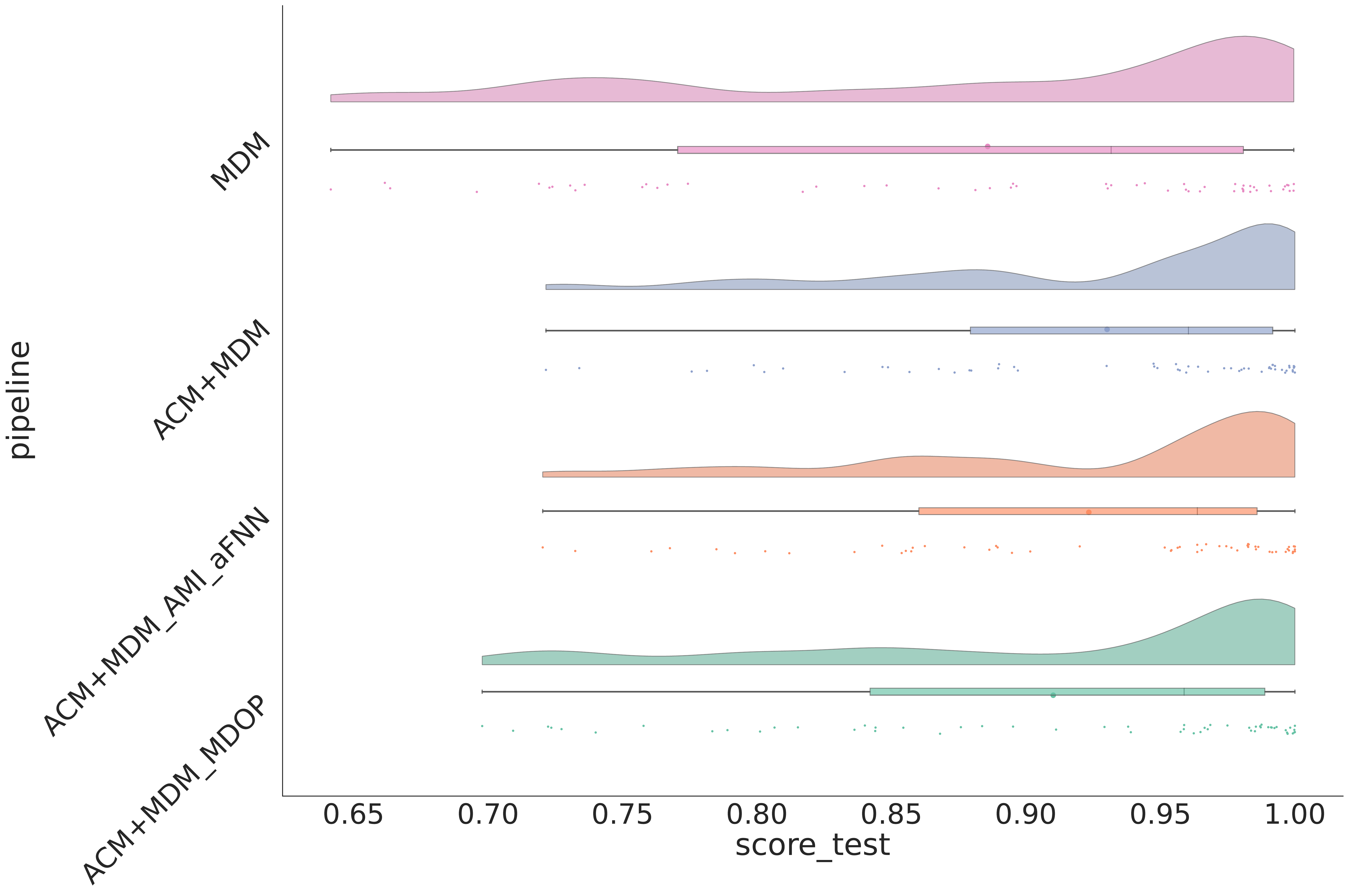}}
            \hfill
     \subfloat[]{%
            \includegraphics[width=0.45\linewidth]{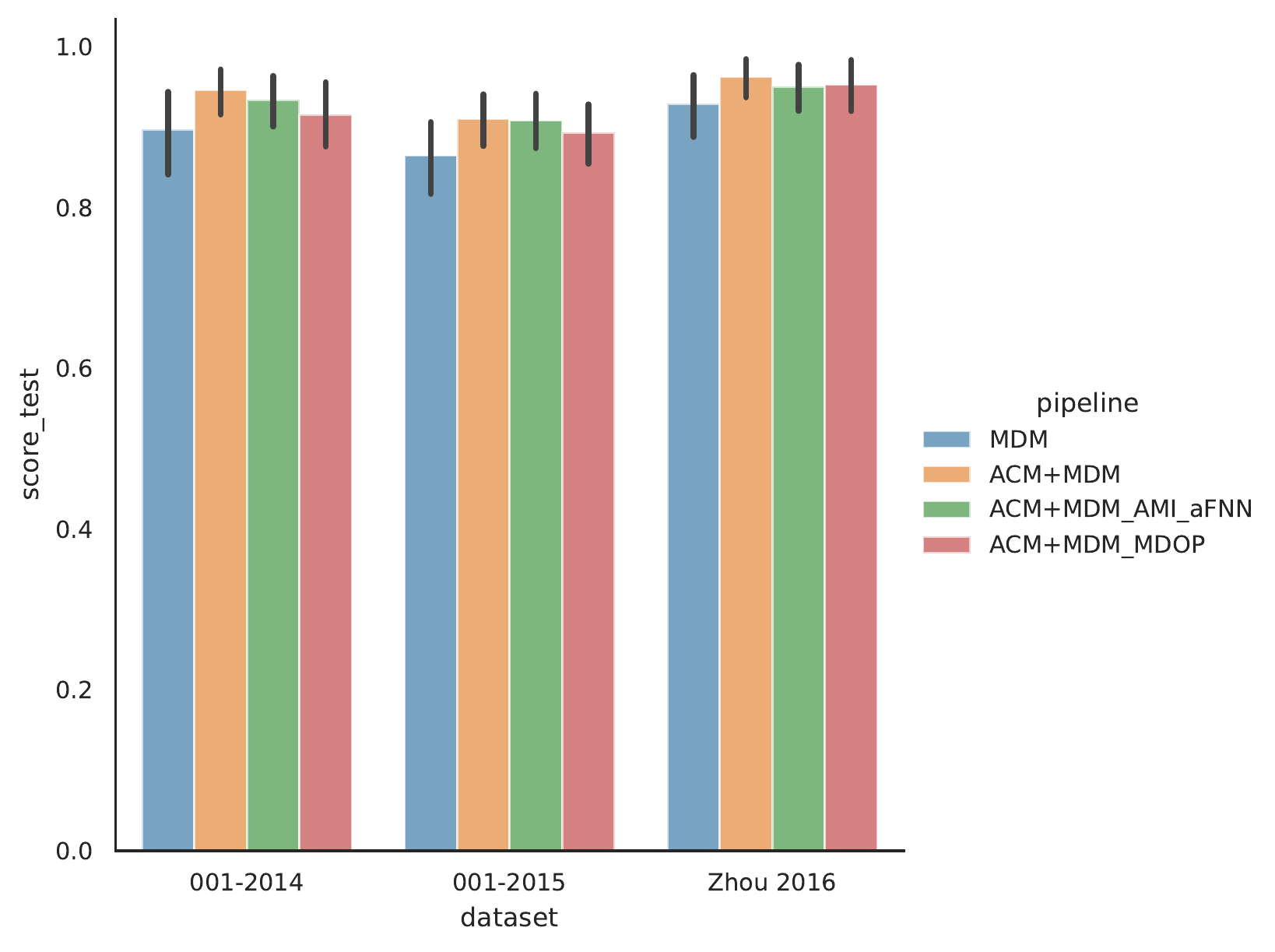}}
    \\
    \subfloat[]{%
        \includegraphics[width=0.5\linewidth]{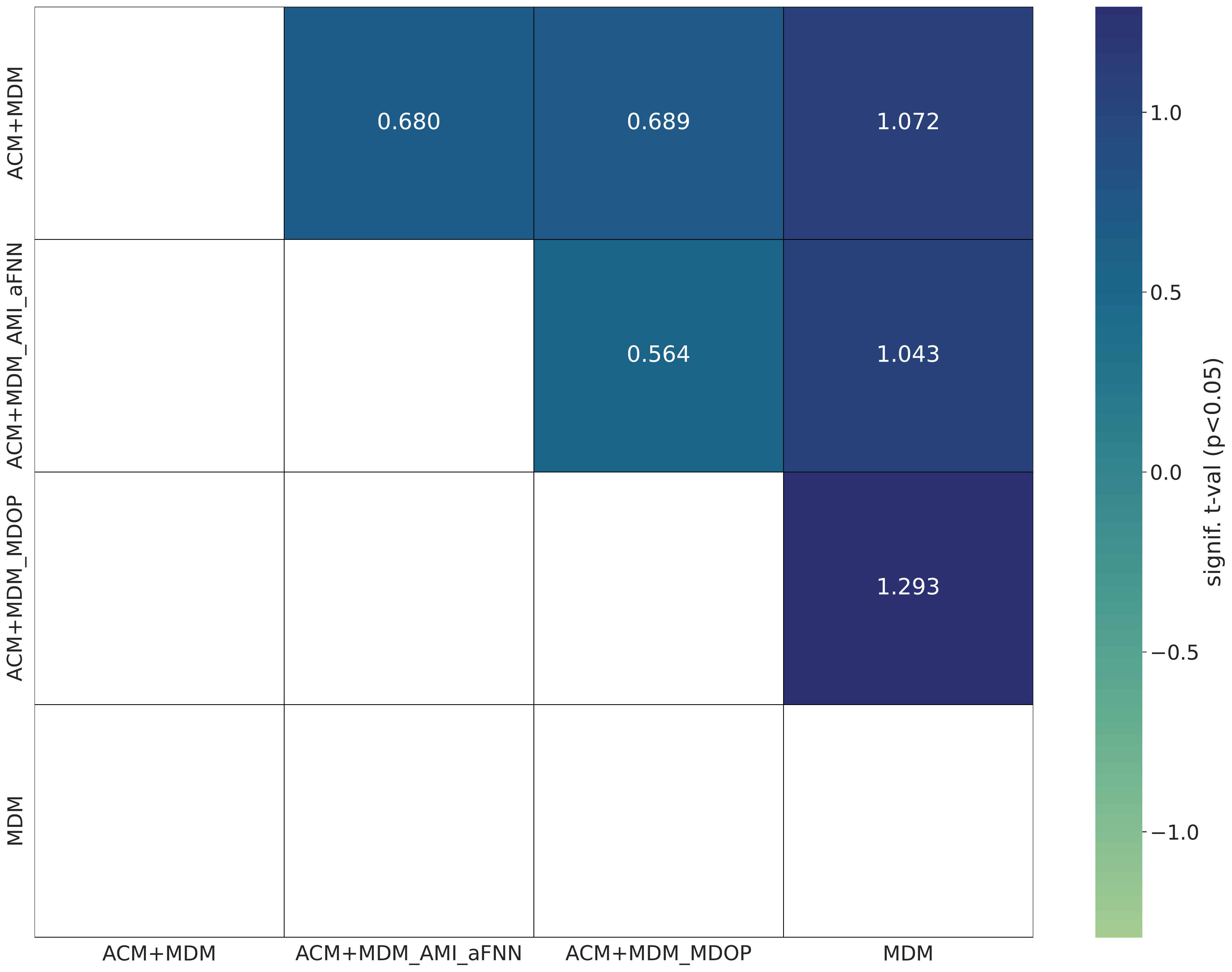}}
        \\
   \subfloat[]{%
            \includegraphics[width=0.30\linewidth]{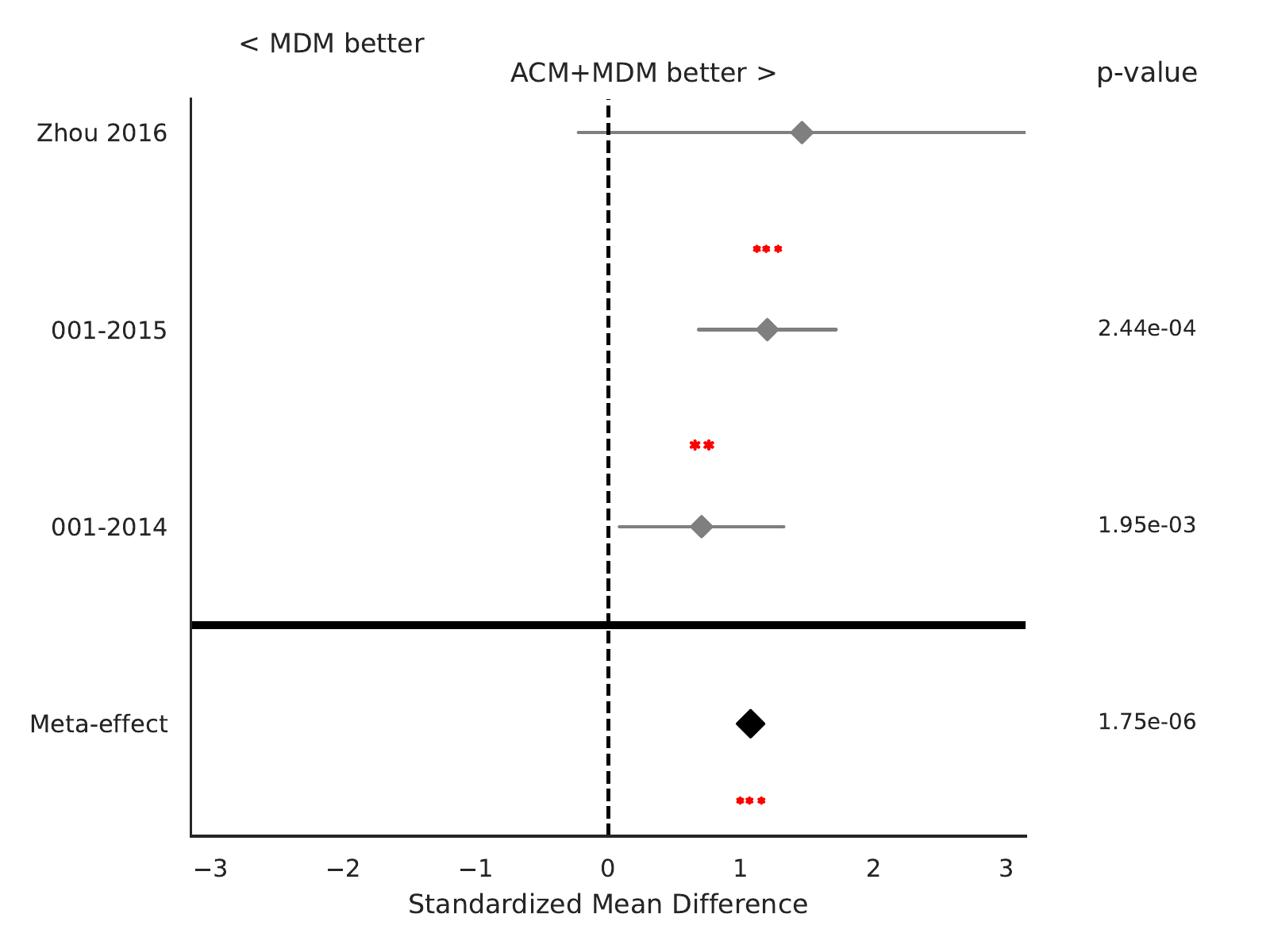}}
            \hfill
   \subfloat[]{%
            \includegraphics[width=0.30\linewidth]{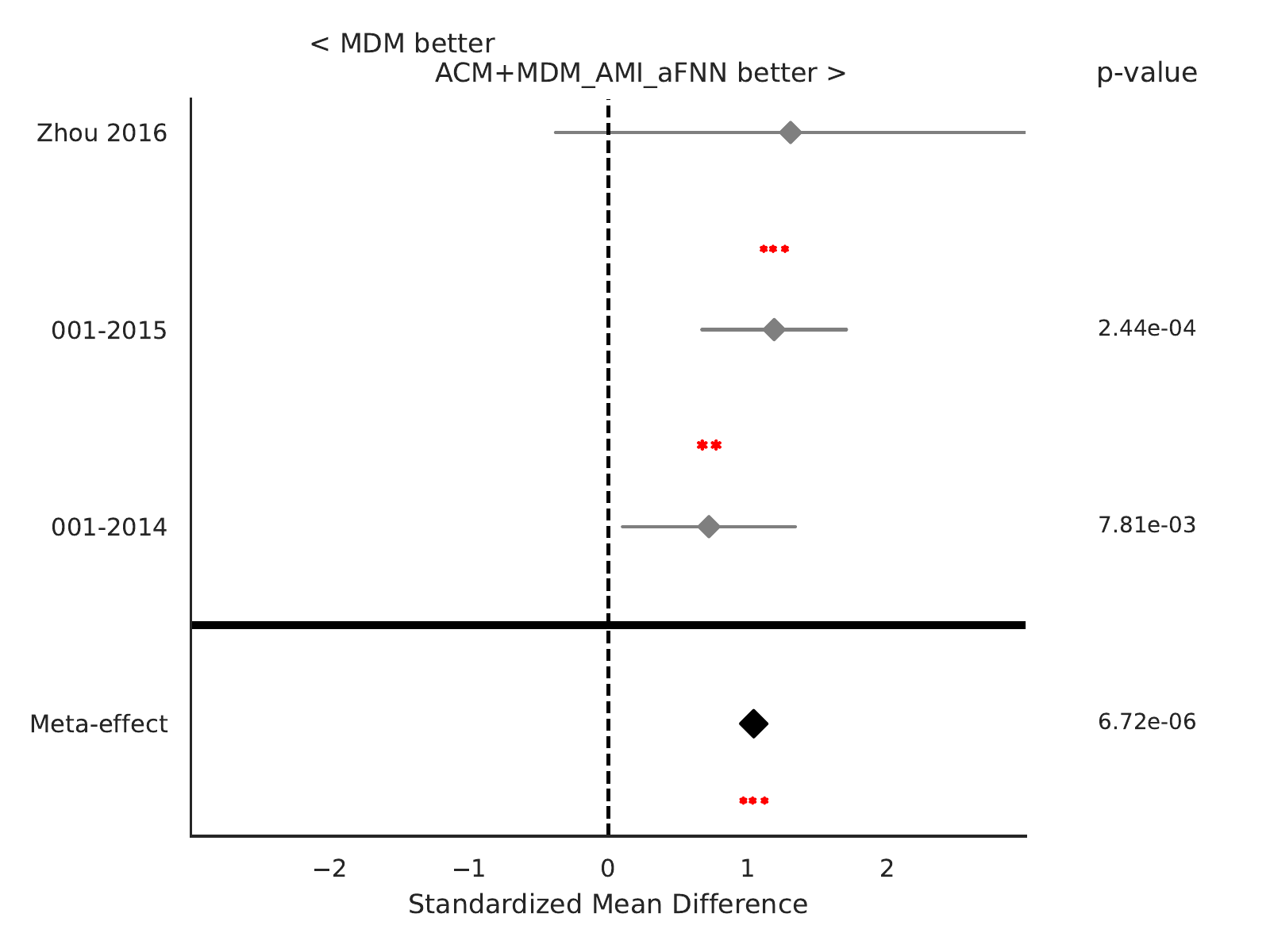}}
            \hfill
   \subfloat[]{%
            \includegraphics[width=0.30\linewidth]{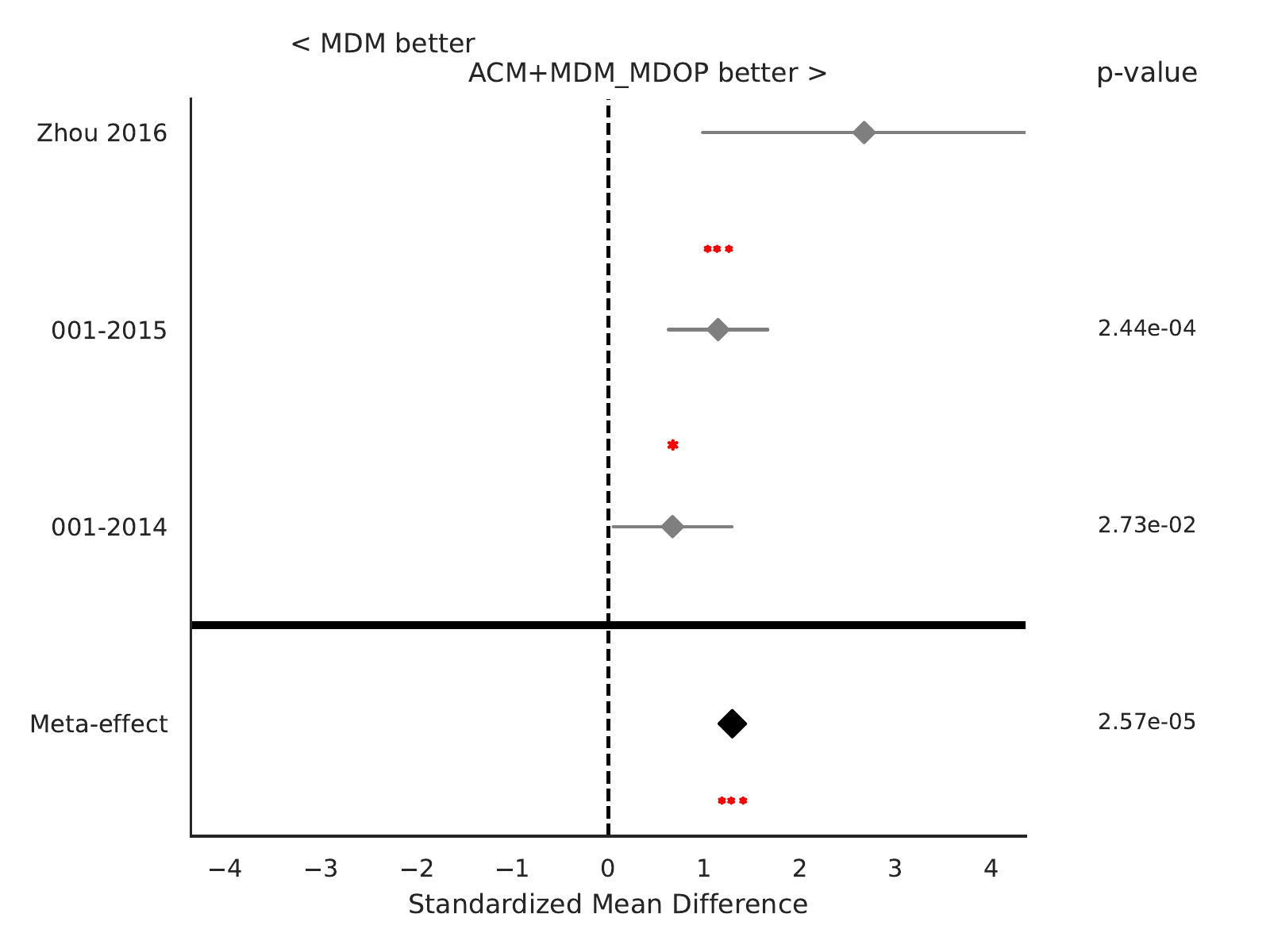}}
            \hfill

    \caption{Result for right hand vs feet classification using the MDM algorithm, using cross-session evaluation. (a) show the rain clouds plots for each pipeline, showing the distribution of the score of every subject. (b) show the bar plot of the score withe the error of the different pipeline and for every dataset considered. (c) show the meta analysis of the different pipeline considered. This plot the significance that the algorithm on the y-axis is better than the one on the x-axis. The color represents the significance level of the difference of accuracy, in terms of t-values, and we show only the significant interactions ($p < 0.05$). (d) (e) (f) show the meta analysis of the standard MDM algorithm against the augmented covariance method with the selection of the hyper-parameter based on grid search, traditional and unified Takens approach respectively. We show the standardized mean differences, while p-values are computed as one-tailed Wilcoxon signed-rank test for the hypothesis given as title of the plot and the gray bar  denote $95\%$ interval. Here, * stands for $p < 0.05$, ** for $p < 0.01$, and *** for $p < 0.001$.
    }
    \label{fig:MDM-rf-crosssession}
\end{figure*}

\begin{figure*}[ht]  
    \centering
    \centering
     \subfloat[]{%
            \includegraphics[width=0.45\linewidth]{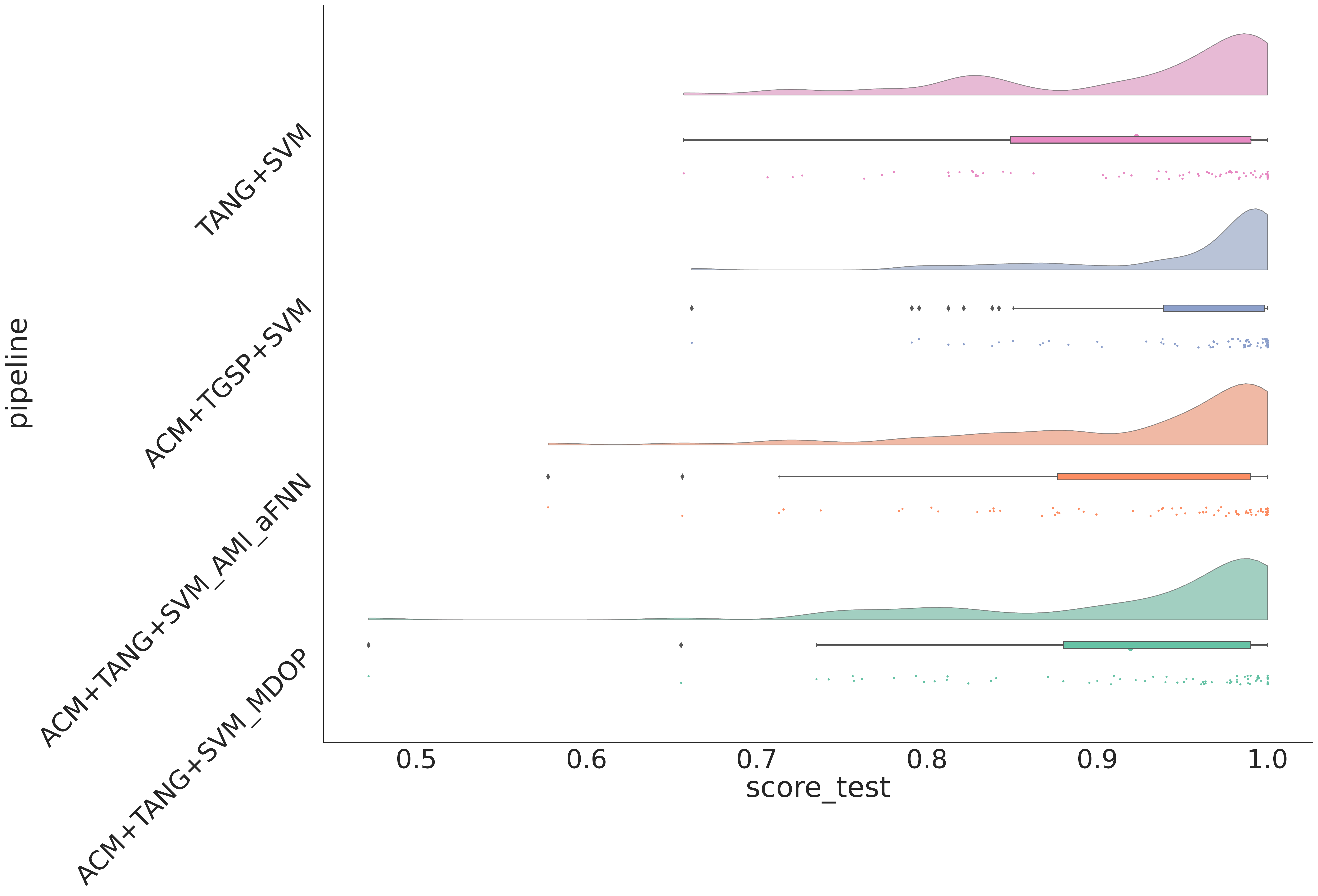}}
            \hfill
     \subfloat[]{%
            \includegraphics[width=0.45\linewidth]{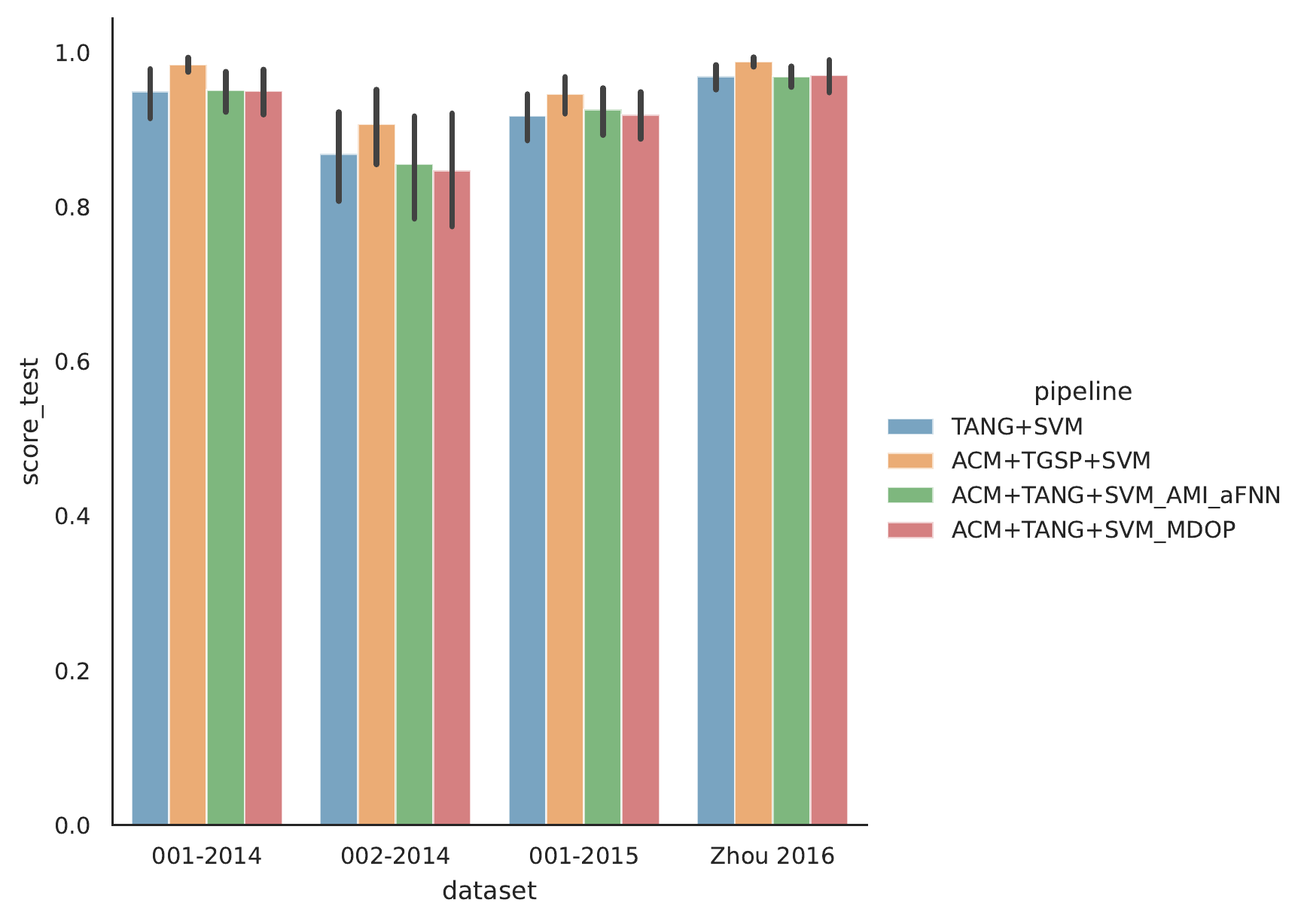}}
    \\
    \subfloat[]{%
        \includegraphics[width=0.5\linewidth]{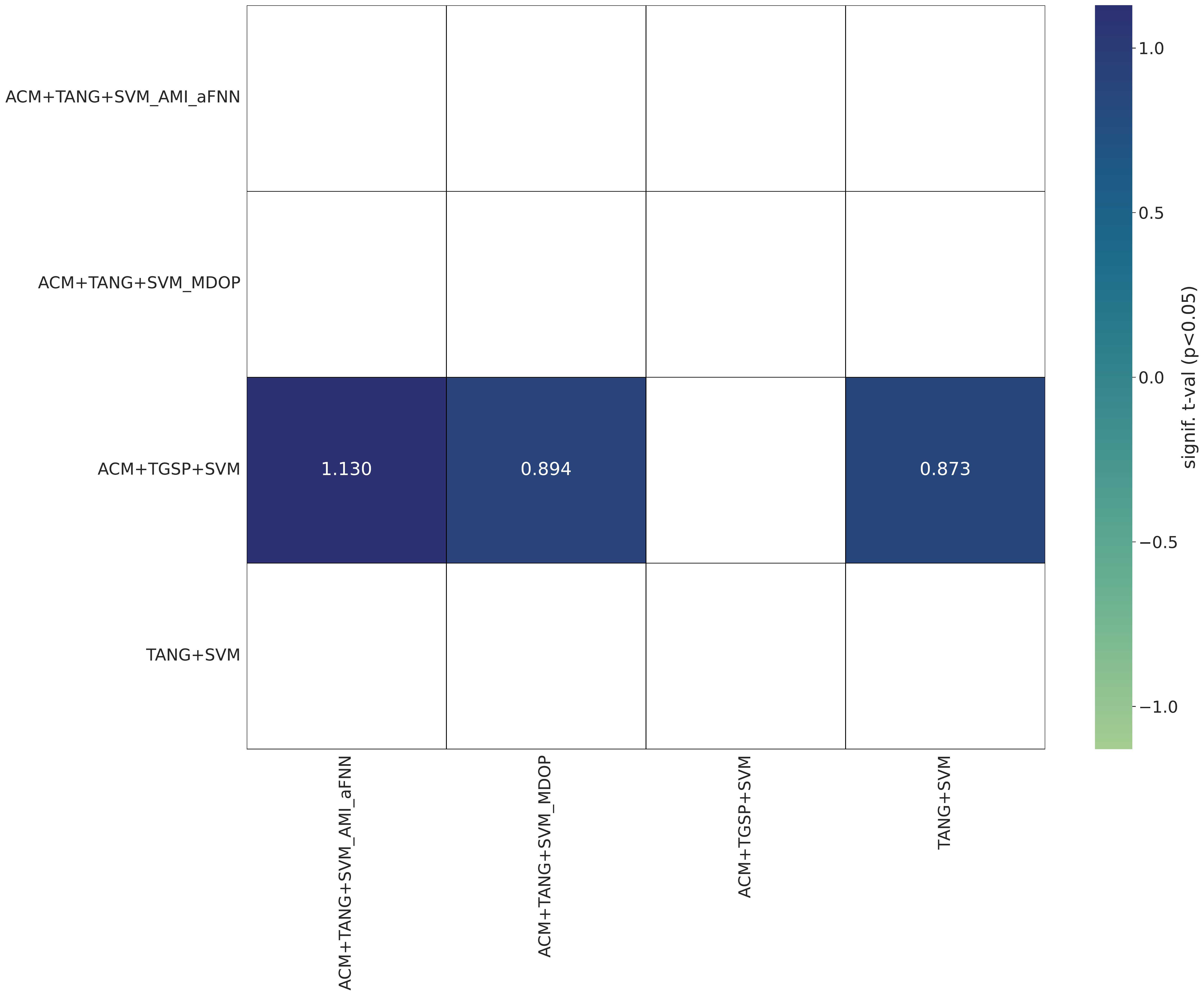}}
        \\
   \subfloat[]{%
            \includegraphics[width=0.30\linewidth]{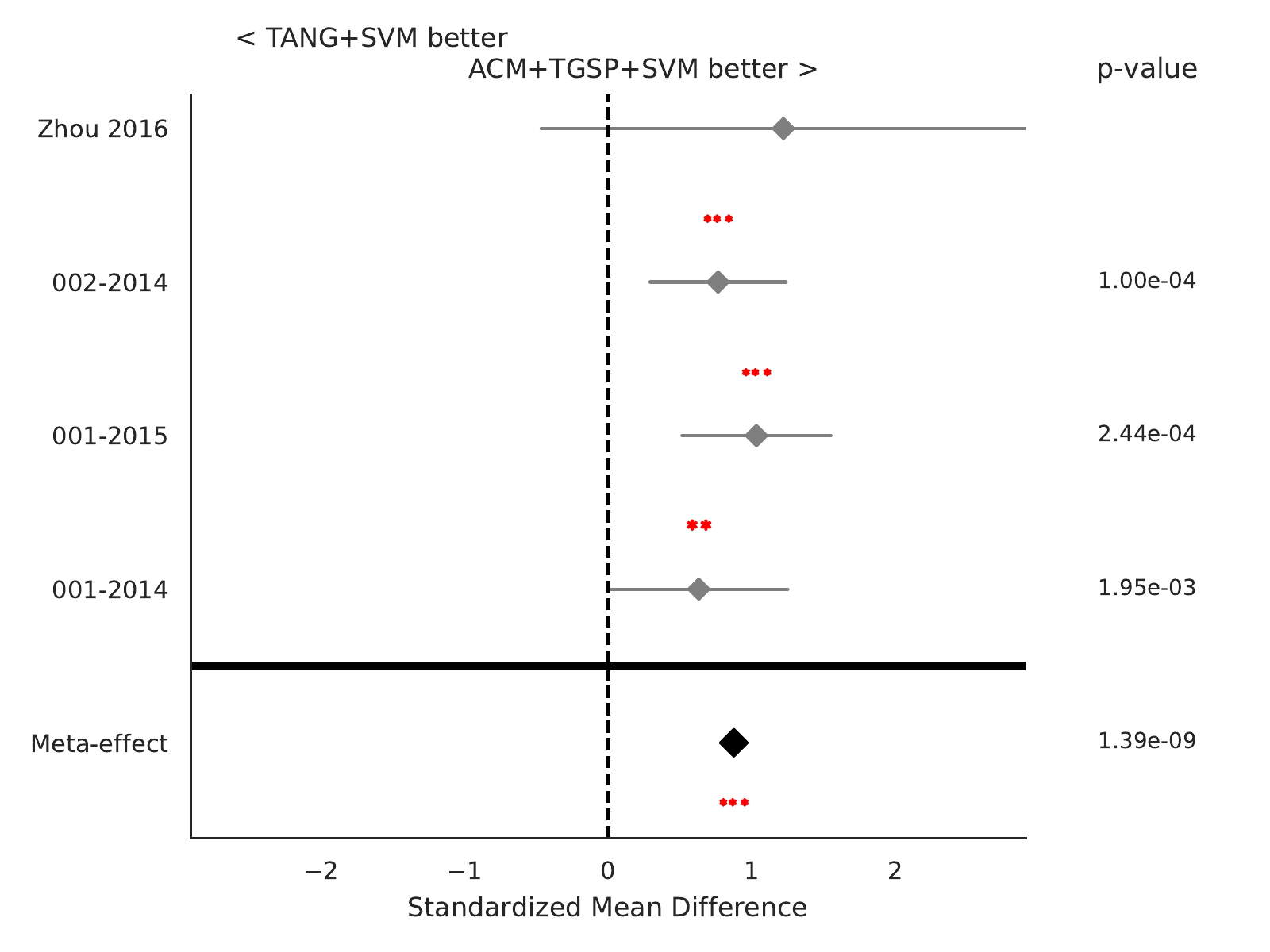}}
            \hfill
   \subfloat[]{%
            \includegraphics[width=0.30\linewidth]{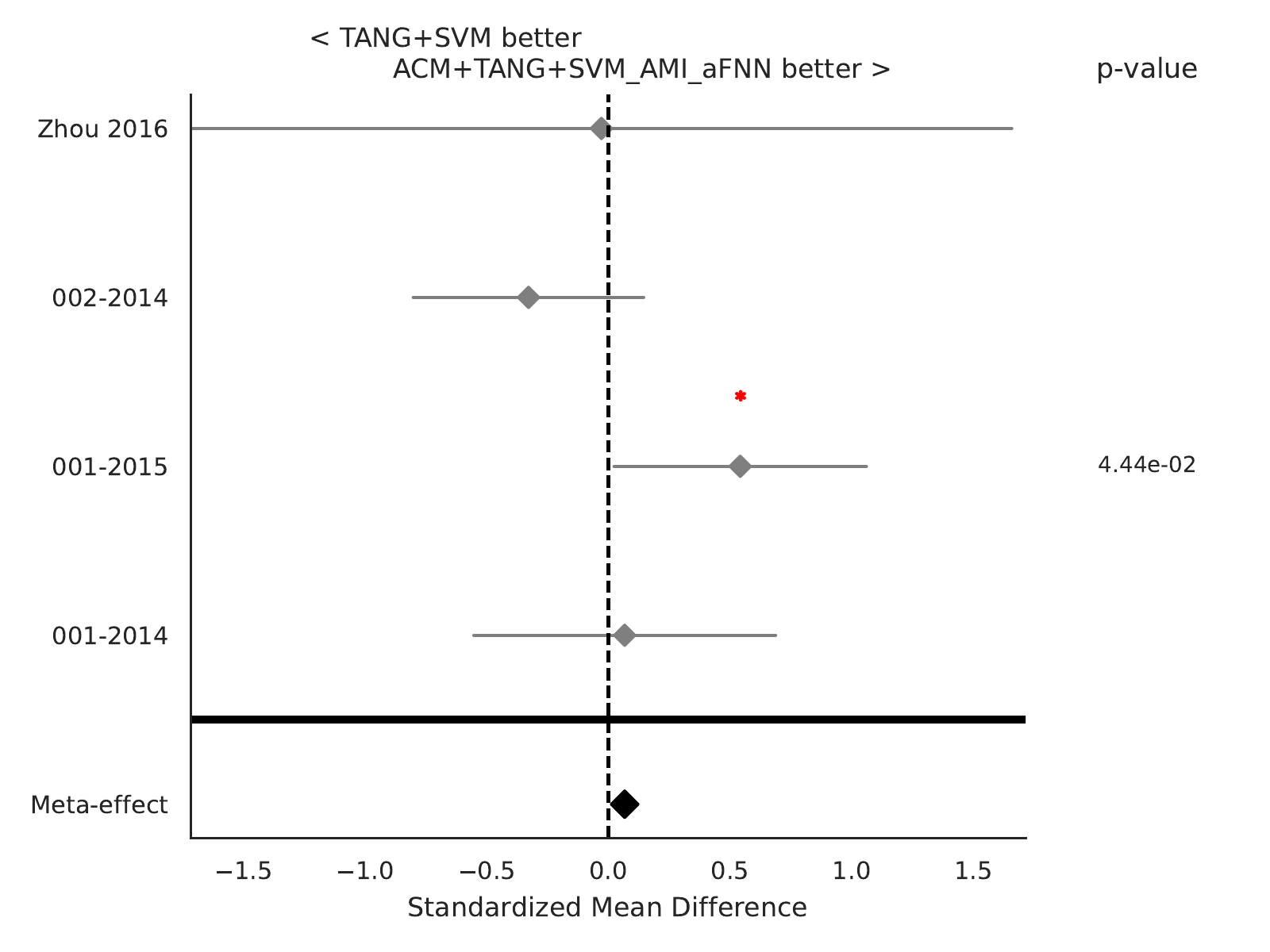}}
            \hfill
   \subfloat[]{%
            \includegraphics[width=0.30\linewidth]{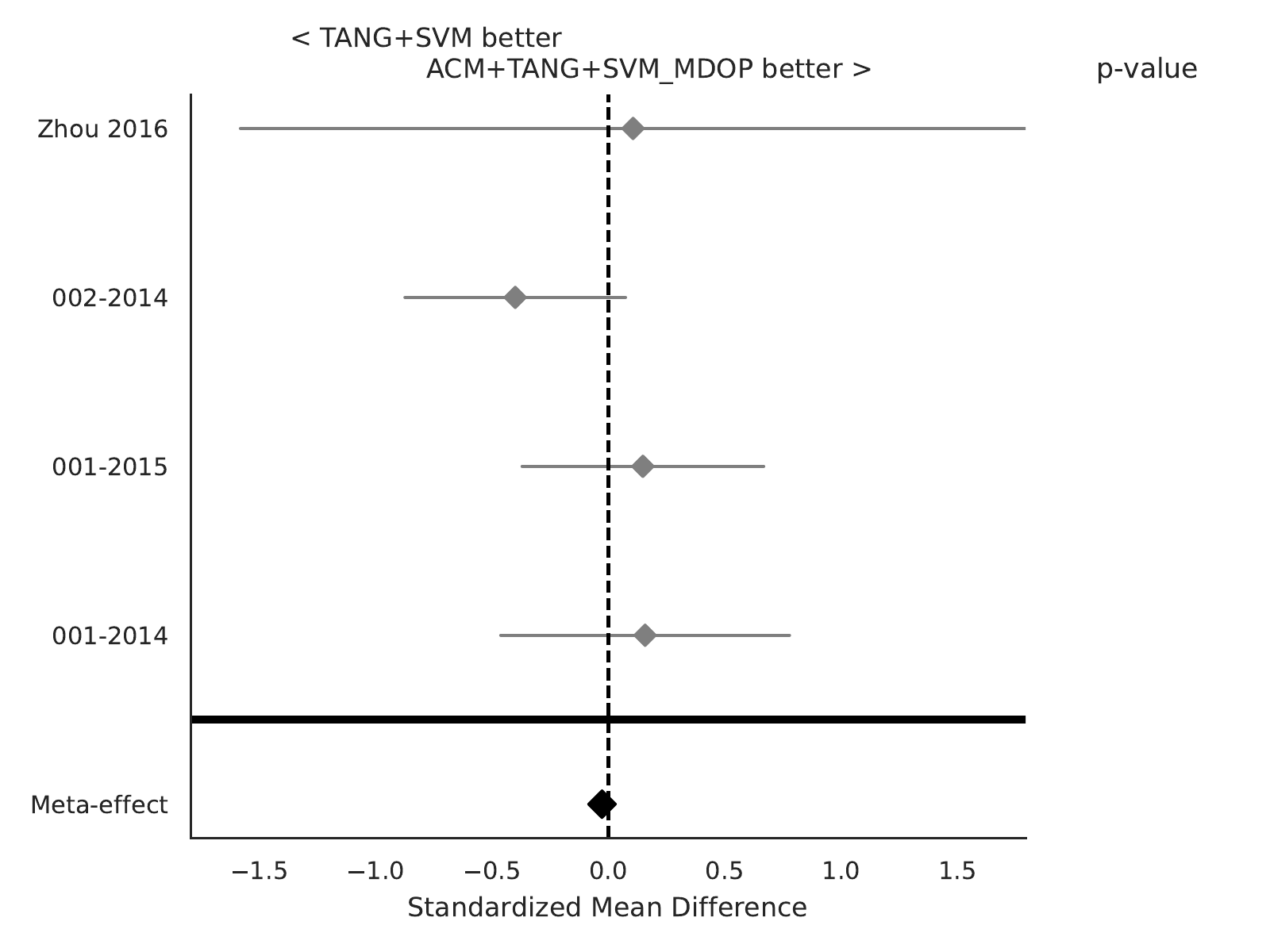}}
            \hfill
            
    \caption{Result for right hand vs feet classification using the TANG algorithm, using withing-session evaluation. (a) show the rain clouds plots for each pipeline, showing the distribution of the score of every subject. (b) show the bar plot of the score withe the error of the different pipeline and for every dataset considered. (c) show the meta analysis of the different pipeline considered. This plot the significance that the algorithm on the y-axis is better than the one on the x-axis. The color represents the significance level of the difference of accuracy, in terms of t-values, and we show only the significant interactions ($p < 0.05$). (d) (e) (f) show the meta analysis of the standard TANG algorithm against the augmented covariance method with the selection of the hyper-parameter based on grid search, traditional and unified Takens approach respectively. We show the standardized mean differences, while p-values are computed as one-tailed Wilcoxon signed-rank test for the hypothesis given as title of the plot and the gray bar  denote $95\%$ interval. Here, * stands for $p < 0.05$, ** for $p < 0.01$, and *** for $p < 0.001$.
    }
    \label{fig:TANG+SVM-rf-whithinsession}
\end{figure*}

\begin{figure*}[ht]  
    \centering
    \centering
     \subfloat[]{%
            \includegraphics[width=0.45\linewidth]{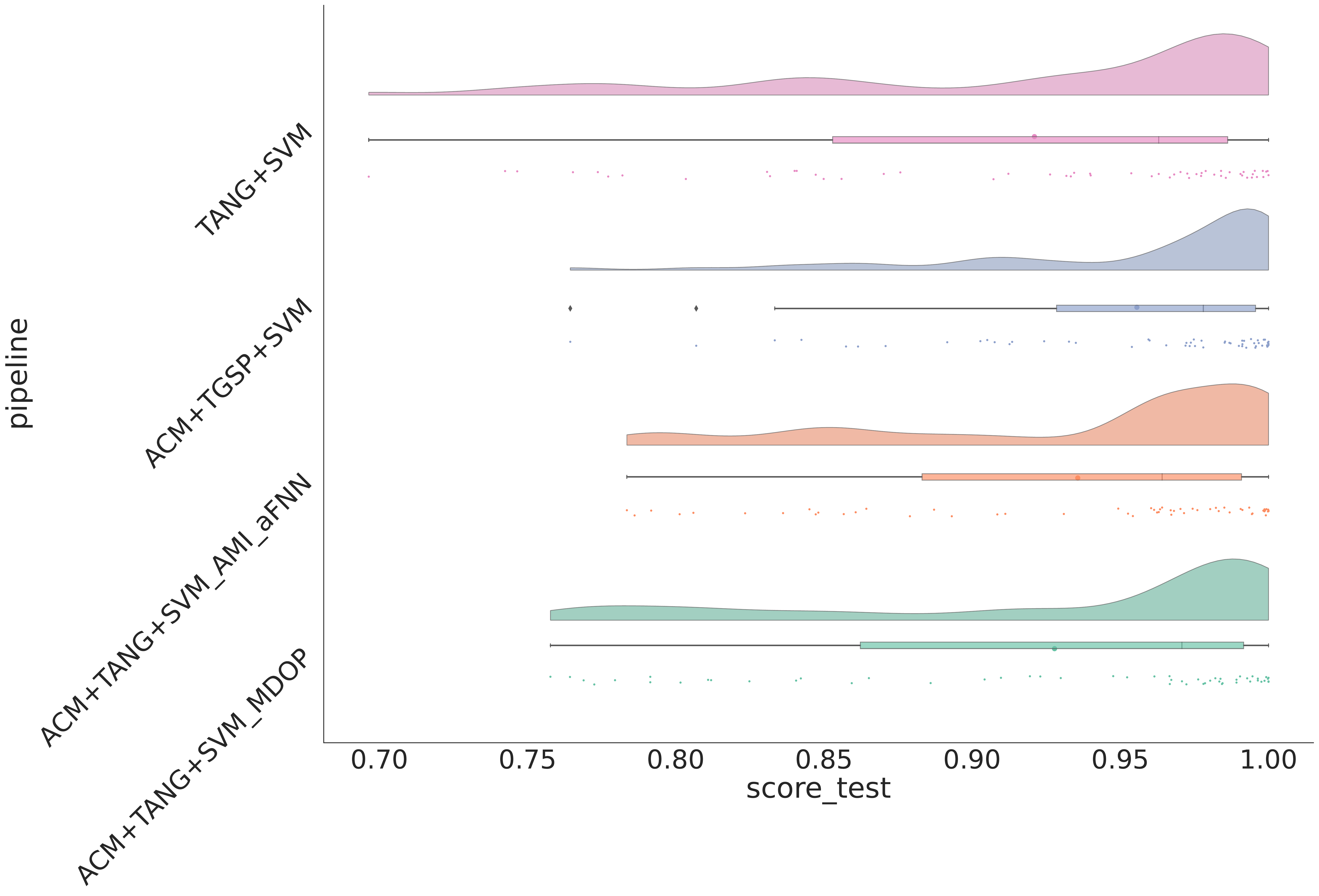}}
            \hfill
     \subfloat[]{%
            \includegraphics[width=0.45\linewidth]{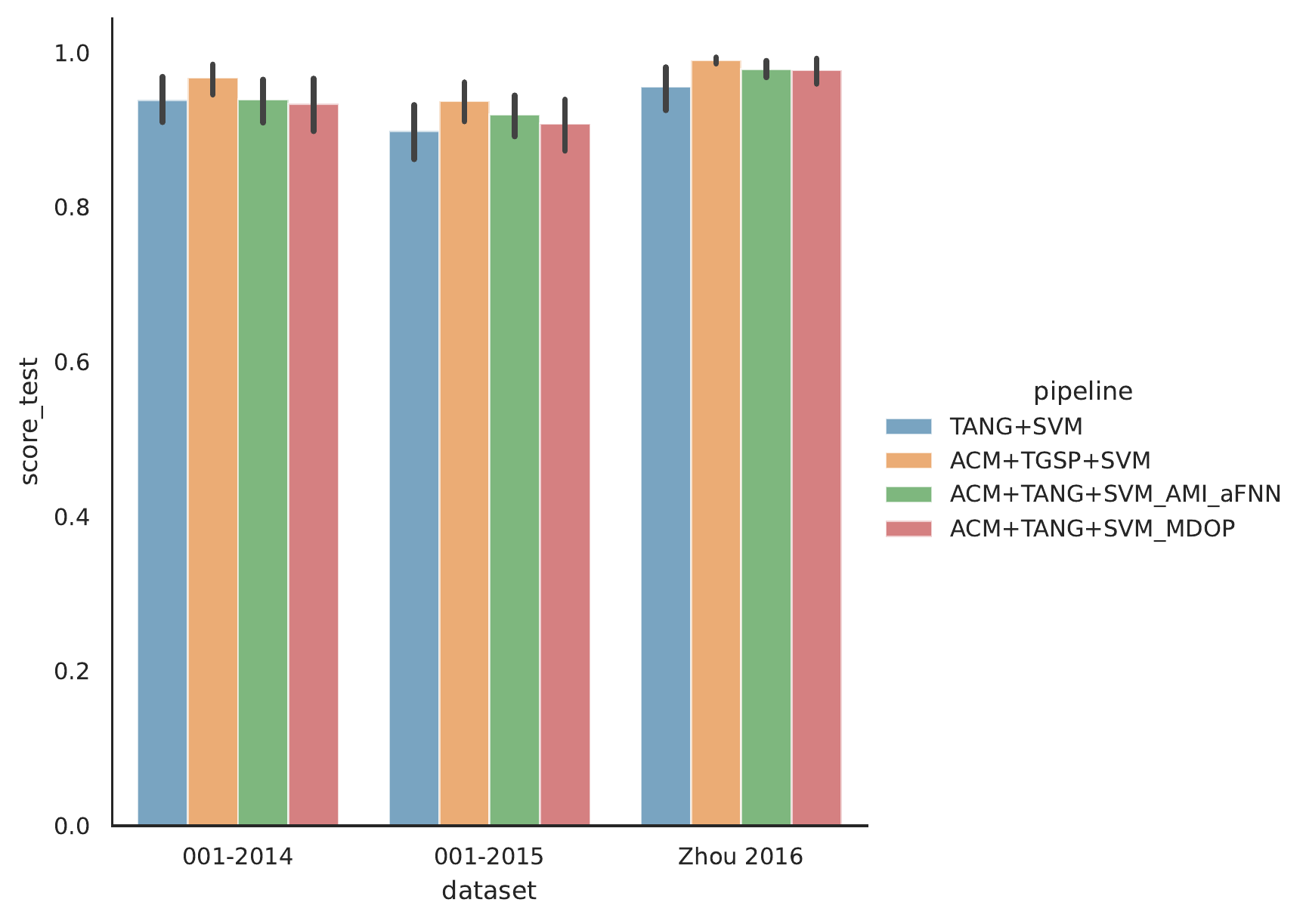}}
    \\
    \subfloat[]{%
        \includegraphics[width=0.5\linewidth]{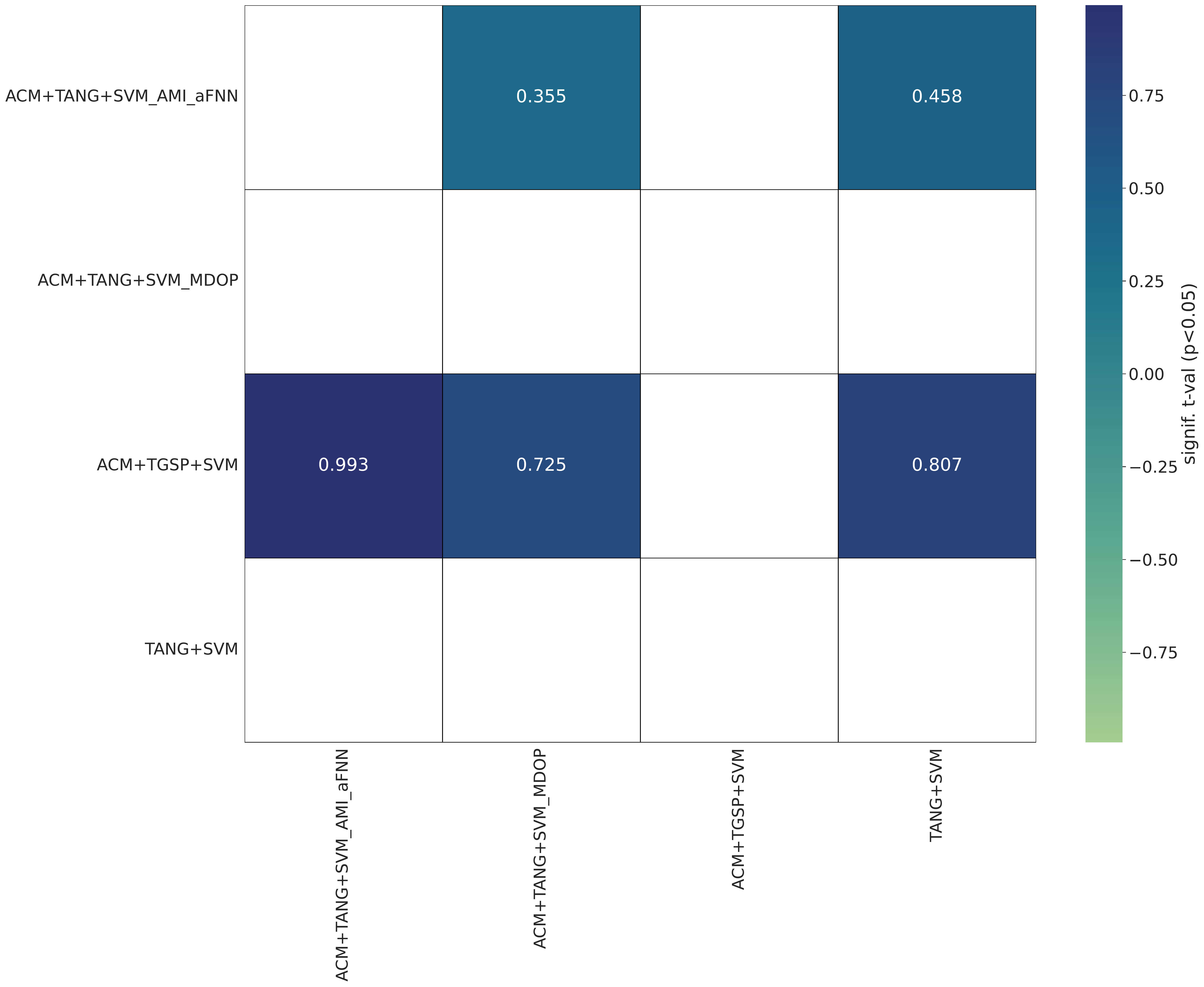}}
        \\
   \subfloat[]{%
            \includegraphics[width=0.30\linewidth]{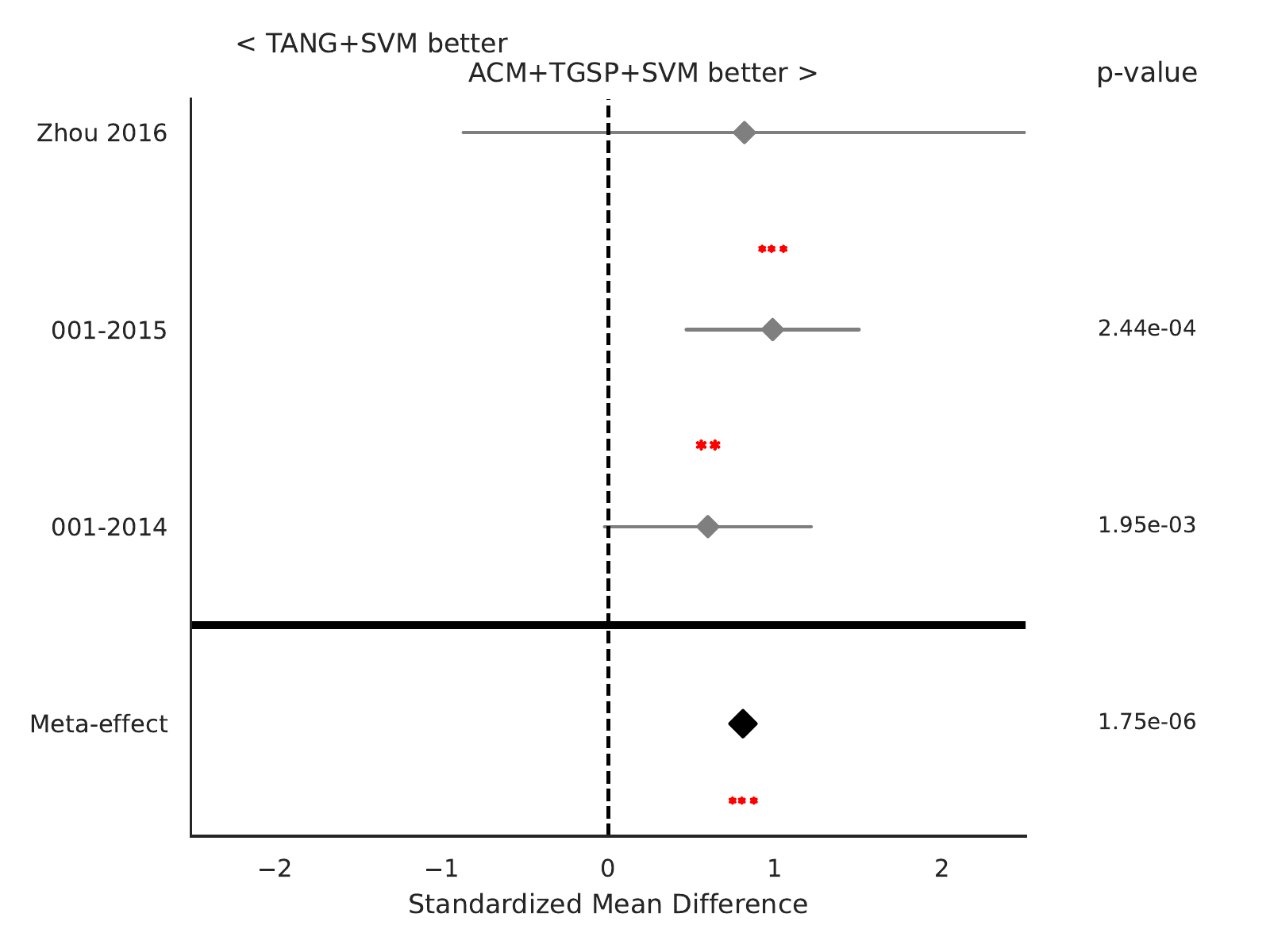}}
            \hfill
   \subfloat[]{%
            \includegraphics[width=0.30\linewidth]{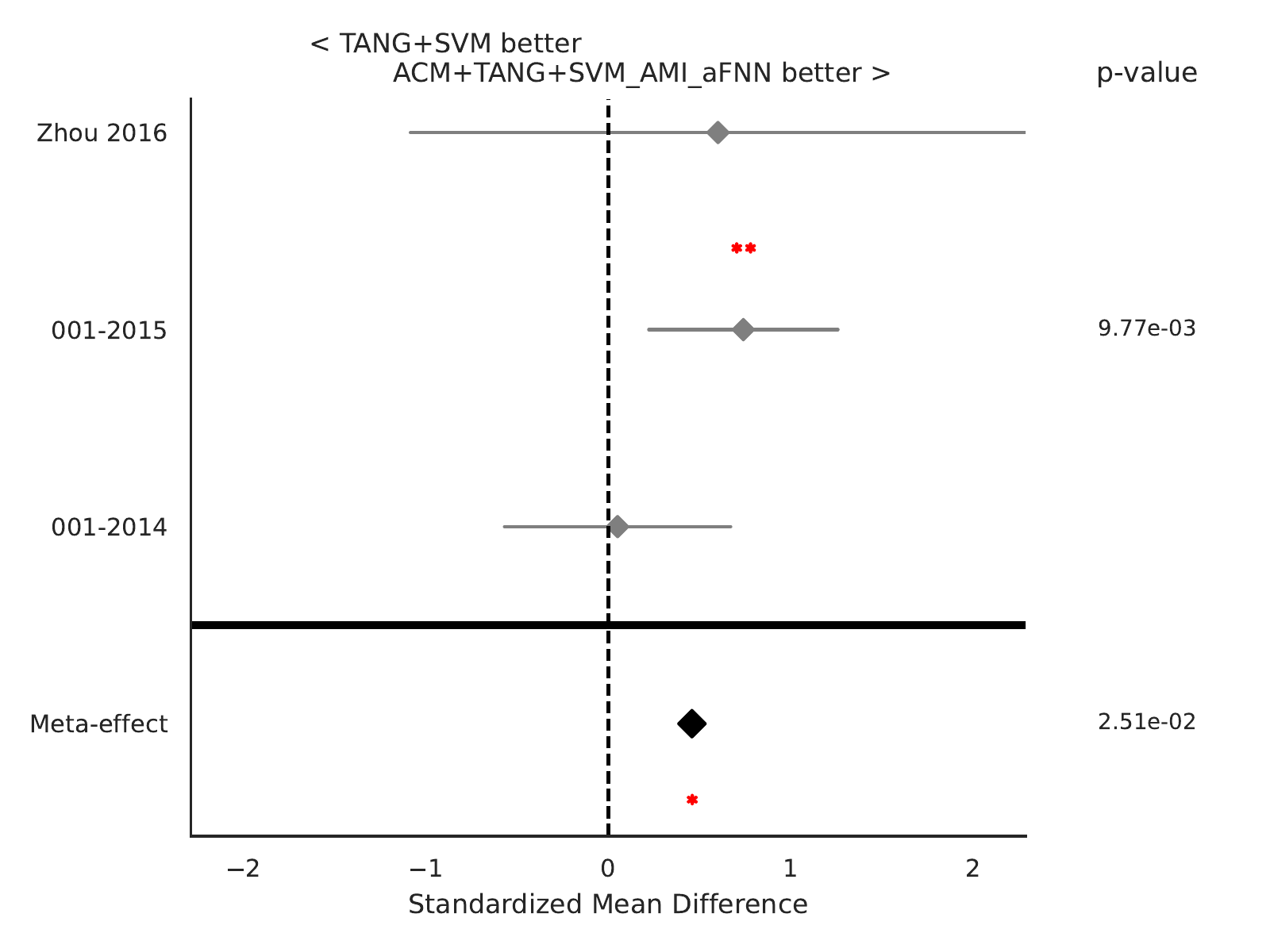}}
            \hfill
   \subfloat[]{%
            \includegraphics[width=0.30\linewidth]{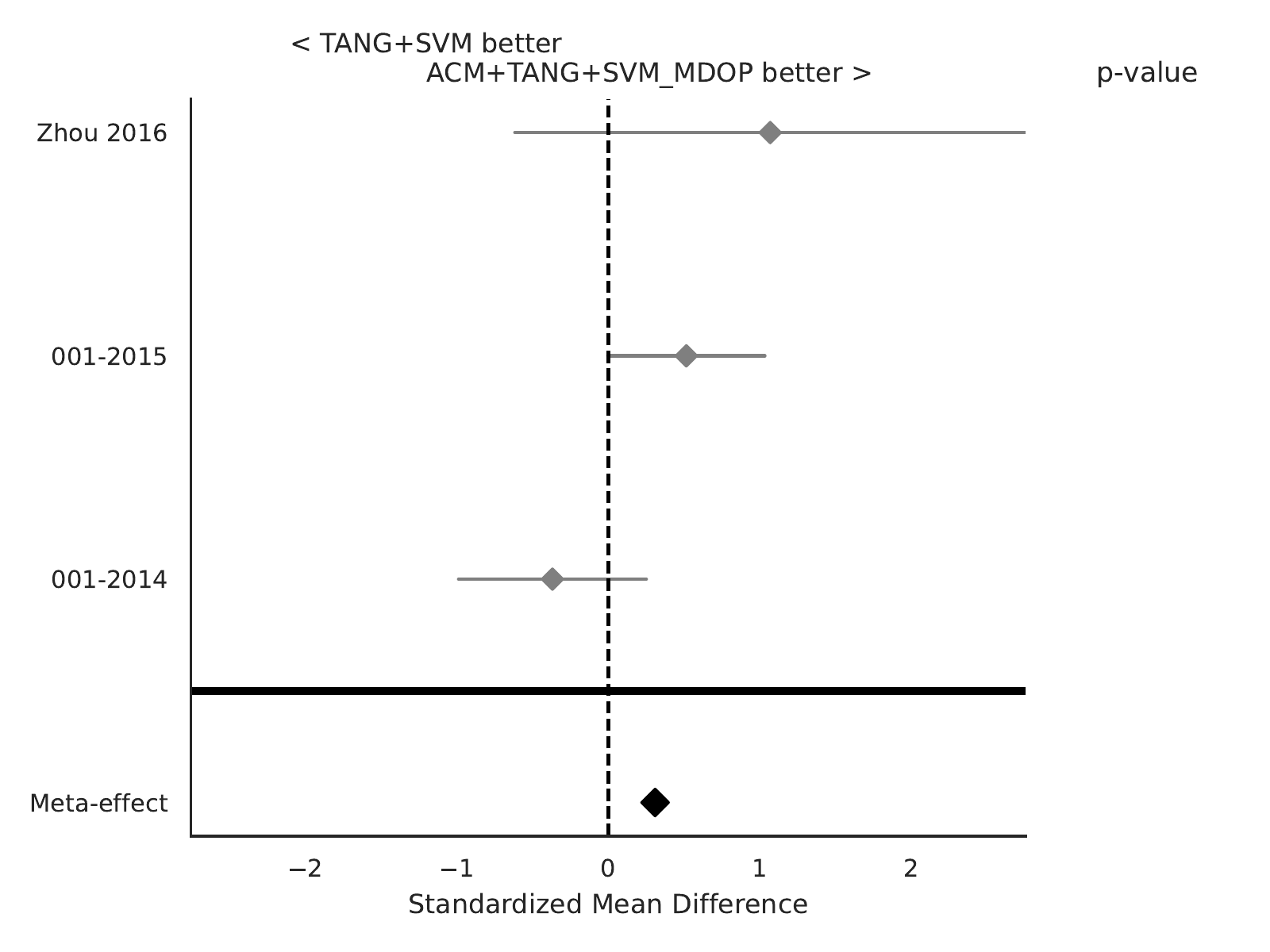}}
            \hfill

    \caption{Result for right hand vs feet classification using the TANG algorithm, using cross-session evaluation. (a) show the rain clouds plots for each pipeline, showing the distribution of the score of every subject. (b) show the bar plot of the score withe the error of the different pipeline and for every dataset considered. (c) show the meta analysis of the different pipeline considered. This plot the significance that the algorithm on the y-axis is better than the one on the x-axis. The color represents the significance level of the difference of accuracy, in terms of t-values, and we show only the significant interactions ($p < 0.05$). (d) (e) (f) show the meta analysis of the standard TANG algorithm against the augmented covariance method with the selection of the hyper-parameter based on grid search, traditional and unified Takens approach respectively. We show the standardized mean differences, while p-values are computed as one-tailed Wilcoxon signed-rank test for the hypothesis given as title of the plot and the gray bar  denote $95\%$ interval. Here, * stands for $p < 0.05$, ** for $p < 0.01$, and *** for $p < 0.001$.
    }
    \label{fig:TANG+SVM-rf-crosssession}
\end{figure*}

\begin{figure*}[ht]  
    \centering
    \centering
     \subfloat[]{%
            \includegraphics[width=0.45\linewidth]{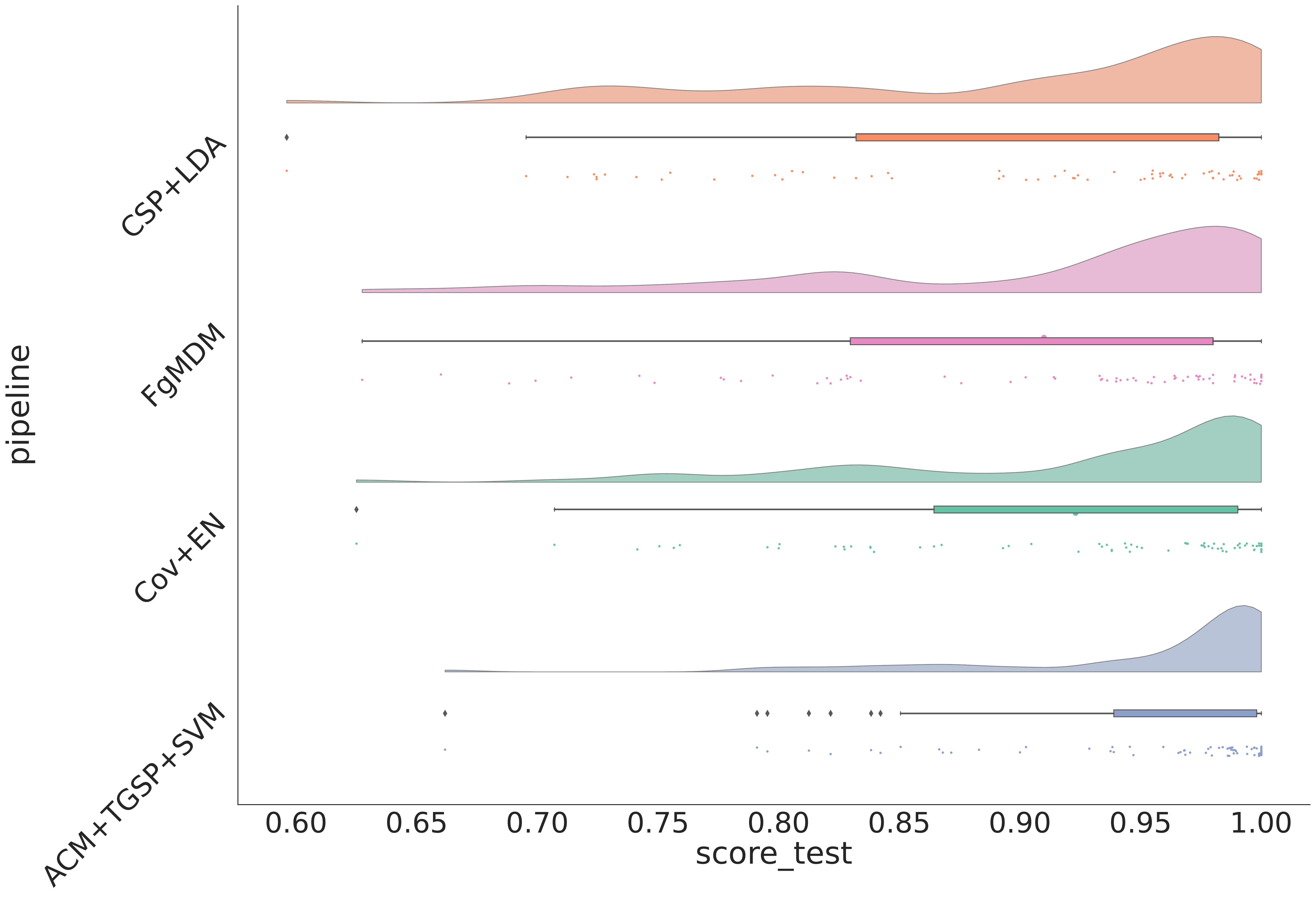}}
            \hfill
     \subfloat[]{%
            \includegraphics[width=0.45\linewidth]{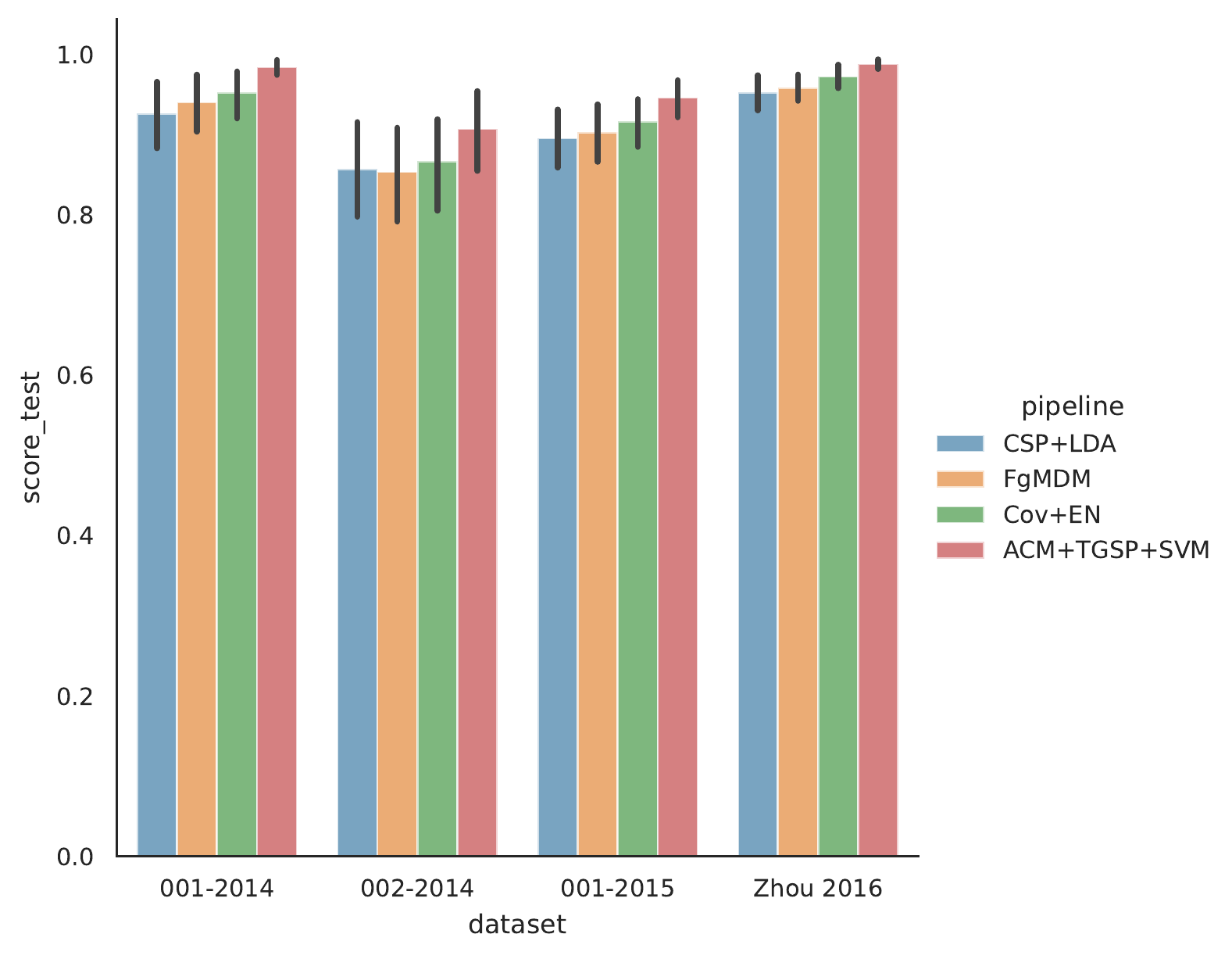}}
    \\
    \subfloat[]{%
        \includegraphics[width=0.5\linewidth]{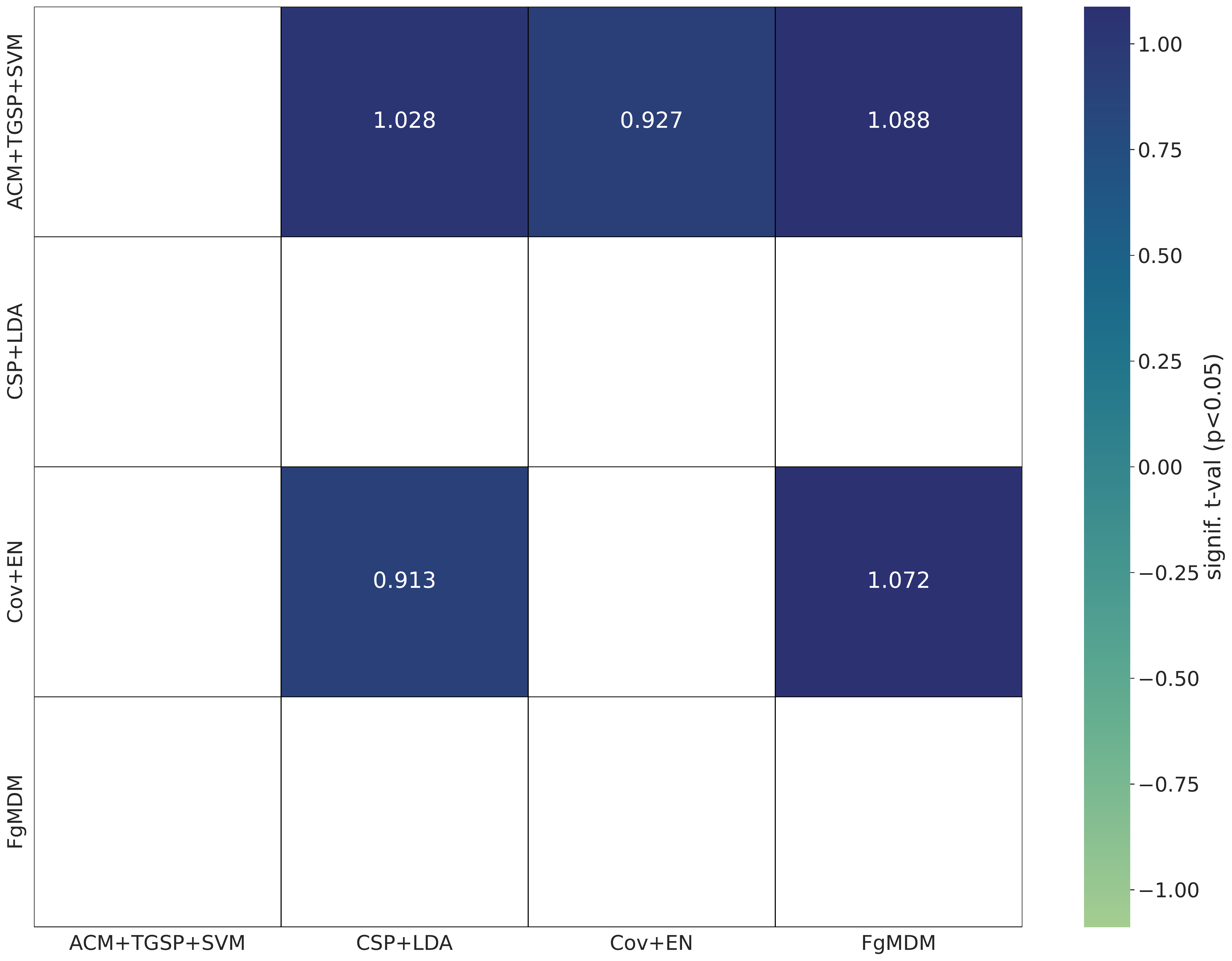}}
        \\
   \subfloat[]{%
            \includegraphics[width=0.30\linewidth]{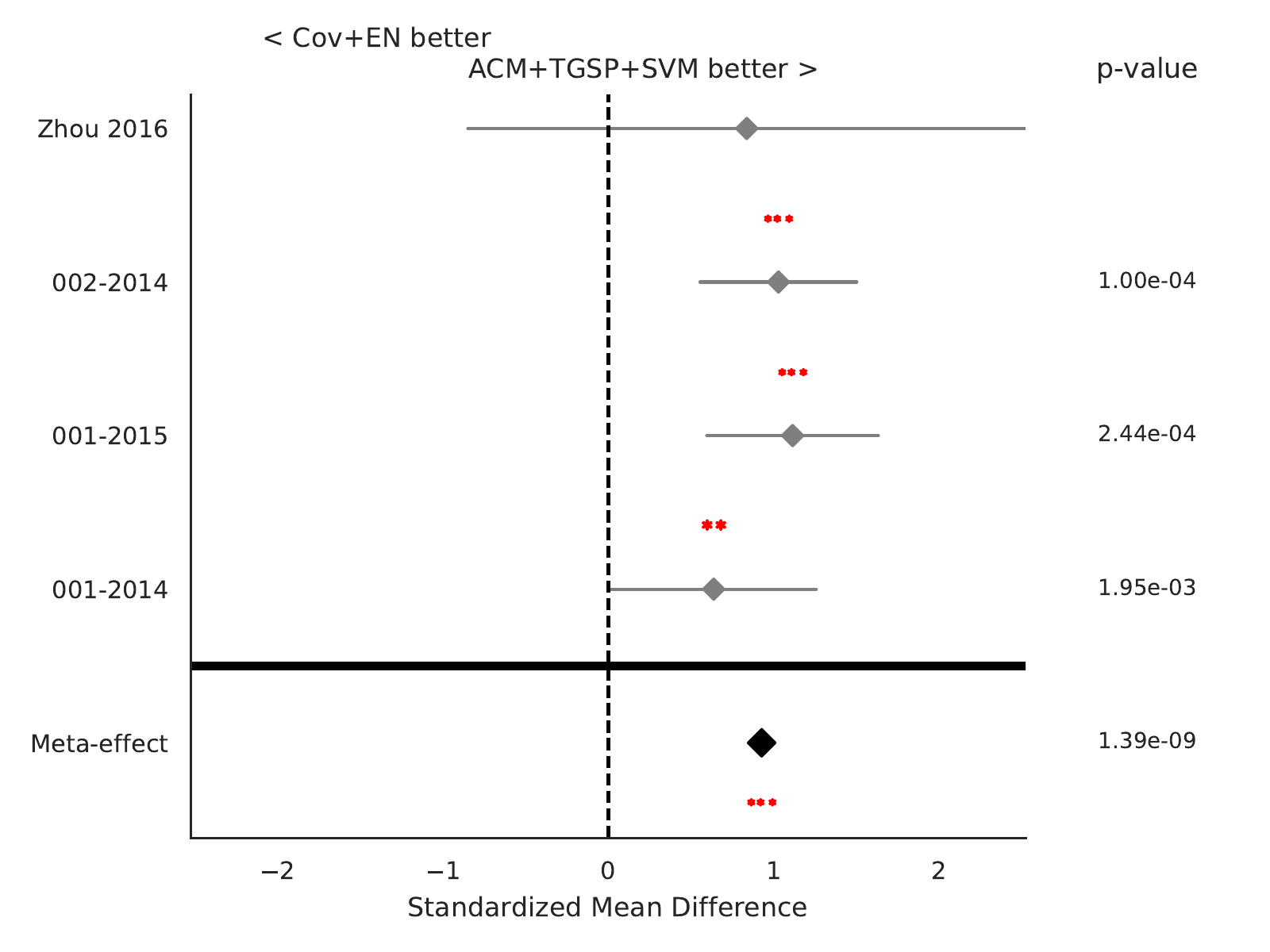}}
            \hfill
   \subfloat[]{%
            \includegraphics[width=0.30\linewidth]{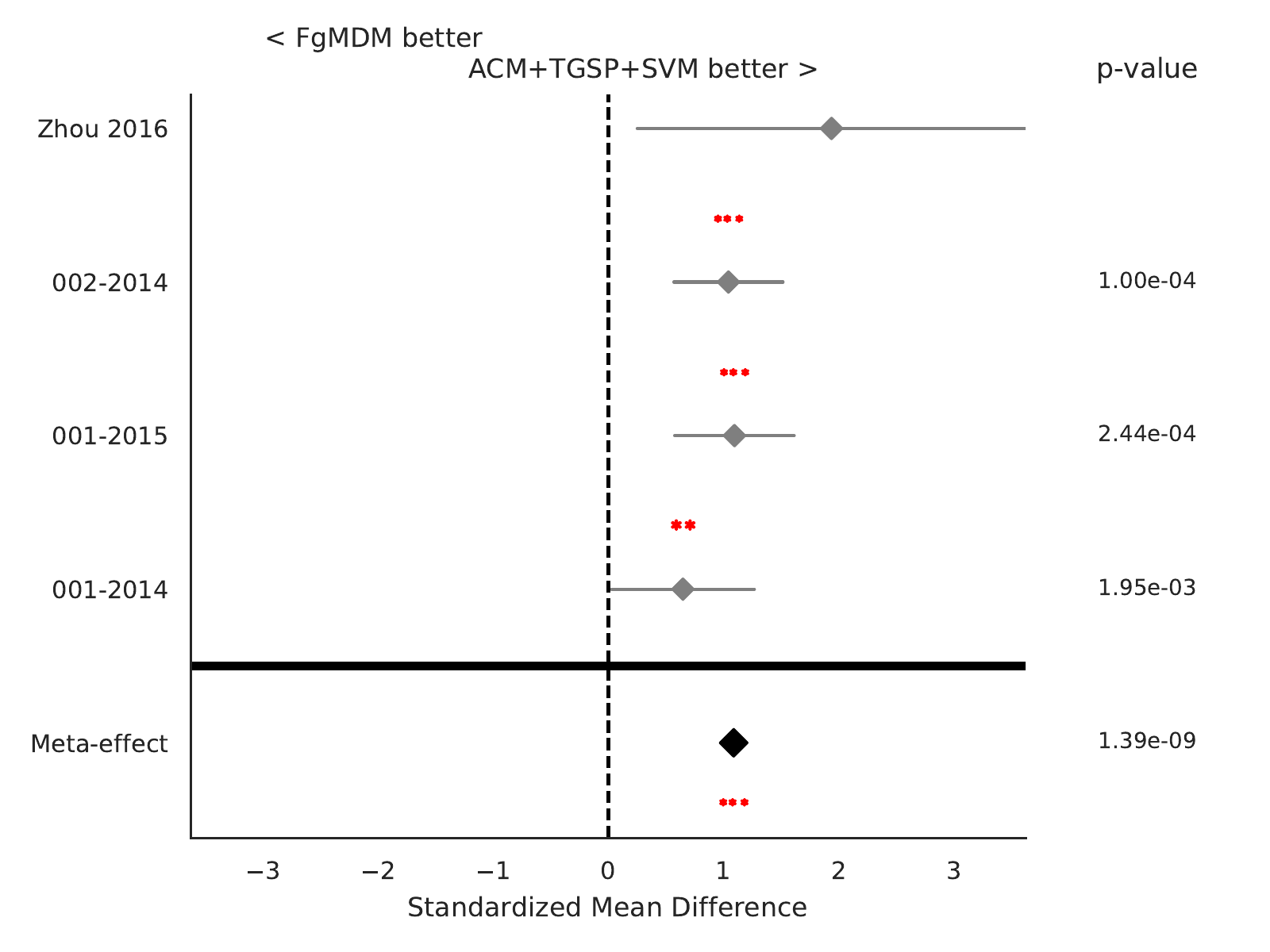}}
            \hfill
   \subfloat[]{%
            \includegraphics[width=0.30\linewidth]{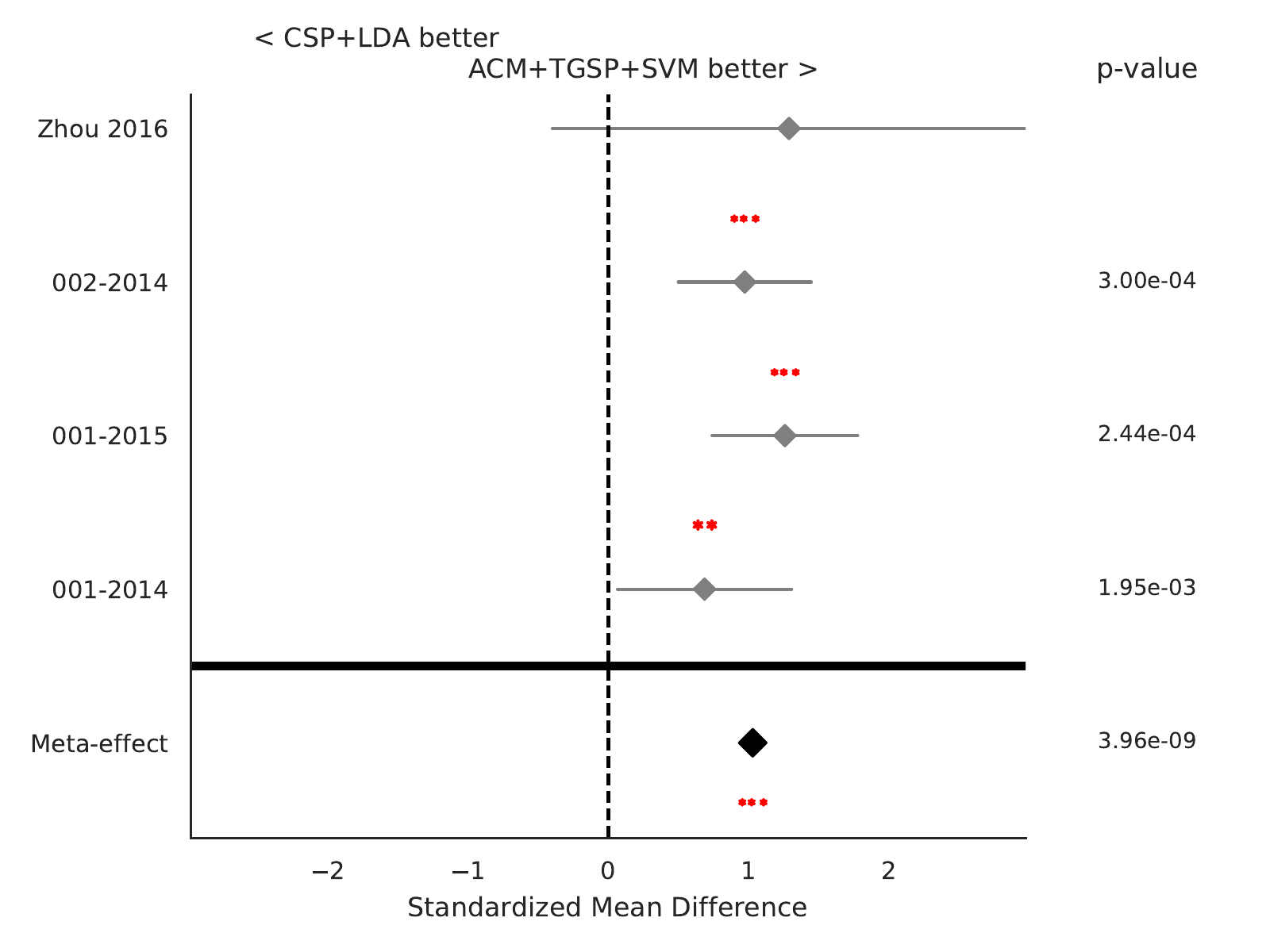}}
            \hfill

    \caption{Result for right hand vs feet classification, using withing-session evaluation. (a) show the rain clouds plots for each pipeline, showing the distribution of the score of every subject. (b) show the bar plot of the score withe the error of the different pipeline and for every dataset considered. (c) show the meta analysis of the different pipeline considered. This plot the significance that the algorithm on the y-axis is better than the one on the x-axis. The color represents the significance level of the difference of accuracy, in terms of t-values, and we show only the significant interactions ($p < 0.05$). (d) (e) (f) show the meta analysis of augmented method with SVM against the state of the art. We show the standardized mean differences, while p-values are computed as one-tailed Wilcoxon signed-rank test for the hypothesis given as title of the plot and the gray bar  denote $95\%$ interval. Here, * stands for $p < 0.05$, ** for $p < 0.01$, and *** for $p < 0.001$.
    }
    \label{fig:TANG+SVM-rf-whithinsession-stateart}
\end{figure*}

\begin{figure*}[ht]  
    \centering
    \centering
     \subfloat[]{%
            \includegraphics[width=0.45\linewidth]{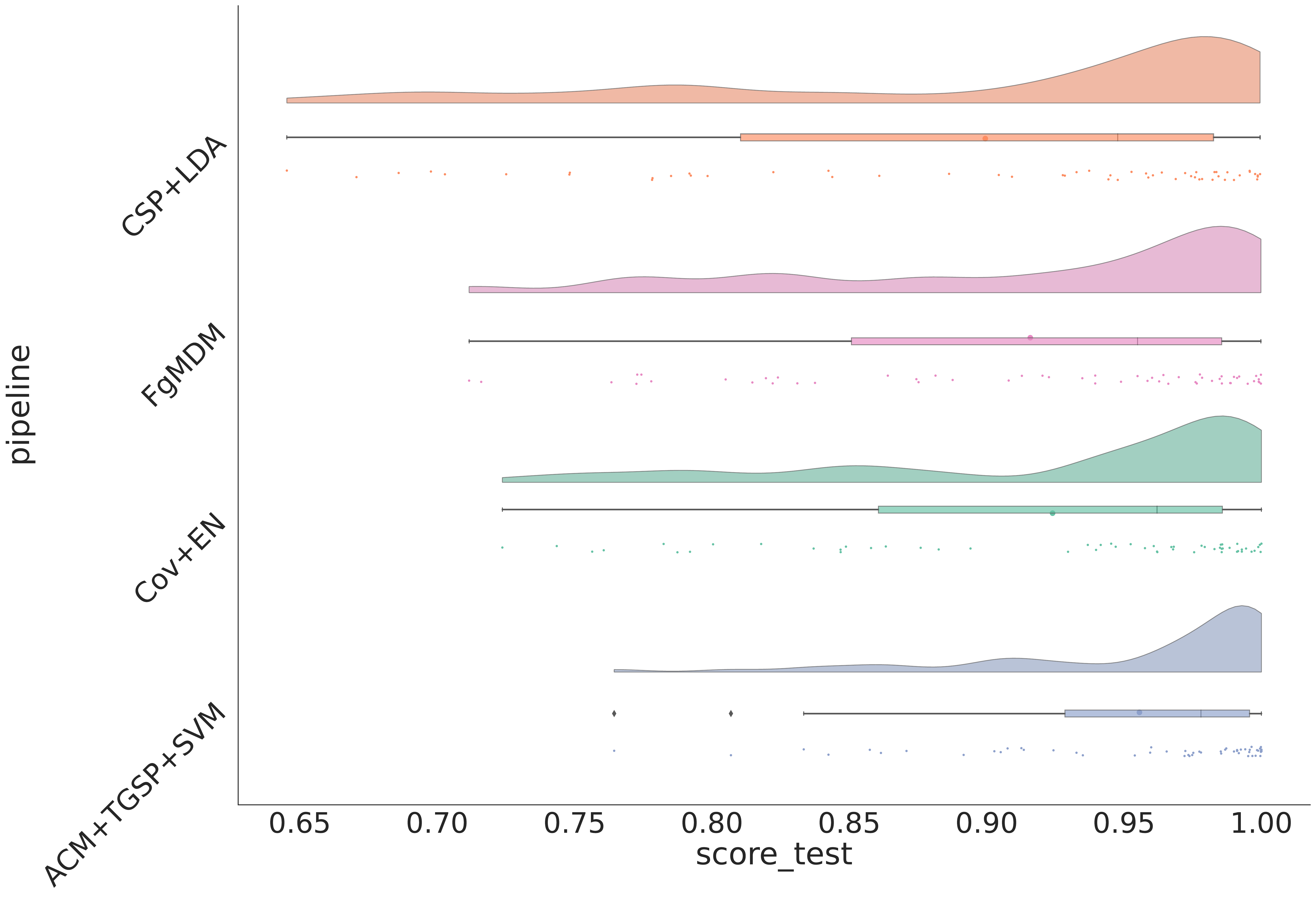}}
            \hfill
     \subfloat[]{%
            \includegraphics[width=0.45\linewidth]{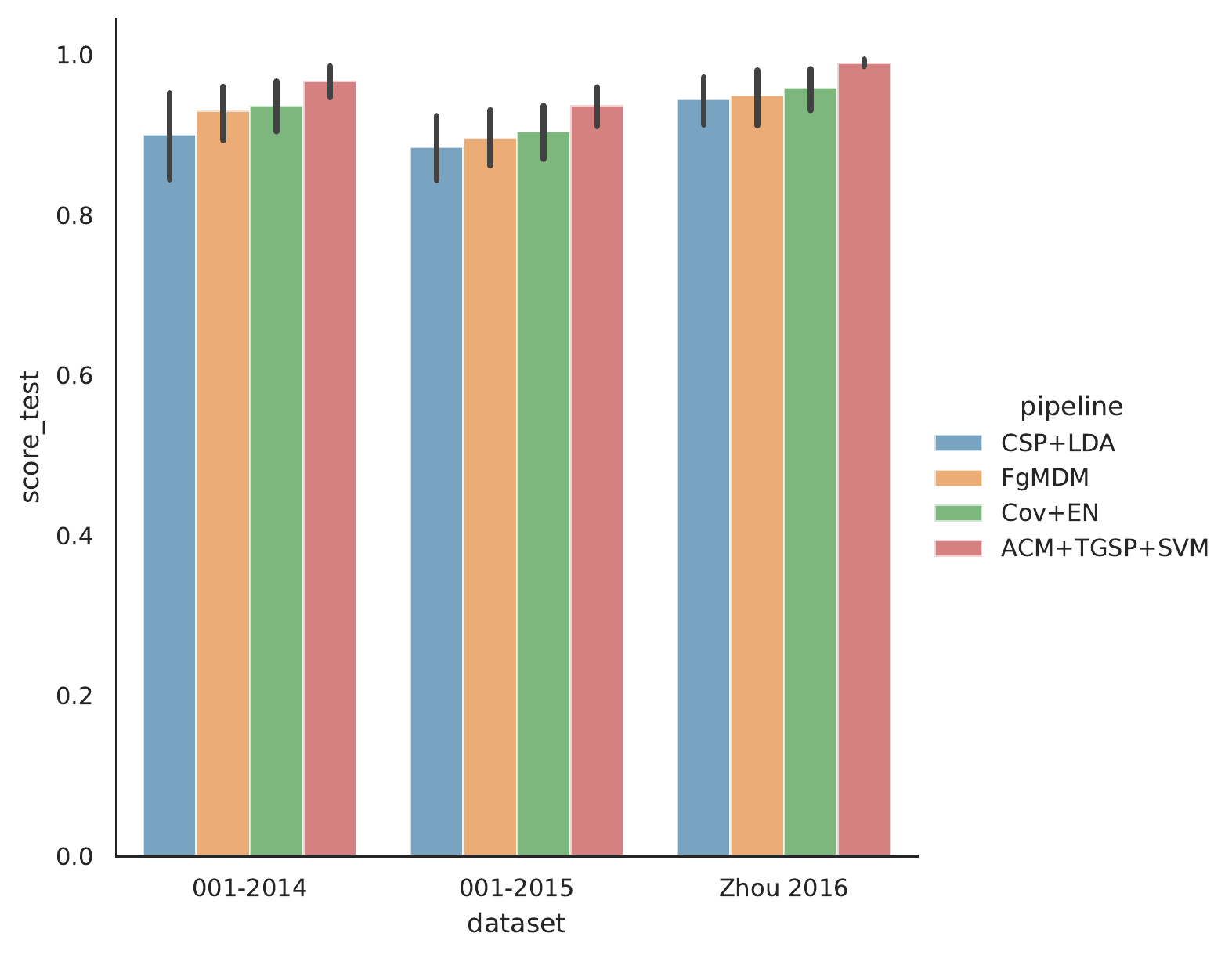}}
    \\
    \subfloat[]{%
        \includegraphics[width=0.5\linewidth]{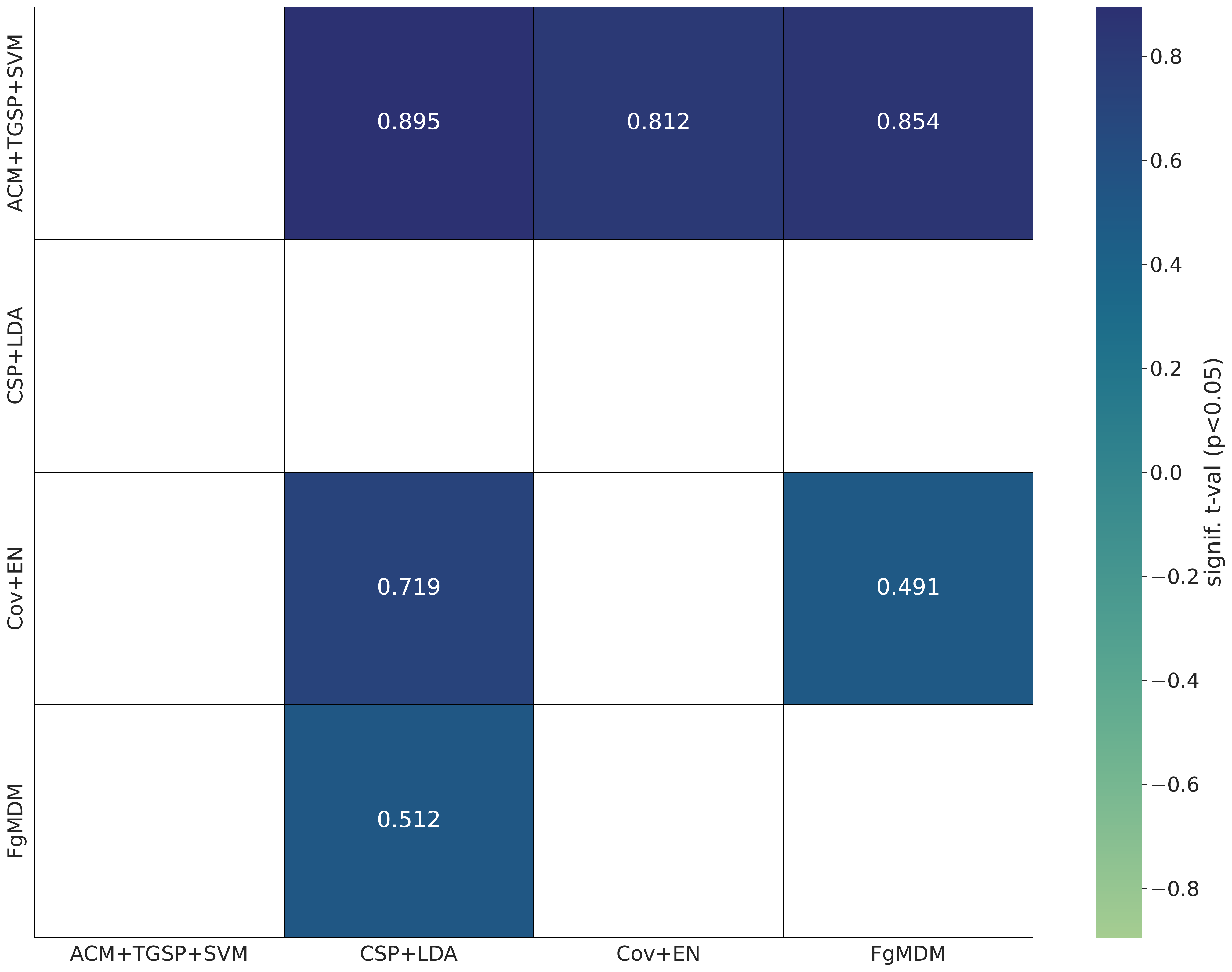}}
        \\
   \subfloat[]{%
            \includegraphics[width=0.30\linewidth]{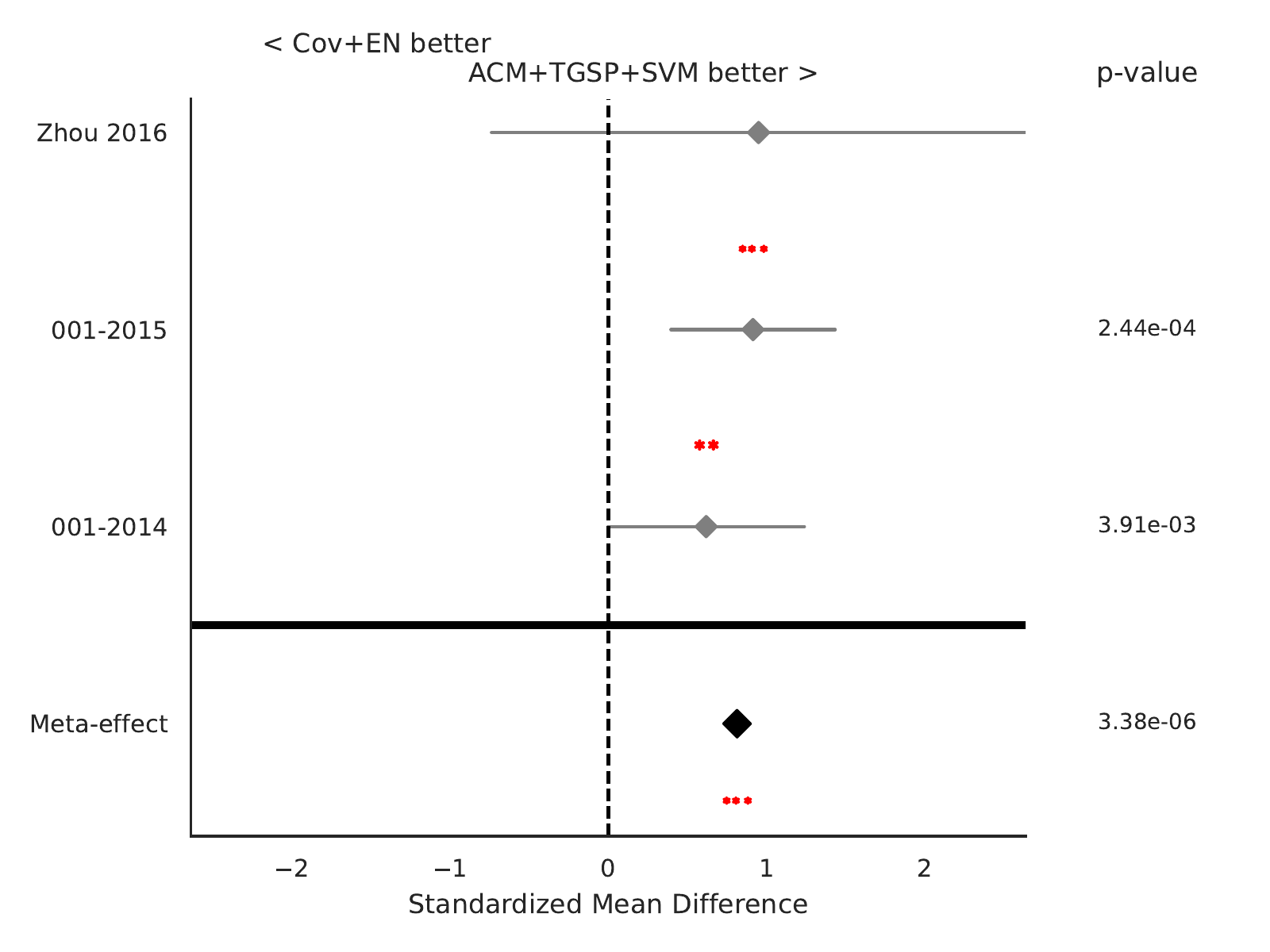}}
            \hfill
   \subfloat[]{%
            \includegraphics[width=0.30\linewidth]{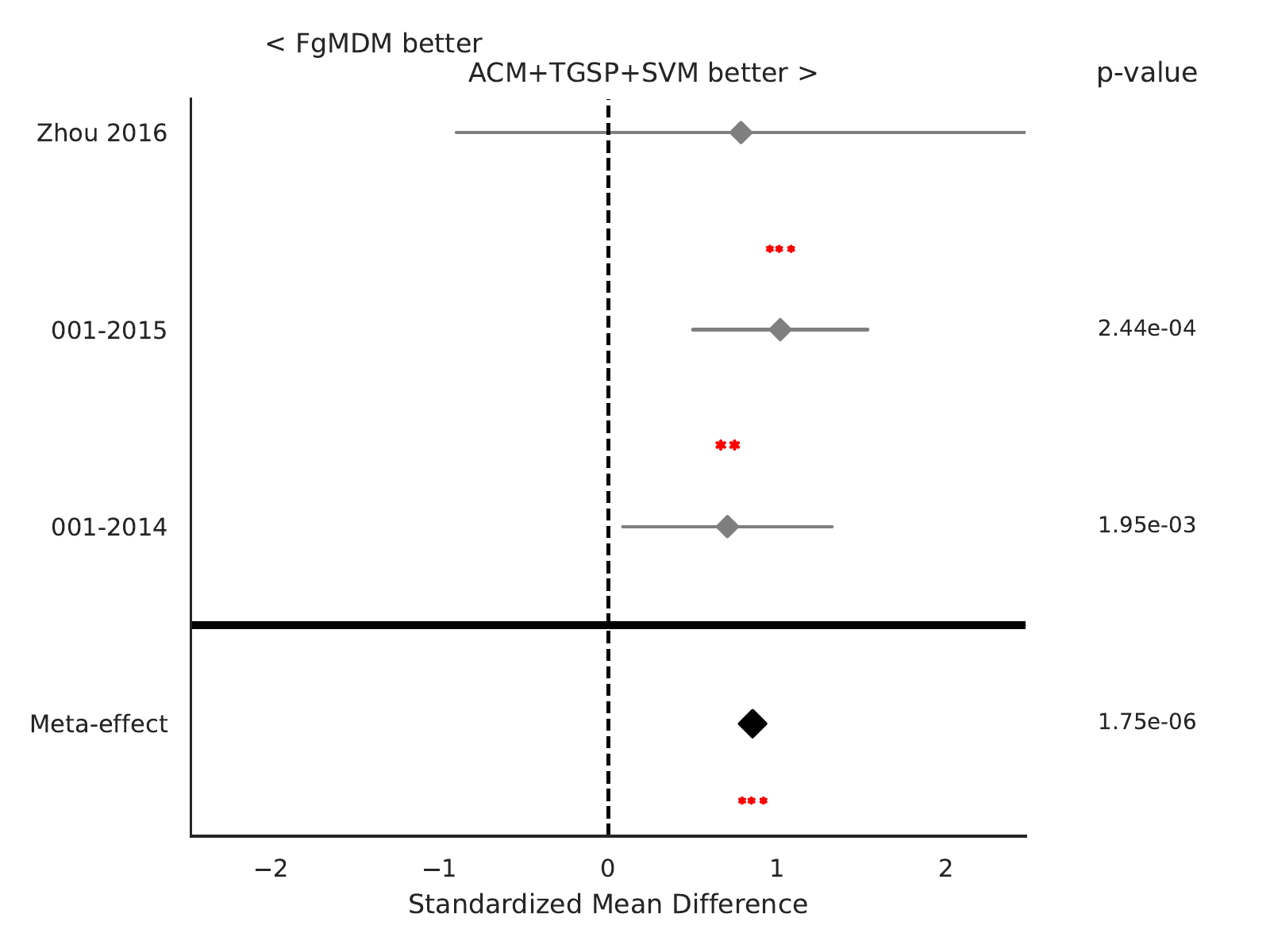}}
            \hfill
   \subfloat[]{%
            \includegraphics[width=0.30\linewidth]{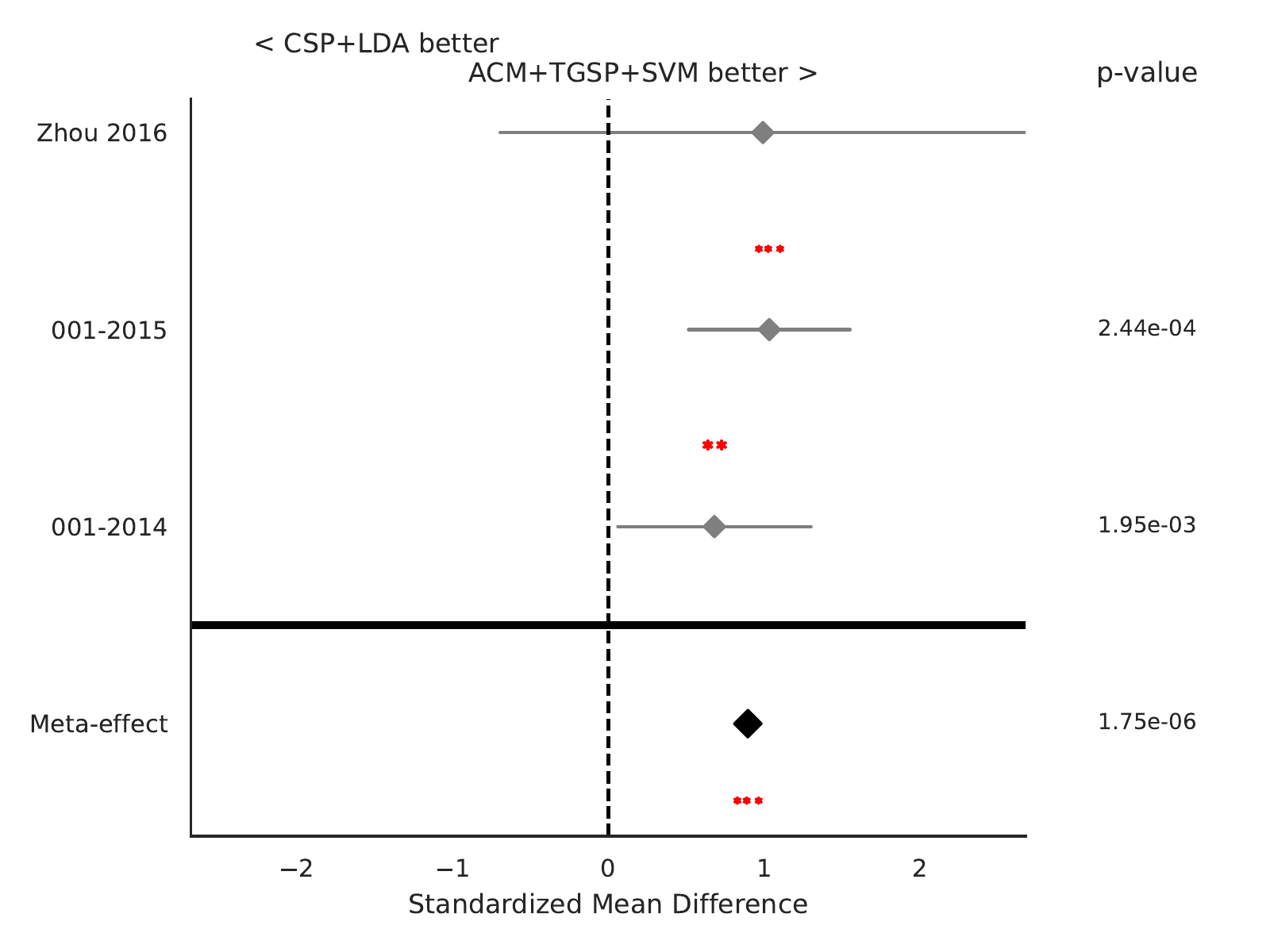}}
            \hfill

    \caption{Result for right hand vs feet classification, using cross-session evaluation. (a) show the rain clouds plots for each pipeline, showing the distribution of the score of every subject. (b) show the bar plot of the score withe the error of the different pipeline and for every dataset considered. (c) show the meta analysis of the different pipeline considered. This plot the significance that the algorithm on the y-axis is better than the one on the x-axis. The color represents the significance level of the difference of accuracy, in terms of t-values, and we show only the significant interactions ($p < 0.05$). (d) (e) (f) show the meta analysis of augmented method with SVM against the state of the art. We show the standardized mean differences, while p-values are computed as one-tailed Wilcoxon signed-rank test for the hypothesis given as title of the plot and the gray bar  denote $95\%$ interval. Here, * stands for $p < 0.05$, ** for $p < 0.01$, and *** for $p < 0.001$.
    }
    \label{fig:TANG+SVM-rf-crosssession-stateart}
\end{figure*}

\begin{figure*}[ht]  
    \centering
    \centering
     \subfloat[]{%
            \includegraphics[width=0.45\linewidth]{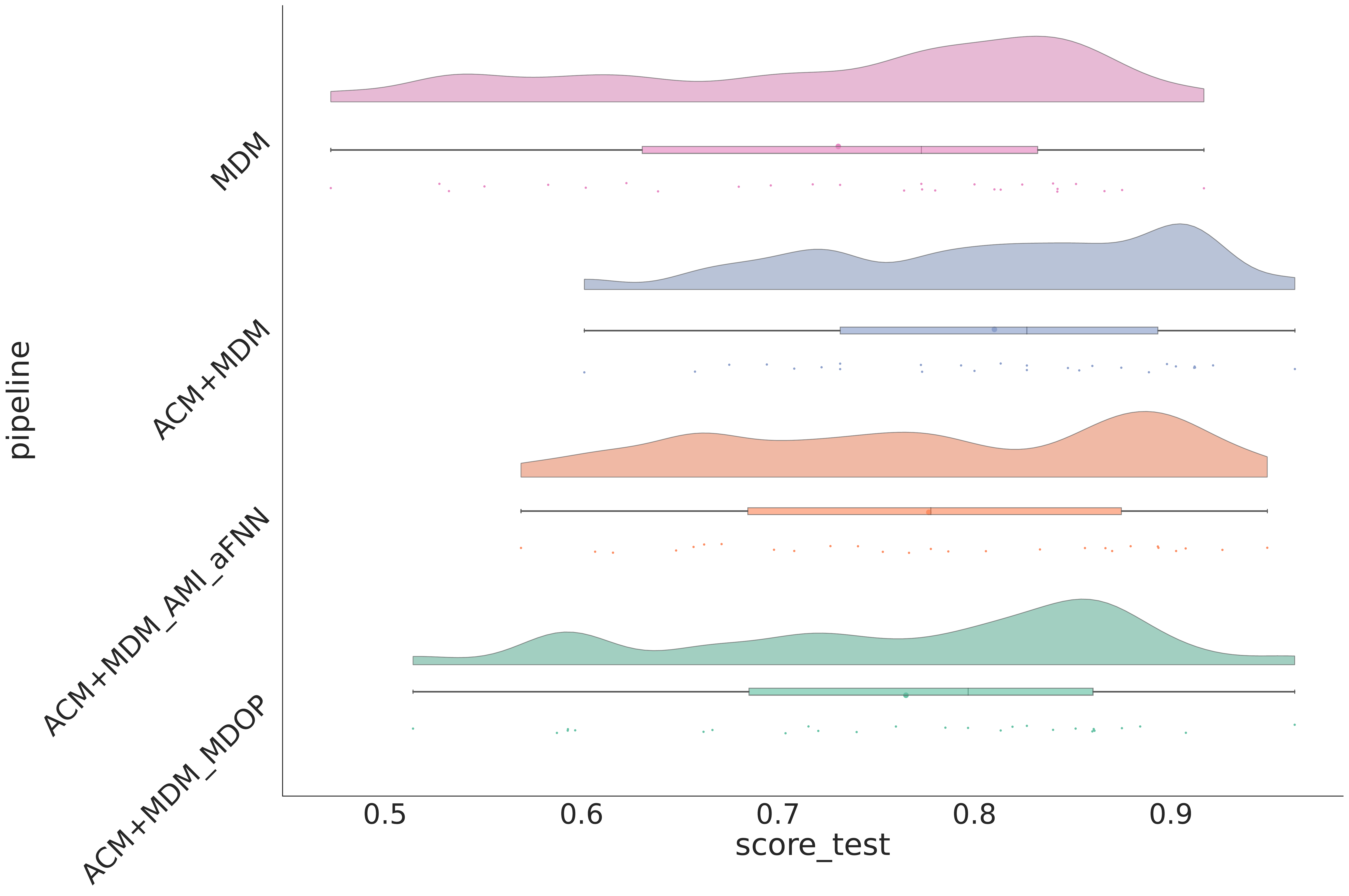}}
            \hfill
     \subfloat[]{%
            \includegraphics[width=0.45\linewidth]{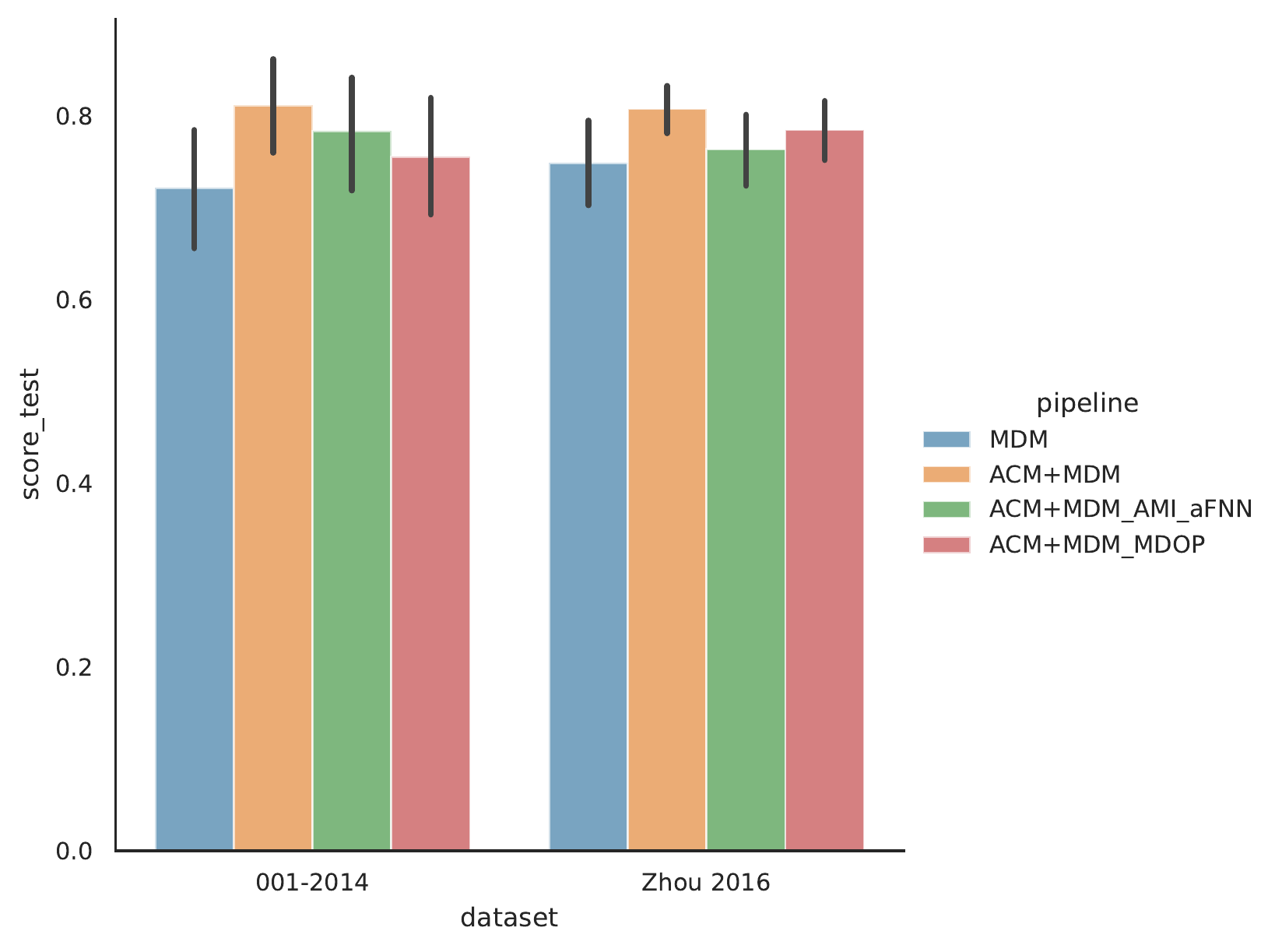}}
    \\
    \subfloat[]{%
        \includegraphics[width=0.5\linewidth]{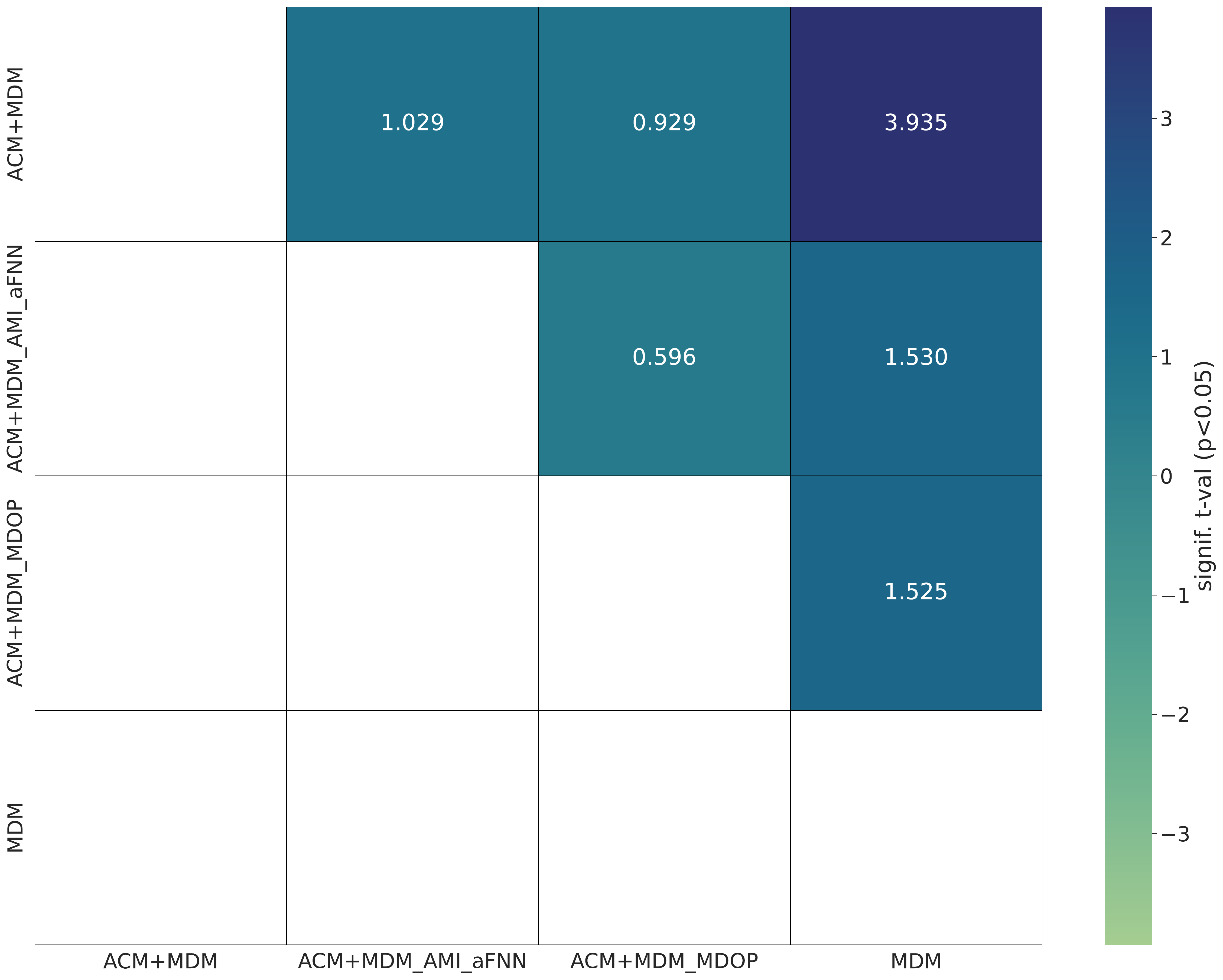}}
        \\
   \subfloat[]{%
            \includegraphics[width=0.30\linewidth]{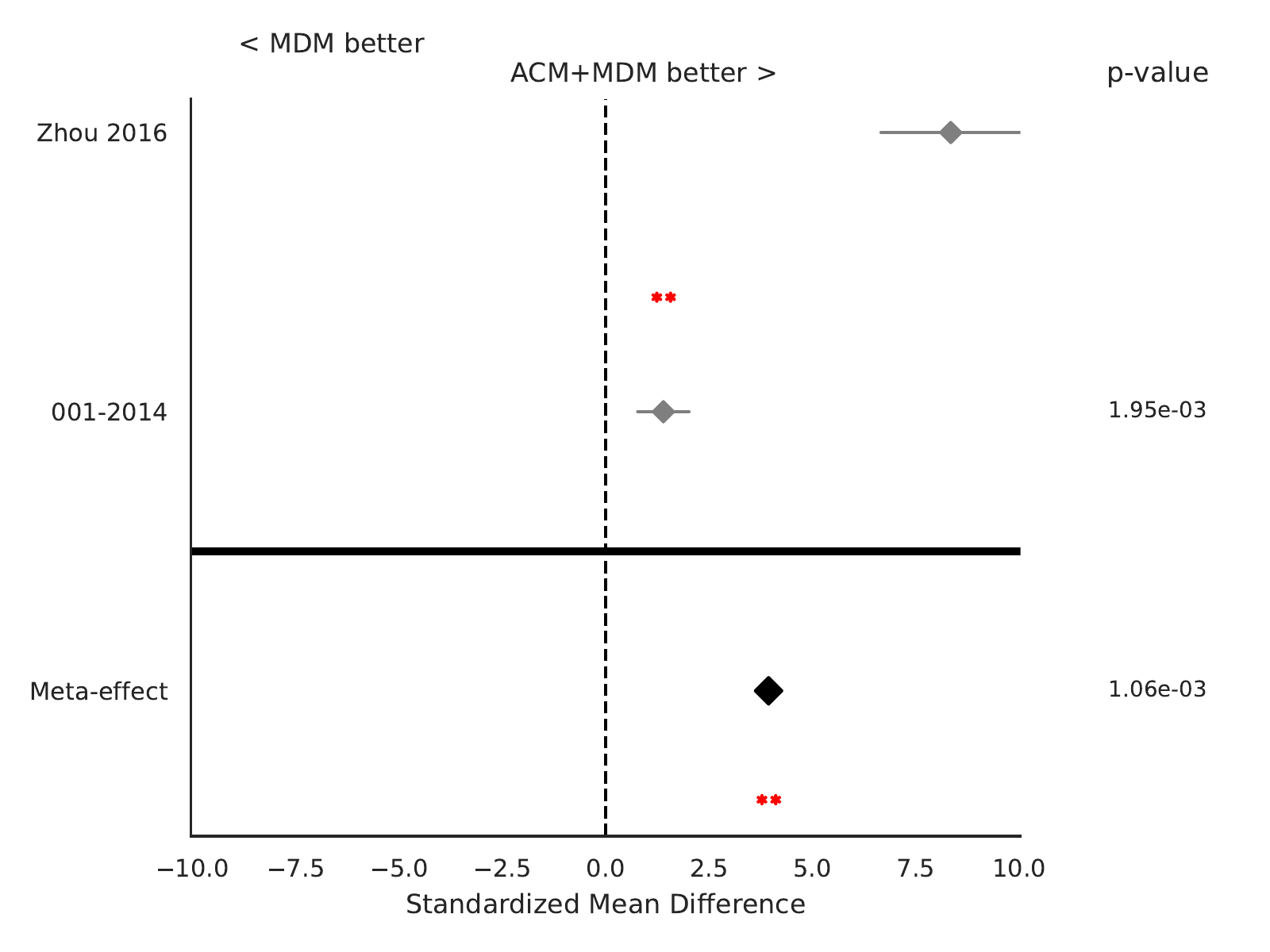}}
            \hfill
   \subfloat[]{%
            \includegraphics[width=0.30\linewidth]{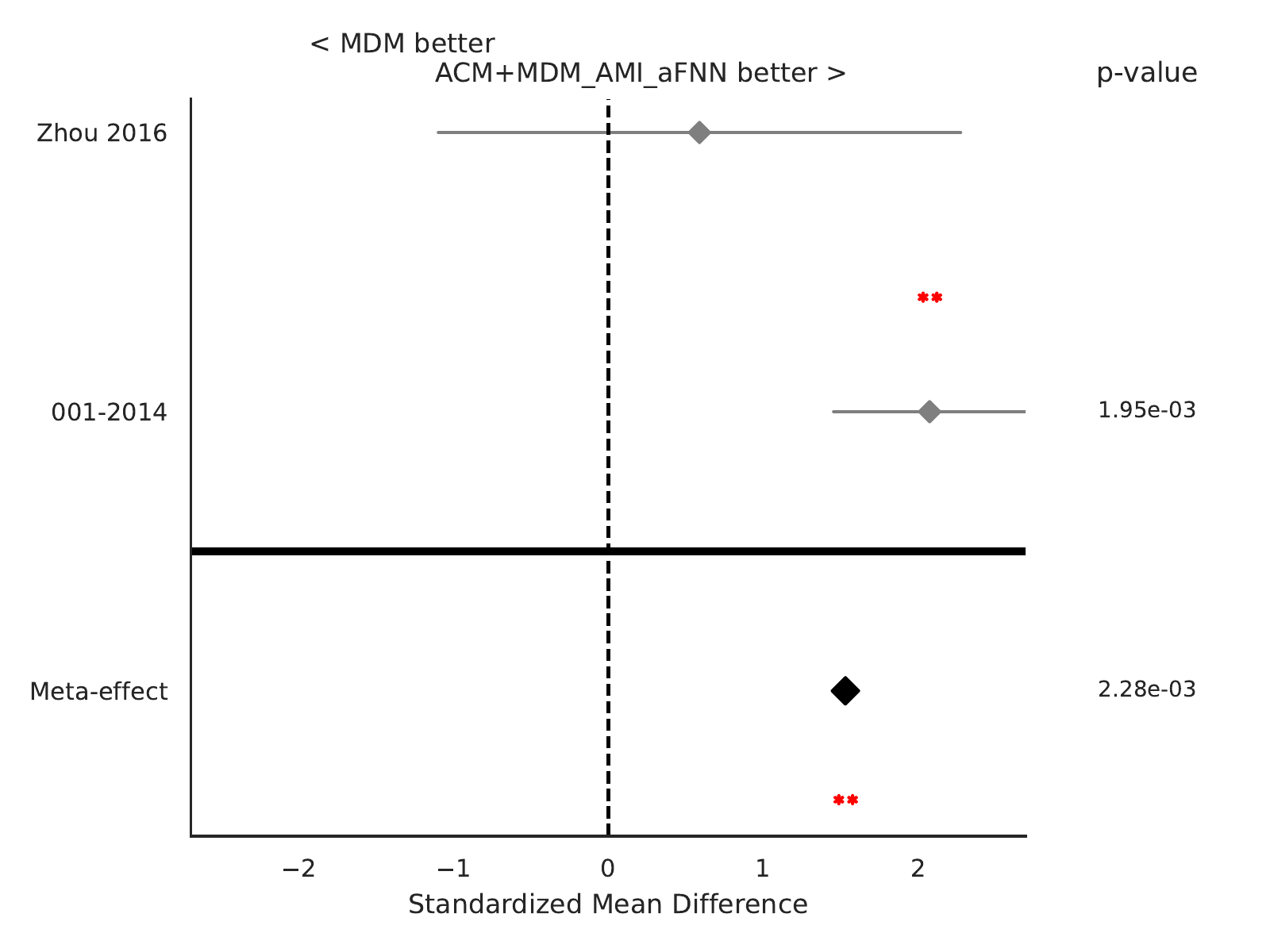}}
            \hfill
   \subfloat[]{%
            \includegraphics[width=0.30\linewidth]{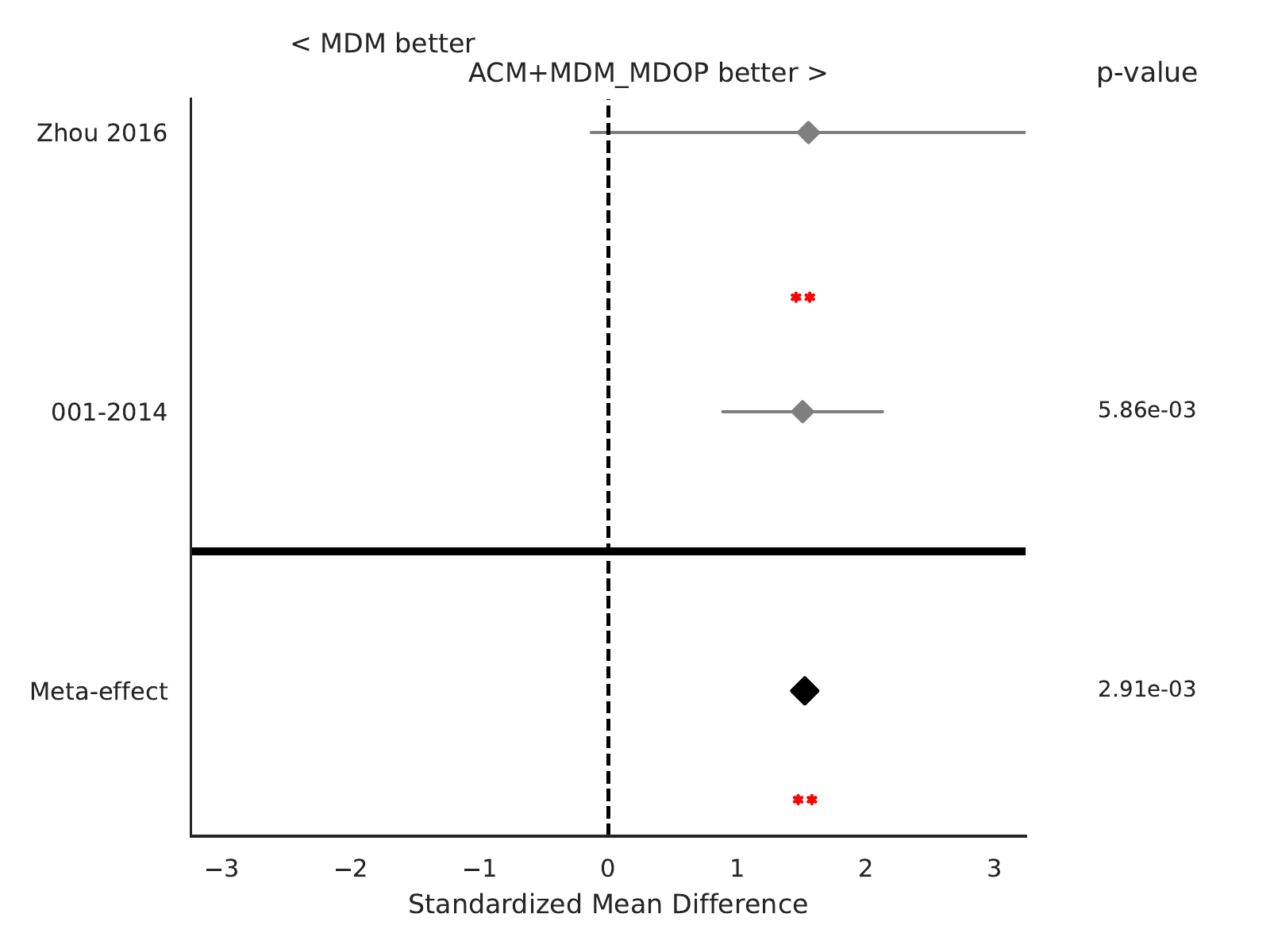}}
            \hfill

    \caption{Result for right hand vs left hand vs feet classification using the MDM algorithm, using withing-session evaluation. (a) show the rain clouds plots for each pipeline, showing the distribution of the score of every subject. (b) show the bar plot of the score withe the error of the different pipeline and for every dataset considered. (c) show the meta analysis of the different pipeline considered. This plot the significance that the algorithm on the y-axis is better than the one on the x-axis. The color represents the significance level of the difference of accuracy, in terms of t-values, and we show only the significant interactions ($p < 0.05$). (d) (e) (f) show the meta analysis of the standard MDM algorithm against the augmented covariance method with the selection of the hyper-parameter based on grid search, traditional and unified Takens approach respectively. We show the standardized mean differences, while p-values are computed as one-tailed Wilcoxon signed-rank test for the hypothesis given as title of the plot and the gray bar  denote $95\%$ interval. Here, * stands for $p < 0.05$, ** for $p < 0.01$, and *** for $p < 0.001$.
    }
    \label{fig:MDM-3class-whithinsession}
\end{figure*}

\begin{figure*}[ht]  
    \centering
    \centering
     \subfloat[]{%
            \includegraphics[width=0.45\linewidth]{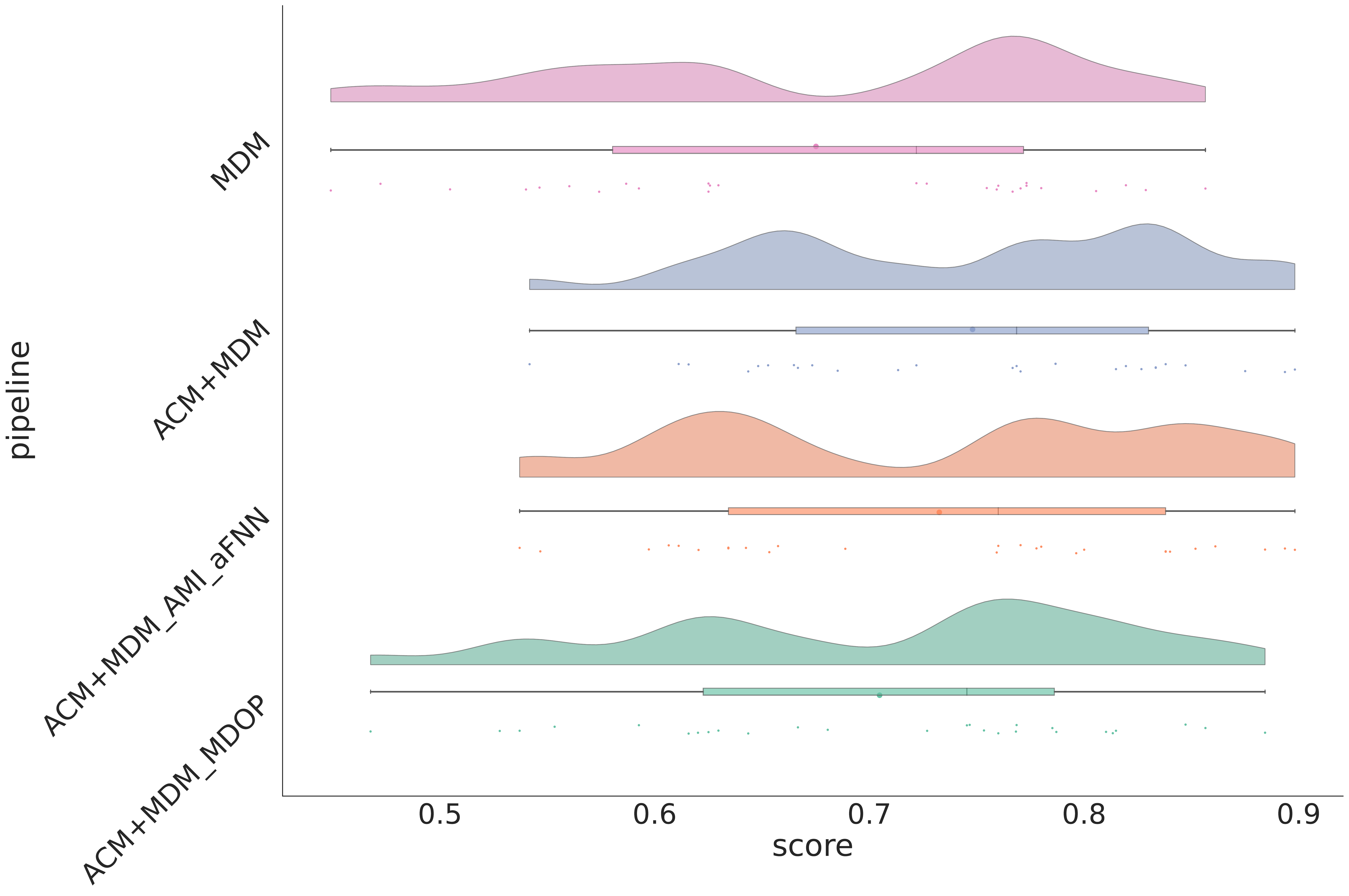}}
            \hfill
     \subfloat[]{%
            \includegraphics[width=0.45\linewidth]{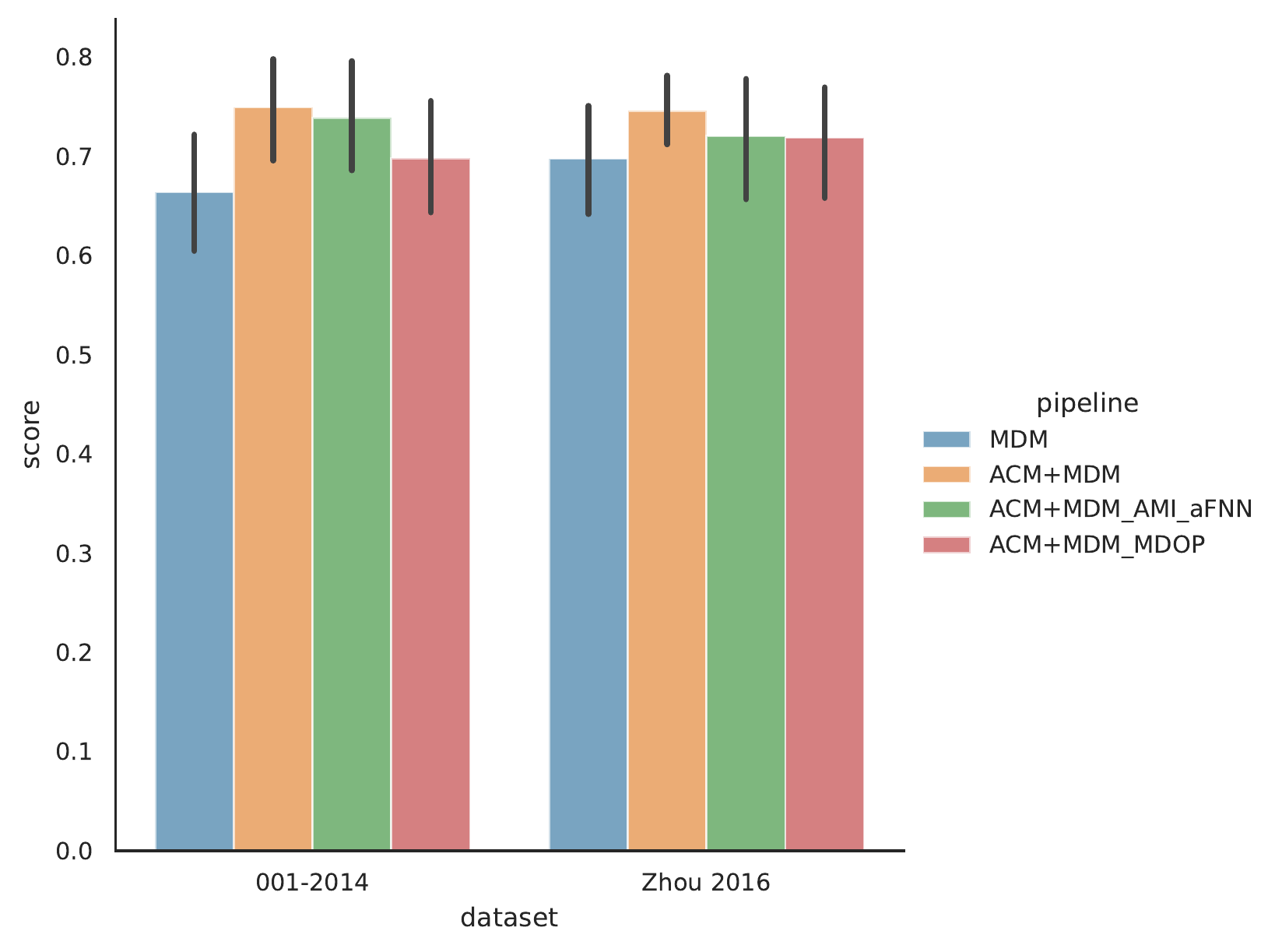}}
    \\
    \subfloat[]{%
        \includegraphics[width=0.5\linewidth]{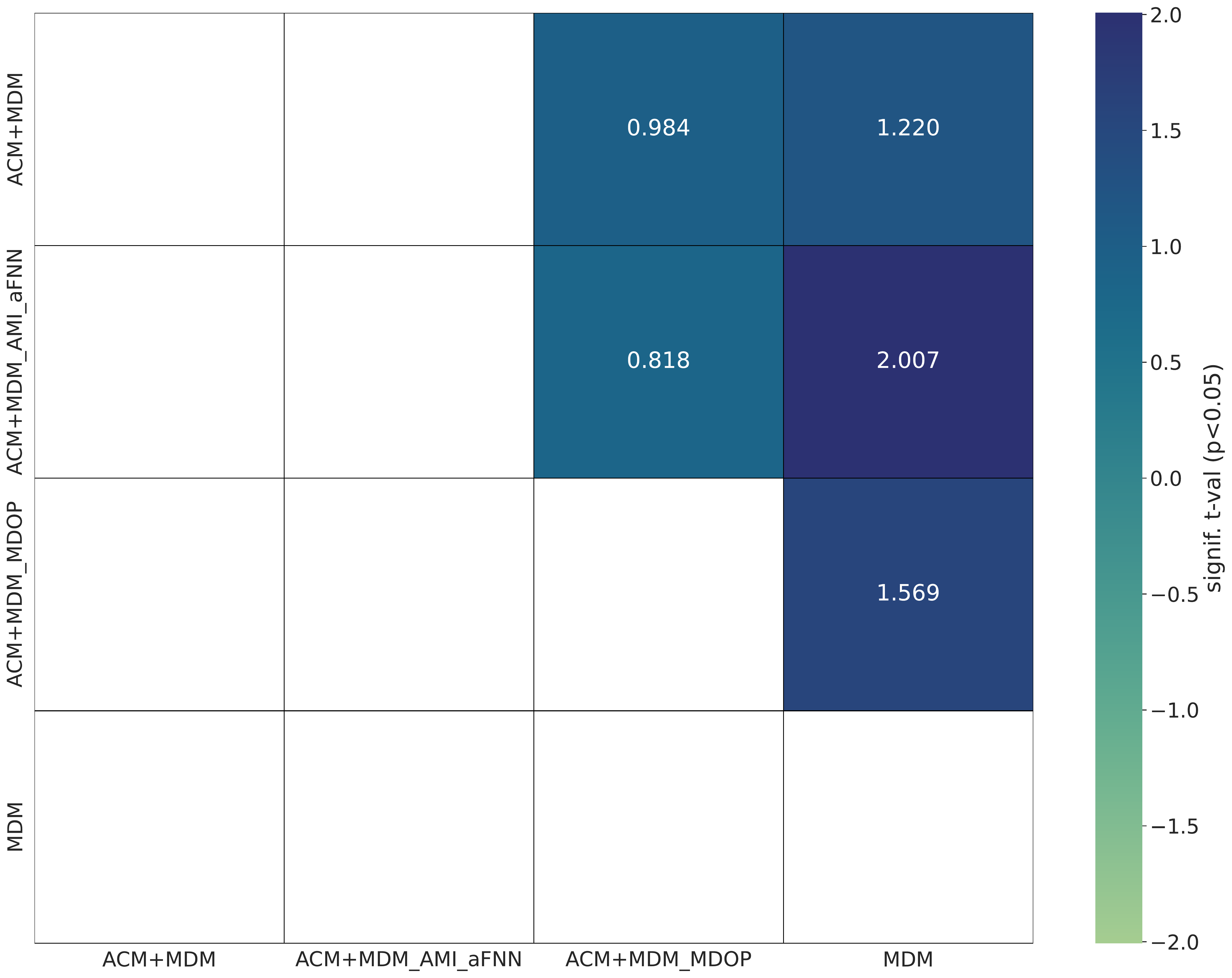}}
        \\
   \subfloat[]{%
            \includegraphics[width=0.30\linewidth]{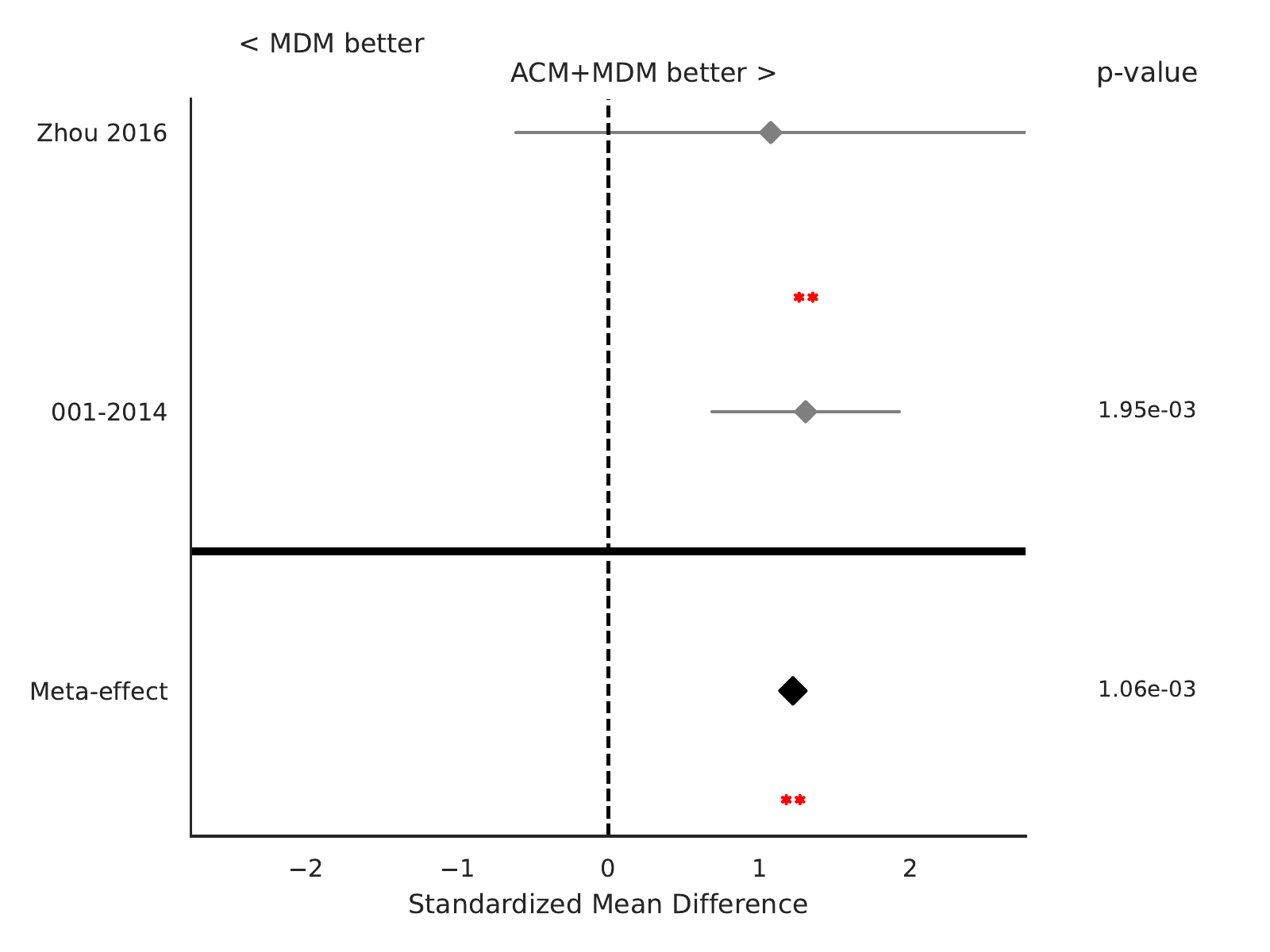}}
            \hfill
   \subfloat[]{%
            \includegraphics[width=0.30\linewidth]{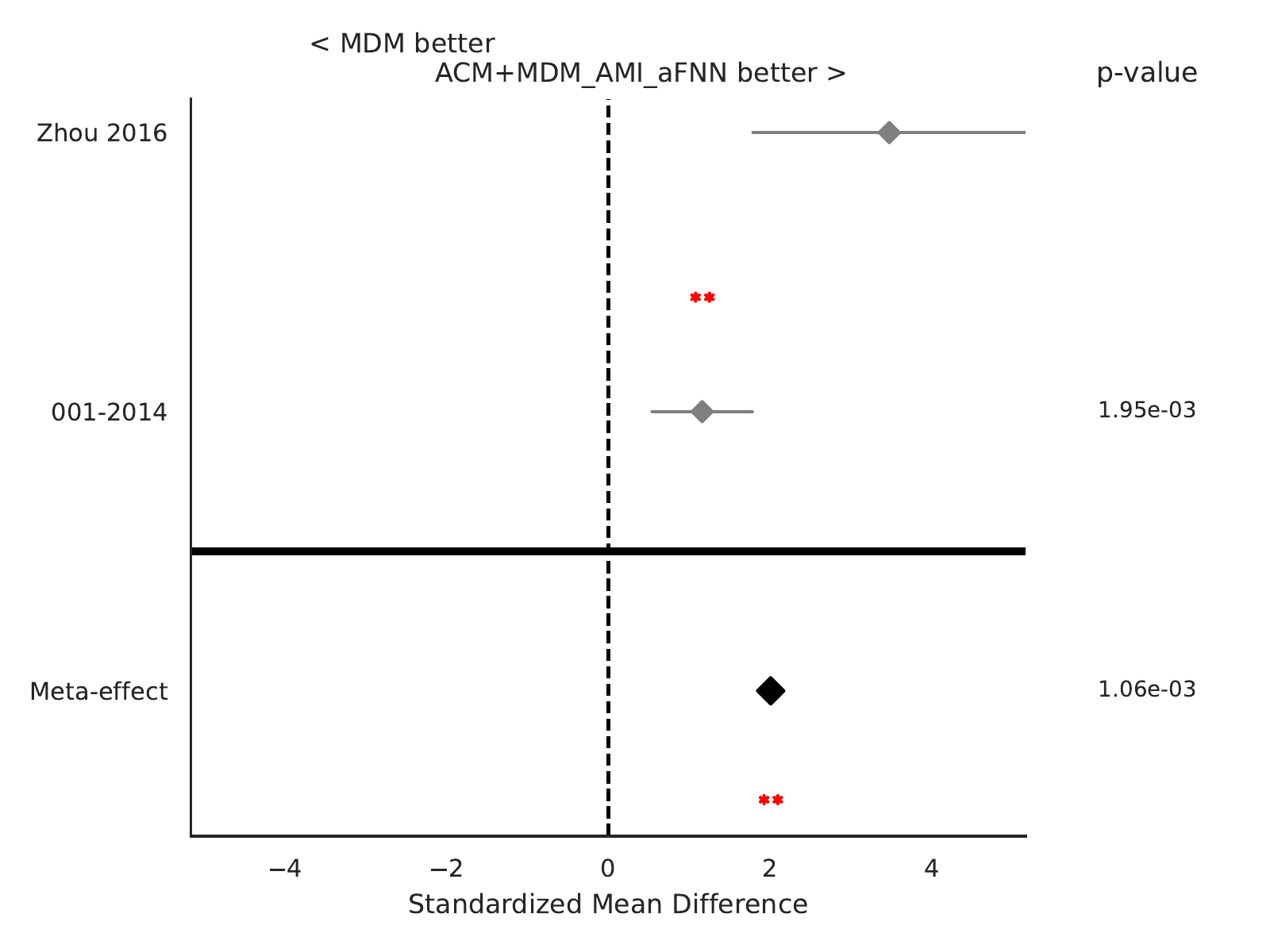}}
            \hfill
   \subfloat[]{%
            \includegraphics[width=0.30\linewidth]{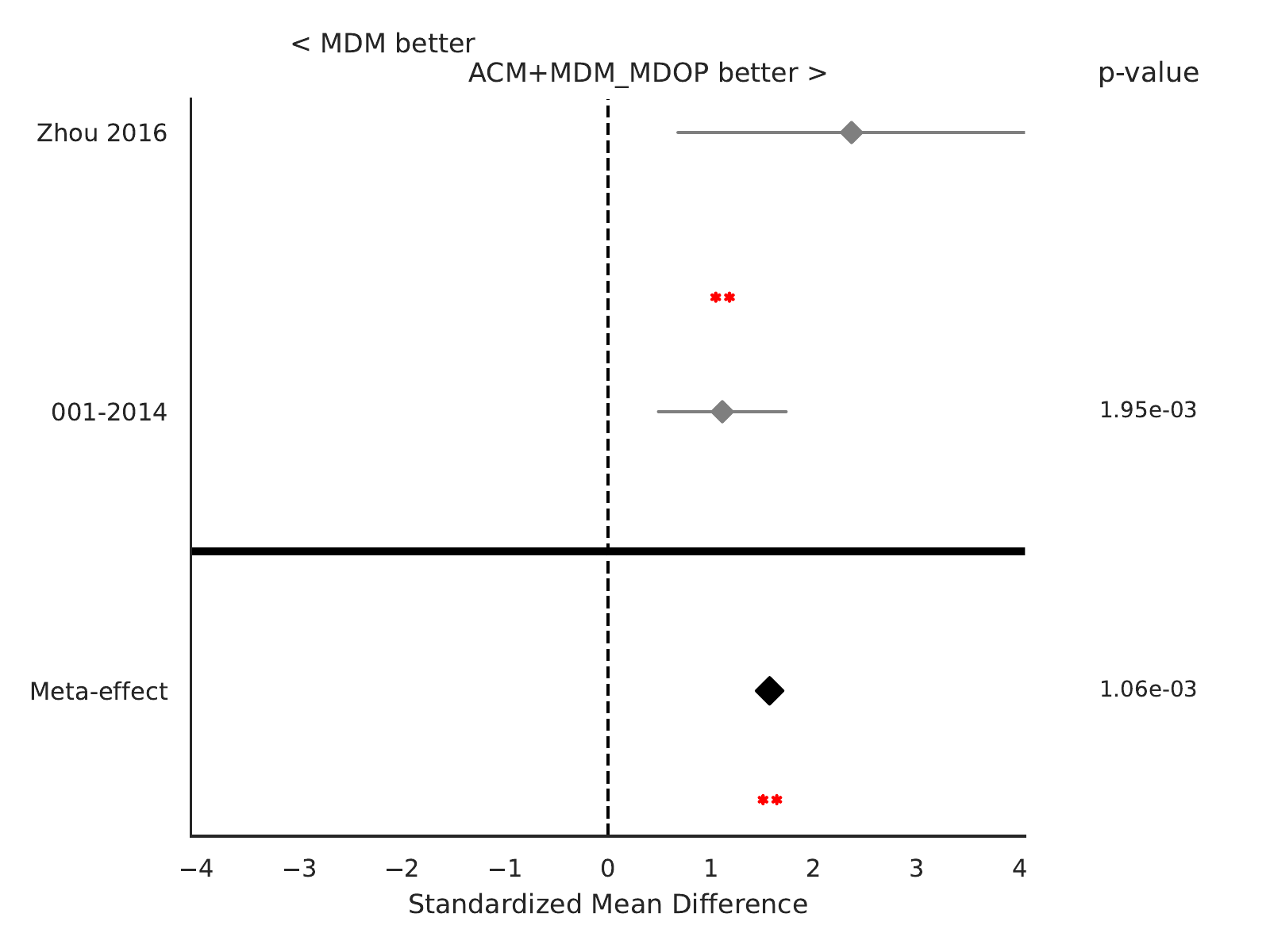}}
            \hfill

    \caption{Result for right hand vs left hand vs feet classification using the MDM algorithm, using cross-session evaluation. (a) show the rain clouds plots for each pipeline, showing the distribution of the score of every subject. (b) show the bar plot of the score withe the error of the different pipeline and for every dataset considered. (c) show the meta analysis of the different pipeline considered. This plot the significance that the algorithm on the y-axis is better than the one on the x-axis. The color represents the significance level of the difference of accuracy, in terms of t-values, and we show only the significant interactions ($p < 0.05$). (d) (e) (f) show the meta analysis of the standard MDM algorithm against the augmented covariance method with the selection of the hyper-parameter based on grid search, traditional and unified Takens approach respectively. We show the standardized mean differences, while p-values are computed as one-tailed Wilcoxon signed-rank test for the hypothesis given as title of the plot and the gray bar  denote $95\%$ interval. Here, * stands for $p < 0.05$, ** for $p < 0.01$, and *** for $p < 0.001$.
    }
    \label{fig:MDM-3class-crosssession}
\end{figure*}

\begin{figure*}[ht]  
    \centering
    \centering
     \subfloat[]{%
            \includegraphics[width=0.45\linewidth]{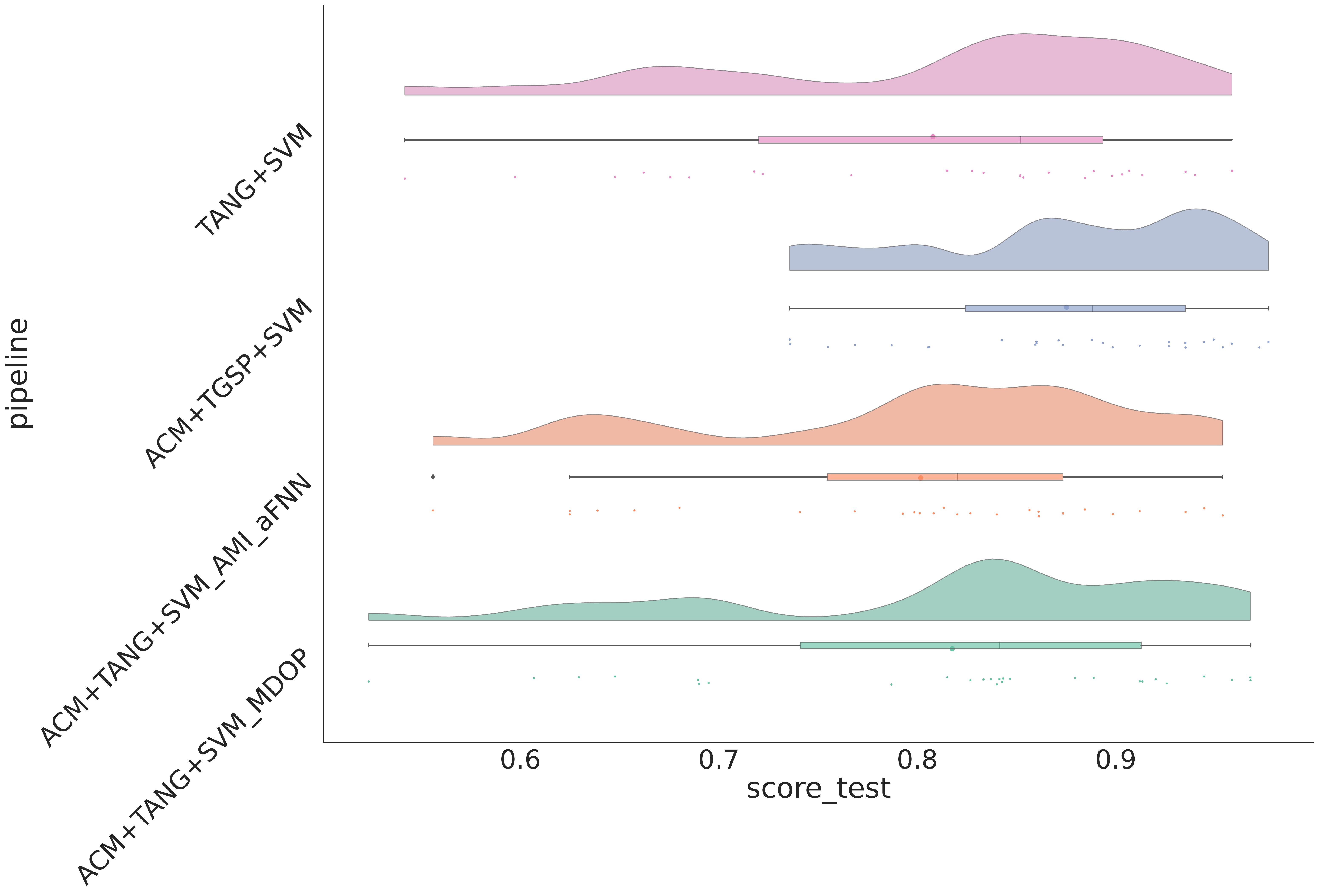}}
            \hfill
     \subfloat[]{%
            \includegraphics[width=0.45\linewidth]{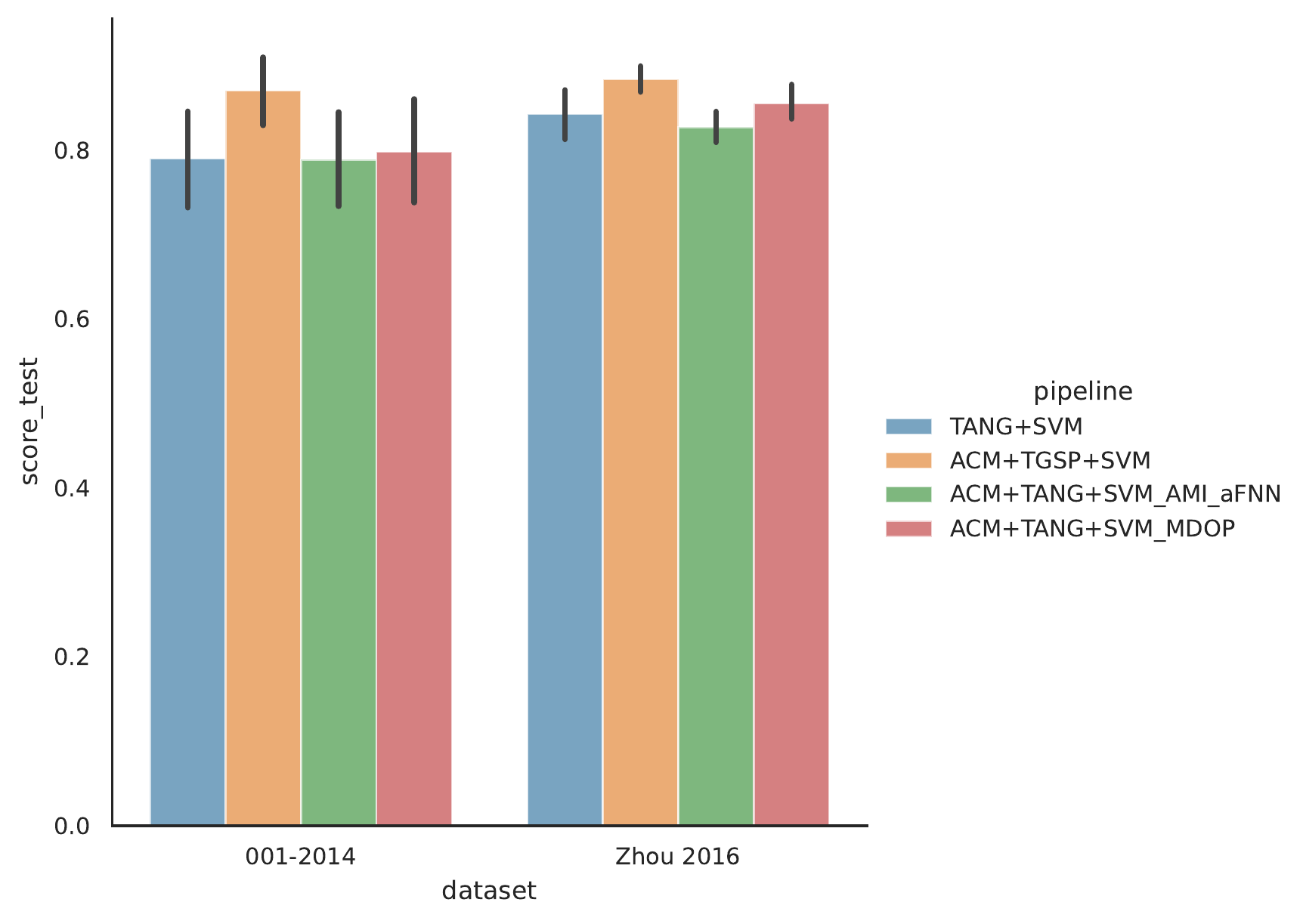}}
    \\
    \subfloat[]{%
        \includegraphics[width=0.5\linewidth]{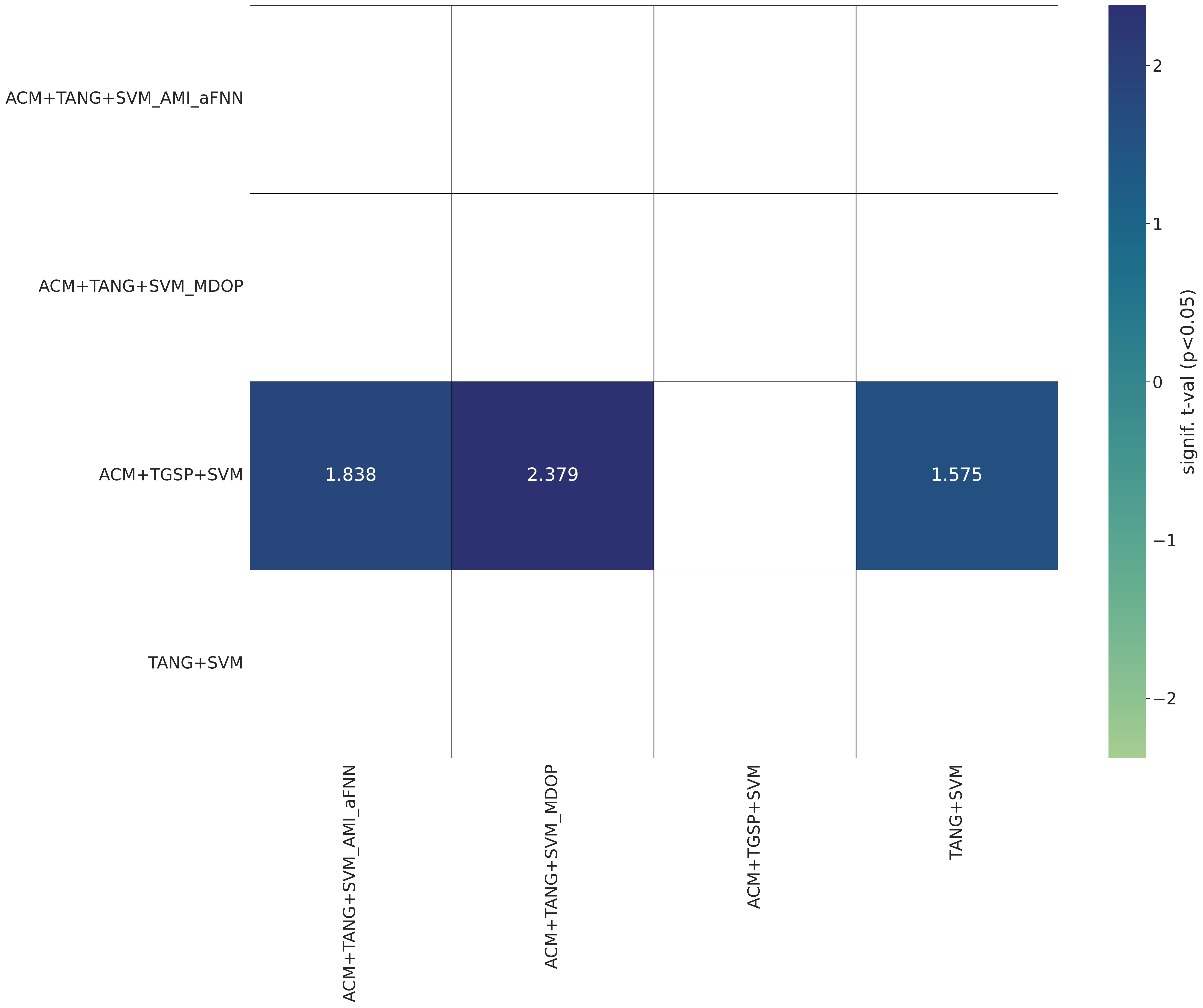}}
        \\
   \subfloat[]{%
            \includegraphics[width=0.30\linewidth]{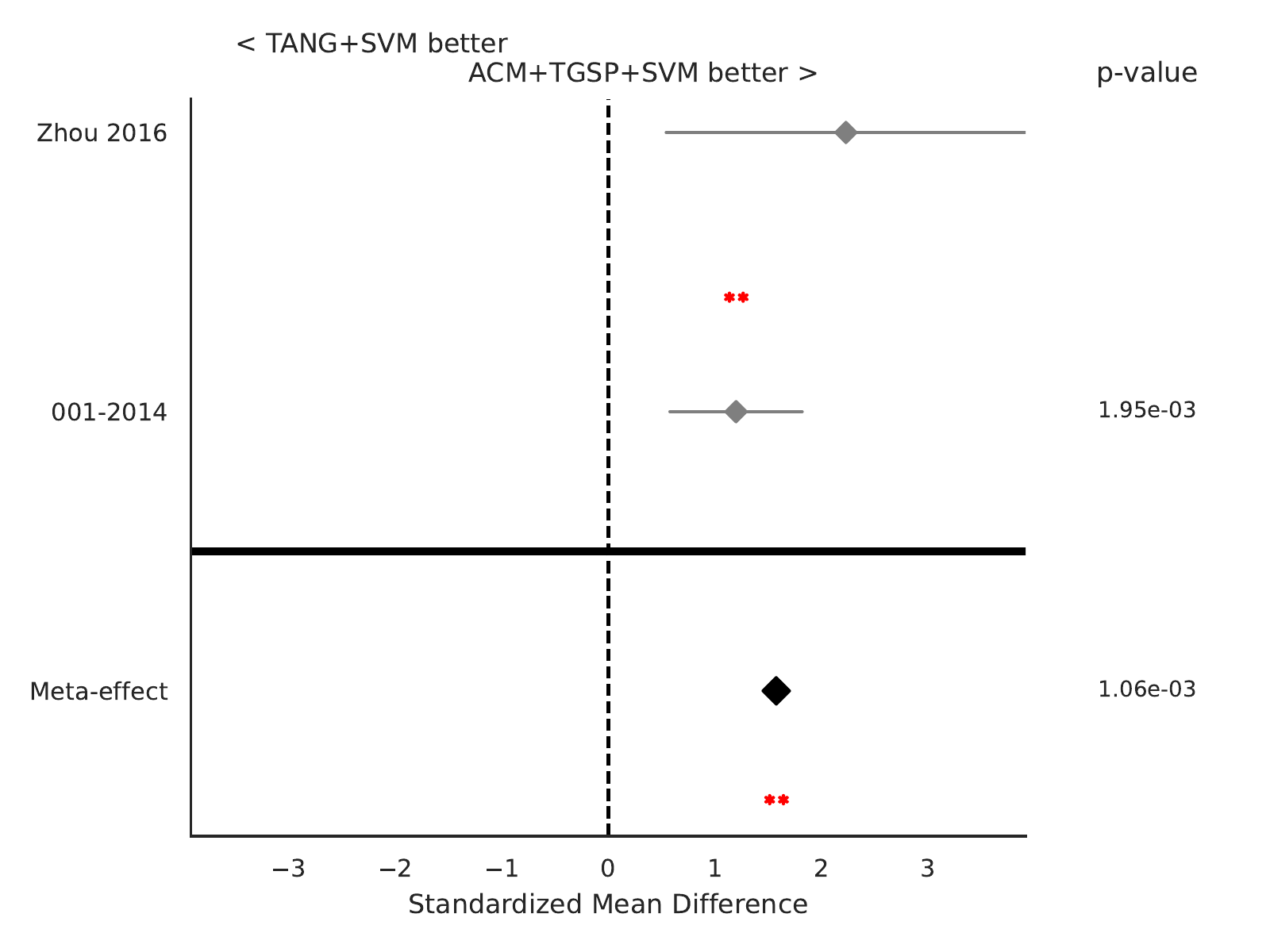}}
            \hfill
   \subfloat[]{%
            \includegraphics[width=0.30\linewidth]{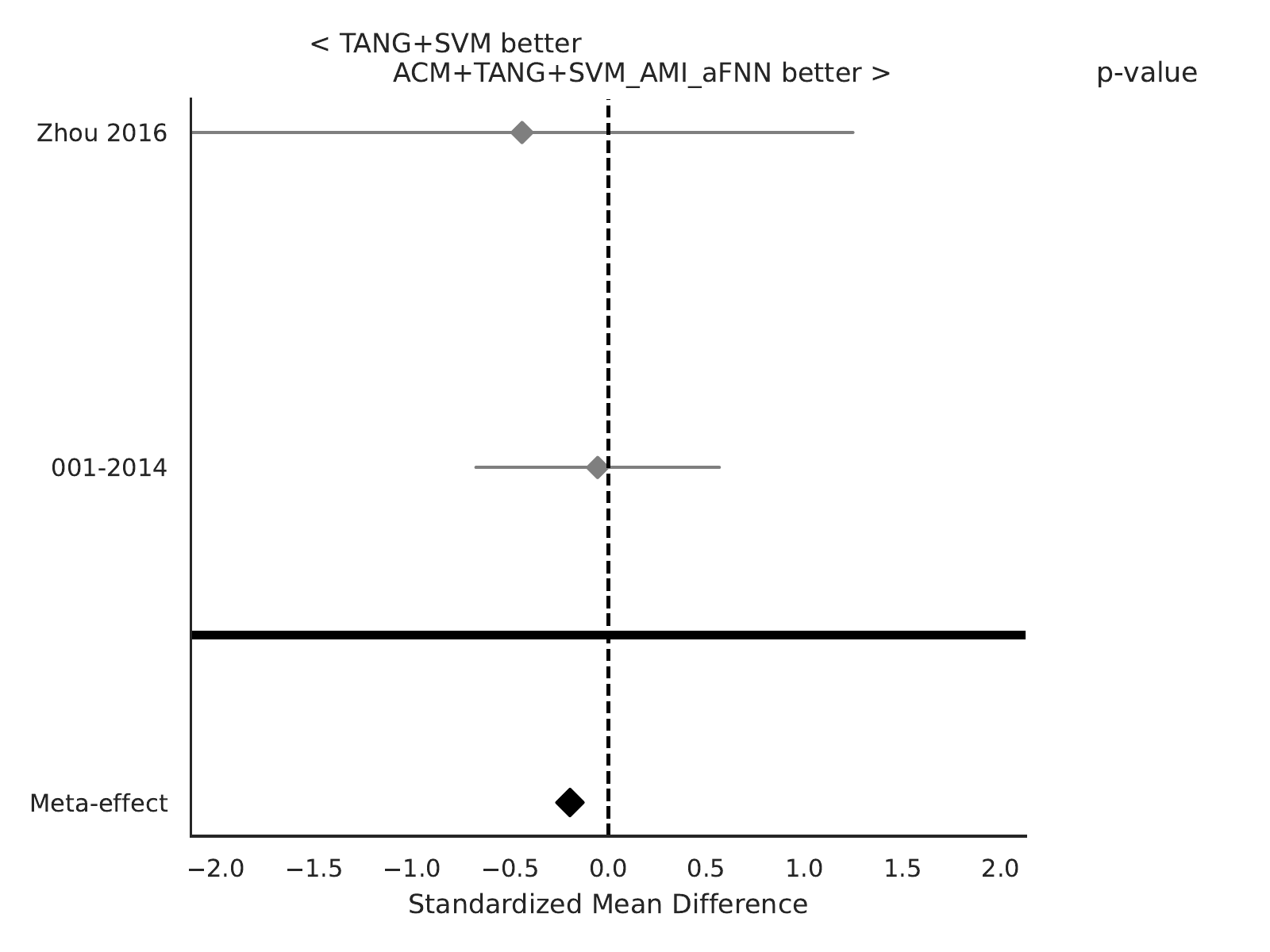}}
            \hfill
   \subfloat[]{%
            \includegraphics[width=0.30\linewidth]{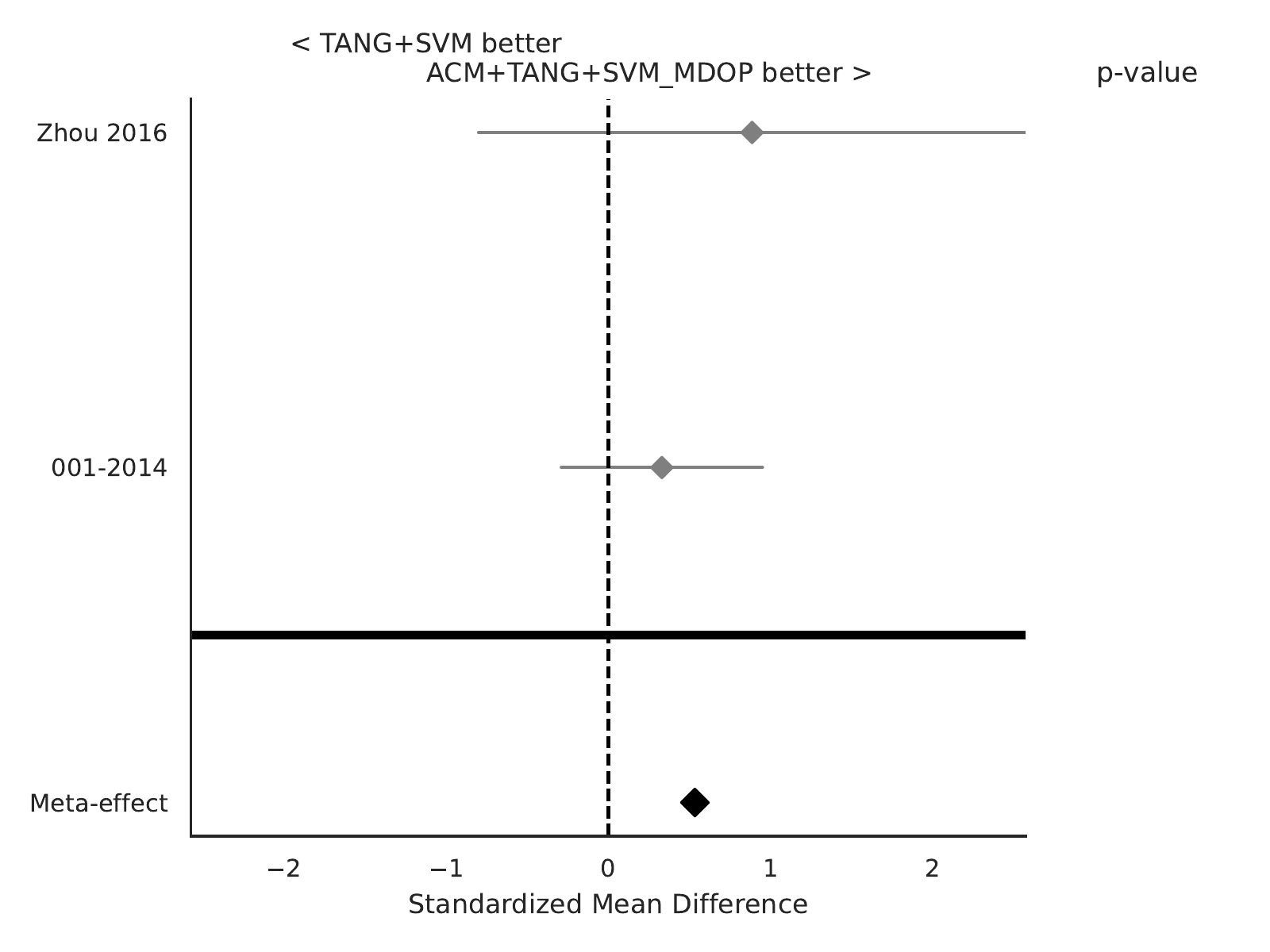}}
            \hfill

    \caption{Result for right hand vs left hand vs feet classification using the TANG algorithm, using withing-session evaluation. (a) show the rain clouds plots for each pipeline, showing the distribution of the score of every subject. (b) show the bar plot of the score withe the error of the different pipeline and for every dataset considered. (c) show the meta analysis of the different pipeline considered. This plot the significance that the algorithm on the y-axis is better than the one on the x-axis. The color represents the significance level of the difference of accuracy, in terms of t-values, and we show only the significant interactions ($p < 0.05$). (d) (e) (f) show the meta analysis of the standard TANG algorithm against the augmented covariance method with the selection of the hyper-parameter based on grid search, traditional and unified Takens approach respectively. We show the standardized mean differences, while p-values are computed as one-tailed Wilcoxon signed-rank test for the hypothesis given as title of the plot and the gray bar  denote $95\%$ interval. Here, * stands for $p < 0.05$, ** for $p < 0.01$, and *** for $p < 0.001$.
    }
    \label{fig:TANG+SVM-3class-whithinsession}
\end{figure*}

\begin{figure*}[ht]  
    \centering
    \centering
     \subfloat[]{%
            \includegraphics[width=0.45\linewidth]{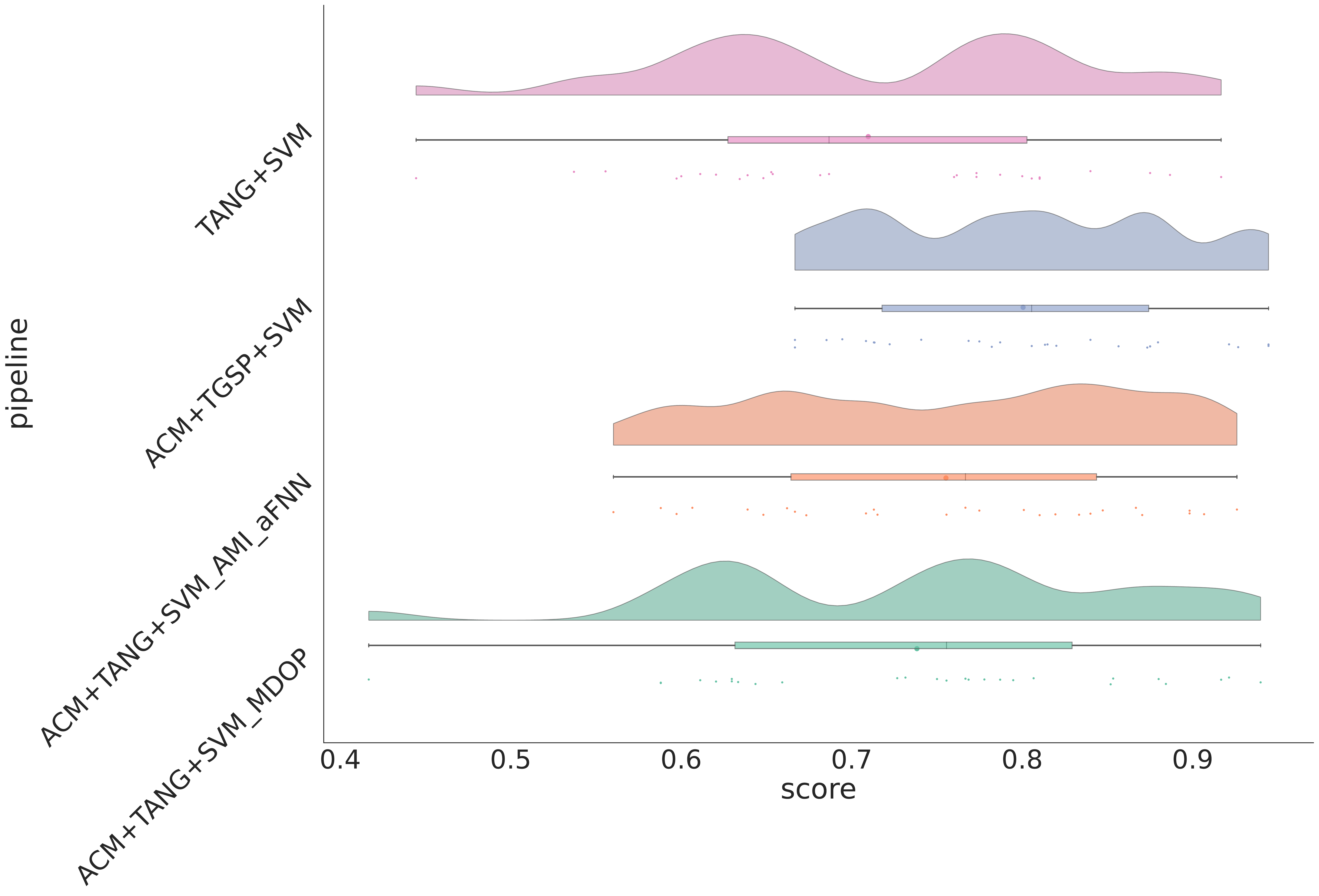}}
            \hfill
     \subfloat[]{%
            \includegraphics[width=0.45\linewidth]{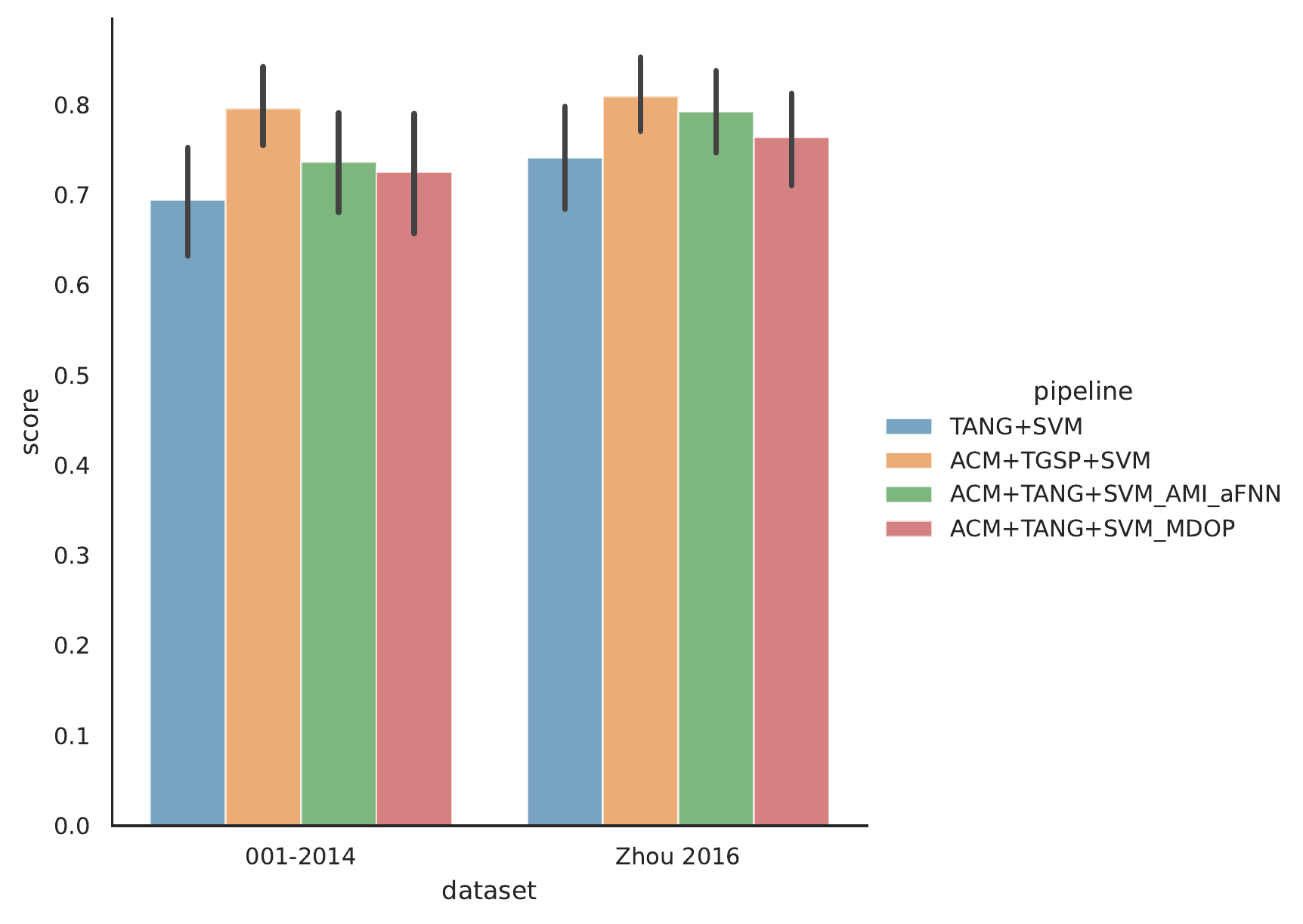}}
    \\
    \subfloat[]{%
        \includegraphics[width=0.5\linewidth]{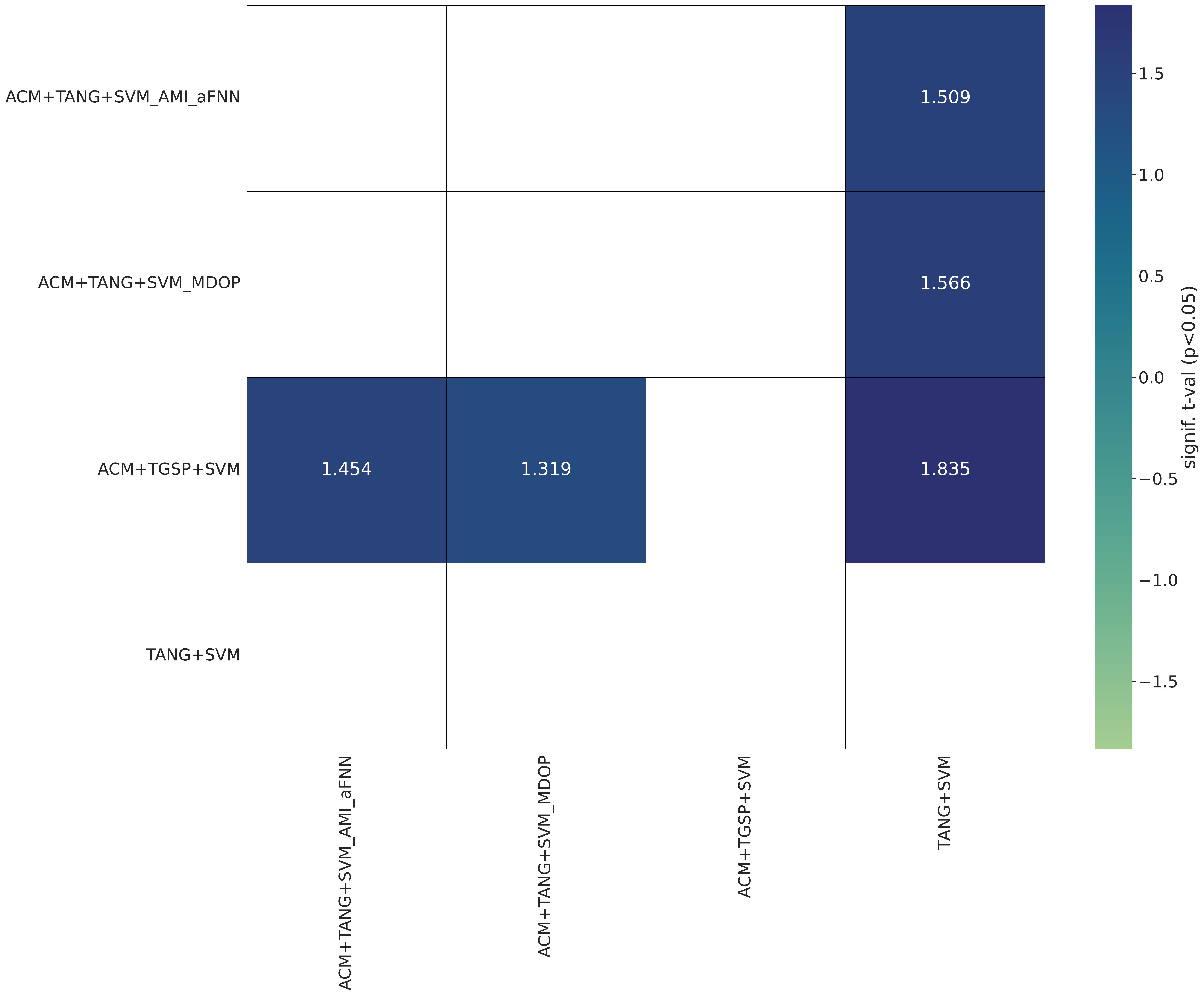}}
        \\
   \subfloat[]{%
            \includegraphics[width=0.30\linewidth]{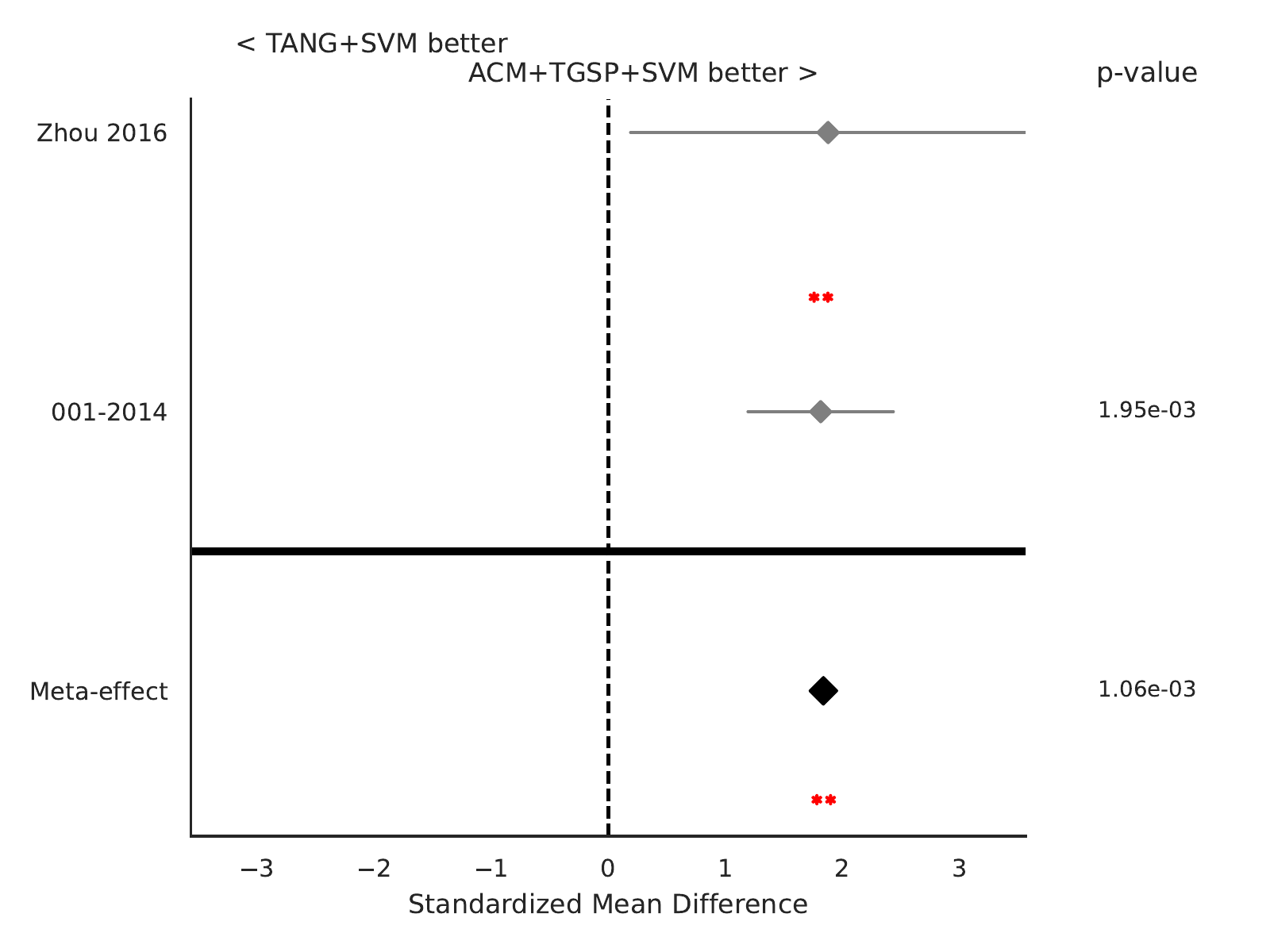}}
            \hfill
   \subfloat[]{%
            \includegraphics[width=0.30\linewidth]{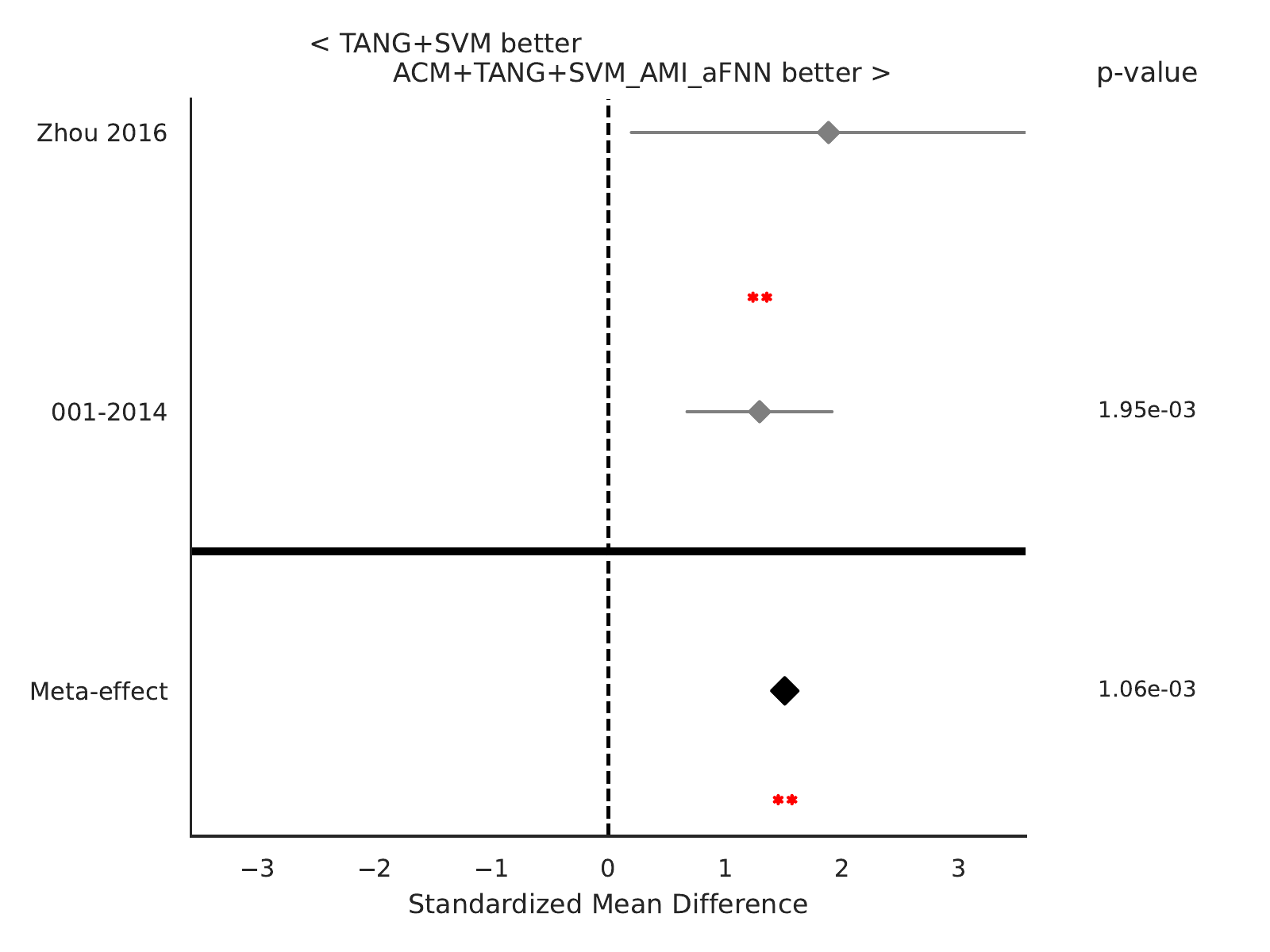}}
            \hfill
   \subfloat[]{%
            \includegraphics[width=0.30\linewidth]{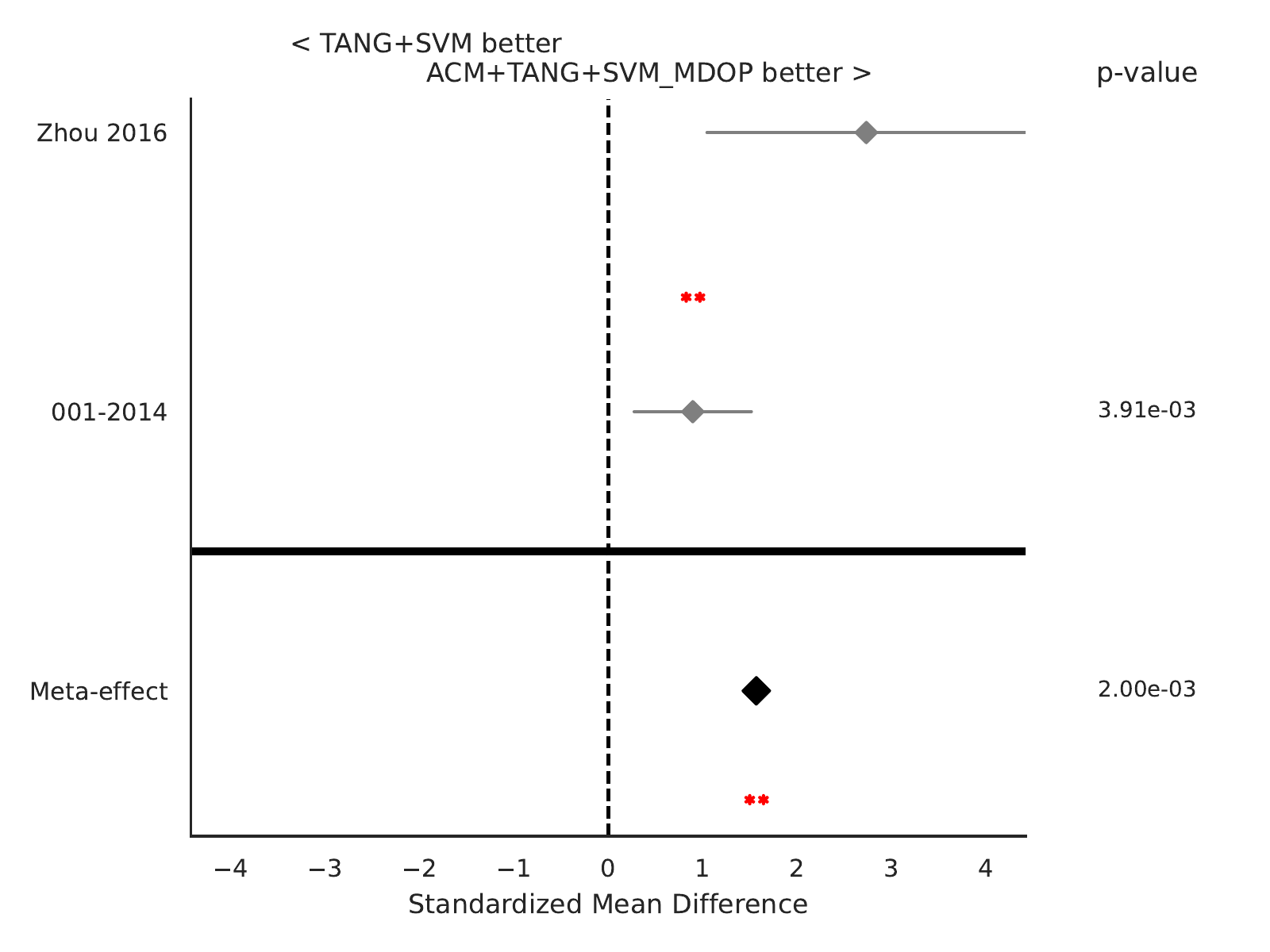}}
            \hfill

    \caption{Result for right hand vs left hand vs feet classification using the TANG algorithm, using cross-session evaluation. (a) show the rain clouds plots for each pipeline, showing the distribution of the score of every subject. (b) show the bar plot of the score withe the error of the different pipeline and for every dataset considered. (c) show the meta analysis of the different pipeline considered. This plot the significance that the algorithm on the y-axis is better than the one on the x-axis. The color represents the significance level of the difference of accuracy, in terms of t-values, and we show only the significant interactions ($p < 0.05$). (d) (e) (f) show the meta analysis of the standard TANG algorithm against the augmented covariance method with the selection of the hyper-parameter based on grid search, traditional and unified Takens approach respectively. We show the standardized mean differences, while p-values are computed as one-tailed Wilcoxon signed-rank test for the hypothesis given as title of the plot and the gray bar  denote $95\%$ interval. Here, * stands for $p < 0.05$, ** for $p < 0.01$, and *** for $p < 0.001$.
    }
    \label{fig:TANG+SVM-3class-crosssession}
\end{figure*}

\begin{figure*}[ht]  
    \centering
    \centering
     \subfloat[]{%
            \includegraphics[width=0.45\linewidth]{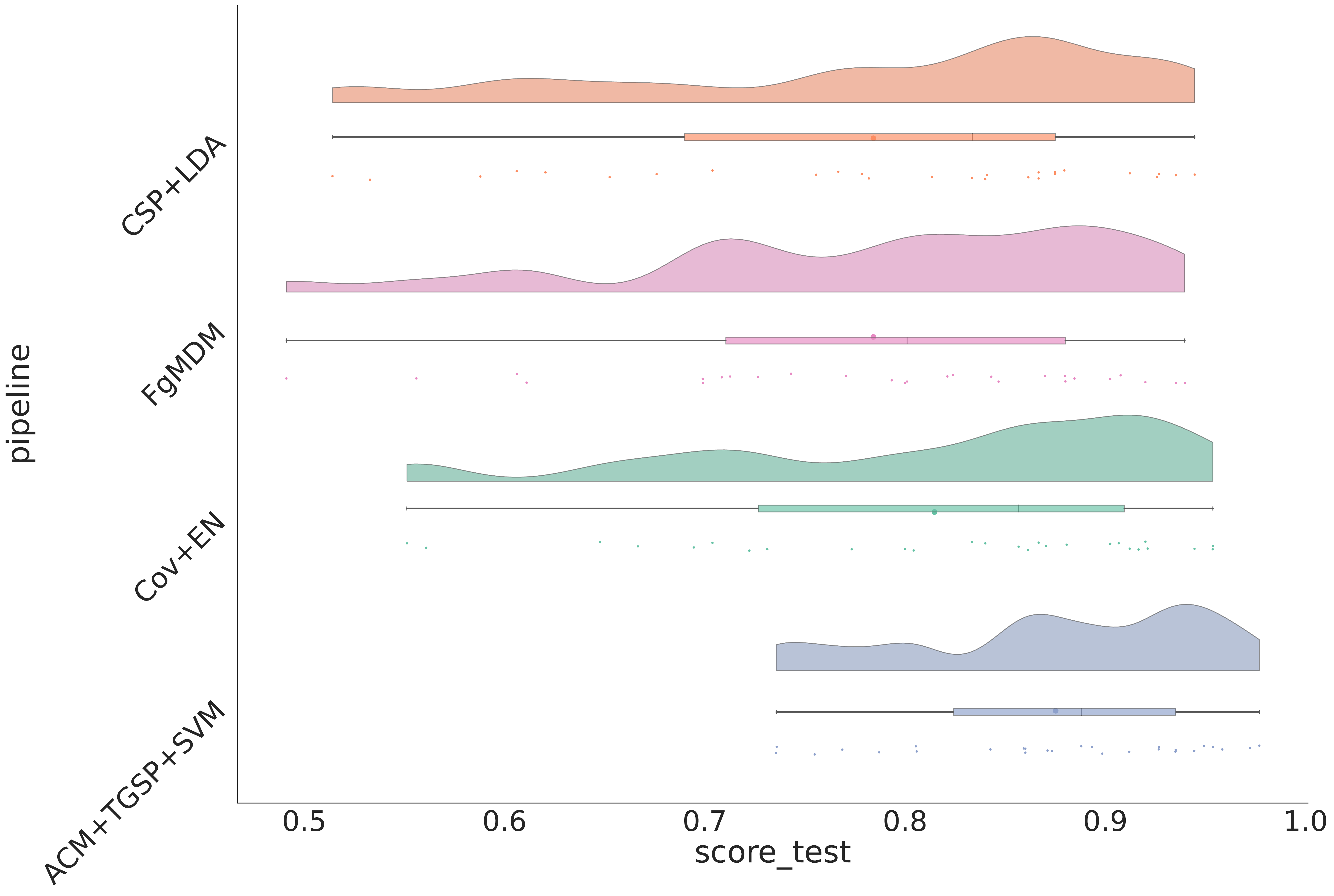}}
            \hfill
     \subfloat[]{%
            \includegraphics[width=0.45\linewidth]{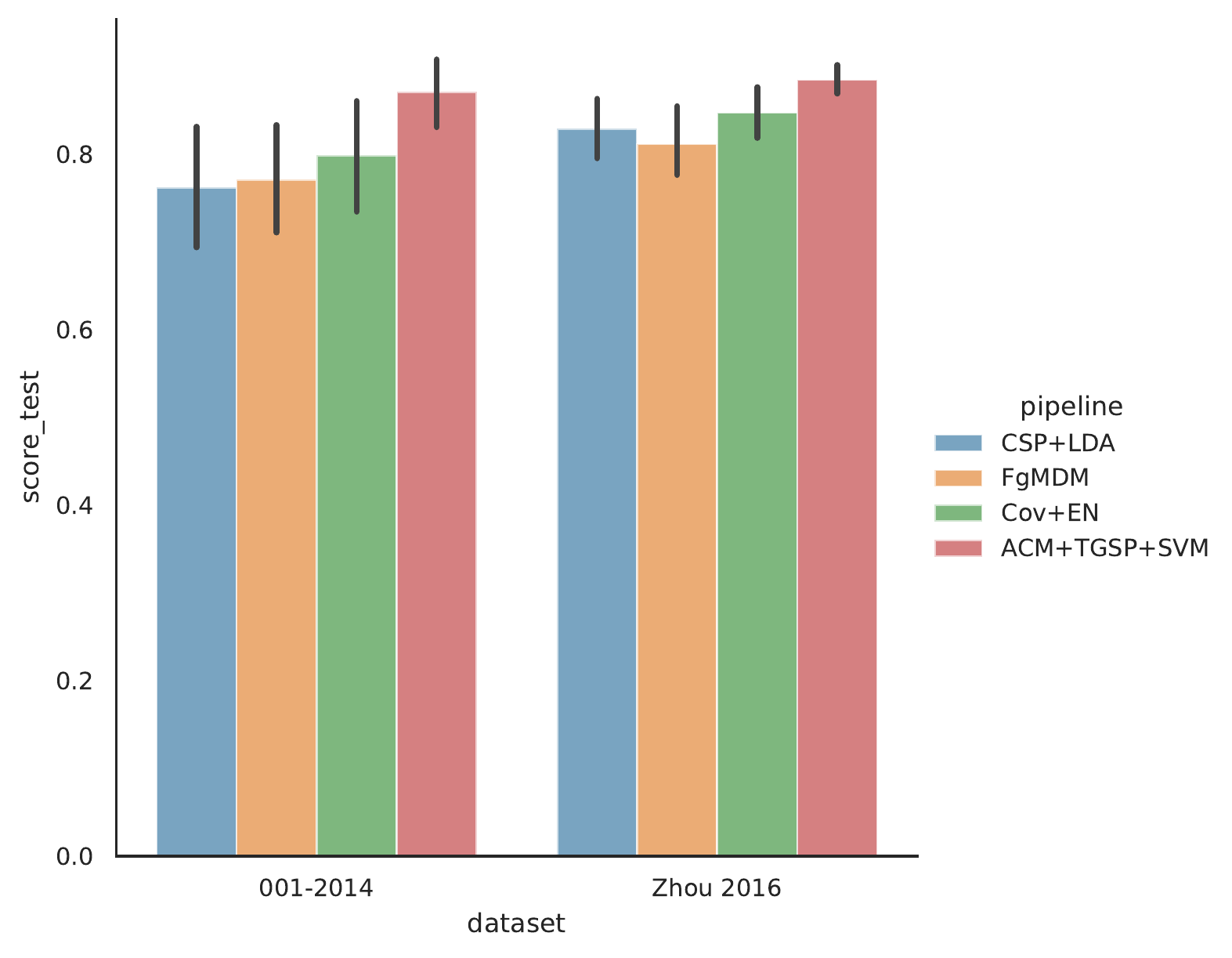}}
    \\
    \subfloat[]{%
        \includegraphics[width=0.5\linewidth]{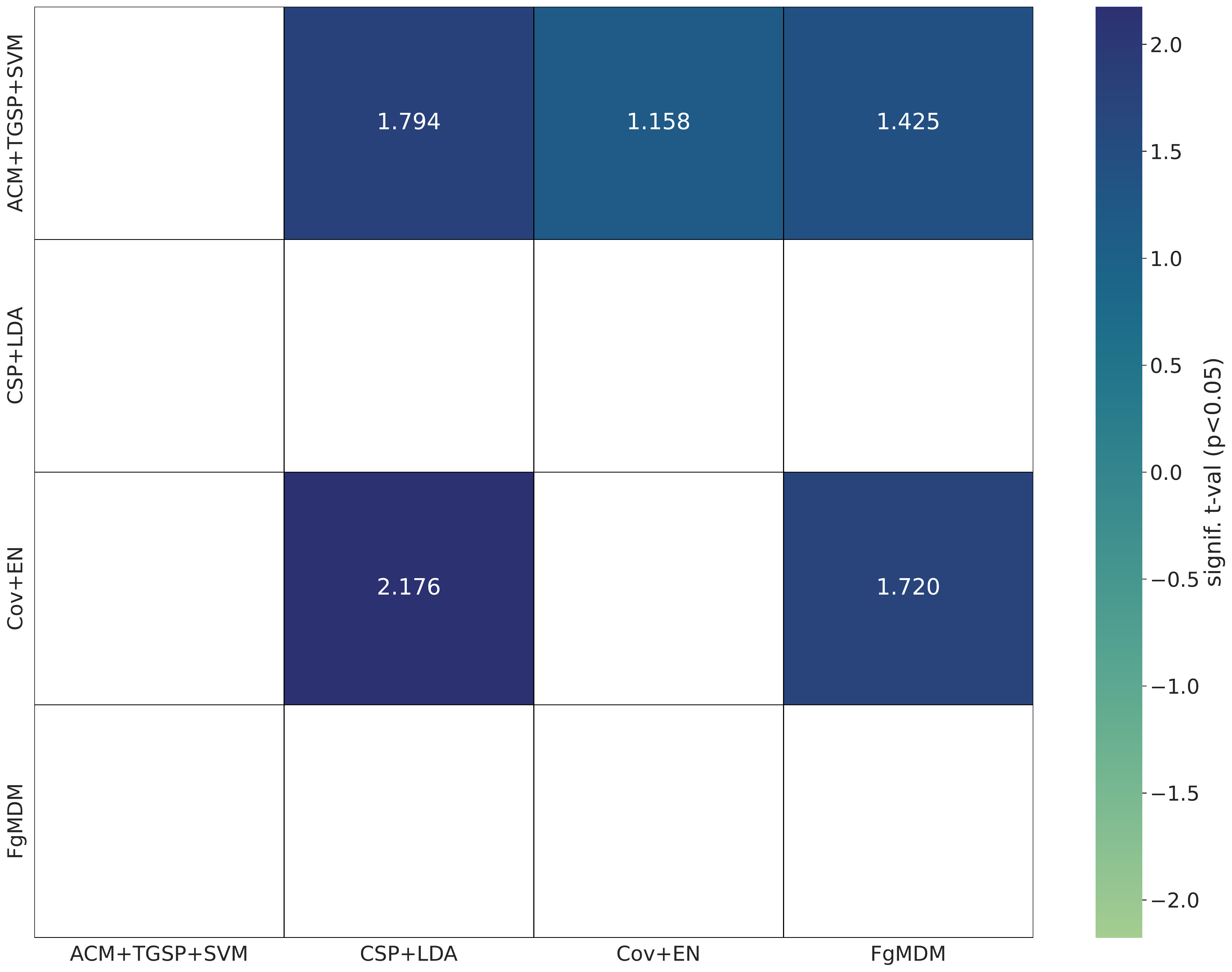}}
        \\
   \subfloat[]{%
            \includegraphics[width=0.30\linewidth]{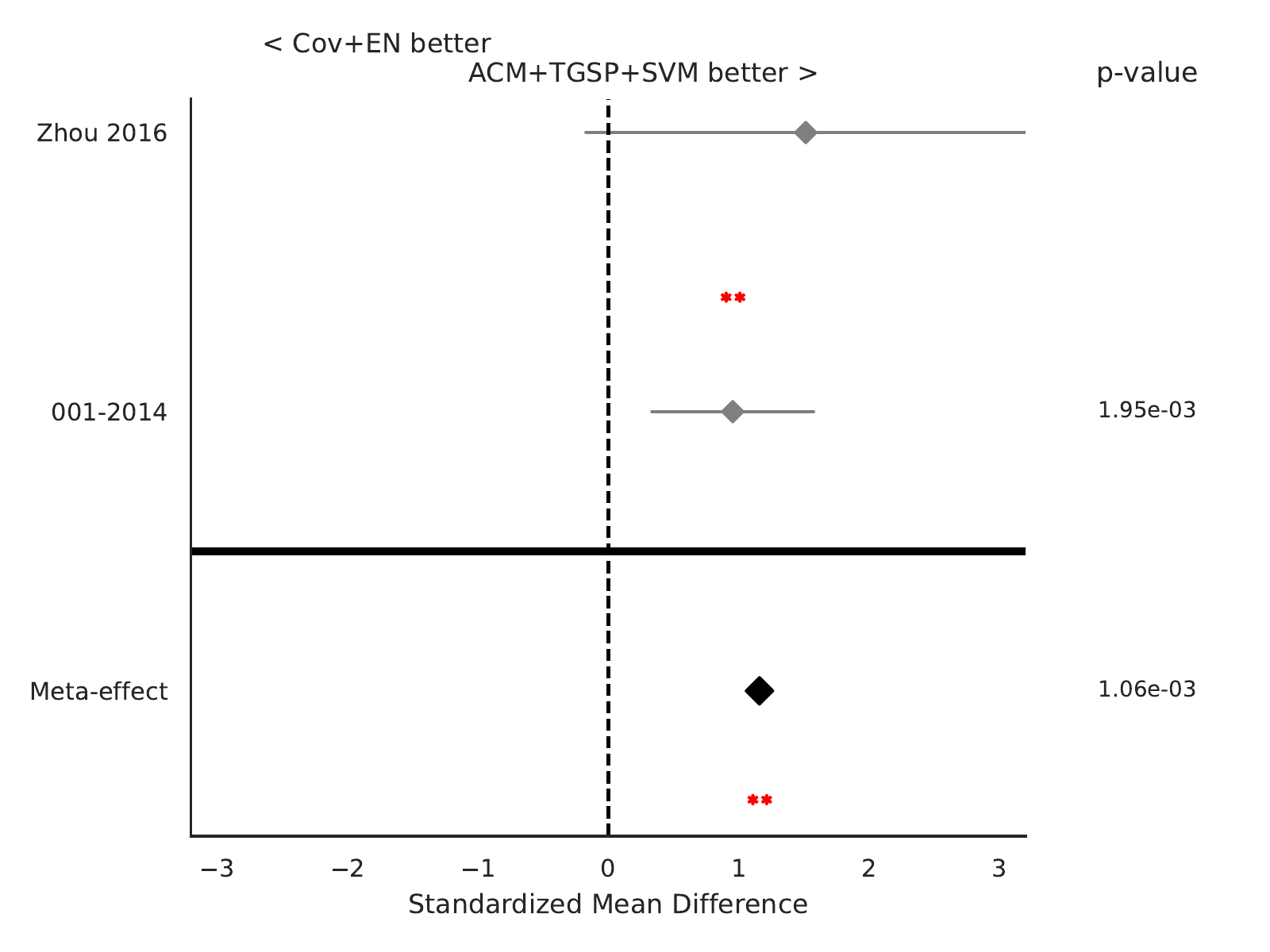}}
            \hfill
   \subfloat[]{%
            \includegraphics[width=0.30\linewidth]{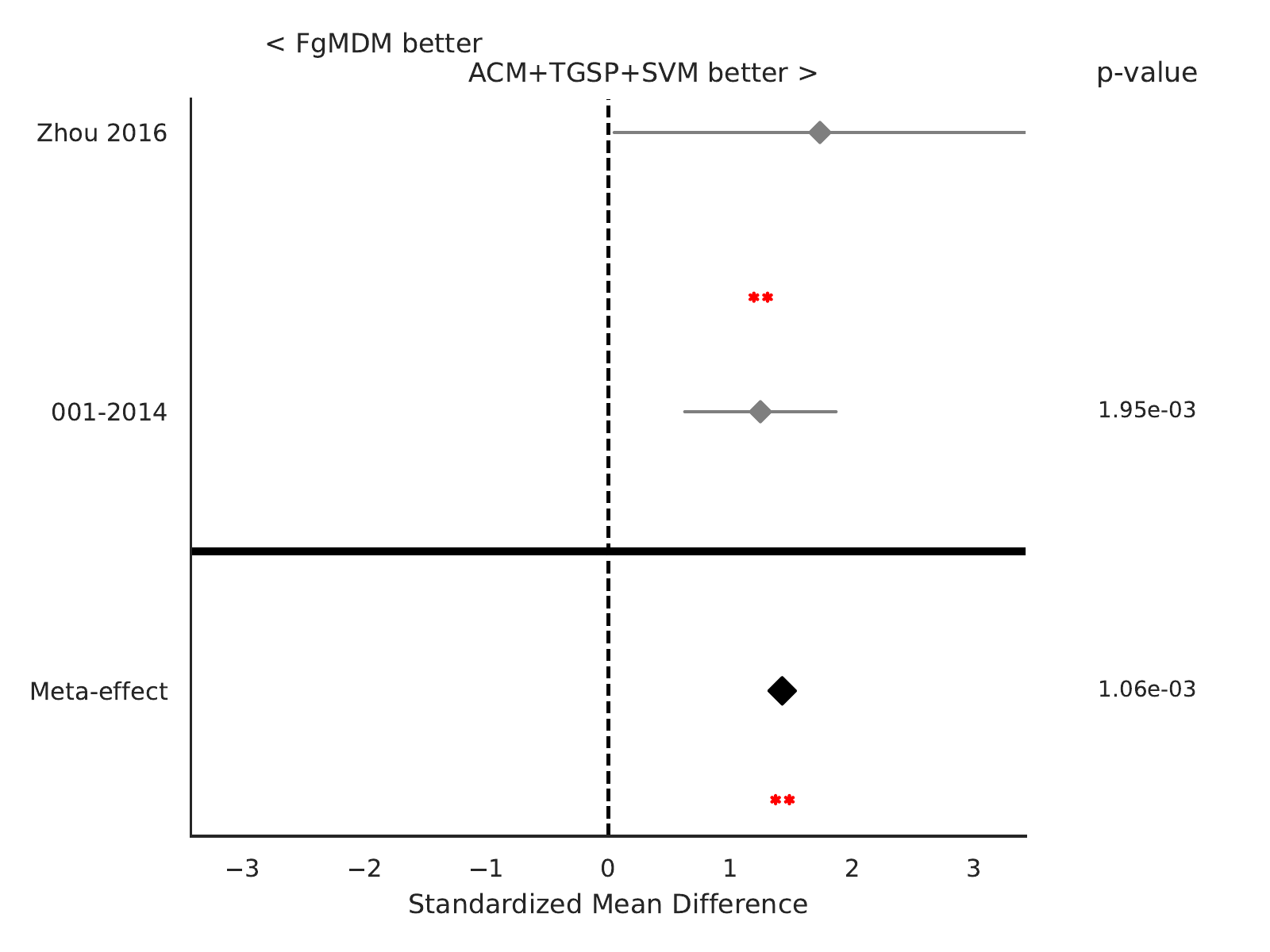}}
            \hfill
   \subfloat[]{%
            \includegraphics[width=0.30\linewidth]{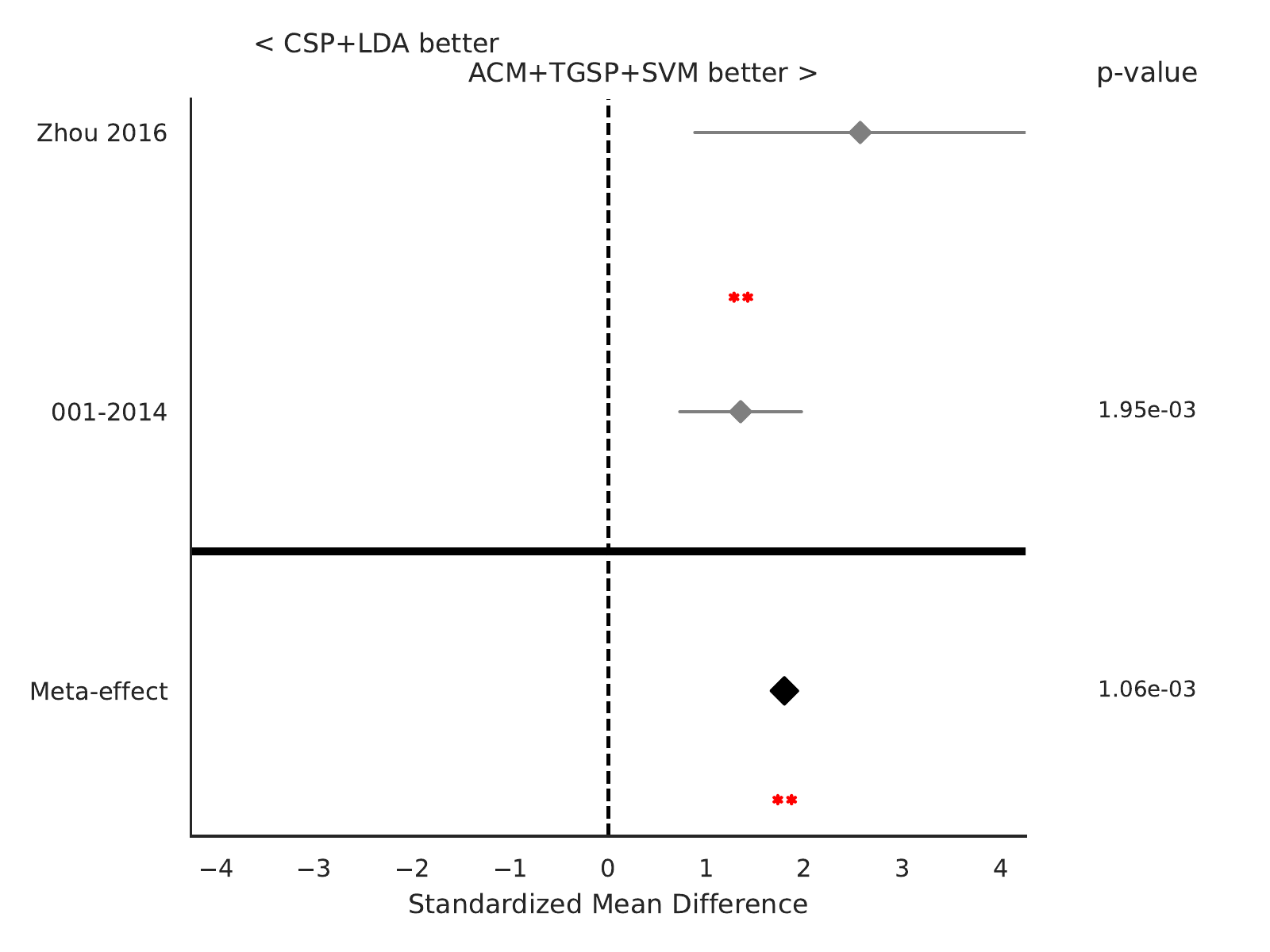}}
            \hfill

    \caption{Result for right hand vs left hand vs feet classification, using withing-session evaluation. (a) show the rain clouds plots for each pipeline, showing the distribution of the score of every subject. (b) show the bar plot of the score withe the error of the different pipeline and for every dataset considered. (c) show the meta analysis of the different pipeline considered. This plot the significance that the algorithm on the y-axis is better than the one on the x-axis. The color represents the significance level of the difference of accuracy, in terms of t-values, and we show only the significant interactions ($p < 0.05$). (d) (e) (f) show the meta analysis of augmented method with SVM against the state of the art. We show the standardized mean differences, while p-values are computed as one-tailed Wilcoxon signed-rank test for the hypothesis given as title of the plot and the gray bar  denote $95\%$ interval. Here, * stands for $p < 0.05$, ** for $p < 0.01$, and *** for $p < 0.001$.
    }
    \label{fig:TANG+SVM-3class-whithinsession-stateart}
\end{figure*}


\begin{figure*}[ht]  
    \centering
    \centering
     \subfloat[]{%
            \includegraphics[width=0.45\linewidth]{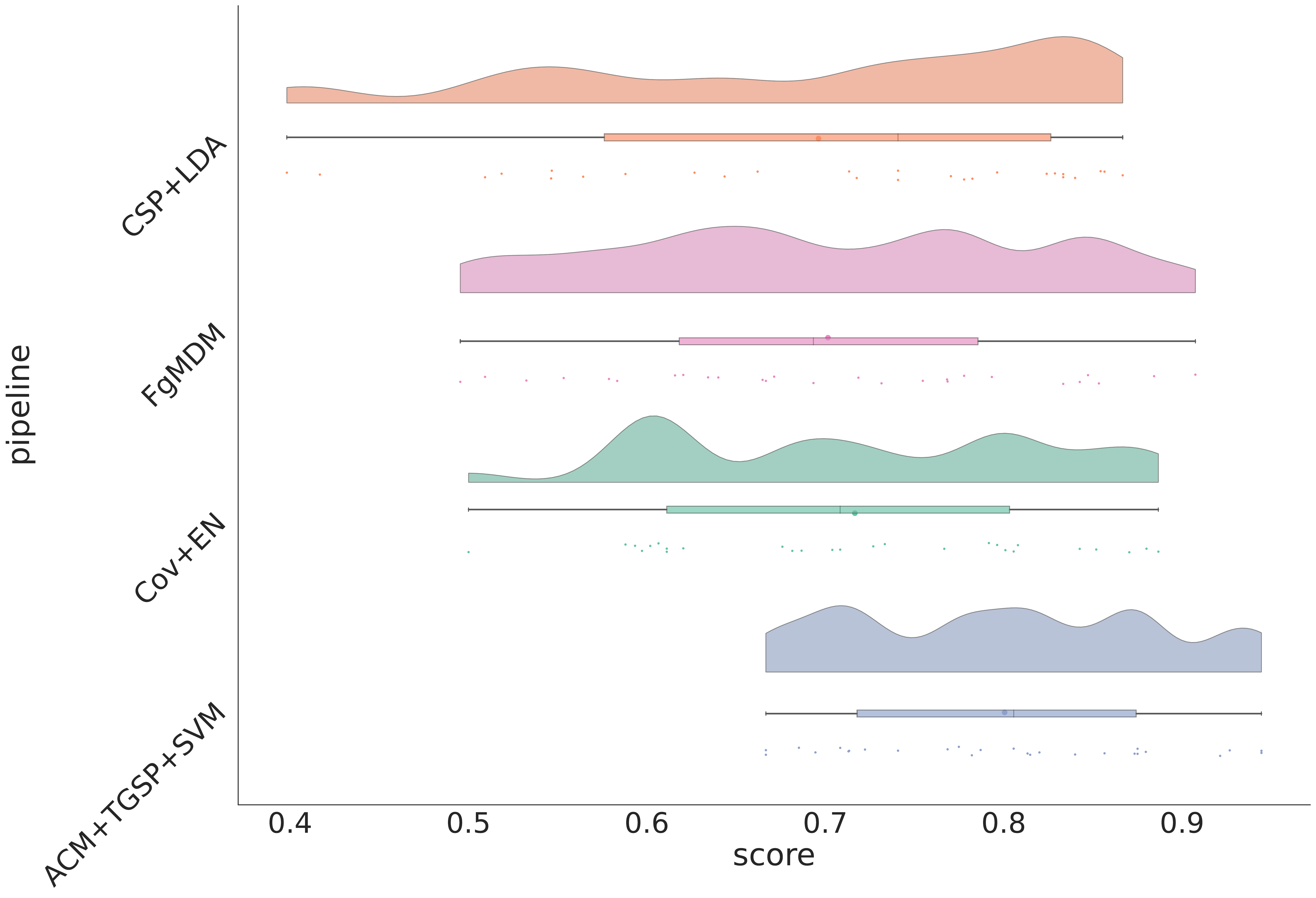}}
            \hfill
     \subfloat[]{%
            \includegraphics[width=0.45\linewidth]{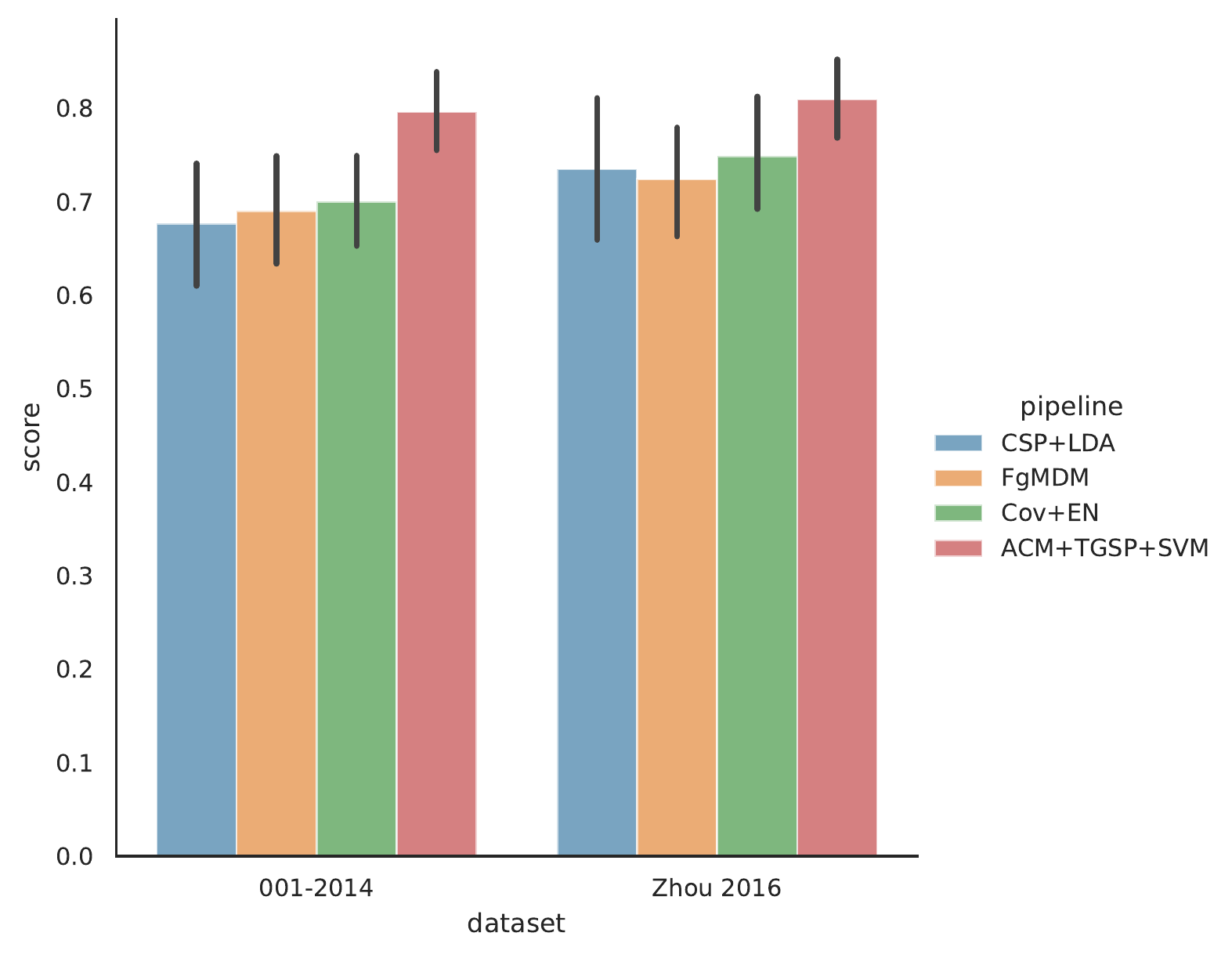}}
    \\
    \subfloat[]{%
        \includegraphics[width=0.5\linewidth]{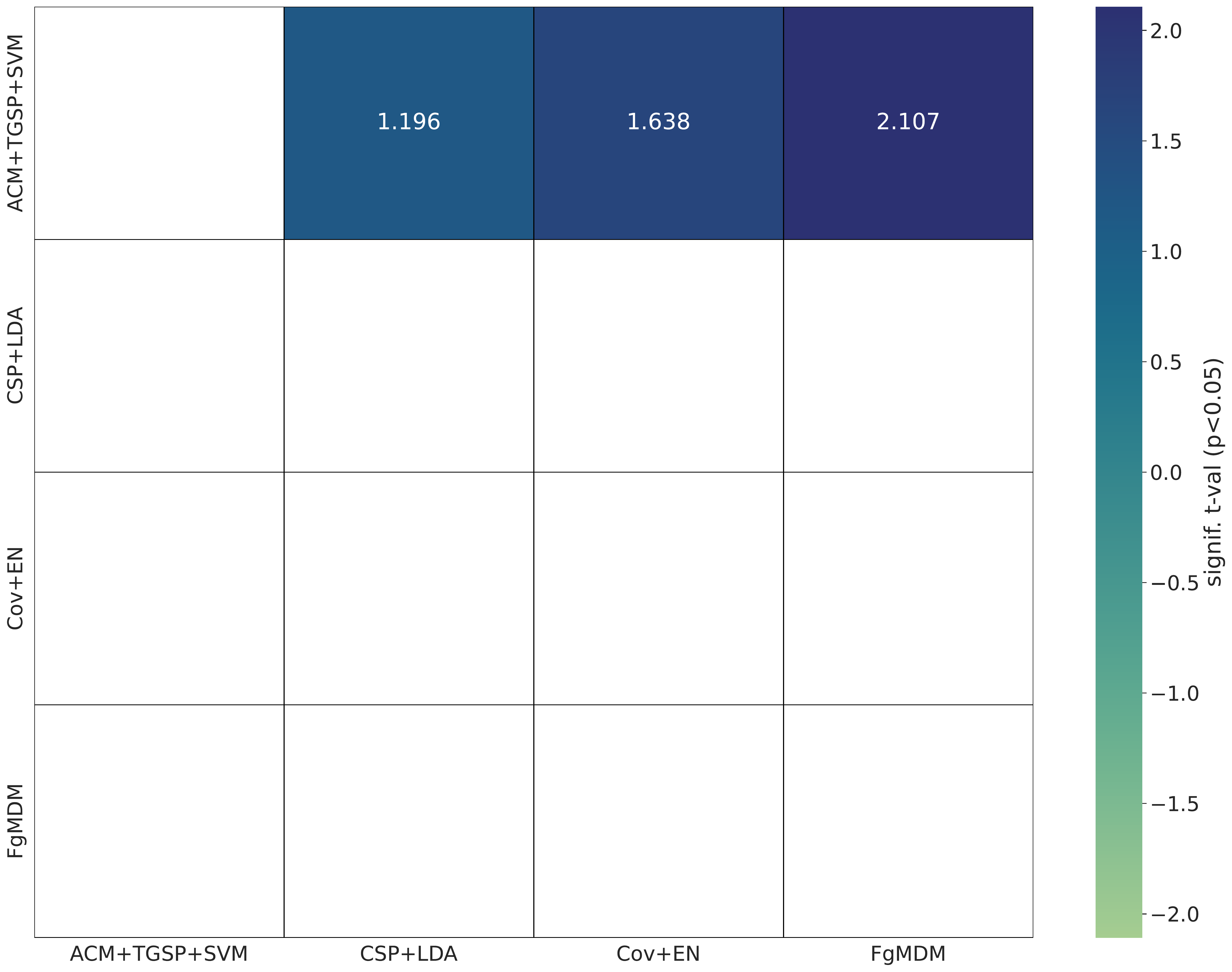}}
        \\
   \subfloat[]{%
            \includegraphics[width=0.30\linewidth]{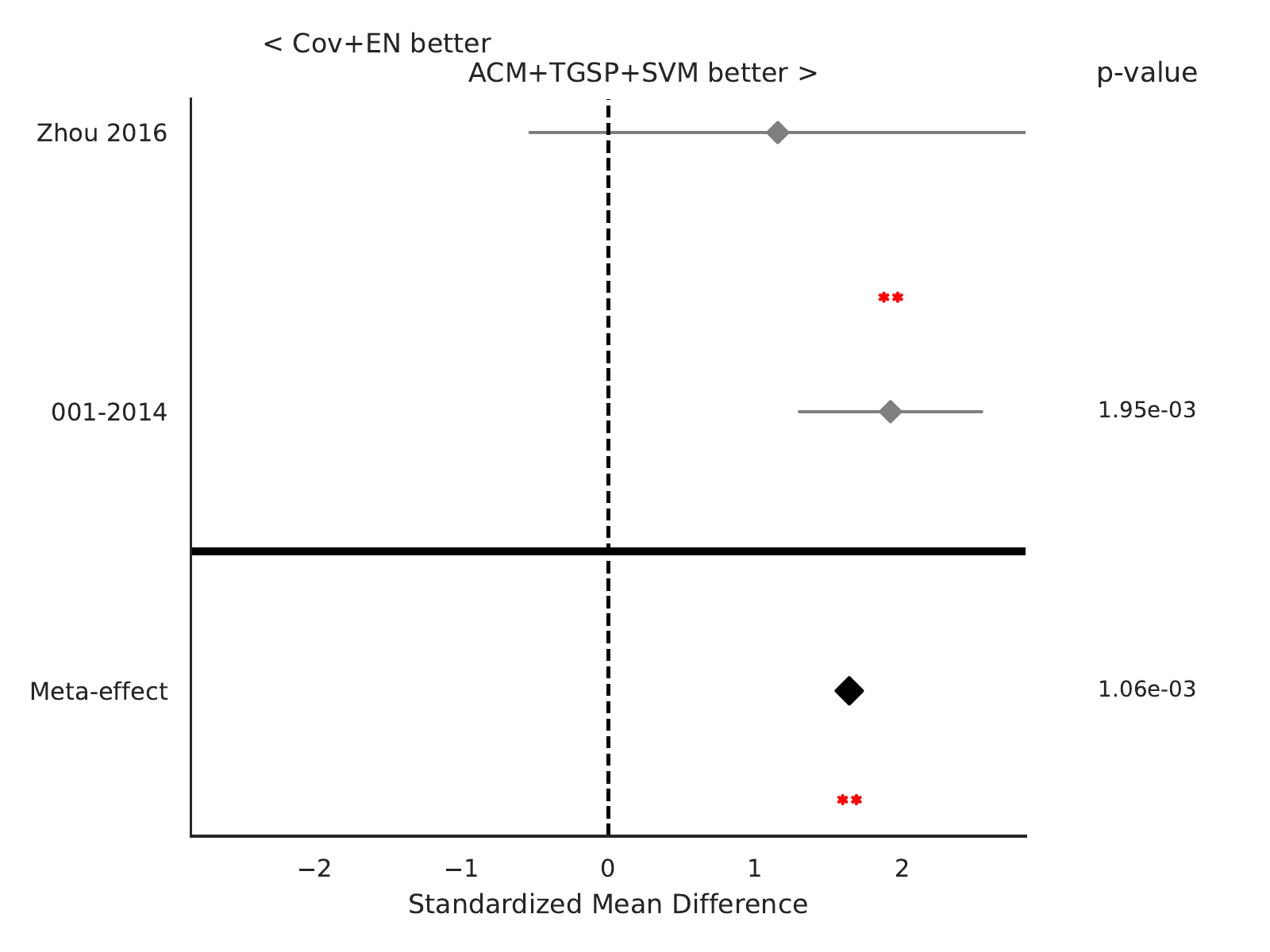}}
            \hfill
   \subfloat[]{%
            \includegraphics[width=0.30\linewidth]{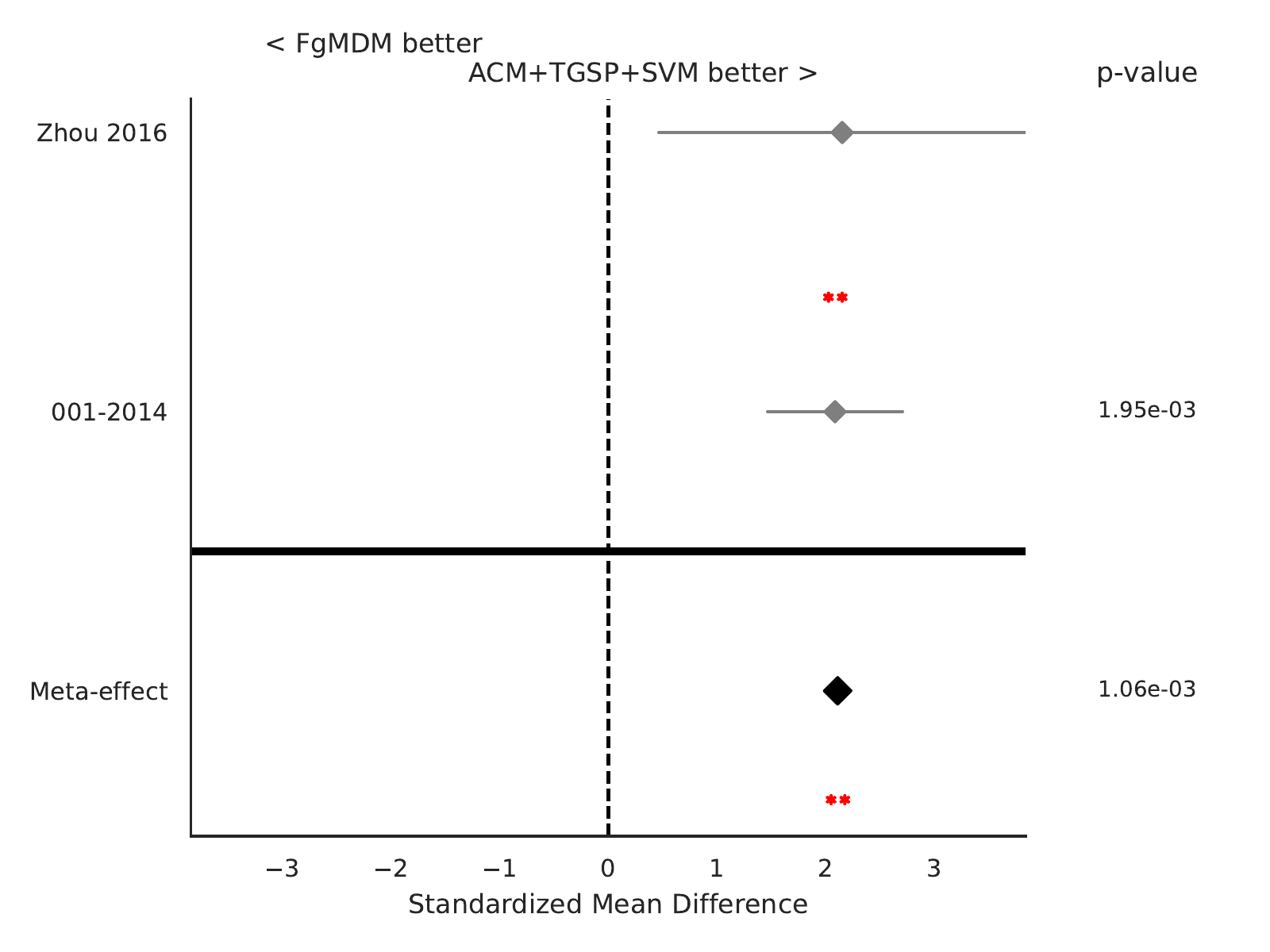}}
            \hfill
   \subfloat[]{%
            \includegraphics[width=0.30\linewidth]{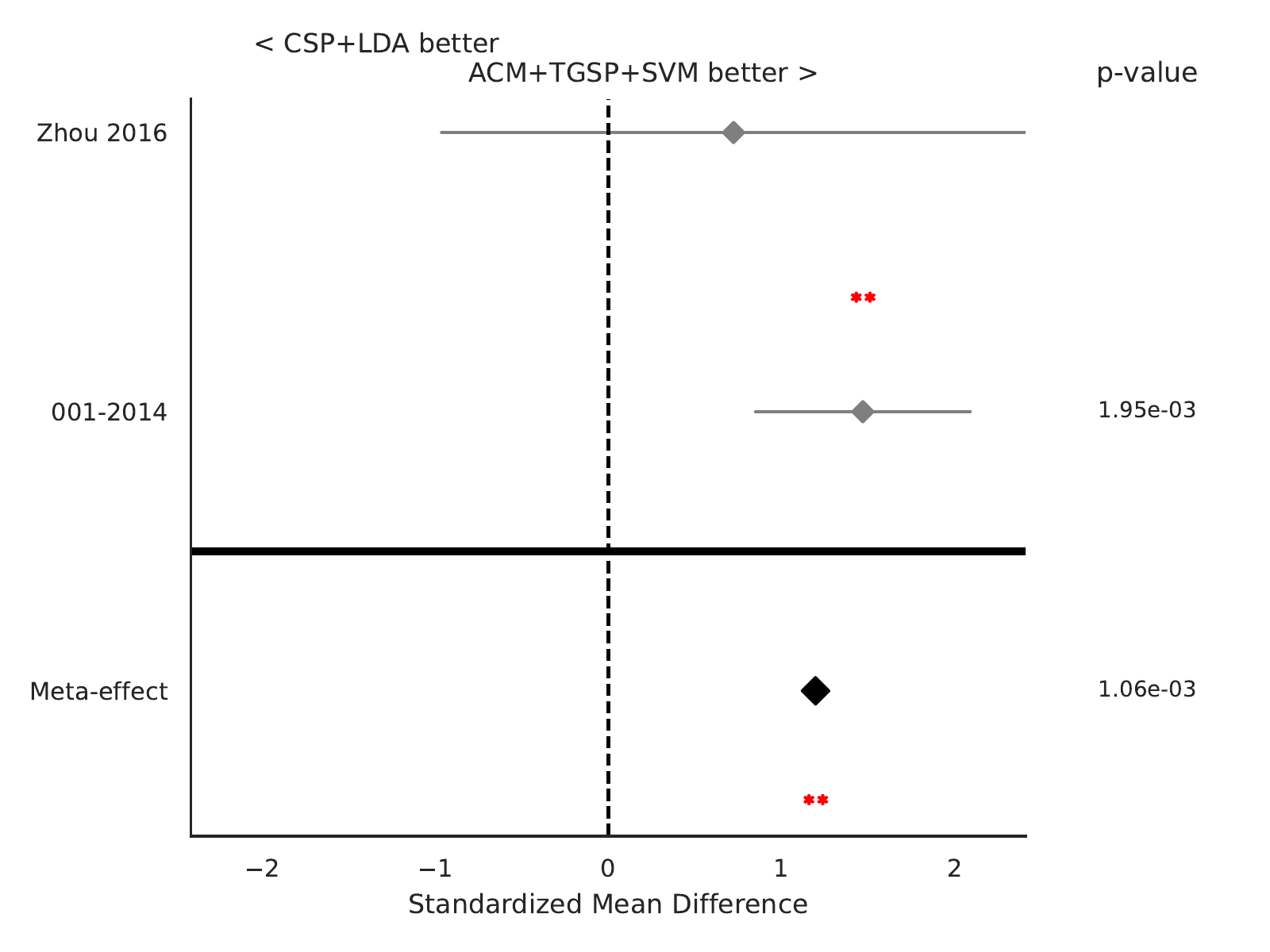}}
            \hfill

    \caption{Result for right hand vs left hand vs feet classification, using cross-session evaluation. (a) show the rain clouds plots for each pipeline, showing the distribution of the score of every subject. (b) show the bar plot of the score withe the error of the different pipeline and for every dataset considered. (c) show the meta analysis of the different pipeline considered. This plot the significance that the algorithm on the y-axis is better than the one on the x-axis. The color represents the significance level of the difference of accuracy, in terms of t-values, and we show only the significant interactions ($p < 0.05$). (d) (e) (f) show the meta analysis of augmented method with SVM against the state of the art. We show the standardized mean differences, while p-values are computed as one-tailed Wilcoxon signed-rank test for the hypothesis given as title of the plot and the gray bar  denote $95\%$ interval. Here, * stands for $p < 0.05$, ** for $p < 0.01$, and *** for $p < 0.001$.
    }
    \label{fig:TANG+SVM-3class-crosssession-stateart}
\end{figure*}

\end{document}